\documentclass[11pt]{amsart}
\usepackage{amsmath,amssymb}
\usepackage{array}
\usepackage{booktabs}
\usepackage{graphicx}
\usepackage[margin=1.15in]{geometry}
\usepackage[colorlinks=true,linkcolor=blue,citecolor=blue,urlcolor=blue]{hyperref}
\hypersetup{pdftitle={Towards the componentwise three-soliton conjecture: cones, collisions, soft momenta and graphs},
  pdfauthor={},
  pdfsubject={Hirota bilinear method, N-soliton conditions, three-soliton condition}}

\numberwithin{equation}{section}

\newtheorem{theorem}{Theorem}[section]
\newtheorem{lemma}[theorem]{Lemma}
\newtheorem{proposition}[theorem]{Proposition}
\newtheorem{corollary}[theorem]{Corollary}
\newtheorem{conjecture}[theorem]{Conjecture}
\newtheorem{mainthm}{Theorem}

\theoremstyle{definition}

\newtheorem{example}[theorem]{Example}
\theoremstyle{remark}
\newtheorem{remark}[theorem]{Remark}

\newcommand{\C}{\mathbb{C}}
\newcommand{\Q}{\mathbb{Q}}
\newcommand{\Z}{\mathbb{Z}}
\newcommand{\PP}{\mathbb{P}}

\newcommand{\bp}{\mathbf{p}}
\newcommand{\sg}{\sigma}
\newcommand{\eps}{\varepsilon}
\newcommand{\Lam}{\Lambda}
\newcommand{\LL}{\mathfrak{L}}
\newcommand{\cS}{\mathcal{S}}
\newcommand{\cL}{\mathcal{L}}
\newcommand{\cQ}{\mathcal{Q}}
\newcommand{\cT}{\mathcal{T}}
\newcommand{\cV}{\mathcal{V}}
\newcommand{\ord}{\operatorname{ord}}
\newcommand{\wt}{\operatorname{wt}}
\newcommand{\Span}{\operatorname{span}}
\newcommand{\Cat}{\operatorname{Cat}}
\newcommand{\TI}{T_{\mathrm{I}}}
\newcommand{\fF}{\mathfrak{F}}
\newcommand{\cP}{\mathcal{P}}
\newcommand{\cD}{\mathcal{D}}
\newcommand{\fm}{\mathfrak{m}}
\newcommand{\cR}{\mathcal{R}}
\newcommand{\PB}{\mathsf{B}}
\newcommand{\PC}{\mathsf{C}}
\newcommand{\PD}{\mathsf{D}}
\newcommand{\PBh}{\widehat{\mathsf{B}}}
\newcommand{\PDh}{\widehat{\mathsf{D}}}
\newcommand{\Esix}{\mathsf{E}_6}
\newcommand{\Etwo}{\mathsf{E}_2}
\allowdisplaybreaks[2]

\begin{document}

\title[Towards the componentwise three-soliton conjecture]{Towards the componentwise
three-soliton conjecture:\\ cones, collisions, soft momenta and graphs}

\author{Shi Zhenhua}

\subjclass[2020]{Primary 37K10; Secondary 37K40, 35Q53, 14H50, 13P15}
\keywords{Hirota bilinear method; three-soliton condition; $N$-soliton condition; dispersion
variety; soft momentum; cubic graph; quintic graph; BKP hierarchy; Hietarinta's equation;
Newton polygon; soliton resonance; translated hyperbola; computer-assisted proof}

\begin{abstract}
The componentwise three-soliton conjecture of Part~I states that if an even polynomial $P$ with
$P(0)=0$ satisfies Hirota's three-soliton condition on a non-hyperplane component of
multiplicity one of its dispersion variety, then it satisfies the $N$-soliton condition there
for every $N$. We prove restrictions on components carrying this condition and classify the
polynomials satisfying it on several families of components, with no degree bound.
Restrictions: if the top homogeneous part of the equation of the component has a nonlinear
irreducible factor, then $P$ is a multiple of a quadratic form; the order of $P$ at the origin
is tied to the multiplicities of the component and of the doubled component; a component
through the origin has odd multiplicity; the condition passes to all limits of the rescaled
component, and the Newton polygon of a plane component has edges of only ten directions.
Classifications: on the cubic graph $v=u^3$, where the four-soliton condition holds exactly in
multiplicity one; on the graphs $v=q(u)$ with $\deg q\ge3$, where $P$ exists only if $q$ is an
odd cubic or an odd quintic, so the conjecture holds for all graphs of polynomials; and for
three families of Hietarinta's list, where each irreducible member is, up to a constant factor,
the only such polynomial on its dispersion variety. All three satisfy the $N$-soliton condition
for every $N$, so the conjecture holds for them: two belong to the BKP hierarchy, and for
Hietarinta's quartic equation, known to satisfy it for $N\le4$, we prove it by fusing resonant
momenta and counting zeros on its curve. For translated hyperbolas we exclude the one through
the origin and all $P$ of degree at most eight. The solutions of multiplicity one on graphs are
known bilinear equations; new is that there are no others. The conjecture remains open. 
\end{abstract}

\maketitle

\section{Introduction}\label{sec:intro}

\subsection{The componentwise three-soliton conjecture}
Let $P$ be an even polynomial in $s$ variables with $P(0)=0$. Hirota's bilinear equation
$P(D)f\cdot f=0$ \cite{Hirota1971,Hirota2004} has a one-soliton solution for every momentum $p$
on the dispersion variety $V(P)=\{p\in\C^s:\ P(p)=0\}$ and a two-soliton solution for every pair
of momenta $p_1,p_2\in V(P)$ with $P(p_1+p_2)\neq0$, whereas the existence of $N$-soliton
solutions of Hirota's form is governed by the polynomials
\begin{equation}\label{eq:SN}
S_N[P](p_1,\dots,p_N):=\sum_{\sg\in\{\pm1\}^N}P\Bigl(\sum_{i=1}^N\sg_ip_i\Bigr)
\prod_{1\le i<j\le N}\sg_i\sg_j\,P(\sg_ip_i-\sg_jp_j)
\end{equation}
of \cite[(2.3)]{PartI}: for nondegenerate momenta $p_1,\dots,p_N\in V(P)$ the $N$-soliton ansatz
solves the equation for all phases if and only if $S_M[P]$ vanishes on every subfamily of
$M\le N$ of the momenta \cite{Hirota2004,HietarintaI,Hietarinta2004,Zhang2026}. Following
\cite[Definition~2.1]{PartI}, we say that $P$ satisfies the \emph{$N$-soliton condition}
($N$SC) on an irreducible component $V_0$ of $V(P)$ if $S_N[P]$ vanishes identically on
$V_0^{\,N}$; for $N=3$ this is the \emph{three-soliton condition} (3SC), which since
Hietarinta's searches
\cite{HietarintaI,HietarintaII,HietarintaIII,HietarintaIV,Hietarinta1991} has served as the
standard algebraic test for integrability in the sense of Hirota.

Whether the 3SC implies the $N$SC for all $N$ is a classical question: that the restrictions
imposed by the solutions with two and three solitons usually suffice for all $N$ was observed
early \cite[p.~58]{Ramani1981}, and for bilinear equations
of KdV type it is generally conjectured that the 3SC implies all $N$SC
\cite[Section~3.2]{Zhang2026}. In its
naive form the implication fails: the family $D_x(D_x^3+\alpha_1D_t+\alpha_2D_y)^{2n+1}$ of G\"urses and Pekcan
\cite{GursesPekcan2025} and, more generally, the families $L\,(D_x^3+\Lam)^{2n+1}$ studied in
the first paper of this series \cite{PartI} satisfy the 3SC on the principal component
$D_x^3+\Lam=0$ for every $n$ \cite[Theorem~5.1]{PartI}, but fail the four-soliton condition
for every $n\ge1$ \cite[Theorem~6.10]{PartI}; and a squarefree polynomial may satisfy the 3SC
but not the 4SC on an affine hyperplane component \cite[Proposition~10.1]{PartI}. In the first
family the component has multiplicity $2n+1\ge3$, in the second example it is a hyperplane.
This led to the following formulation.

\begin{conjecture}[{\cite[Conjecture~1.1]{PartI}}]\label{conj:main}
Let $P$ be an even polynomial with $P(0)=0$, and let $V_0$ be an irreducible component of the
dispersion variety $V(P)$ which is not a hyperplane and along which $P$ vanishes to order one.
If $P$ satisfies the three-soliton condition on $V_0$, then it satisfies the $N$-soliton
condition on $V_0$ for every $N\ge3$.
\end{conjecture}

The conjecture concerns one named component: all momenta lie on $V_0$, mixed products of
different components are a separate matter \cite[Section~7]{PartI}, and $V_0$ need not pass
through the origin. If $C$ is an irreducible polynomial with $V_0=V(C)$, the hypothesis on the
order of vanishing means that $C$ divides $P$ and $C^2$ does not; we then call $C$ a
\emph{simple factor} of $P$ and $V_0$ a \emph{simple component}.

\subsection{What was known}\label{subsec:known}
\emph{Finitely many conditions.} Newell and Zeng \cite{NewellZeng1986} studied the conditions
$S_N=0$ for weighted homogeneous polynomials $P$ in $D_{t_1},D_{t_3},D_{t_5},\dots$, where
$D_{t_j}$ has weight $j$, on the curve $(k,-k^3,k^5,\dots)$. Extending Hirota's proof for the
Korteweg--de Vries equation \cite{Hirota1971}, they showed the following: if $P$ has weight $2L$
and its two-soliton phase factor on this curve is that of a polynomial of weight $2M\le2L$,
then the $N$-soliton condition is the vanishing of a polynomial of degree $2L+MN(N-1)$ in
$k_1,\dots,k_N$ which, if the conditions with fewer solitons hold, is divisible by an explicit
polynomial of degree $3N^2-N$. Hence the conditions with $3N^2-N>2L+MN(N-1)$ follow from the
preceding ones; for $M=3$ these are the conditions with $N>L$. They also determined all such
polynomials of weight six which satisfy the conditions for all $N$
\cite[Section~IV]{NewellZeng1986}. We use this argument, in a form which allows phase factors
that are not homogeneous, for the polynomials on the cubic and quintic graphs
(Proposition~\ref{prop:sextic}, Lemma~\ref{lem:quintic}).

\emph{Hietarinta's search.} Hietarinta \cite{HietarintaI} formulated the three-soliton
condition as the membership of $S_3[P]$ in the ideal of $V(P)^3$ --- in the terminology of
\cite[Definition~2.1]{PartI}, the 3SC on the whole dispersion variety --- and carried out a
search, by computer algebra, for the polynomials in two variables of degree at most eight which
satisfy it, with extensions to more variables \cite[Table~IV]{HietarintaI}. Apart from
polynomials whose dispersion variety is a union of hyperplanes, the resulting list consists of
four families: the Kadomtsev--Petviashvili polynomial $D_x^4-4D_xD_t+3D_y^2$, with the
Korteweg--de Vries and Boussinesq equations among its reductions; the polynomial
$D_x^3D_t+aD_x^2+D_tD_y$, with Ito's equation \cite{Ito1980} and the shallow water wave equation
of Hirota and Satsuma \cite{HirotaSatsuma1976}, which belongs to the BKP hierarchy with an
additional time variable of weight $-1$ \cite[end of Section~9]{JimboMiwa1983}; the first equation
$D_x^6+5D_x^3D_t-5D_t^2+D_xD_y$ of the BKP hierarchy \cite{DJKM1982,JimboMiwa1983}, with the
Sawada--Kotera equation \cite{CaudreyDoddGibbon1976,SawadaKotera1974} and Ramani's equation
$D_x^6+5D_x^3D_t-5D_t^2$ \cite{Ramani1981}; and one equation which was new,
$D_xD_t(D_x^2+\sqrt3\,D_xD_t+D_t^2)+aD_x^2+bD_xD_t+cD_t^2$. For these four families the
existence of four-soliton solutions and the Painlev\'e property are reported in
\cite{GrammaticosRamaniHietarinta1990}.

\emph{The condition on one component.} The condition studied in \cite{PartI} and in this paper
concerns the triples of momenta on a single component of $V(P)$ and is weaker than the
condition on the whole variety. An example is the family $D_x^6+D_x^3D_t+kD_t^2$, for which
Ramani \cite[(3)]{Ramani1981} states that there are no three-soliton solutions unless
$k=-\frac15$. For $k\neq0,\frac14$ its dispersion variety consists of two cubic graphs. On the
whole variety the 3SC holds only for Ramani's value $k=-\frac15$ \cite[(19)]{HietarintaI}; on
one of the two components it holds also for $k=-2$, where the polynomial is $P_6$ of
\eqref{eq:Kbasis} below and satisfies all $N$SC on that component
(Remark~\ref{rem:cubic-scope}(iv)). Up to the sign of $t$, this case is the example of
\cite[Appendix~A]{Ramani1981}, where the bilinear equation is attributed to Hirota: the
associated nonlinear equation is satisfied by every solution of the potential Korteweg--de
Vries equation and therefore has $N$-soliton solutions for all $N$ in which all solitary waves
belong to one branch of the dispersion relation; the solitary waves of the other branch do not
form such solutions for $N\ge3$, and there are solutions which those of the first branch
cannot cross. Ramani concludes that multi-soliton solutions have to be required for every
combination of solitary waves, which is the condition on the whole variety. Such polynomials
arise when the
equations of a hierarchy are combined in such a way that only two of its variables remain:
$P_6$ is the limiting member without the term $D_{t_1}D_{t_5}$ of the family of weight six with
the phase factor of the Korteweg--de Vries equation in \cite[(4.9)]{NewellZeng1986}, and the
solutions of weighted degree ten in Theorem~\ref{thm:E} below are obtained in this way from the
BKP hierarchy (Remark~\ref{rem:bkp}).

\emph{The first paper.}
In \cite{PartI} the conjecture was proved, together with a complete classification of the
polynomials satisfying the 3SC, in the following classes: graph-type dispersions with a linear
cofactor, $P=L\,(q(D_{\mathbf x})+D_t)$ \cite[Theorem~5.7]{PartI}; all irreducible even quadric
components, with an arbitrary cofactor \cite[Theorem~9.15]{PartI}; parabolic components
\cite[Theorem~9.17]{PartI}; the dispersions $F(\ell)+Q$, which include the
Kadomtsev--Petviashvili surface \cite[Theorem~9.18]{PartI}; and the cubic graph $v=u^3$ under
the additional hypothesis that the two-soliton phase factor belongs to the one-parameter family
\begin{equation}\label{eq:Aalpha}
A_\alpha(s,t)=\frac{(s-t)^2}{(s+t)^2}\cdot\frac{s^2+\alpha st+t^2}{s^2-\alpha st+t^2},\qquad\alpha\in\C
\end{equation}
\cite[Theorem~9.19 and (9.10)]{PartI}. Among the cases left open in
\cite[Section~10]{PartI} were: the cubic graph without a hypothesis on the phase factor (phase
factors outside \eqref{eq:Aalpha} do occur); the translated hyperbolas $uv+au+bv+c=0$ with
$ab\neq0$, the only irreducible quadratic components not classified in \cite{PartI}, where
polynomials of degree at most six were excluded \cite[Proposition~9.20]{PartI}; and components
of higher degree, where the only general information was that the top homogeneous part of the
equation of the component cannot contain a quadratic factor of rank at least three unless $P$
is a quadratic form \cite[Corollary~9.16]{PartI}. For the hyperbola $uv=1$ the $N$-soliton
conditions were derived in \cite[Lemma~9.6, Theorems~9.7 and~9.11]{PartI} from the
three-soliton condition by a reduction of $S_N$ to $S_{N-1}$ at the pairs of momenta whose
difference lies on the curve; this fusion argument is used again in
Section~\ref{subsec:resonances}.

\subsection{Results}
We prove eight theorems, which play three roles. Theorems~\ref{thm:A}, \ref{thm:B}
and~\ref{thm:C} are general restrictions on a component carrying the 3SC.
Theorems~\ref{thm:D}, \ref{thm:E} and~\ref{thm:G} classify the polynomials $P$ satisfying the
3SC on the graphs of polynomials and prove the conjecture for these components;
Theorem~\ref{thm:G} also restricts the branches of a plane curve at infinity and the Newton
polygon of its equation. Theorem~\ref{thm:F} restricts the translated hyperbolas, the only
quadratic components left unclassified in \cite{PartI}, and Theorem~\ref{thm:H} proves the
conjecture for the irreducible polynomials of Hietarinta's list: each of them is, up to a
constant factor, the only polynomial satisfying the 3SC on its dispersion variety, and it
satisfies the $N$-soliton conditions for all $N$; for Hietarinta's equation the second
statement is, to our knowledge, new. Throughout, $P$ denotes a nonzero even polynomial on $\C^s$ with $P(0)=0$.

\begin{mainthm}[Cones and top homogeneous parts; Theorem~\ref{thm:cone},
Corollary~\ref{cor:top}]\label{thm:A}
Let $C$ be an irreducible factor of $P$ such that the top homogeneous part of $C$ has an
irreducible factor $F$ of degree at least two; for instance, let $C=F$ be homogeneous of degree
at least two. If $P$ satisfies the \textup{3SC} on $V(C)$, then $\deg F=2$ and
$P=\lambda F$ with $\lambda\neq0$. No hypothesis on the multiplicity of $C$ in $P$ is made.
Conversely, for an irreducible quadratic form $F$ the polynomial $\lambda F$ satisfies the
$N$\textup{SC} on $V(F)$ for every $N$.
\end{mainthm}

Thus Conjecture~\ref{conj:main} holds for every component which is a cone with vertex at the
origin, and whenever $\deg P>2$ the top homogeneous part of the equation of a component
carrying the 3SC is a product of linear forms. This extends \cite[Theorem~9.15(i) and
Corollary~9.16]{PartI} from quadratic factors to arbitrary nonlinear factors. Since every
binary form is a product of linear forms, Theorem~\ref{thm:A} is a statement about at least
three variables and gives nothing for plane curves.

For a component $V=V(C)$ of $V(P)$ we write $\mu$ for the multiplicity of $C$ in $P$,
$d:=\ord_0P$ for the order of vanishing of $P$ at the origin and, for a positive integer $r$,
$\nu_r$ for the multiplicity of the dilated hypersurface $rV=\{rp:\ p\in V\}$ in $P$
($\nu_r=0$ if $rV\not\subset V(P)$, and $\nu_1=\mu$).

\begin{mainthm}[Collision orders; Theorem~\ref{thm:collision}]\label{thm:B}
Let $V$ be a component of $V(P)$ which is not a hyperplane, of multiplicity $\mu$, on which
$P$ satisfies the \textup{3SC}. If $\mu$ is odd, then
\[
d-2\nu_2\in\{\mu-1,\ \mu+1\},\qquad\text{and}\qquad d-2\nu_2=\mu-1\ \Longrightarrow\ \nu_3=1 .
\]
If $\mu$ is even, then $d-2\nu_2\in\{\mu-2,\ \mu\}$, with $\nu_3=2$ in the first case and
$\nu_3=0$ in the second. In particular $d-2\nu_2\in\{0,2\}$ for a simple component.
\end{mainthm}

Thus, if $V$ is simple and $2V$ is not a component of $V(P)$, then $P$ has a nonzero
quadratic part. Theorem~\ref{thm:B} is obtained by letting three momenta collide along a
curved analytic arc on $V$. In the branch $d=2\nu_2$ of a simple component we also obtain an
identity for doubled momenta (Lemma~\ref{lem:doubled}) and a criterion excluding this branch
(Proposition~\ref{prop:additive}). Even multiplicities do not occur for components through the
origin (Theorem~\ref{thm:C}), but they occur on the hyperbola $uv=1$
(Remark~\ref{rem:collision-scope}); the statement for even $\mu$ is used in the proof of
Theorem~\ref{thm:G}(iv).

\begin{mainthm}[Soft momenta; Theorem~\ref{thm:soft}, Corollary~\ref{cor:origin}]\label{thm:C}
Let $V=V(C)$ be a component of $V(P)$ which is not a hyperplane and passes through the origin,
and suppose that $P$ satisfies the \textup{3SC} on $V$. Then:
\begin{enumerate}
\item[\textup{(a)}] the multiplicity $\mu$ is odd;
\item[\textup{(b)}] $\nu_2\le1$;
\item[\textup{(c)}] the soft form \eqref{eq:soft}, which is bilinear in the second-order jets
of $P$ at $p_1+p_2$ and $p_1-p_2$, vanishes for all $p_1,p_2\in V$ and all arc directions of
$V$ at the origin.
\end{enumerate}
If moreover $V$ is not a cone, then $d=\mu+1+2\nu_2$; the origin is a smooth point of $V$, or
$\mu=1$ and the origin is a point of multiplicity two; and if $V\neq-V$, then $\mu=1$, the
origin is a smooth point of $V$, and $P=C\,C^-\,(C_2C_2^-)^{\nu_2}K$ with $K(0)\neq0$, where
$C^-(z)=C(-z)$, $C_2(z)=C(z/2)$ and $C_2^-(z)=C_2(-z)$.
\end{mainthm}

Theorem~\ref{thm:C} is obtained by letting one of the three momenta tend to the origin
along $V$; we call it a soft momentum. For a simple component through the origin which is not a
cone it shows that the branch $d=2\nu_2$ of Theorem~\ref{thm:B} does not occur. It excludes,
for example, every component of even multiplicity through the origin, every component with a
point of multiplicity at least three at the origin which is not a cone, and every component
with a singular point at the origin which is not centrally symmetric, such as the cuspidal
cubic $v^2=u^3$.

For the cubic graph let $C:=u^3-v$, where $u,v$ are two of the coordinates of $\C^s$, and, as
in \cite[Theorem~9.19]{PartI},
\begin{equation}\label{eq:Kbasis}
\begin{aligned}
P_4&=uC,\qquad P_6=C(u^3+2v),\qquad P_8=u^2C(u^3-4v),\\
P_{10}&=uC(u^3-4v)(u^3+2v),\qquad P_{12}=C(u^3-4v)(u^6+28u^3v+16v^2).
\end{aligned}
\end{equation}

\begin{mainthm}[The cubic graph in arbitrary degree; Theorems~\ref{thm:cubic},
\ref{thm:cubic-all} and~\ref{thm:fourth}]\label{thm:D}
A nonzero even polynomial $P$ on $\C^s$ satisfies $S_3[P]=0$ identically on $V(C)^3$ if and
only if $P=C^{2n}P_0$ with $n\ge0$ and $P_0$ one of the following:
\begin{enumerate}
\item[\textup{(a)}] a nonzero linear combination of $P_4,P_6,P_8,P_{10},P_{12}$;
\item[\textup{(b)}] $P_0=C\cdot L$ with $L$ a nonzero linear form on $\C^s$;
\item[\textup{(c)}] $P_0=C\,\bigl[\alpha u+\beta(u^3+rv)\bigr]$ with $\beta\neq0$ and
$r^2-7r+1=0$.
\end{enumerate}
If $C$ is a simple factor of $P$, that is, if $n=0$, then $P$ satisfies the $N$\textup{SC} on
$V(C)$ for every $N\ge3$. If $n\ge1$, then $P$ does not satisfy the \textup{4SC} on $V(C)$.
\end{mainthm}

No bound on the degree of $P$, no restriction on the number of variables and no hypothesis on
the phase factors is imposed; in two variables the weighted degree of a solution with simple
factor $C$ ($\wt u=1$, $\wt v=3$) turns out to be at most $12$. For $n=0$ the polynomials of
the theorem are not new as bilinear equations: family (a) and the polynomials $vC$ and
$C(u^3+rv)$ are those of \cite[Theorem~9.19]{PartI}, and $C(u^3+rv)$ is Ramani's equation
\cite{Ramani1981};
family (b) is the case $n=0$ of the families $L\,(D_x^3+\Lam)^{2n+1}$ of \cite{PartI}; and
family (c) is, up to linear changes of variables, the set of the reducible members of the
family $D_x^6+5D_x^3D_t-5D_t^2+aD_x^2+bD_xD_t$ of \cite{HietarintaI}, a reduction of the first
equation of the BKP hierarchy (Remark~\ref{rem:cubic-scope}(iv)). The new statements are that
there are no other solutions, in any degree, the description of the solutions of higher
multiplicity, and the failure of the four-soliton condition for all of them. The phase factors of family (b) for general $L$, and those of family (c) with
$\alpha\neq0$, lie outside \eqref{eq:Aalpha}. Since a linear change of variables maps
$V(\ell_1^3-\ell_2)$ to $V(C)$, we obtain:

\begin{corollary}[Corollary~\ref{cor:conj-cubic}]\label{cor:intro}
Conjecture~\ref{conj:main} holds for every component of the form $V_0=V(\ell_1^3-\ell_2)$ with
$\ell_1,\ell_2$ linearly independent linear forms, that is, for the dispersion relation
$D_x^3+\Lam=0$ of \cite{PartI} with an arbitrary cofactor.
\end{corollary}

Thus on the cubic graph the four-soliton condition holds exactly for the solutions of
multiplicity one, in accordance with the hypothesis of Conjecture~\ref{conj:main}. This is a
property of the cubic graph: on the hyperbola $uv=1$ a component of multiplicity two can carry
all $N$SC \cite[Remark~9.12]{PartI}. For
$P=L\,C^{2n+1}$ this is \cite[Theorem~6.10]{PartI}; for the other polynomials $C^{2n}P_0$ the
proof evaluates $S_4$ at one quadruple of points of the curve, for all multiplicities at once
(Theorem~\ref{thm:fourth}).

For the graphs of higher degree the answer is different.

\begin{mainthm}[Graphs of degree at least four; Theorems~\ref{thm:monomial}
and~\ref{thm:graphs}]\label{thm:E}
Let $u,v$ be two of the coordinates of $\C^s$, let $q\in\C[u]$ be a polynomial of degree
$m\ge4$, and suppose that $q(u)-v$ divides $P$ and that $P$ satisfies the \textup{3SC} on
$V(q(u)-v)$. Then $m=5$, the polynomial $q$ is odd, $q=a_5u^5+a_3u^3+a_1u$, and
\[
P=\bigl(q(u)-v\bigr)\,\bigl[\alpha u+\beta\,(a_5u^5+2a_3u^3+4v)\bigr],\qquad(\alpha,\beta)\neq(0,0).
\]
Conversely, for every odd quintic $q$ these polynomials are even, have $q(u)-v$ as a simple
factor, and satisfy the $N$\textup{SC} on $V(q(u)-v)$ for every $N\ge3$.
\end{mainthm}

No bound on the degree of $P$, on the number of variables or on the multiplicity of the
component is assumed. For $\beta=0$ the solutions are the family $D_x(q(D_x)+D_t)$ of
\cite[Theorem~5.7]{PartI}: the Sawada--Kotera equation
\cite{CaudreyDoddGibbon1976,SawadaKotera1974} for $a_3=0$,
and in general its combination with the Korteweg--de Vries equation, a reduction of the first
equation of the BKP hierarchy \cite{DJKM1982,Hirota1989}. The polynomials with $\beta\neq0$
have weighted degree ten and the same two-soliton phase factor \eqref{eq:quintic-phase}. They
are not new bilinear equations either: under the same reduction they are the linear
combinations of the equations of degrees ten and eight of the BKP hierarchy, as listed by Jimbo
and Miwa \cite[Appendix~3]{JimboMiwa1983}, from which the higher variables of the hierarchy
disappear (Remark~\ref{rem:bkp}). What Theorem~\ref{thm:E} adds is completeness: on a graph of
degree at least four no other polynomial, of whatever degree, satisfies the 3SC. Special cases
of the nonexistence were known: for the equations $D_x(D_t+D_x+D_x^{2n+1})f\cdot f=0$ with
$n\ge3$ only one- and two-soliton solutions of Hirota's form were found in
\cite[(3.10)--(3.12)]{CaudreyDoddGibbon1976}, and according to \cite[(1), (2)]{Ramani1981},
where this is attributed to Hirota, the equations $D_x(D_t+D_x^{2n+1})f\cdot f=0$ with $n\ge3$
and $D_t(D_t+D_x^{2n+1})f\cdot f=0$ with $n\ge2$ have no three-soliton solutions. The
polynomials with $\beta\neq0$ satisfy the 3SC on $V(q(u)-v)$ but not on their second
component (Remark~\ref{rem:bkp}(iii)); they are outside the scope of the search of
\cite{HietarintaI}, which was limited to degree eight and to the condition on the whole
dispersion variety.
Together with Corollary~\ref{cor:intro} we obtain:

\begin{corollary}[Corollary~\ref{cor:consequences}]\label{cor:intro-graphs}
Conjecture~\ref{conj:main} holds for every component of the form $V_0=V(\ell_2-q(\ell_1))$
with $\ell_1,\ell_2$ linearly independent linear forms and $q$ a polynomial which is an odd
cubic or has degree at least four. For $\deg q\ge4$ a polynomial satisfying the
\textup{3SC} on $V_0$ exists only if $q$ is an odd quintic.
\end{corollary}

The graphs of degree two are parabolas, for which the conjecture was proved in
\cite[Theorem~9.17]{PartI}; the cubic graphs which are not odd are the subject of
Theorem~\ref{thm:G} below. As a by-product we find that a plane curve through the origin which
is smooth there and carries the 3SC has contact of order two, three or five with its tangent
line, and of order two if it is not centrally symmetric (Corollary~\ref{cor:consequences}(b)).

For the translated hyperbolas $uv+au+bv+c=0$, $ab\neq0$, $c\neq ab$, a scaling of $u$ and $v$
gives the normal form
\begin{equation}\label{eq:Ch}
C_h:=uv+u+v+h,\qquad h=\frac{c}{ab}\neq1 .
\end{equation}

\begin{mainthm}[Translated hyperbolas; Theorems~\ref{thm:hzero} and~\ref{thm:algebraic},
Proposition~\ref{prop:degree8}]\label{thm:F}
Let $u,v$ be two of the coordinates of $\C^s$. Suppose that $C_h$ divides $P$ for some
$h\neq1$ and that $P$ satisfies the \textup{3SC} on $V(C_h)$. Then $h\neq0$, $h$ is an algebraic number, and $\deg P\ge10$. More precisely, for
every $D$ the set of parameters $h$ admitting such a $P$ with $\deg P\le D$ is contained in a
finite subset of $\overline{\Q}\setminus\{0\}$.
\end{mainthm}

We do not know whether any such $P$ exists. Together with \cite{PartI}, Theorem~\ref{thm:F}
reduces the quadratic case of Conjecture~\ref{conj:main} to the hyperbolas $C_h$ with
$h\in\overline{\Q}\setminus\{0,1\}$ and polynomials $P$ of degree at least ten.

The last two theorems rest on the following observation (Lemma~\ref{lem:limits}). For an
integral weight vector $\omega$ let $G^\omega$ denote the part of a polynomial $G$ of highest
$\omega$-weight. If $S_N[P]$ vanishes on $V(C)^N$, then $S_N[P^\omega]$ vanishes on
$V(C^\omega)^N$ for every $\omega$, and the $N$ momenta may lie on different components of
$V(C^\omega)$. For a plane curve the varieties $V(C^\omega)$ describe its branches at infinity
and at the origin.

\begin{mainthm}[Cubic graphs; branches at infinity; Newton polygons;
Theorems~\ref{thm:cubic-graphs}, \ref{thm:branches} and~\ref{thm:monomial-curves},
Corollaries~\ref{cor:all-graphs} and~\ref{cor:newton}]\label{thm:G}
\textup{(i)} Let $u,v$ be two of the coordinates of $\C^s$ and $q\in\C[u]$ a polynomial of
degree three. If $S_3[P]$ vanishes identically on $V(q(u)-v)^3$, then $q$ is odd.

\textup{(ii)} Let $\ell_1,\ell_2$ be linearly independent linear forms on $\C^s$ and $q$ a
polynomial of degree at least two. A polynomial $P$ satisfying the \textup{3SC} on
$V_0=V(\ell_2-q(\ell_1))$ exists if and only if $q=a_2u^2+a_1u$, or $q$ is an odd cubic, or $q$
is an odd quintic; and Conjecture~\textup{\ref{conj:main}} holds for $V_0$.

\textup{(iii)} Let $C\in\C[u,v]$ be irreducible, and suppose that $P$ is divisible by $C$ and
satisfies the \textup{3SC} on $V(C)$. If for the weights $\wt u=1$, $\wt v=m\ge2$ the top
weighted part of $C$ is divisible by $u^m-cv$ with $c\neq0$, then $m\in\{2,3,5\}$, and for
$m=5$ the curve $V(C)$ is the graph of an odd quintic.

\textup{(iv)} Let $C$ and $P$ be as in \textup{(iii)}. Then $C$ is not of the form
$u^av^b-c$ with $c\neq0$ and coprime integers $a,b\ge1$, $(a,b)\neq(1,1)$, and every edge of
the Newton polygon of $C$ is parallel to one of the vectors $(1,0)$, $(0,1)$, $(1,1)$,
$(1,-1)$, $(m,-1)$, $(1,-m)$ with $m\in\{2,3,5\}$.
\end{mainthm}

Part (i) includes the cubic graphs which do not pass through the origin. Its proof uses the
classification of Theorem~\ref{thm:D} for all multiplicities, and the first, respectively the
third, order of the expansion of $S_3$ along the scaling which contracts the graph to $v=u^3$.
With Theorems~\ref{thm:D} and~\ref{thm:E} and \cite[Theorem~9.17]{PartI}, part (ii) settles the
components which are graphs of polynomials. In (iii) the hypothesis means that the curve has a
branch at infinity which is asymptotic to the graph $v=u^m/c$; for $m=3$ we obtain a
classification when the top weighted part of $C$ contains two different cubic graphs, or a
cubic graph and the line $v=0$ (Theorem~\ref{thm:branches}(c),(d)), which leads to the next
theorem. In (iv) the edges with a direction $(a,b)$, $a,b\ge1$, belong to the branches which
approach an asymptote through the origin; such a branch has to approach its asymptote like a
branch of a hyperbola. The curves $u^av^b=c$ are excluded by letting one momentum tend to
infinity on the curve, which gives an identity for the two remaining momenta, and by letting
these collide; the result is then compared with Theorem~\ref{thm:B} for all multiplicities.
Let
\begin{gather*}
\PB_{a,b,c}:=X^3T+aX^2+bXT+cT^2,\qquad\PD_{a,b}:=X^6+5X^3T-5T^2+aX^2+bXT,\\
\PC_{a,b,c}:=XT\,(X^2+\sqrt3\,XT+T^2)+aX^2+bXT+cT^2,\\
\PBh_a:=X^3T+aX^2+TY,\qquad\PDh:=X^6+5X^3T-5T^2+XY .
\end{gather*}
These are the polynomials (I\,B), (I\,D) and (I\,C) of Hietarinta's list
\cite[Section~IV.E and Table~IV]{HietarintaI}, in two and in three variables
(Section~\ref{subsec:known}); $\PC_{a,b,c}$ is the equation found in \cite{HietarintaI}.

\begin{mainthm}[The equations of Hietarinta's list; Theorems~\ref{thm:rigidity}
and~\ref{thm:hietarinta-NSC}, Proposition~\ref{prop:NSC-BD}, Corollaries~\ref{cor:hietarinta}
and~\ref{cor:hietarinta-conj}]\label{thm:H}
Let $X,T$, respectively $X,T,Y$, be coordinates of $\C^s$, and let $C$ be one of the polynomials
\[
\PB_{a,b,c}\ (ac\neq0),\qquad\PD_{a,b}\ (45a+b^2\neq0),\qquad\PBh_a\ (a\neq0),\qquad\PDh,
\qquad\PC_{a,b,c}\ \text{irreducible}.
\]
Then $C$ is irreducible, and if $P$ is divisible by $C$ and satisfies the \textup{3SC} on
$V(C)$, then $P$ is a constant multiple of $C$. Each of the five polynomials satisfies the
$N$\textup{SC} on its dispersion variety for every $N$; hence
Conjecture~\textup{\ref{conj:main}} holds for these components. Moreover, Hietarinta's
equation $\PC_{a,b,c}$ satisfies the $N$-soliton condition on its whole dispersion variety for
every $N$ and for all values of $a,b,c$, the reducible members included.
\end{mainthm}

For the Kadomtsev--Petviashvili polynomial the corresponding statements are contained in
\cite[Section~9.4 and Theorem~9.18]{PartI}. Thus, for each of the four families with nonlinear
dispersion in Hietarinta's list, the polynomial itself is, up to a constant factor, the only
polynomial which satisfies the 3SC on its dispersion variety, as long as it is irreducible, and
Conjecture~\ref{conj:main} is proved for all four families. The new statements of
Theorem~\ref{thm:H} are the first one and the $N$-soliton conditions with $N\ge5$ for
Hietarinta's equation.

The proof of the first statement for $\PC_{a,b,c}$ contains a bound for the degree of $P$
which does not use the phase factors: the four hyperbolic branches of the curve at infinity and
\cite[Theorem~9.14]{PartI} give $\deg P\in\{4,12\}$, and degree twelve is excluded by a
condition on three lines through the origin (Lemma~\ref{lem:threelines},
Remark~\ref{rem:degreebound}). The $N$-soliton conditions for $\PBh_a$ and $\PDh$ are not new:
both polynomials belong to the BKP hierarchy, with the momenta
$(p+q,\,p^{-1}+q^{-1},\,p^3+q^3)$, respectively $(p+q,\,p^3+q^3,\,p^5+q^5)$, of its solitons
(Remark~\ref{rem:BD}). We include a proof by the classical induction on $N$, which is short and
treats the two cases in the same way (Lemma~\ref{lem:bkp-sums}).

For Hietarinta's equation the 3SC is the result of \cite{HietarintaI}, and four-soliton
solutions are reported in \cite{GrammaticosRamaniHietarinta1990}; in the survey
\cite[Section~3.2.3]{Zhang2026} its integrability is described as unclear. For an irreducible
member the normalization of the dispersion curve has genus one or two
(Remark~\ref{rem:quartic}(iii)), so that the momenta of the solitons cannot be parametrized
rationally and the classical induction is not available. We use the
fusion identity of \cite[Lemma~9.6]{PartI} instead, which was introduced there for the
hyperbola $uv=1$: if the difference of two momenta $p_1,p_2$ is a zero of $P$ and
$P(p_1+p_2)\neq0$, that is, at a zero of the two-soliton phase factor, the three-soliton
condition gives an identity which, for generic further momenta $p_3,\dots,p_N$, expresses
$S_N[P](p_1,p_2,p_3,\dots)$ as a multiple of $S_{N-1}[P](p_1-p_2,p_3,\dots)$
(Lemma~\ref{lem:fusion}). On Hietarinta's curve a momentum $p_2$ has in general eight such
resonant partners $p_1$, and $-p_2$ has eight others. Together with the coincidences
$p_1=\pm p_2$ this gives $18(N-1)$ zeros
of $S_N$ as a function of one momentum on the curve, where it is the restriction of a
polynomial of degree at most $3N$; by B\'ezout's theorem it vanishes identically for $N\ge4$
(Theorem~\ref{thm:hietarinta-NSC}); by a closedness argument this extends to all values of the
parameters. Thus Hietarinta's equation has $N$-soliton solutions of Hirota's form for every $N$,
unless $a=b=c=0$, where its dispersion variety consists of four lines through the origin
(Section~\ref{subsec:resonances}). What is new in this argument is the count on a curve which is
not rational; in \cite{PartI} the identity was combined with a rational parametrization. The
latter form of the argument shows that the rigidity statement of
Theorem~\ref{thm:H} does not extend to the reducible members: on the cubic curve of the shallow
water wave equation of Hirota and Satsuma, a component of the reducible members
$X\,(X^2T+aX+bT)$ of the family $\PB_{a,b,c}$, there is a second family of solutions of the
3SC, and it satisfies the $N$-soliton conditions for all $N$ (Proposition~\ref{prop:sww}).

\subsection{What is not proved}
Conjecture~\ref{conj:main} remains open. Theorem~\ref{thm:A} says nothing when the top
homogeneous part of $C$ is a product of linear forms, which is always the case for plane
curves. Theorems~\ref{thm:B} and~\ref{thm:C} give necessary conditions only, and
Theorem~\ref{thm:C} says nothing about components which do not pass through the
origin. Theorem~\ref{thm:D} is restricted to linear images of the cubic graph: its proof rests
on an identity (Lemma~\ref{lem:cov}) which is special to this curve.
Theorem~\ref{thm:F} does not decide whether there is a solution of degree at least ten on
$V(C_h)$ for some algebraic $h\neq0,1$. Theorem~\ref{thm:G}(iii),(iv) restricts, but does not
classify, the plane curves with a branch at infinity which is asymptotic to a cubic graph, and
the limits of Lemma~\ref{lem:limits} give no information when they are affine lines, as for the
translated hyperbolas. The rigidity statement of Theorem~\ref{thm:H} concerns only the
irreducible members of its families: for the components of the reducible members which are
neither cubic graphs nor quadrics --- for instance
for the cubic curve of the shallow water wave equation of Hirota and Satsuma, on which
Proposition~\ref{prop:sww} gives a second family of solutions --- we have no classification
of the polynomials satisfying the 3SC. The reduction at resonant momenta which proves the
$N$-soliton conditions for Hietarinta's equation is not a proof of the conjecture for other
components: it gives no information at the zeros of the phase factor which come from another
component of $V(P)$ (Remark~\ref{rem:resonance-scope}). The precise list of open cases is given
in Section~\ref{sec:discussion}.

\subsection{Methods and organization}
Section~\ref{sec:prelim} collects elementary facts: the behaviour of $S_N$ under linear maps,
the dominance of sums and differences of points of a non-hyperplane component, a normalized
form of the 3SC, and a degeneration principle (Lemma~\ref{lem:degeneration}) by which the 3SC
passes to limits of rescaled polynomials on limits of rescaled components.
Section~\ref{sec:cones} proves Theorem~\ref{thm:A}: a tangential degeneration at a general
point reduces an arbitrary nonlinear cone to the polynomials $\xi^h(\xi\zeta+\eta^2)^m$ in
three variables, which are treated by an explicit expansion. Section~\ref{sec:collisions}
proves Theorem~\ref{thm:B}.

Section~\ref{sec:soft} proves Theorem~\ref{thm:C}. For $p_1,p_2\in V$ the function
$q\mapsto S_3[P](p_1,p_2,q)$ is an even polynomial on $\C^s$ which vanishes on $V$. Its
homogeneous parts of the two lowest degrees, $2\mu$ and $2\mu+2$, are computed explicitly
(Lemma~\ref{lem:taylor}); they must vanish on the arc directions of $V$ at the origin, which
gives the parity of $\mu$ and the identity (c), and the behaviour of (c) when $p_2$ tends to
$p_1$ gives the bound for $\nu_2$.

Section~\ref{sec:cubic} proves Theorem~\ref{thm:D}. Its new
ingredient is the passage to Laurent polynomials in $C$: on the cubic graph the product of the
four values of $C$ occurring in a summand of $S_3$ does not depend on the signs, so that
$S_3[C^mF]$ is a fixed multiple of $S_3[F]$ for every integer $m$ (Lemma~\ref{lem:cov}); a
polynomial whose top weighted part contains a high power of $C$ thereby becomes a Laurent
polynomial whose top part is of one of seven explicit types, and the pole order of the lower
terms is controlled by a double-collision argument (Lemma~\ref{lem:pole}), which reduces the
classification to finitely many linear systems. The same property gives the four-soliton
condition in closed form in the multiplicity: at a quadruple of points of the curve
$S_4[C^mT]$ is a sum of eight terms $A_\sg^mB_\sg$, and at one explicit quadruple one of them
dominates the others for every $m\ge3$ (Theorem~\ref{thm:fourth}).

Section~\ref{sec:graphs} proves Theorem~\ref{thm:E}. There is no covariance in degree at least
four, and the multiplicity is kept as an unknown instead. By the degeneration principle the top
weighted part of $P$ is a weighted homogeneous solution on a monomial graph $v=u^m$, and
Theorem~\ref{thm:C} leaves two candidates for even $m$ and five explicit families for
odd $m$ (Proposition~\ref{prop:candidates}). They are tested with the soft-momentum identity
(c) in the limit in which the second momentum also tends to the origin
(Lemma~\ref{lem:doublesoft}); this gives a sequence of conditions which are polynomial in the
degree $m$, in the multiplicity and in the parameters of the family. Even degrees are
excluded by hand, and so are two of the five families, apart from one solution for $m=5$. For
$m=5$ and $m=7$ the three remaining families are treated by Gr\"obner bases, which give the
second solution for $m=5$ and none for $m=7$, and for odd $m\ge9$ they are excluded uniformly
in $m$ and $\mu$: one further
condition comes from the collision of the two momenta (Lemma~\ref{lem:star}), the unknowns
are eliminated by resultants, and the resulting polynomials in $m$ and $2^{m-1}$ are shown to
have no zeros by an elementary size estimate and finitely many exact evaluations
(Proposition~\ref{prop:uniform}). That the quintic solutions satisfy all $N$SC follows from a
divisibility property of $S_N$ and a comparison of degrees (Lemma~\ref{lem:quintic}), which is
the argument of \cite{NewellZeng1986}; their relation to the BKP hierarchy is the subject of
Remark~\ref{rem:bkp}.

Section~\ref{sec:hyperbola} proves
Theorem~\ref{thm:F}; the algebraicity of $h$ follows from the exclusion of $h=0$ by a
properness argument.

Section~\ref{sec:limits} proves Theorems~\ref{thm:G} and~\ref{thm:H}. For a cubic graph
$v=u^3+a_2u^2+a_0$ the scaling which contracts it to $v=u^3$ displaces the curve to the first
order in the scaling parameter if $a_2\neq0$, and to the third order if $a_2=0\neq a_0$,
whereas an even polynomial has only parts of even relative weight; hence the term of this order
of $S_3$ involves only the top weighted part of $P$, which is one of the polynomials $C^\mu T$
of Section~\ref{sec:cubic}, and its non-vanishing is a finite computation, uniform in the
multiplicity $\mu$ by the covariance (Lemma~\ref{lem:firstorder}). For the equations of
Hietarinta's list the top parts of $P$ with respect to several weight vectors are compared:
for (I\,D) the two cubic graphs at infinity admit only one common solution of
Theorem~\ref{thm:D}; for (I\,B) the cubic branch leaves the top parts $v(u^3-cv)^\mu$, and the
hyperbolic branch, with the classification on the hyperbola $uv=1$ of
\cite[Theorem~9.14]{PartI}, gives $\mu=1$; for (I\,C) there are four hyperbolic branches, which
bound the multiplicities of the four asymptotic directions in the top homogeneous part of $P$
from below, and three lines through the origin, on which the top homogeneous part must satisfy
a parity condition (Lemma~\ref{lem:threelines}). The $N$-soliton conditions of
Proposition~\ref{prop:NSC-BD} follow from a factorization of $P$ at the sum of two momenta and
from the vanishing of the relevant alternating sum on the hyperplanes where two parameters
coincide (Lemma~\ref{lem:bkp-sums}). The last three subsections of Section~\ref{sec:limits}
contain the exclusion of the curves $u^av^b=c$ and its consequence for Newton polygons
(Theorem~\ref{thm:monomial-curves}, Corollary~\ref{cor:newton}), the reduction at resonant
momenta with the $N$-soliton conditions for Hietarinta's equation (Lemma~\ref{lem:fusion},
Theorem~\ref{thm:hietarinta-NSC}), and the solutions on the cubic curve of the shallow water
wave equation (Proposition~\ref{prop:sww}). Section~\ref{sec:discussion} states the remaining
cases.

\subsection*{Computer-assisted steps}
Theorems~\ref{thm:A}, \ref{thm:B} and~\ref{thm:C} are proved without computer
assistance. The following steps are finite computations in exact rational, integer or
algebraic arithmetic (no floating point): the evaluation of one constant in
Proposition~\ref{prop:sextic}, the weighted
homogeneous solutions of the weights $5$, $7$, $9$ on the cubic graph
(Proposition~\ref{prop:homog}), the kernels of the linearized condition
(Proposition~\ref{prop:kernel}), the value \eqref{eq:Q2value} in the proof of
Theorem~\ref{thm:cubic}, the integers in the proof of Theorem~\ref{thm:fourth}; the Gr\"obner
bases of the table in Section~\ref{sec:remaining} for
the degrees five and seven, the resultants and evaluations in the proof of
Proposition~\ref{prop:uniform}, and the evaluations in the proofs of Lemma~\ref{lem:quintic}
and Theorem~\ref{thm:graphs}; the finite jet computation in the proof of
Theorem~\ref{thm:hzero}, and the eliminations in the proof of Proposition~\ref{prop:degree8};
the numbers of Table~\ref{tab:critical} in Lemma~\ref{lem:firstorder}; the resultants of
Lemma~\ref{lem:certificate}, and the three-soliton condition for Hietarinta's equation with
indeterminate coefficients, on which Theorem~\ref{thm:hietarinta-NSC} depends
(Remark~\ref{rem:quartic}(i)).
In each case the text states exactly what is computed and how the result is used. Sample
points are used in three ways: to generate necessary conditions; to show that an explicit
quantity is not zero; and to prove an identity by interpolation, when the polynomial in
question is known beforehand to lie in an explicit finite-dimensional space on which the
evaluation at the sample points is injective (see the end of
Appendix~\ref{app:certificates}). The scripts and
their recorded outputs are described in Appendix~\ref{app:certificates}; all of them were
rerun for this paper, and the computation for Proposition~\ref{prop:uniform} was carried out
twice, with independently generated input and two computer algebra systems. The polynomial
identities which relate the results to the literature (Remarks~\ref{rem:cubic-scope}(iv)
and~\ref{rem:bkp}) and those used in Section~\ref{sec:limits} (Lemmas~\ref{lem:threelines}
and~\ref{lem:bkp-sums}(i)) are elementary; they are also checked by scripts. For Hietarinta's equation
the four-soliton condition with indeterminate coefficients and the conditions with
$5\le N\le7$ at sample points of one curve were computed as well
(Remark~\ref{rem:quartic}(i),(ii)); these computations are independent checks of
Theorem~\ref{thm:hietarinta-NSC}. One experiment in floating-point arithmetic is reported in
Remark~\ref{rem:resonance-scope}(iii); nothing depends on it.

\subsection*{Conventions}
A polynomial $P$ on $\C^s$ is \emph{even} if $P(-z)=P(z)$, where all coordinates change sign
simultaneously. Points of $\C^s$ are denoted by $p,q,\dots$; for a polynomial $C$ we write
$V(C)$ for its zero set, $DC(p)$ and $D^2C(p)$ for its first and second differentials, and
$\ord_0C$ for its order of vanishing at the origin. ``Generic'' means: in a nonempty Zariski
open subset. When a component $V(C)$ is fixed, a nonzero even polynomial $P$ which satisfies
the 3SC on $V(C)$ is called a \emph{solution of the 3SC on $V(C)$}, or simply a
\emph{solution}; what is classified in this paper are these polynomials, not the functions $f$
which solve a bilinear equation. References to \cite{PartI} use the numbering of
arXiv:2611.04953v1.

\section{Preliminaries}\label{sec:prelim}

\subsection{The sign sums}\label{subsec:signsums}
Let $P$ be an even polynomial on $\C^s$. The polynomial $S_N[P]$ of \eqref{eq:SN} is
homogeneous of degree $1+\binom N2$ in $P$, that is,
$S_N[\lambda P]=\lambda^{1+N(N-1)/2}S_N[P]$. The summands for $\sg$ and $-\sg$ coincide, so
$S_N[P]$ is twice the sum over the sign vectors with $\sg_1=+1$. For $N=3$ the product of the
signs $\sg_i\sg_j$ over $i<j$ equals $1$, so that
\begin{equation}\label{eq:S3}
S_3[P](p_1,p_2,p_3)=\sum_{\sg\in\{\pm1\}^3}P(\sg_1p_1+\sg_2p_2+\sg_3p_3)
\prod_{1\le i<j\le3}P(\sg_ip_i-\sg_jp_j)
\end{equation}
is a quartic form in $P$. The following observation will be used constantly.

\begin{lemma}[Linear maps]\label{lem:linear}
Let $A\colon\C^r\to\C^s$ be linear and $P$ a polynomial on $\C^s$. Then
$S_N[P\circ A](z_1,\dots,z_N)=S_N[P](Az_1,\dots,Az_N)$. In particular:
\begin{enumerate}
\item[\textup{(i)}] if $A$ is invertible, then $P$ satisfies the $N$\textup{SC} on a component
$V_0$ if and only if $P\circ A$ satisfies it on $A^{-1}(V_0)$, and $P$ is even, respectively
vanishes to order $m$ along $V_0$, if and only if $P\circ A$ is even, respectively vanishes to
order $m$ along $A^{-1}(V_0)$;
\item[\textup{(ii)}] if $S_N[P]$ vanishes on $V_0^{\,N}$ for a subset $V_0\subset\C^s$, then
$S_N[P\circ A]$ vanishes on $A^{-1}(V_0)^N$.
\end{enumerate}
\end{lemma}
\begin{proof}
All arguments of $P$ in \eqref{eq:SN} are linear combinations of the momenta.
\end{proof}

We shall use (ii) for the planes
\begin{equation}\label{eq:planes}
A_{\xi\eta}\colon\C^2\to\C^s,\qquad A_{\xi\eta}(u,v)=(u,v,\xi u+\eta v),\qquad \xi,\eta\in\C^{s-2},
\end{equation}
when $u,v$ are two of the coordinates of $\C^s$ and the component is defined by a polynomial
in $u,v$ alone: every point $(u,v,z)$ with $u\neq0$ lies on such a plane.

\begin{lemma}[Non-hyperplane components]\label{lem:dominance}
Let $C$ be an irreducible polynomial on $\C^s$ such that $V=V(C)$ is not an affine hyperplane,
let $V_{\mathrm{sm}}=\{p\in V:\ DC(p)\neq0\}$ and $T_pV=\ker DC(p)$ for $p\in V_{\mathrm{sm}}$.
\begin{enumerate}
\item[\textup{(a)}] The tangent hyperplanes $T_pV$, $p\in V_{\mathrm{sm}}$, take infinitely many
values.
\item[\textup{(b)}] For generic $p\in V_{\mathrm{sm}}$ the restriction of the quadratic form
$D^2C(p)$ to $T_pV$ is not zero.
\item[\textup{(c)}] For $\alpha,\beta\in\C\setminus\{0\}$ the map $V\times V\to\C^s$,
$(p,q)\mapsto\alpha p+\beta q$, is dominant.
\end{enumerate}
Consequently, for every nonzero polynomial $P$ on $\C^s$ the functions
$(p,q)\mapsto P(p\pm q)$ do not vanish identically on $V\times V$, and for every $N$ the
$N$-tuples $(p_1,\dots,p_N)\in V^N$ with $P(p_i\pm p_j)\neq0$ for all $i<j$ form a nonempty
Zariski open subset of $V^N$.
\end{lemma}
\begin{proof}
Since $V$ is irreducible, $V_{\mathrm{sm}}$ is a connected complex manifold, dense in $V$.
Consider the Gauss map $g\colon V_{\mathrm{sm}}\to\PP((\C^s)^*)$, $g(p)=[DC(p)]$.

(a) If $g$ took only finitely many values, it would be constant, say $g\equiv[\ell]$ with a
linear form $\ell$. Then $\ell$ vanishes on every $T_pV$, so the restriction of $\ell$ to
$V_{\mathrm{sm}}$ has zero differential and is constant, $\ell|_V\equiv c_0$. Thus $V$ is
contained in the hyperplane $\{\ell=c_0\}$; both being irreducible hypersurfaces, they
coincide, contrary to the hypothesis.

(b) For $p\in V_{\mathrm{sm}}$ and $v\in T_pV$ the derivative of $p\mapsto DC(p)$ in the
direction $v$ is the covector $D^2C(p)[v,\cdot\,]$. Hence $dg_p(v)=0$ if and only if
$D^2C(p)[v,\cdot\,]$ is a multiple of $DC(p)$, that is, if and only if $D^2C(p)[v,w]=0$ for
all $w\in T_pV$. The set of points $p\in V_{\mathrm{sm}}$ at which the restriction of
$D^2C(p)$ to $T_pV$ vanishes is Zariski closed in $V_{\mathrm{sm}}$; if it were all of
$V_{\mathrm{sm}}$, the differential of $g$ would vanish identically and $g$ would be constant,
contradicting (a).

(c) At $(p,q)\in V_{\mathrm{sm}}^2$ the image of the differential of the map is
$T_pV+T_qV$, which is $\C^s$ as soon as $T_pV\neq T_qV$. By (a) such a pair exists, so the map
is a submersion at some point and therefore dominant.

For the last statement, the image of a dominant map is Zariski dense, so $P(\alpha p+\beta q)$
is not identically zero on $V\times V$; and a finite intersection of nonempty open subsets of
the irreducible variety $V^N$ is nonempty.
\end{proof}

The tuples in the last statement are the \emph{nondegenerate} ones of \cite{PartI}; for them
all two-soliton phase factors are finite and nonzero. By Lemma~\ref{lem:dominance}(c), applied for a fixed third point,
the sums $\sg_1p_1+\sg_2p_2+\sg_3p_3$ of triples of points of $V$ are Zariski dense in $\C^s$ as
well, so that for generic triples all eight summands of \eqref{eq:S3} are different from zero.
Thus on a non-hyperplane component the 3SC is never satisfied by the vanishing of each summand
separately; it is a cancellation among summands which are nonzero at generic triples. For $p,q\in V$ with $P(p+q)\neq0$ we write
\begin{equation}\label{eq:phase}
A(p,q):=-\frac{P(p-q)}{P(p+q)}
\end{equation}
for the two-soliton phase factor \cite[(2.2)]{PartI}; it is a rational function on
$V\times V$, not identically zero, with $A(q,p)=A(p,q)$.

\begin{lemma}[Normalized three-soliton condition]\label{lem:normalized}
Let $P$ be even and $p_1,p_2,p_3\in V(P)$ with $P(p_i+p_j)\neq0$ for $i<j$, and write
$A_{ij}=A(p_i,p_j)$. Then
\begin{multline}\label{eq:normalized}
S_3[P](p_1,p_2,p_3)=-2\,P(p_1+p_2)P(p_1+p_3)P(p_2+p_3)\,\bigl[P(p_1+p_2+p_3)A_{12}A_{13}A_{23}\\
+P(p_1+p_2-p_3)A_{12}+P(p_1-p_2+p_3)A_{13}+P(-p_1+p_2+p_3)A_{23}\bigr].
\end{multline}
In particular, on a non-hyperplane component $V$ the \textup{3SC} is equivalent to the
vanishing of the bracket as a rational function on $V^3$.
\end{lemma}
\begin{proof}
Take the four sign vectors with $\sg_1=+1$ in \eqref{eq:S3}. By evenness,
$P(\sg_ip_i-\sg_jp_j)$ equals $P(p_i-p_j)=-A_{ij}P(p_i+p_j)$ if $\sg_i=\sg_j$ and $P(p_i+p_j)$
otherwise. For $\sg=(+,+,+)$ this gives $-P(p_1+p_2+p_3)A_{12}A_{13}A_{23}$ times the product
$\Pi:=P(p_1+p_2)P(p_1+p_3)P(p_2+p_3)$; for $(+,+,-)$, $(+,-,+)$, $(+,-,-)$ it gives
$-P(p_1+p_2-p_3)A_{12}\,\Pi$, $-P(p_1-p_2+p_3)A_{13}\,\Pi$ and $-P(p_1-p_2-p_3)A_{23}\,\Pi$,
and $P(p_1-p_2-p_3)=P(-p_1+p_2+p_3)$. The last statement follows from
Lemma~\ref{lem:dominance}.
\end{proof}

\subsection{A degeneration principle}
The following lemma will be applied three times, with a tangential scaling
(Section~\ref{sec:cones}), with the scaling to infinity (Corollary~\ref{cor:top}) and with a
weighted scaling at a point of the component (Theorem~\ref{thm:hzero}).

\begin{lemma}[Degeneration]\label{lem:degeneration}
Let $P$ be a polynomial on $\C^s$, let $V_0\subset\C^s$ and $N\ge2$, and suppose that $S_N[P]$
vanishes on $V_0^{\,N}$. Let $A_\eps\colon\C^r\to\C^s$, $\eps\in\C\setminus\{0\}$, be linear maps
and $a\in\Z$ an integer such that the coefficients of the polynomials
$P_\eps:=\eps^{-a}\,P\circ A_\eps$ are polynomials in $\eps$; let $P_0$ be their value at
$\eps=0$. Let $W\subset\C^r$ be an irreducible variety and $W'\subset W$ a Zariski
dense subset such that for every $z\in W'$ there are points $z(\eps)\in\C^r$, defined for all
sufficiently small $\eps\neq0$, with
\[
A_\eps z(\eps)\in V_0\qquad\text{and}\qquad z(\eps)\to z\quad(\eps\to0).
\]
Then $S_N[P_0]$ vanishes on $W^N$.
\end{lemma}
\begin{proof}
Let $z_1,\dots,z_N\in W'$. By Lemma~\ref{lem:linear} and the homogeneity of $S_N$ in $P$,
\[
S_N[P_\eps]\bigl(z_1(\eps),\dots,z_N(\eps)\bigr)
=\eps^{-a(1+N(N-1)/2)}\,S_N[P]\bigl(A_\eps z_1(\eps),\dots,A_\eps z_N(\eps)\bigr)=0
\]
for small $\eps\neq0$. The coefficients of the polynomial $S_N[P_\eps]$ are polynomials in
$\eps$ with value $S_N[P_0]$ at $\eps=0$, so letting $\eps\to0$ gives
$S_N[P_0](z_1,\dots,z_N)=0$. Since $W'^N$ is Zariski dense in $W^N$, the claim follows.
\end{proof}

\section{Nonlinear cones and top homogeneous parts}\label{sec:cones}

\begin{theorem}[Nonlinear homogeneous components]\label{thm:cone}
Let $P\neq0$ be an even polynomial on $\C^s$ with $P(0)=0$, and let $F$ be an irreducible
homogeneous polynomial of degree at least two which divides $P$. If $S_3[P]$ vanishes
identically on $V(F)^3$, then
\[
\deg F=2\qquad\text{and}\qquad P=\lambda F,\quad\lambda\in\C\setminus\{0\}.
\]
Conversely, if $F$ is an irreducible quadratic form, then $S_N[\lambda F]$ vanishes identically
on $V(F)^N$ for every $N\ge2$.
\end{theorem}

No hypothesis is made on the multiplicity of $F$ in $P$. For quadratic $F$ the theorem is
\cite[Theorem~9.15(i)]{PartI}; the proof below is independent of it. It has three steps: an
explicit obstruction for a model family in three variables, a tangential degeneration which
reduces a homogeneous $P$ to the model, and the passage to the top homogeneous part of a
general $P$.

\subsection{The model family}

\begin{lemma}\label{lem:model}
On $\C^3$ with coordinates $(\xi,\eta,\zeta)$ let $Q:=\xi\zeta+\eta^2$, let $m\ge1$ be an
integer and $h\ge0$ an even integer, and put $R:=\xi^hQ^m$. Then $S_3[R]$ vanishes identically
on $V(Q)^3$ if and only if $(m,h)=(1,0)$.
\end{lemma}
\begin{proof}
Let $B$ be the symmetric bilinear form with $Q(x)=B(x,x)$. For
$p_i=(\xi_i,\eta_i,\zeta_i)\in V(Q)$, $i=1,2,3$, put $B_{ij}:=B(p_i,p_j)$. Since $Q(p_i)=0$,
\[
Q\Bigl(\sum_i\sg_ip_i\Bigr)=2\sum_{i<j}\sg_i\sg_jB_{ij},\qquad
Q(\sg_ip_i-\sg_jp_j)=-2\,\sg_i\sg_jB_{ij},
\]
and $\prod_{i<j}\sg_i\sg_j=1$, so \eqref{eq:S3} gives
\begin{equation}\label{eq:model-normalize}
S_3[R](p_1,p_2,p_3)=(-1)^m2^{4m}\,(B_{12}B_{13}B_{23})^m\,U_{m,h},
\end{equation}
where
\begin{equation}\label{eq:Udef}
U_{m,h}:=\sum_{\sg\in\{\pm1\}^3}\Bigl(\sum_{i<j}\sg_i\sg_jB_{ij}\Bigr)^m
\Bigl(\sum_i\sg_i\xi_i\Bigr)^h\prod_{i<j}(\sg_i\xi_i-\sg_j\xi_j)^h .
\end{equation}
The variety $V(Q)^3$ is irreducible, and $B_{12}B_{13}B_{23}$ does not vanish identically on
it: for the points $(1,t,-t^2)\in V(Q)$ one has
$B\bigl((1,t,-t^2),(1,t',-t'^2)\bigr)=-\tfrac12(t-t')^2$. Hence $S_3[R]$ vanishes on $V(Q)^3$
if and only if $U_{m,h}$ does. (This cancellation has to precede the specialization below, in
which $B_{13}=0$.)

\emph{The case $h>0$.} Let $p:=(1,0,0)$ and $q:=(1,1,-1)$; both lie on $V(Q)$, and
$B(p,p)=0$, $b_0:=B(p,q)=-\tfrac12$. For $(p_1,p_2,p_3)=(p,\eps q,tp)\in V(Q)^3$ we have
$B_{12}=\eps b_0$, $B_{13}=0$, $B_{23}=\eps tb_0$ and $(\xi_1,\xi_2,\xi_3)=(1,\eps,t)$. The
summands of \eqref{eq:Udef} for $\sg$ and $-\sg$ coincide because $h$ is even; writing
$\sg=(1,\varsigma,\rho)$ we obtain
\begin{equation}\label{eq:Uexp}
U_{m,h}=2\,\eps^mb_0^m\sum_{\varsigma,\rho=\pm1}\varsigma^m(1+\rho t)^m\,\psi_{\varsigma\rho}(\eps)^h,
\end{equation}
where
\[
\psi_{\varsigma\rho}(\eps):=(1+\varsigma\eps+\rho t)(1-\varsigma\eps)(1-\rho t)(\varsigma\eps-\rho t).
\]
Here $\psi_{\varsigma\rho}(0)=-\rho t(1-t^2)$ and
\[
\frac{\psi_{\varsigma\rho}'(0)}{\psi_{\varsigma\rho}(0)}
=\varsigma\Bigl(\frac1{1+\rho t}-1-\frac1{\rho t}\Bigr)
=-\frac{\varsigma\,(1+\rho t+t^2)}{\rho t\,(1+\rho t)},
\]
so that, $h$ being even,
$\psi_{\varsigma\rho}(\eps)^h=t^h(1-t^2)^h\bigl[1-h\varsigma\eps\,\frac{1+\rho t+t^2}{\rho t(1+\rho t)}+O(\eps^2)\bigr]$.
These are identities of rational functions of $t$; the coefficients extracted below are
polynomials in $t$.

If $m$ is even, the coefficient of $\eps^m$ in $U_{m,h}$ is
\[
4\,b_0^m\,t^h(1-t^2)^h\bigl[(1+t)^m+(1-t)^m\bigr]\neq0 .
\]
If $m$ is odd, then $\varsigma^m=\varsigma$, the coefficient of $\eps^m$ vanishes, and the
coefficient of $\eps^{m+1}$ is
\[
-4h\,b_0^m\,t^{h-1}\,\Phi_{m,h}(t),\qquad
\Phi_{m,h}(t):=\sum_{\rho=\pm1}\rho\,(1+\rho t)^{m+h-1}(1-\rho t)^h(1+\rho t+t^2),
\]
where we used $(1-t^2)^h=(1+\rho t)^h(1-\rho t)^h$ and $1/\rho=\rho$. Now
$\Phi_{m,h}(t)=\varphi(t)-\varphi(-t)$ with $\varphi(x)=(1+x)^{m+h-1}(1-x)^h(1+x+x^2)$ and
$\varphi'(0)=(m+h-1)-h+1=m$, so $\Phi_{m,h}(t)=2mt+O(t^3)\neq0$. In both cases $U_{m,h}$ does not
vanish identically on $V(Q)^3$.

\emph{The case $h=0$.} The points $(1,0,0)$, $(0,0,2)$, $(1,\mathrm i\sqrt2,2)$ lie on $V(Q)$
and satisfy $B_{12}=B_{13}=B_{23}=1$. There
$\sum_{i<j}\sg_i\sg_jB_{ij}$ equals $3$ for the two constant sign vectors and $-1$ for the six
others, so $U_{m,0}=2\cdot3^m+6(-1)^m$, which is nonzero for every $m\ge2$. For $m=1$,
$U_{1,0}=\sum_\sg\sum_{i<j}\sg_i\sg_jB_{ij}=0$ identically.
\end{proof}

\subsection{Tangential degeneration: homogeneous \texorpdfstring{$P$}{P}}
Assume first that $P$ is homogeneous of degree $d$; $d$ is even because $P$ is even. Write
\[
P=F^mH,\qquad e:=\deg F\ge2,\qquad m\ge1,\qquad F\nmid H,
\]
so that $H$ is homogeneous of degree $d-me$. Since $F$ is irreducible of degree at least two,
$V(F)$ is not a hyperplane, and by Lemma~\ref{lem:dominance}(b) we can choose a smooth point
$p\neq0$ of $V(F)$ with $H(p)\neq0$ at which the restriction of $D^2F(p)$ to $T_pV(F)$ is not
zero. By polarization there is $v\in T_pV(F)$ with $D^2F(p)[v,v]\neq0$; choose $w\in\C^s$ with
$DF(p)[w]\neq0$, and put
\[
\kappa_1:=DF(p)[w]\neq0,\qquad\kappa_2:=\tfrac12D^2F(p)[v,v]\neq0 .
\]
By Euler's identities $DF(p)[p]=eF(p)=0$ and $D^2F(p)[p,\cdot\,]=(e-1)DF(p)$, so
$p\in T_pV(F)$ and $D^2F(p)[p,p]=0$; hence $v\notin\C p$, and since $w\notin T_pV(F)$, the
vectors $p,v,w$ are linearly independent. Define the injective linear maps
\[
A_\eps\colon\C^3\to\C^s,\qquad A_\eps(\xi,\eta,\zeta):=\xi p+\eps\eta v+\eps^2\zeta w\qquad(\eps\neq0).
\]
By Taylor's formula and the homogeneity $D^kF(\xi p)=\xi^{e-k}D^kF(p)$ of the derivatives,
\[
F\bigl(\xi p+y\bigr)=\sum_{k=1}^{e}\frac{\xi^{e-k}}{k!}D^kF(p)[y,\dots,y],\qquad y=\eps\eta v+\eps^2\zeta w .
\]
The term $k=1$ is $\eps^2\xi^{e-1}\kappa_1\zeta$ because $DF(p)[v]=0$, the term $k=2$ is
$\eps^2\xi^{e-2}\kappa_2\eta^2+O(\eps^3)$, and the terms with $k\ge3$ are $O(\eps^3)$. Hence, as
an identity of polynomials in $\eps,\xi,\eta,\zeta$,
\begin{equation}\label{eq:tangent-deg}
F\bigl(A_\eps(\xi,\eta,\zeta)\bigr)=\eps^2\,\xi^{e-2}K(\xi,\eta,\zeta)+O(\eps^3),\qquad
K:=\kappa_1\xi\zeta+\kappa_2\eta^2 ,
\end{equation}
and likewise $H(A_\eps(\xi,\eta,\zeta))=\xi^{d-me}H(p)+O(\eps)$. Consequently the coefficients of
$P_\eps:=\eps^{-2m}P\circ A_\eps$ are polynomials in $\eps$, with
\begin{equation}\label{eq:Pzero}
P_0=H(p)\,\xi^{\,d-2m}K^m .
\end{equation}
The exponent $d-2m$ is even, and nonnegative because $d\ge me\ge2m$.

We apply Lemma~\ref{lem:degeneration} with $V_0=V(F)$, $W=V(K)\subset\C^3$ (irreducible, since
$K$ has rank three) and $W'=\{(\xi,\eta,\zeta)\in V(K):\ \xi\neq0\}$. Let $z=(\xi,\eta,\zeta)\in W'$.
By \eqref{eq:tangent-deg}, $\Phi(\eps,\zeta'):=\eps^{-2}F(A_\eps(\xi,\eta,\zeta'))$ is a polynomial in
$(\eps,\zeta')$ with $\Phi(0,\zeta')=\xi^{e-2}K(\xi,\eta,\zeta')$; thus $\Phi(0,\zeta)=0$ and
$\partial_{\zeta'}\Phi(0,\zeta)=\kappa_1\xi^{e-1}\neq0$. By the implicit function theorem there
is an analytic function $\zeta(\eps)$ with $\zeta(0)=\zeta$ and $\Phi(\eps,\zeta(\eps))=0$, and
$z(\eps):=(\xi,\eta,\zeta(\eps))$ satisfies $z(\eps)\to z$ and $A_\eps z(\eps)\in V(F)$. Note that the
lifted points lie on the given component $V(F)$, not merely on $V(P)$. Lemma~\ref{lem:degeneration}
now shows that $S_3[P_0]$ vanishes on $V(K)^3$.

The linear change of coordinates $(\xi,\eta,\zeta)\mapsto(\xi,\kappa_2^{-1/2}\eta,\kappa_1^{-1}\zeta)$
transforms $K$ into $Q=\xi\zeta+\eta^2$ and $P_0$ into $H(p)\,\xi^{d-2m}Q^m$. By
Lemmas~\ref{lem:linear}(i) and~\ref{lem:model}, $(m,d-2m)=(1,0)$. Hence $m=1$, $d=2$, and
then $e=2$ and $H$ is constant: $P=\lambda F$ with $\deg F=2$.

\subsection{General \texorpdfstring{$P$}{P}, and the converse}
Now let $P$ be as in Theorem~\ref{thm:cone}, of degree $d$, with top homogeneous part $P_d$.
Since $P$ is even, $d$ is even and $P_d$ is even; since $F$ is homogeneous and divides $P$, it
divides every homogeneous component of $P$, in particular $P_d$, and $d\ge2$. For
$p_1,p_2,p_3\in V(F)$ and $t\in\C$ the points $tp_i$ lie on $V(F)$, so
$S_3[P](tp_1,tp_2,tp_3)=0$ for all $t$; the coefficient of $t^{4d}$ of this polynomial in $t$
is $S_3[P_d](p_1,p_2,p_3)$. Thus $P_d$ satisfies the hypotheses of the theorem, and by the
homogeneous case $d=2$ and $P_d=\lambda F$. An even polynomial of degree two with $P(0)=0$ is a
quadratic form, so $P=P_d=\lambda F$.

For the converse let $F(x)=B(x,x)$ be a quadratic form and $p_1,\dots,p_N\in V(F)$. Then
$\sg_i\sg_jF(\sg_ip_i-\sg_jp_j)=-2B(p_i,p_j)$ does not depend on $\sg$, so
\[
S_N[F](p_1,\dots,p_N)=\prod_{i<j}\bigl(-2B(p_i,p_j)\bigr)\cdot\sum_{\sg}2\sum_{i<j}\sg_i\sg_jB(p_i,p_j)=0
\qquad(N\ge2).
\]
This completes the proof of Theorem~\ref{thm:cone}. \qed

\subsection{Top homogeneous parts}

\begin{corollary}[Top homogeneous parts]\label{cor:top}
Let $P\neq0$ be an even polynomial on $\C^s$ with $P(0)=0$, let $C$ be an irreducible factor
of $P$, and suppose that $S_3[P]$ vanishes identically on $V(C)^3$. If the top homogeneous
part of $C$ has an irreducible factor $F$ of degree at least two, then $\deg F=2$,
$P=\lambda F$ and $C=\lambda'F$ with $\lambda,\lambda'\neq0$.

Equivalently: if $\deg P>2$, then the top homogeneous part of the equation of every
irreducible component of $V(P)$ on which $P$ satisfies the \textup{3SC} is a product of linear
forms.
\end{corollary}
\begin{proof}
Let $e=\deg C$, $d=\deg P$, and let $C_e$, $P_d$ be the top homogeneous parts. Put
$A_\eps:=\eps^{-1}\mathrm{id}$, so that $P_\eps(z):=\eps^dP(z/\eps)=P_d+\eps P_{d-1}+\dots$ has
value $P_d$ at $\eps=0$, and similarly $C_\eps(z):=\eps^eC(z/\eps)=C_e+\eps C_{e-1}+\cdots$. Write
$C_e=F^\rho G$ with $F\nmid G$, and let $W'$ be the set of smooth points $z$ of $W:=V(F)$ with
$G(z)\neq0$. For $z\in W'$ choose $w$ with $DF(z)[w]\neq0$. The polynomial
$t\mapsto C_e(z+tw)$ has a zero of order exactly $\rho$ at $t=0$, so by the continuity of the
roots of a polynomial (or Rouch\'e's theorem), for all sufficiently small $\eps$ the polynomial
$t\mapsto C_\eps(z+tw)$ has a root $t_\eps$ with $t_\eps\to0$ as $\eps\to0$; no single-valued
analytic choice is needed. Then $z(\eps):=z+t_\eps w\to z$ and $A_\eps z(\eps)=z(\eps)/\eps\in V(C)$.
By Lemma~\ref{lem:degeneration} (with $a=-d$), $S_3[P_d]$ vanishes on $V(F)^3$. Since $C$
divides $P$, $C_e$ divides $P_d$, so $F\mid P_d$; moreover $P_d$ is even and $d\ge2$.
Theorem~\ref{thm:cone} gives $\deg F=2$, $d=2$ and $P_d=\lambda F$; as before, $P=P_d=\lambda F$,
and $C\mid\lambda F$ gives $C=\lambda'F$.
\end{proof}

\begin{remark}\label{rem:cone-scope}
(i) Corollary~\ref{cor:top} extends \cite[Corollary~9.16]{PartI}, where $F$ was a quadratic
form. For example, a translated cone $F(z-z_0)=0$ over a nonlinear irreducible hypersurface
carries the 3SC only if it coincides with the cone $F=0$, and then only for $P=\lambda F$ with
$F$ quadratic.

(ii) Every homogeneous polynomial in two variables is a product of linear forms over $\C$, and
an irreducible homogeneous polynomial of degree at least two needs at least three variables.
Hence Theorem~\ref{thm:cone} and Corollary~\ref{cor:top} are statements about three or more
variables. They give no information when the top homogeneous part splits into linear forms:
the cubic graph $u^3-v$ and the hyperbolas $uv+u+v+h$ have the top parts $u^3$ and $uv$, and
the Kadomtsev--Petviashvili quartic $k^4+3l^2-4kw$ has the top part $k^4$.
\end{remark}

\section{Collision orders}\label{sec:collisions}

In this section and the next, $P\neq0$ is an even polynomial on $\C^s$ with $P(0)=0$, $C$ is an
irreducible factor of $P$, the hypersurface $V=V(C)$ is not a hyperplane, and $S_3[P]$ vanishes
identically on $V^3$. No hypothesis on the multiplicity of $C$ in $P$ is made at this point.
For a positive integer $r$ put
\[
rV:=\{rp:\ p\in V\}=V(C_r),\qquad C_r(z):=C(z/r),
\]
and let $\nu_r\ge0$ be the multiplicity of $C_r$ in $P$. We write $\mu:=\nu_1\ge1$ for the
multiplicity of $C$ itself, $P=C_r^{\nu_r}K_r$ with $C_r\nmid K_r$, and $H:=K_1$, so that
$P=C^\mu H$. Let
\[
d:=\ord_0P
\]
be the order of vanishing of $P$ at the origin (not the degree), and let $P_d\neq0$ be the
homogeneous part of $P$ of lowest degree; $d$ is even and $d\ge2$.

\begin{theorem}[Collision orders]\label{thm:collision}
Under these hypotheses:

\textup{(a)} If the multiplicity $\mu$ is odd, then
\begin{equation}\label{eq:collision-order}
d-2\nu_2\in\{\mu-1,\ \mu+1\},
\end{equation}
and if $d-2\nu_2=\mu-1$, then $\nu_3=1$. In particular, for a simple component
\textup{(}$\mu=1$\textup{)}: $d-2\nu_2\in\{0,2\}$; if $d=2\nu_2$, then $\nu_3=1$; and if $2V$ is
not a component of $V(P)$, then $d=2$, that is, $P$ has a nonzero quadratic part.

\textup{(b)} If $\mu$ is even, then $d-2\nu_2\in\{\mu-2,\ \mu\}$; moreover $\nu_3=2$ if
$d-2\nu_2=\mu-2$, and $\nu_3=0$ if $d-2\nu_2=\mu$.

In all cases $d-2\nu_2\le\mu+1$.
\end{theorem}

The multiplicity is odd for every component through the origin
(Theorem~\ref{thm:soft}(a) below), and $\mu=1$ in the situation of
Conjecture~\ref{conj:main}.

\begin{proof}
\emph{Step 1: a curved arc.} We choose a smooth point $p\in V$ and a tangent vector
$v\in T_pV$ such that
\begin{equation}\label{eq:arc-conditions}
\kappa:=D^2C(p)[v,v]\neq0,\qquad P_d(v)\neq0,\qquad K_r(rp)\neq0\quad(r=1,2,3).
\end{equation}
This is possible. The third condition holds for generic $p$ because $C_r\nmid K_r$. The tangent
bundle $\{(p,v):\ p\in V_{\mathrm{sm}},\ v\in T_pV\}$ is an irreducible variety, and the first
condition defines a nonempty open subset of it by Lemma~\ref{lem:dominance}(b). So does the
second: otherwise $P_d$ would vanish on $T_pV=\ker DC(p)$ for every $p\in V_{\mathrm{sm}}$, hence
would be divisible by the linear form $DC(p)$ for every such $p$, and by
Lemma~\ref{lem:dominance}(a) these are infinitely many pairwise non-proportional forms. Let
$\varphi$ be an analytic arc on $V$ with $\varphi(0)=p$, $\varphi'(0)=v$. Differentiating
$C(\varphi(t))=0$ twice at $t=0$ gives
\begin{equation}\label{eq:second-order}
DC(p)[\varphi''(0)]=-\kappa .
\end{equation}
For pairwise distinct $x_1,x_2,x_3\in\C$ and small $\eps$ we consider the points
\[
a_i:=\varphi(\eps x_i)=p+\eps x_iv+\tfrac12\eps^2x_i^2\varphi''(0)+O(\eps^3)\in V .
\]

\emph{Step 2: three expansions.} If $y=\eps\xi v+\eps^2\eta\,\varphi''(0)+O(\eps^3)$, then by
$C(p)=0$, $DC(p)[v]=0$ and \eqref{eq:second-order},
\[
C(p+y)=DC(p)[y]+\tfrac12D^2C(p)[y,y]+O(\eps^3)=\kappa\bigl(\tfrac12\xi^2-\eta\bigr)\eps^2+O(\eps^3).
\]
Applying this to $\tfrac12(a_i+a_j)$, $\tfrac13(a_1+a_2+a_3)$ and $a_i+a_j-a_k$, for which
$(\xi,\eta)$ equals $\bigl(\tfrac12(x_i+x_j),\tfrac14(x_i^2+x_j^2)\bigr)$,
$\bigl(\tfrac13\sum x_l,\tfrac16\sum x_l^2\bigr)$ and
$\bigl(x_i+x_j-x_k,\tfrac12(x_i^2+x_j^2-x_k^2)\bigr)$, we obtain
\begin{align}
C_2(a_i+a_j)=C\bigl(\tfrac12(a_i+a_j)\bigr)&=-\tfrac{\kappa}{8}\,(x_i-x_j)^2\,\eps^2+O(\eps^3),\label{eq:paircoll}\\
C_3(a_1+a_2+a_3)=C\bigl(\tfrac13(a_1+a_2+a_3)\bigr)&=-\tfrac{\kappa}{9}\,\Delta\,\eps^2+O(\eps^3),\label{eq:triplecoll}\\
C(a_i+a_j-a_k)&=\kappa\,(x_i-x_k)(x_j-x_k)\,\eps^2+O(\eps^3),\label{eq:mixedcoll}
\end{align}
where $\Delta:=x_1^2+x_2^2+x_3^2-x_1x_2-x_1x_3-x_2x_3$ and $\{i,j,k\}=\{1,2,3\}$.

\emph{Step 3: the orders of the terms.} Take $x_1,x_2,x_3$ pairwise distinct with $\Delta\neq0$.
Since $a_i-a_j=\eps(x_i-x_j)v+O(\eps^2)$ and $P_d(v)\neq0$,
$P(a_i-a_j)=P_d(v)(x_i-x_j)^d\eps^d\,(1+O(\eps))$. By \eqref{eq:paircoll} and
$K_2(2p)\neq0$, $P(a_i+a_j)=\bigl(-\tfrac\kappa8\bigr)^{\nu_2}(x_i-x_j)^{2\nu_2}\eps^{2\nu_2}K_2(2p)\,(1+O(\eps))$.
Hence the phase factors \eqref{eq:phase} satisfy
\begin{equation}\label{eq:A-asymptotic}
A(a_i,a_j)=c_0\,(x_i-x_j)^m\,\eps^m\,(1+O(\eps)),\qquad m:=d-2\nu_2,
\end{equation}
with a constant $c_0\neq0$ which does not depend on $i,j$ or on the $x_l$. The integer $m$ is
even; a priori it may be negative. Similarly, by \eqref{eq:triplecoll}, \eqref{eq:mixedcoll}
and $K_3(3p)K_1(p)\neq0$,
\begin{align*}
P(a_1+a_2+a_3)&=\bigl(-\tfrac\kappa9\bigr)^{\nu_3}\Delta^{\nu_3}\eps^{2\nu_3}K_3(3p)\,(1+O(\eps)),\\
P(a_i+a_j-a_k)&=\kappa^\mu\,\bigl[(x_i-x_k)(x_j-x_k)\bigr]^\mu\,\eps^{2\mu}K_1(p)\,(1+O(\eps)).
\end{align*}

\emph{Step 4: comparison.} By Lemma~\ref{lem:normalized} the bracket in \eqref{eq:normalized}
vanishes at $(a_1,a_2,a_3)$. Its first term is
\begin{equation}\label{eq:Tplus}
P(a_1+a_2+a_3)A_{12}A_{13}A_{23}=c_1\,\Delta^{\nu_3}\prod_{i<j}(x_i-x_j)^m\;\eps^{3m+2\nu_3}\,(1+O(\eps)),\qquad c_1\neq0,
\end{equation}
and the sum of the other three terms is
\begin{equation}\label{eq:Tmix}
c_0\,\kappa^\mu K_1(p)\,\Psi(x)\,\eps^{m+2\mu}+O(\eps^{m+2\mu+1}),\qquad
\Psi(x):=\sum_{k=1}^3\bigl[(x_i-x_k)(x_j-x_k)\bigr]^\mu(x_i-x_j)^m ,
\end{equation}
where $\{i,j,k\}=\{1,2,3\}$. Let first $\mu$ be odd. Since $m$ is even,
\begin{equation}\label{eq:Psi}
\Psi(0,1,t)=t^\mu(1-t)^m+t^m(1-t)^\mu-t^\mu(1-t)^\mu=[t(1-t)]^m\,\psi_n(t)
\end{equation}
with the odd integer $n:=\mu-m$ and $\psi_n(t):=t^n+(1-t)^n-[t(1-t)]^n$. Now $\psi_{-1}=0$, and $\psi_n\neq0$ for every odd $n\neq-1$:
for $n\le-3$, $[t(1-t)]^{-n}\psi_n=(1-t)^{-n}+t^{-n}-1\neq0$, and for $n\ge1$, $\psi_n$ is a
polynomial of degree $2n$.

If $n=-1$, then $m=\mu+1$. Let $n\neq-1$. Then $\Psi\not\equiv0$, so for generic $x$ the sum
\eqref{eq:Tmix} has the exact order $\eps^{m+2\mu}$ and must be cancelled by \eqref{eq:Tplus}.
Hence $3m+2\nu_3=m+2\mu$, that is, $\nu_3=n$; in particular $n\ge1$. Moreover the leading
coefficients cancel: $c\,\Delta^n\prod_{i<j}(x_i-x_j)^m+\Psi(x)\equiv0$ with a constant $c$. At
$x=(0,1,t)$ this reads
\begin{equation}\label{eq:negative-contra}
c\,(t^2-t+1)^n+\psi_n(t)=0 .
\end{equation}
For odd $n\ge3$ we have $\psi_n(t)=t^{2n}-nt^{2n-1}+\binom n2t^{2n-2}-\cdots$, because
$t^n+(1-t)^n$ has degree $n-1<2n-2$, whereas
$(t^2-t+1)^n=t^{2n}-nt^{2n-1}+\bigl(\binom n2+n\bigr)t^{2n-2}-\cdots$; so \eqref{eq:negative-contra}
is impossible. Hence $n=1$: $m=\mu-1$ and $\nu_3=1$. This proves \eqref{eq:collision-order} and
the statement on $\nu_3$. For $\mu=1$ and $\nu_2=0$ we get $d\in\{0,2\}$, and $d\ge2$.

\emph{Step 5: even multiplicity.} Let now $\mu$ be even. Steps~1--3 and the formulas
\eqref{eq:Tplus}, \eqref{eq:Tmix} do not depend on the parity of $\mu$, but in place of
\eqref{eq:Psi} we have
\begin{align*}
\Psi(0,1,t)&=t^\mu(1-t)^m+t^m(1-t)^\mu+t^\mu(1-t)^\mu=[t(1-t)]^m\,\psi^+_n(t),\\
\psi^+_n(t)&:=t^n+(1-t)^n+[t(1-t)]^n ,
\end{align*}
with the even integer $n:=\mu-m$. For real $t$ with $0<t<1$ the three terms are positive, also
if $m<0$; hence $\Psi\not\equiv0$ for every $n$. As in Step~4, the sum \eqref{eq:Tmix} must be
cancelled by \eqref{eq:Tplus}, so that $\nu_3=n\ge0$ and $c\,(t^2-t+1)^n+\psi^+_n(t)=0$ with a
constant $c$. Now $\psi^+_0=3$ and $\psi^+_2=(t^2-t+1)^2$, whereas for even $n\ge4$
\[
\psi^+_n(t)=t^{2n}-nt^{2n-1}+\tbinom n2t^{2n-2}-\cdots,\qquad
(t^2-t+1)^n=t^{2n}-nt^{2n-1}+\bigl(\tbinom n2+n\bigr)t^{2n-2}-\cdots
\]
are not proportional. Hence $n\in\{0,2\}$: either $m=\mu$ and $\nu_3=0$, or $m=\mu-2$ and
$\nu_3=2$. This proves (b), and the last statement follows from (a) and (b).
\end{proof}

\begin{remark}\label{rem:collision-scope}
The expansion \eqref{eq:A-asymptotic} holds for an arc satisfying \eqref{eq:arc-conditions}
and generic $x_i$; it is not a statement about every arc or every direction of collision. The
vanishing of $\psi_{-1}$ is precisely the cancellation that the necessary condition
\eqref{eq:collision-order} allows in the case $m=\mu+1$, which is the case of all classical
examples (for instance $\mu=1$, $d=2$, $\nu_2=0$ for the Korteweg--de Vries, Sawada--Kotera and
Kadomtsev--Petviashvili equations). The case $m=\mu-1$ occurs for quadric cones, where $\mu=1$,
$2V=V$ and $d=2$, $\nu_2=\nu_3=1$ (Theorem~\ref{thm:cone}); we know no other example. For
components through the origin which are not cones it does not occur
(Corollary~\ref{cor:origin}). Even multiplicities do not occur for components through the
origin (Theorem~\ref{thm:soft}(a)), but they do occur otherwise: the polynomials
$(au^2+bv^2)(uv-1)^2(uv-3)$ of \cite[Remark~9.12]{PartI} satisfy the 3SC, and in fact all
$N$SC, on the hyperbola $uv=1$, with $\mu=2$, $d=2$ and $\nu_2=\nu_3=0$, which is the case
$d-2\nu_2=\mu$ of part~(b). We know no example of the case $d-2\nu_2=\mu-2$. Part~(b) is
used in Section~\ref{subsec:monomial-curves}.
\end{remark}

\subsection{Dilates, and the branch \texorpdfstring{$d=2\nu_2$}{d = 2 nu2} for simple components}
We record two facts for later use. The first concerns the dilated hypersurfaces $rV$ when $V$
is not a cone, that is, when $C$ is not homogeneous; cones are covered by
Theorem~\ref{thm:cone}. The second concerns the branch $d=2\nu_2$ of
Theorem~\ref{thm:collision} for a simple component. For components through the origin which
are not cones this branch is excluded in Section~\ref{sec:soft}; for the remaining components
Lemma~\ref{lem:doubled} and Proposition~\ref{prop:additive} are the only information we have.

\begin{lemma}\label{lem:dilates}
If $C$ is not homogeneous, the hypersurfaces $rV$, $r=1,2,3,\dots$, are pairwise distinct. If
moreover $-rV=r'V$ for some positive integers $r,r'$, then $r=r'$ and $V=-V$.
\end{lemma}
\begin{proof}
If $\rho V=V$ for a real number $\rho>0$, $\rho\neq1$, then $C(\rho z)=\lambda C(z)$ for some
$\lambda$; comparing homogeneous components $C_j$ gives $\rho^jC_j=\lambda C_j$, and since the
numbers $\rho^j$ are pairwise distinct, $C$ has only one homogeneous component. This proves the
first statement. If $-rV=r'V$, then $V=-\rho V$ with $\rho=r'/r$, hence $V=\rho^2V$, so
$\rho=1$.
\end{proof}

\begin{lemma}[Doubled momenta]\label{lem:doubled}
In the situation of Theorem~\ref{thm:collision}, suppose that $\mu=1$ and $d=2\nu_2$. Then
\begin{equation}\label{eq:doubled}
P(2p+q)\,P(p-q)^2+P(2p-q)\,P(p+q)^2=0\qquad\text{for all }p,q\in V .
\end{equation}
\end{lemma}
\begin{proof}
Let $p,v,\varphi$ be as in the proof of Theorem~\ref{thm:collision}, let $x_1\neq x_2$,
$a_i=\varphi(\eps x_i)$, and let $q\in V$ be generic. By \eqref{eq:A-asymptotic} with $m=0$,
$A(a_1,a_2)\to c_0\neq0$ as $\eps\to0$, while $A(a_i,q)\to A(p,q)$. In the bracket of
\eqref{eq:normalized} for $(a_1,a_2,q)$, the terms $P(a_1-a_2+q)A(a_1,q)$ and
$P(-a_1+a_2+q)A(a_2,q)$ tend to zero because $P(q)=0$, and the limit of the remaining two terms
is $c_0\bigl[P(2p+q)A(p,q)^2+P(2p-q)\bigr]$. Multiplying by $P(p+q)^2$ gives
\eqref{eq:doubled} for generic $(p,q)\in V^2$, hence for all.
\end{proof}

The identity \eqref{eq:doubled} must not be read as a three-soliton condition on $2V$. In three
or more variables the branch $d=2\nu_2$ can often be excluded as follows.

\begin{proposition}[Additive triples]\label{prop:additive}
In the situation of Theorem~\ref{thm:collision}, suppose that $\mu=1$ and that there are smooth points
$p,q,r\in V$ with $r=p+q$ such that the covectors $DC(p)$, $DC(q)$, $DC(r)$ are linearly
independent. Then $d=2\nu_2+2$.
\end{proposition}
\begin{proof}
Suppose that $d=2\nu_2$, so that \eqref{eq:doubled} holds. Consider
\[
X:=\{(a,b)\in\C^s\times\C^s:\ C(a)=C(b)=C(a+b)=0\}.
\]
The differentials of the three equations at $(p,q)$ are $(DC(p),0)$, $(0,DC(q))$ and
$(DC(r),DC(r))$; they are linearly independent, so $X$ is a complex manifold near $(p,q)$ with
tangent space $\{(y,y'):\ DC(p)[y]=DC(q)[y']=DC(r)[y+y']=0\}$. Let
$\alpha,\beta\in\C$ with $\alpha\beta(\alpha-\beta)\neq0$. We claim that the map
$\mu_{\alpha\beta}\colon X\to\C^s$, $(a,b)\mapsto\alpha a+\beta b$, is a submersion at $(p,q)$.
Indeed, if a covector $\lambda$ annihilates the image of its differential, then
$(\alpha\lambda,\beta\lambda)$ is a combination of the three differentials:
\[
\alpha\lambda=x\,DC(p)+z\,DC(r),\qquad\beta\lambda=y\,DC(q)+z\,DC(r).
\]
Eliminating $\lambda$ gives $\beta x\,DC(p)-\alpha y\,DC(q)+(\beta-\alpha)z\,DC(r)=0$, hence
$x=y=z=0$ and $\lambda=0$.

On $X$ we have $P(a+b)=0$, so \eqref{eq:doubled} reduces to $P(2a+b)\,P(a-b)^2=0$ on $X$. Since
$\mu_{1,-1}$ is a submersion at $(p,q)$, the function $P(a-b)$ does not vanish identically on
any neighbourhood of $(p,q)$ in $X$; as $X$ is a connected manifold near $(p,q)$, $P(2a+b)$
vanishes identically there. But $\mu_{2,1}$ is a submersion at $(p,q)$ as well, so $P$ would
vanish on a nonempty open subset of $\C^s$, a contradiction.
\end{proof}

\begin{example}\label{ex:additive}
Let $C=w-k^3-l^3$ on $\C^3$. The points $p=(1,-1,0)$, $q=(2,3,35)$ and $p+q=(3,2,35)$ lie on
$V(C)$, and the covectors $DC=(-3k^2,-3l^2,1)$ at these points, $(-3,-3,1)$, $(-12,-27,1)$,
$(-27,-12,1)$, have determinant $-495$. By Proposition~\ref{prop:additive}, every even
polynomial $P$ with simple factor $C$ which satisfies the 3SC on $V(C)$ has $\ord_0P=2\nu_2+2$.
This surface passes through the origin, so the same conclusion follows from
Corollary~\ref{cor:origin}. The translate $C'=w-k^3-l^3-24$ does not pass through the origin;
there the points $(1,1,26)$, $(1,2,33)$ and their sum $(2,3,59)$ lie on $V(C')$, the three
covectors $(-3,-3,1)$, $(-3,-12,1)$, $(-12,-27,1)$ have determinant $-81$, and
Proposition~\ref{prop:additive} gives $\ord_0P=2\nu_2+2$ for every even $P$ with simple factor
$C'$ satisfying the 3SC on $V(C')$. For plane curves three covectors are always linearly
dependent, and the proposition does not apply.
\end{example}

\section{Soft momenta}\label{sec:soft}
The setting and the notation of Section~\ref{sec:collisions} remain in force: $P=C^\mu H$ with
$C\nmid H$, the hypersurface $V=V(C)$ is not a hyperplane, and $S_3[P]$ vanishes identically on
$V^3$. In Section~\ref{sec:collisions} two or three momenta collide at a general point of $V$.
Here we suppose that $V$ passes through the origin and let one of the three momenta tend to
the origin along $V$; we call it a \emph{soft momentum}. The first two nontrivial orders of the
expansion of $S_3$ in the soft momentum give the parity of $\mu$, a bilinear identity for all
pairs of momenta on $V$, and a bound for $\nu_2$.

\subsection{Statements}
If $0\in V$, an \emph{arc direction} of $V$ is a vector $e\neq0$ for which there is a convergent
power series $\gamma(\eps)=\eps^ke+O(\eps^{k+1})$, $k\ge1$, with $C(\gamma(\eps))\equiv0$; the
set of arc directions is denoted by $\cT(V)$. If $0$ is a smooth point of $V$, then
$\cT(V)=T_0V\setminus\{0\}$. For $z,e\in\C^s$ let
\[
\pi_j(z;e):=\tfrac1{j!}D_e^jP(z),\qquad\text{so that}\quad P(z+\tau e)=\sum_{j\ge0}\pi_j(z;e)\,\tau^j ;
\]
for $z\in V$ we have $\pi_j(z;e)=0$ if $j<\mu$, and $\pi_\mu(z;e)=(D_eC(z))^\mu H(z)$. For
$p_1,p_2\in V$ put $P^\pm:=P(p_1\pm p_2)$, $D_e^jP^\pm:=(D_e^jP)(p_1\pm p_2)$,
$a_j:=\pi_{\mu+j}(p_1;e)$, $b_j:=\pi_{\mu+j}(p_2;e)$, and define the \emph{soft form}
\begin{multline}\label{eq:soft}
\fF(p_1,p_2;e):=a_0b_0\bigl(P^-D_e^2P^+-P^+D_e^2P^-\bigr)-2(a_1b_0+a_0b_1)\,P^-D_eP^+\\
-2(a_0b_1-a_1b_0)\,P^+D_eP^-+4a_1b_1\,P^+P^- .
\end{multline}

\begin{theorem}[Soft-momentum identities]\label{thm:soft}
In the setting of Section~\textup{\ref{sec:collisions}} suppose that $0\in V$. Then:
\begin{enumerate}
\item[\textup{(a)}] the multiplicity $\mu$ is odd;
\item[\textup{(b)}] $\fF(p_1,p_2;e)=0$ for all $p_1,p_2\in V$ and all $e\in\cT(V)$;
\item[\textup{(c)}] $\nu_2\le1$: the doubled hypersurface $2V$ has multiplicity at most one in $P$.
\end{enumerate}
\end{theorem}

Where $a_0b_0\neq0$, \textup{(b)} reads, with $\Theta_1:=2a_1/a_0$ and $\Theta_2:=2b_1/b_0$,
\begin{equation}\label{eq:softTheta}
P^-D_e^2P^+-P^+D_e^2P^--(\Theta_1+\Theta_2)\,P^-D_eP^++(\Theta_1-\Theta_2)\,P^+D_eP^-+\Theta_1\Theta_2\,P^+P^-=0,
\end{equation}
and $\Theta(z):=2\pi_{\mu+1}(z;e)/\pi_\mu(z;e)=\mu\,D_e^2C(z)/D_eC(z)+2\,D_eH(z)/H(z)$ for $z\in V$.
Since $P(z\mp\eps e)=(\mp\eps)^\mu\pi_\mu(z;e)\,(1\mp\tfrac12\Theta(z)\eps+O(\eps^2))$, the function
$-\Theta$ is the linear coefficient of the two-soliton phase factor at a soft momentum:
$-P(z-\eps e)/P(z+\eps e)=1-\Theta(z)\eps+O(\eps^2)$ for odd $\mu$.

\begin{corollary}[Components through the origin]\label{cor:origin}
In the setting of Section~\textup{\ref{sec:collisions}} suppose that $0\in V$ and that $V$ is not a cone, and let $e_0:=\ord_0C$ be the multiplicity of the origin on $V$.
Then:
\begin{enumerate}
\item[\textup{(1)}] $\mu$ is odd, $\nu_2\in\{0,1\}$, and $d=\mu+1+2\nu_2$;
\item[\textup{(2)}] $e_0=1$, or $\mu=1$ and $e_0=2$;
\item[\textup{(3)}] if $V\neq-V$, then $\mu=1$, $e_0=1$, and
$P=C\,C^-\,(C_2C_2^-)^{\nu_2}K$ with $K(0)\neq0$, where $C^-(z):=C(-z)$ and $C_2^-(z):=C_2(-z)$.
\end{enumerate}
In particular, for $\mu=1$ the branch $d=2\nu_2$ of Theorem~\textup{\ref{thm:collision}} does not occur for a
component through the origin which is not a cone.
\end{corollary}
For cones Theorem~\ref{thm:cone} applies. The proofs of Theorem~\ref{thm:soft} and of
Corollary~\ref{cor:origin} occupy the rest of this section. A further consequence of the
identity \textup{(b)}, needed in Section~\ref{sec:graphs}, is Lemma~\ref{lem:star}.

\subsection{The expansion}\label{subsec:expansion}
In this subsection $S_3[P]=0$ on $V^3$ is assumed only where stated. For $p_1,p_2,q\in\C^s$ put
\begin{equation}\label{eq:Fdef}
F(q):=\sum_{\sg_1,\sg_2=\pm1}P(\sg_1p_1+\sg_2p_2+q)\,P(\sg_1p_1-\sg_2p_2)\,P(\sg_1p_1-q)\,P(\sg_2p_2-q).
\end{equation}

\begin{lemma}\label{lem:F}
$F(-q)=F(q)$ and $S_3[P](p_1,p_2,q)=2F(q)$.
\end{lemma}
\begin{proof}
For $N=3$ the product of the signs in $S_3$ equals one. The summand of $S_3[P](p_1,p_2,q)$ for
$(\sg_1,\sg_2,+1)$ is the summand of \eqref{eq:Fdef} for $(\sg_1,\sg_2)$, and the summand for
$(\sg_1,\sg_2,-1)$ is, by the evenness of $P$ applied to each of the four factors, the summand of
\eqref{eq:Fdef} for $(-\sg_1,-\sg_2)$. In the same way the summand of $F(-q)$ for
$(\sg_1,\sg_2)$ is the summand of $F(q)$ for $(-\sg_1,-\sg_2)$.
\end{proof}

\begin{lemma}[Taylor coefficients]\label{lem:taylor}
Let $p_1,p_2\in V$ and $e\in\C^s$. As polynomials in $\eps$,
\[
F(\eps e)=\begin{cases}
4\,P^+P^-\,\pi_\mu(p_1;e)\,\pi_\mu(p_2;e)\,\eps^{2\mu}+O(\eps^{2\mu+2}),&\mu\ \text{even},\\[2pt]
\fF(p_1,p_2;e)\,\eps^{2\mu+2}+O(\eps^{2\mu+4}),&\mu\ \text{odd}.
\end{cases}
\]
Consequently, if $F=\sum_jF_j$ is the decomposition of the polynomial $F$ into homogeneous parts,
then $F_j=0$ for $j<2\mu$ and for all odd $j$, $F_{2\mu}(e)=2(1+(-1)^\mu)P^+P^-\pi_\mu(p_1;e)\pi_\mu(p_2;e)$,
and $F_{2\mu+2}(e)=\fF(p_1,p_2;e)$ if $\mu$ is odd.
\end{lemma}
\begin{proof}
Let $\sg=\pm1$ and $z\in V$. By evenness,
\[
P(\sg z-\eps e)=P(z-\sg\eps e)=\sum_{j\ge0}\pi_{\mu+j}(z;e)\,(-\sg\eps)^{\mu+j},
\]
hence
\[
P(\sg_1p_1-\eps e)\,P(\sg_2p_2-\eps e)=(\sg_1\sg_2)^\mu\eps^{2\mu}\,
\bigl[a_0-\sg_1\eps a_1+\eps^2a_2-\cdots\bigr]\bigl[b_0-\sg_2\eps b_1+\eps^2b_2-\cdots\bigr].
\]
Also by evenness, $P(\sg(p_1\pm p_2)+\eps e)=P^\pm+\sg\eps D_eP^\pm+\tfrac12\eps^2D_e^2P^\pm+O(\eps^3)$,
and $P(\sg_1p_1-\sg_2p_2)$ equals $P^-$ if $\sg_1=\sg_2$ and $P^+$ otherwise. The sum over
$\sg=\pm1$ of a power series in $\sg\eps$ is twice its even part. Thus the two terms of
\eqref{eq:Fdef} with $\sg_1=\sg_2$ contribute
\begin{multline*}
2P^-\eps^{2\mu}\Bigl\{P^+a_0b_0+\eps^2\bigl[\tfrac12a_0b_0D_e^2P^+-(a_1b_0+a_0b_1)D_eP^+\\
+(a_2b_0+a_1b_1+a_0b_2)P^+\bigr]\Bigr\}+O(\eps^{2\mu+4}),
\end{multline*}
and the two terms with $\sg_1=-\sg_2$ contribute
\begin{multline*}
2(-1)^\mu P^+\eps^{2\mu}\Bigl\{P^-a_0b_0+\eps^2\bigl[\tfrac12a_0b_0D_e^2P^-+(a_0b_1-a_1b_0)D_eP^-\\
+(a_2b_0-a_1b_1+a_0b_2)P^-\bigr]\Bigr\}+O(\eps^{2\mu+4}).
\end{multline*}
For even $\mu$ the leading terms add up to $4P^+P^-a_0b_0\eps^{2\mu}$. For odd $\mu$ they cancel,
the terms with $a_2$ and $b_2$ cancel, and the coefficient of $\eps^{2\mu+2}$ is \eqref{eq:soft}.
The last statement follows because $F(\eps e)=\sum_j\eps^jF_j(e)$ for every $e$ and $F$ is even.
\end{proof}

\begin{lemma}[Arc directions]\label{lem:arcs}
Let $0\in V$. There is $e\in\cT(V)$ such that $D_eC$ does not vanish identically on $V$. If $0$
is a smooth point of $V$, every $e\in T_0V$ outside a proper linear subspace has this property.
\end{lemma}
\begin{proof}
Let $L$ be the linear space of all $w\in\C^s$ with $D_wC\equiv0$. Then $C$ is constant along the
cosets of $L$, that is, $C$ is the pull-back of a polynomial on $\C^s/L$. If $\dim\C^s/L\le1$,
the irreducible polynomial $C$ would be constant or a polynomial of degree one in a linear form,
and $V$ would be empty or a hyperplane. Hence $\dim\C^s/L\ge2$, and we can choose a
two-dimensional linear subspace $\Pi$ with $\Pi\cap L=0$ and $C|_\Pi\not\equiv0$. The plane curve
germ $\{C|_\Pi=0\}$ at the origin has a branch, which by the Newton--Puiseux theorem is
parametrized, in suitable linear coordinates of $\Pi$, by $\eps\mapsto(\eps^k,\sum_{j\ge k}c_j\eps^j)$;
this is an arc in $V$ with $e=(1,c_k)\in\Pi\setminus\{0\}$. Since $e\notin L$, the polynomial
$D_eC$ is not zero; its degree is smaller than $\deg C$, so it is not divisible by $C$ and does
not vanish identically on $V$. If $0$ is a smooth point, $\cT(V)=T_0V\setminus\{0\}$, and
$T_0V\subset L$ is impossible, because then $C$ would be a polynomial in the linear form $DC(0)$.
\end{proof}

\begin{proof}[Proof of Theorem~\ref{thm:soft}\,\textup{(a), (b)}]
Let $p_1,p_2\in V$ and $e\in\cT(V)$, with an arc $\gamma(\eps)=\eps^ke+O(\eps^{k+1})$ in $V$. By
the hypothesis and Lemma~\ref{lem:F}, $F(\gamma(\eps))=0$ for all small $\eps$. By
Lemma~\ref{lem:taylor}, $F=F_{2\mu}+(\text{terms of degree}>2\mu)$, so
$F(\gamma(\eps))=\eps^{2\mu k}F_{2\mu}(e)+O(\eps^{2\mu k+1})$ and therefore $F_{2\mu}(e)=0$. If
$\mu$ were even, this would say $P(p_1+p_2)P(p_1-p_2)\pi_\mu(p_1;e)\pi_\mu(p_2;e)=0$ for all
$p_1,p_2\in V$. Choose $e$ as in Lemma~\ref{lem:arcs}. Then $\pi_\mu(\,\cdot\,;e)=(D_eC)^\mu H$
does not vanish identically on $V$, and neither do $P(p_1\pm p_2)$ on $V^2$ by
Lemma~\ref{lem:dominance}; since $V^2$ is irreducible, this is a contradiction. Hence $\mu$ is odd.
Then $F=F_{2\mu+2}+(\text{terms of degree}\ge2\mu+4)$, so
$F(\gamma(\eps))=\eps^{(2\mu+2)k}\fF(p_1,p_2;e)+O(\eps^{(2\mu+2)k+1})$, which proves (b).
\end{proof}

\begin{proof}[Proof of Theorem~\ref{thm:soft}\,\textup{(c)}]
Write $P=C_2^{\nu_2}K_2$ with $C_2\nmid K_2$, and let $P_d$ be the homogeneous part of $P$ of
lowest degree. Choose $e$ as in Lemma~\ref{lem:arcs}, and choose, as in Step~1 of the proof of
Theorem~\ref{thm:collision}, a smooth point $p\in V$ and $v\in T_pV$ with
\[
\kappa:=D^2C(p)[v,v]\neq0,\quad P_d(v)\neq0,\quad K_2(2p)\neq0,\quad\theta_e:=DC(p)[e]\neq0,\quad H(p)\neq0 ;
\]
each condition defines a nonempty Zariski open subset of the irreducible variety of pairs
$(p,v)$. Let $\varphi$ be an analytic arc on $V$ with $\varphi(0)=p$, $\varphi'(0)=v$, and take
$p_1=p$, $p_2=\varphi(y)$. As $y\to0$ we have $P^-=P(-yv+O(y^2))=P_d(v)\,y^d(1+O(y))$,
$D_eP^-=O(y^{d-1})$, $D_e^2P^-=O(y^{d-2})$, and by \eqref{eq:paircoll},
$C_2(p_1+p_2)=C(\tfrac12(p_1+p_2))=-\tfrac\kappa8y^2+O(y^3)$, while
$D_eC_2(p_1+p_2)=\tfrac12\theta_e+O(y)$. Suppose that $\nu_2\ge2$. Then $P^+=O(y^{2\nu_2})$,
$D_eP^+=O(y^{2\nu_2-2})$ and
\[
D_e^2P^+=\nu_2(\nu_2-1)\bigl(-\tfrac\kappa8\bigr)^{\nu_2-2}\bigl(\tfrac12\theta_e\bigr)^2K_2(2p)\,y^{2\nu_2-4}\,(1+O(y)).
\]
Moreover $a_0=\theta_e^\mu H(p)\neq0$, $b_0\to a_0$, and $a_1,b_1$ are bounded. In
\eqref{eq:soft} the term $a_0b_0P^-D_e^2P^+$ then has the exact order $y^{d+2\nu_2-4}$, and all
other terms are $O(y^{d+2\nu_2-2})$. This contradicts (b); hence $\nu_2\le1$.
\end{proof}

\begin{lemma}[Collision condition for $\nu_2=1$]\label{lem:star}
In the setting of Section~\textup{\ref{sec:collisions}} let $0\in V$, $e\in\cT(V)$, and suppose that
$\nu_2=1$; write $P=C_2K_2$ and let $P_d$ be the homogeneous part of $P$ of lowest degree. Let
$p$ be a smooth point of $V$ and $w\in T_pV$ such that
\[
P_d(w)\neq0,\qquad K_2(2p)\neq0,\qquad D_eC(p)\neq0,\qquad H(p)\neq0 .
\]
Then, with $\kappa:=D^2C(p)[w,w]$,
\begin{equation}\label{eq:star-general}
\Theta(p)=\frac{D_e^2C(p)}{4\,D_eC(p)}+\frac{D_eK_2(2p)}{K_2(2p)}+\frac{\kappa\,D_e^2P_d(w)}{8\,D_eC(p)\,P_d(w)} .
\end{equation}
\end{lemma}
\begin{proof}
Let $\varphi$ be an analytic arc on $V$ with $\varphi(0)=p$, $\varphi'(0)=w$, and take $p_1=p$,
$p_2=\varphi(y)$. Since $d=\ord_0P$ is even, we have, as $y\to0$, $P^-=P_d(w)\,y^d(1+O(y))$,
$D_eP^-=O(y^{d-1})$ and $D_e^2P^-=D_e^2P_d(w)\,y^{d-2}+O(y^{d-1})$. By \eqref{eq:paircoll},
$C_2(p_1+p_2)=C(\tfrac12(p_1+p_2))=-\tfrac\kappa8y^2+O(y^3)$; moreover
$D_eC_2(p_1+p_2)=\tfrac12D_eC(p)+O(y)$ and $D_e^2C_2(p_1+p_2)=\tfrac14D_e^2C(p)+O(y)$. Hence
\begin{gather*}
P^+=-\tfrac\kappa8K_2(2p)\,y^2+O(y^3),\qquad D_eP^+=\tfrac12D_eC(p)\,K_2(2p)+O(y),\\
D_e^2P^+=\tfrac14D_e^2C(p)\,K_2(2p)+D_eC(p)\,D_eK_2(2p)+O(y).
\end{gather*}
Finally $b_j=a_j+O(y)$. In \eqref{eq:soft} the terms $-2(a_0b_1-a_1b_0)P^+D_eP^-$ and
$4a_1b_1P^+P^-$ are $O(y^{d+2})$, and the coefficient of $y^d$ in $\fF(p,\varphi(y);e)$ is
\begin{multline*}
a_0^2P_d(w)\bigl[\tfrac14D_e^2C(p)\,K_2(2p)+D_eC(p)\,D_eK_2(2p)\bigr]+\tfrac\kappa8\,a_0^2\,K_2(2p)\,D_e^2P_d(w)\\
-2a_0a_1\,D_eC(p)\,K_2(2p)\,P_d(w).
\end{multline*}
It vanishes by Theorem~\ref{thm:soft}(b). Here $a_0=(D_eC(p))^\mu H(p)\neq0$ and
$\Theta(p)=2a_1/a_0$, so division by $a_0^2\,D_eC(p)\,K_2(2p)\,P_d(w)$ gives
\eqref{eq:star-general}.
\end{proof}
The right-hand side of \eqref{eq:star-general} does not change if $C_2$ is multiplied by a
constant.

\begin{remark}\label{rem:soft}
(i) Only the vanishing of $S_3[P]$ at triples $(p_1,p_2,q)$ with $q$ near the origin is used.
If $0\notin V$, the expansion is not available.

(ii) Treating $\Theta_1,\Theta_2$ as constants, \eqref{eq:softTheta} is equivalent to
$\bigl(D_e-\tfrac12(\Theta_1+\Theta_2)\bigr)^2P^+/P^+=\bigl(D_e-\tfrac12(\Theta_1-\Theta_2)\bigr)^2P^-/P^-$.

(iii) On the cubic graph the three-soliton condition transfers to the odd cofactor $H=P/C$ by
the covariance \eqref{eq:M} of Section~\ref{sec:cubic}, and the identity \eqref{eq:diamond}
proved there is the counterpart of \eqref{eq:softTheta} for $H$. Lemma~\ref{lem:nu2}(ii) is
Theorem~\ref{thm:soft}(c) for the cubic graph, and the parity statement of
Lemma~\ref{lem:divis} is Theorem~\ref{thm:soft}(a). The proofs above do not use the covariance.
\end{remark}

\subsection{Components through the origin}
\begin{proof}[Proof of Corollary~\ref{cor:origin}]
By Theorem~\ref{thm:soft}, $\mu$ is odd and $\nu_2\le1$. Since $V$ is not a cone, the
polynomials $C$, $C_2$, $C_3$ are pairwise non-associated (Lemma~\ref{lem:dilates}), and each of them
vanishes to order $e_0$ at the origin.

(1) By Theorem~\ref{thm:collision}, $d-2\nu_2\in\{\mu-1,\mu+1\}$. If $d=2\nu_2+\mu-1$, then
$\nu_3=1$, so $C^\mu C_2^{\nu_2}C_3$ divides $P$ and $d\ge e_0(\mu+\nu_2+1)\ge\mu+\nu_2+1$, which
gives $\nu_2\ge2$, a contradiction. Hence $d=\mu+1+2\nu_2$.

(2) Since $C^\mu C_2^{\nu_2}$ divides $P$, $e_0(\mu+\nu_2)\le\mu+1+2\nu_2$. If $e_0\ge2$, this
gives $\mu\le1$, and then $e_0(1+\nu_2)\le2+2\nu_2$ gives $e_0=2$.

(3) Since $P$ is even, $-V$ and $-2V$ have the multiplicities $\mu$ and $\nu_2$. If $V\neq-V$,
the hypersurfaces $V,-V,2V,-2V$ are pairwise distinct (Lemma~\ref{lem:dilates}), so
$2e_0(\mu+\nu_2)\le\mu+1+2\nu_2$. Hence $\mu=1$ and $e_0=1$, and
$K:=P/\bigl(CC^-(C_2C_2^-)^{\nu_2}\bigr)$ satisfies $\ord_0K=d-2-2\nu_2=0$.
\end{proof}
\section{The cubic graph}\label{sec:cubic}

Throughout this section $u,v$ are two of the coordinates of $\C^s$, $s\ge2$,
\[
C:=u^3-v,\qquad\bp(t):=(t,t^3),\qquad\wt u:=1,\quad\wt v:=3 ,
\]
$P_4,\dots,P_{12}$ are the polynomials \eqref{eq:Kbasis}, and $r$ denotes a root of
$r^2-7r+1=0$. The ordinary degree of a monomial $u^iv^j$ is $i+j$ and its weight is $i+3j$;
the indices of $P_4,\dots,P_{12}$ are weights.

\subsection{Statements}

\begin{theorem}[Two variables, arbitrary degree]\label{thm:cubic}
Let $P\in\C[u,v]\setminus\{0\}$ be even, and let $C$ be a simple factor of $P$. Then $P$
satisfies the \textup{3SC} on $V(C)$ if and only if $P$ belongs to one of the following
families:
\begin{enumerate}
\item[\textup{(a)}] $P\in\Span(P_4,P_6,P_8,P_{10},P_{12})$;
\item[\textup{(b)}] $P=C\,(\alpha u+\gamma v)$ with $(\alpha,\gamma)\neq(0,0)$;
\item[\textup{(c)}] $P=C\,\bigl[\alpha u+\beta(u^3+rv)\bigr]$ with $\beta\neq0$ and $r^2-7r+1=0$.
\end{enumerate}
Every such $P$ satisfies the $N$\textup{SC} on $V(C)$ for all $N\ge3$. In particular the
weighted degree of $P$ is at most $12$.
\end{theorem}

The families overlap: $\alpha P_4$ belongs to (a) and (b). The hypothesis $P(0)=0$ is not
needed; it is a consequence.

\begin{theorem}[All solutions, any number of variables]\label{thm:cubic-all}
Let $P\neq0$ be an even polynomial on $\C^s$. Then $S_3[P]$ vanishes identically on $V(C)^3$
if and only if
\[
P=C^{2n}P_0,\qquad n\ge0,
\]
where $P_0$ is one of the following: \textup{(a)} a nonzero element of
$\Span(P_4,\dots,P_{12})\subset\C[u,v]$; \textup{(b)} $P_0=C\cdot L$ with $L\neq0$ a linear form
on $\C^s$; \textup{(c)} $P_0=C\,[\alpha u+\beta(u^3+rv)]$ with $\beta\neq0$ and $r^2-7r+1=0$.
In each case $C$ is a simple factor of $P_0$, and $P_0$ satisfies the $N$\textup{SC} on $V(C)$
for every $N\ge3$.
\end{theorem}

\begin{corollary}\label{cor:conj-cubic}
Let $\ell_1,\ell_2$ be linearly independent linear forms on $\C^s$ and $V_0:=V(\ell_1^3-\ell_2)$.
Then Conjecture~\ref{conj:main} holds for $V_0$: if $P$ is an even polynomial which vanishes to
order one along $V_0$ and satisfies the \textup{3SC} on $V_0$, then $P$ satisfies the
$N$\textup{SC} on $V_0$ for every $N\ge3$. These $P$ are exactly the polynomials
\textup{(a)}, \textup{(b)}, \textup{(c)} of Theorem~\ref{thm:cubic-all} in the variables
$u=\ell_1$, $v=\ell_2$. In the notation of \cite{PartI}, where $\ell_1=D_x$ and
$\ell_2=-\Lam$, family \textup{(b)} consists of the polynomials $L\,(D_x^3+\Lam)$ of
\cite[Theorem~4.1]{PartI}.
\end{corollary}

\begin{remark}[Scope]\label{rem:cubic-scope}
(i) Theorem~\ref{thm:cubic-all} describes the solutions of the 3SC for every multiplicity of
the component --- the multiplicity is necessarily odd --- but for $n\ge1$ it asserts only the
3SC. In fact the 4SC fails for every $n\ge1$ (Theorem~\ref{thm:fourth} below); for
$P=L\,C^{2n+1}$ this is \cite[Theorem~6.10]{PartI}. The polynomials with $n\ge1$ are outside
the hypotheses of Conjecture~\ref{conj:main}.

(ii) With the phase factor prescribed in the family \eqref{eq:Aalpha},
\cite[Theorem~9.19]{PartI} found family (a) (with $\alpha=0$ in \eqref{eq:Aalpha}), the
polynomial $vC$ and the polynomials $C(u^3+rv)$. Theorem~\ref{thm:cubic} shows that, in two
variables and without any hypothesis on the phase factor, the only further solutions with
simple factor $C$ are the deformations by $\alpha uC$ in families (b) and (c), whose phase
factors \eqref{eq:deformedphase} are not homogeneous in $(s,t)$.

(iii) Nothing is claimed for components that are not linear images of the cubic graph; the
proof depends on the identity \eqref{eq:M} below.

(iv) \emph{Known equations.} For $n=0$ the polynomials of Theorem~\ref{thm:cubic-all} are not
new as bilinear equations; the new statement is the completeness of the list. Family (a) is that
of \cite[Theorem~9.19]{PartI}, and $P_4$ is the Korteweg--de Vries polynomial. Family (b) is
the class $L\,(D_x^3+\Lam)$ of \cite{PartI}; in two variables it consists, up to linear changes
of variables, of the Korteweg--de Vries polynomial and of the polynomials
$D_t(D_x^3+\lambda D_x+D_t)$, $\lambda\in\C$ \cite[Remark~4.2]{PartI}, of which $\lambda=0$ is
Ito's equation \cite{Ito1980}. For the weighted homogeneous sextics $C(\beta u^3+\gamma v)$ the
three factors of the condition \eqref{eq:sextic-condition} below correspond to Ito's equation
($\beta=0$), to $P_6$ ($\gamma=2\beta$) and to Ramani's equation \cite[(11)]{Ramani1981}
($\gamma=r\beta$): with $u=X$ and $v=(r-6)T$,
\[
(u^3-v)(u^3+rv)=X^6+5X^3T-5T^2 .
\]
For these sextics \eqref{eq:sextic-condition} is the limiting case of the condition
\cite[(4.8)]{NewellZeng1986} of Newell and Zeng for the polynomials
$D_{t_1}D_{t_5}+aD_{t_3}^2+bD_{t_1}^3D_{t_3}+cD_{t_1}^6$ in which the first term disappears.
In the family $D_x^6+D_x^3D_t+kD_t^2$ of \cite[(3)]{Ramani1981} and \cite[(19)]{HietarintaI},
$P_6$ and Ramani's equation are the members $k=-2$ and $k=-\frac15$. For $k\neq0,\frac14$ the dispersion variety of this
family consists of the two cubic graphs $D_t=\rho_iD_x^3$, where $\rho_1+\rho_2=-1/k$ and
$\rho_1\rho_2=1/k$, and by Proposition~\ref{prop:sextic} the 3SC holds on the first of them if
and only if $\rho_1=-2\rho_2$ or $\rho_1^2+7\rho_1\rho_2+\rho_2^2=0$, that is, if and only if
$k=-2$ and $\rho_1=1$, or $k=-\frac15$. Thus the 3SC holds on both components for
$k=-\frac15$, on exactly one component for $k=-2$, and on no component otherwise; on the whole
dispersion variety it holds only for $k=-\frac15$ \cite{HietarintaI,Ramani1981}. The member
$k=-2$ is, after the change of sign of $t$, the equation $(2D_t^2+D_tD_x^3-D_x^6)f\cdot f=0$ of
\cite[Appendix~A]{Ramani1981}, that is, $P_6(D_x,-D_t)f\cdot f=0$. With
$w:=2\,\partial_x\log f$, its nonlinear form \cite[(A-1)]{Ramani1981} is obtained by applying the
operator $2\partial_t-\partial_x^3-9w_x\partial_x+3w_{xx}$ to the left-hand side of the potential
Korteweg--de Vries equation $w_t+w_{xxx}+3w_x^2=0$, which is the nonlinear form of
$P_4(D_x,-D_t)f\cdot f=0$; hence every solution of the latter solves the former, as stated in
\cite{Ramani1981}.
For three momenta $(a,-\frac12a^3)$, $(b,-\frac12b^3)$, $(c,-\frac12c^3)$ on the second
component $v=-\frac12u^3$ of $V(P_6)$ the value of $S_3[P_6]$ is
\[
-6561\,a^4b^4c^4\,(a^2-b^2)^2(a^2-c^2)^2(b^2-c^2)^2 ,
\]
and $S_3[P_6]$ does not vanish identically on the triples with momenta on both components
either. Finally, the substitution
$u=X$, $v=(r-6)\bigl(T-\tfrac\alpha5X\bigr)$ transforms the polynomial $C\,[\alpha u+u^3+rv]$
of family (c) into
\[
X^6+5X^3T-5T^2+aX^2+bXT,\qquad b=(8-r)\,\alpha,\qquad a=-\frac{b^2}{45},
\]
and $X^6+5X^3T-5T^2+aX^2+bXT$ is reducible if and only if $a=-b^2/45$, because its
discriminant with respect to $T$ is $X^2(45X^4+10bX^2+b^2+20a)$. Hence family (c) is, up to
linear changes of variables, the set of the reducible members of the family
$D_x^6+5D_x^3D_t-5D_t^2+aD_x^2+bD_xD_t$ of \cite[Section~IV.E, (I\,D2)]{HietarintaI}, which is
the reduction $D_5=a'D_1+b'D_3$ of the first equation $D_1^6-5D_1^3D_3-5D_3^2+9D_1D_5$ of the
BKP hierarchy \cite{DJKM1982,JimboMiwa1983}; Ramani's equation itself is the reduction $D_5=0$
\cite[Sections~1 and~7]{DJKM1982}. For the irreducible members of this family see
Theorem~\ref{thm:rigidity} and Proposition~\ref{prop:NSC-BD}. The identities of this remark are elementary; they
are also verified by the script of Group~VIII (Appendix~\ref{app:certificates}).
\end{remark}

\subsection{Coordinates, Laurent polynomials and the covariance}\label{subsec:laurent}
We write $c$ for the polynomial $C=u^3-v$ when it is used as a coordinate, so that $(u,c)$ are
coordinates on the $(u,v)$-plane with $\wt c=3$, and the reflection $(u,v)\mapsto(-u,-v)$ is
$(u,c)\mapsto(-u,-c)$. Let
\[
\LL:=\C[u,c,c^{-1}]\supset\C[u,c]=\C[u,v].
\]
An element $F\in\LL$ is \emph{odd} (\emph{even}) if $F(-u,-c)=-F(u,c)$ (respectively
$=F(u,c)$); it is \emph{weighted homogeneous of weight $w$} if it is a combination of monomials
$u^ic^j$ with $i+3j=w$ ($i\ge0$, $j\in\Z$); its \emph{pole order} is the largest $b\ge0$ such
that $c^{-b}$ occurs in $F$. A weighted homogeneous element is odd if and only if its weight is
odd. For $z=(z_1,z_2,z_3)\in\C^3$ put
\[
\Sigma(z):=\bp(z_1)+\bp(z_2)+\bp(z_3),\qquad d_{ab}(z):=\bp(z_a)-\bp(z_b)\quad(1\le a<b\le3).
\]
Since $(z_1+z_2+z_3)^3-z_1^3-z_2^3-z_3^3=3(z_1+z_2)(z_1+z_3)(z_2+z_3)$ and
$(z_a-z_b)^3-(z_a^3-z_b^3)=-3z_az_b(z_a-z_b)$, these points are, in the coordinates $(u,c)$,
\begin{equation}\label{eq:points}
\Sigma(z)=\bigl(z_1+z_2+z_3,\ 3(z_1+z_2)(z_1+z_3)(z_2+z_3)\bigr),\qquad
d_{ab}(z)=\bigl(z_a-z_b,\ -3z_az_b(z_a-z_b)\bigr).
\end{equation}
For $F\in\LL$ and $k=(k_1,k_2,k_3)$ we define the rational function
\begin{equation}\label{eq:cS}
\cS[F](k):=\sum_{\sg\in\{\pm1\}^3}F(\Sigma(\sg k))\,F(d_{12}(\sg k))\,F(d_{13}(\sg k))\,F(d_{23}(\sg k)),
\end{equation}
where $\sg k:=(\sg_1k_1,\sg_2k_2,\sg_3k_3)$.
Since $\sg_a\bp(k_a)=\bp(\sg_ak_a)$, comparison with \eqref{eq:S3} shows that for every
polynomial $P\in\C[u,v]$
\[
\cS[P](k)=S_3[P]\bigl(\bp(k_1),\bp(k_2),\bp(k_3)\bigr).
\]
Hence, for even $P\in\C[u,v]$,
\begin{equation}\label{eq:3SC-cubic}
P\ \text{satisfies the 3SC on }V(C)\iff\cS[P]=0 .
\end{equation}
If $F$ is weighted homogeneous of weight $w$, its values at the points \eqref{eq:points} are
homogeneous of degree $w$ in $z$. Therefore, if $F=\sum_wF_w$ is the decomposition of
$F\in\LL$ into weighted homogeneous parts, $\cS[F]$ is a finite sum of rational functions
which are homogeneous in $k$ of different degrees, and each of them vanishes if $\cS[F]=0$; in
particular the top weighted part $F^{\mathrm{top}}$ of $F$ then satisfies $\cS[F^{\mathrm{top}}]=0$.

\begin{lemma}[Covariance]\label{lem:cov}
For all $F\in\LL$ and all $m\in\Z$,
\begin{equation}\label{eq:M}
\cS[c^mF]=M^m\,\cS[F],\qquad M(k):=-81\,k_1^2k_2^2k_3^2\,(k_1^2-k_2^2)(k_1^2-k_3^2)(k_2^2-k_3^2).
\end{equation}
More generally, if $p_1,p_2,p_3\in V(C)\subset\C^s$ have the $u$-coordinates $k_1,k_2,k_3$, then
for every polynomial $F$ on $\C^s$ and every integer $m\ge0$
\[
S_3[C^mF](p_1,p_2,p_3)=M(k)^m\,S_3[F](p_1,p_2,p_3).
\]
\end{lemma}
\begin{proof}
By \eqref{eq:points} the product of the values of $c$ at the four points
$\Sigma(z),d_{12}(z),d_{13}(z),d_{23}(z)$ equals
\[
3\prod_{a<b}(z_a+z_b)\cdot(-27)\prod_az_a^2\prod_{a<b}(z_a-z_b)=-81\prod_az_a^2\prod_{a<b}(z_a^2-z_b^2),
\]
which for $z=\sg k$ is $M(k)$, independently of $\sg$. Hence each summand of \eqref{eq:cS} for
$c^mF$ is $M^m$ times the corresponding summand for $F$. For the second statement note that
the $(u,v)$-coordinates of $\sg_ip_i$ are $\bp(\sg_ik_i)$ and that $C$ depends on $u,v$ only, so
the same computation applies to the summands of \eqref{eq:S3}.
\end{proof}

\begin{lemma}[Divisibility and parity of the multiplicity]\label{lem:divis}
Let $Q\neq0$ be an even polynomial on $\C^s$ such that $S_3[Q]$ vanishes on $V(C)^3$. Then $C$
divides $Q$, and the multiplicity of $C$ in $Q$ is odd.
\end{lemma}
\begin{proof}
The origin lies on $V(C)$, and for $p_1,p_2\in V(C)$ evenness gives
\[
S_3[Q](p_1,p_2,0)=8\,Q(p_1)\,Q(p_2)\,Q(p_1+p_2)\,Q(p_1-p_2).
\]
By Lemma~\ref{lem:dominance} the functions $Q(p_1\pm p_2)$ do not vanish identically on
$V(C)^2$. Since the coordinate ring of $V(C)^2$ is a domain, $Q$ vanishes on $V(C)$, that is,
$C\mid Q$. Let $m$ be the multiplicity of $C$ in $Q$. By Lemma~\ref{lem:cov},
$S_3[Q/C^m]=M^{-m}S_3[Q]=0$ on the dense subset of $V(C)^3$ where $M\neq0$, hence on $V(C)^3$.
If $m$ were even, $Q/C^m$ would be a nonzero even polynomial satisfying the hypothesis of the
lemma and not divisible by $C$, a contradiction.
\end{proof}

Consequently, in two variables the polynomials $P$ of Theorem~\ref{thm:cubic} are exactly the
products $P=cH$ with
\begin{equation}\label{eq:Hcond}
H\in\C[u,c]\ \text{odd},\qquad H(u,0)\not\equiv0,\qquad\cS[H]=0 ;
\end{equation}
here $H(u,0)$ is the restriction of $H$ to $V(C)$, and $\cS[H]=M^{-1}\cS[P]$ by \eqref{eq:M}.

\subsection{Sufficiency}\label{subsec:sufficiency}
The $N$-soliton conditions for the polynomials of Theorem~\ref{thm:cubic} are obtained by a
divisibility argument and a comparison of degrees. This is Hirota's proof for the
Korteweg--de Vries equation \cite{Hirota1971}, which Newell and Zeng
\cite[Section~III]{NewellZeng1986} extended to weighted homogeneous polynomials with a
prescribed phase factor; see Section~\ref{subsec:known}. We give the argument in full, because
for $\alpha\neq0$ the polynomials and the phase factors below are not weighted homogeneous.

\begin{proposition}[The sextic family]\label{prop:sextic}
Let $P=C\,(\alpha u+\beta u^3+\gamma v)$ with $(\alpha,\beta,\gamma)\neq(0,0,0)$. Then $P$
satisfies the \textup{3SC} on $V(C)$ if and only if
\begin{equation}\label{eq:sextic-condition}
\beta\,(2\beta-\gamma)\,(\beta^2-7\beta\gamma+\gamma^2)=0 .
\end{equation}
If \eqref{eq:sextic-condition} holds, then $C$ is a simple factor of $P$ and $P$ satisfies the
$N$\textup{SC} on $V(C)$ for every $N\ge3$.
\end{proposition}
\begin{proof}
Put
\[
R(s,t):=(s-t)^2\bigl[\alpha+(\beta+\gamma)(s^2+t^2)+(\gamma-2\beta)st\bigr].
\]
Since $C(\bp(s)-\bp(t))=-3st(s-t)$ and
$\alpha(s-t)+\beta(s-t)^3+\gamma(s^3-t^3)=(s-t)[\alpha+(\beta+\gamma)(s^2+t^2)+(\gamma-2\beta)st]$,
\begin{equation}\label{eq:sextic-pair}
P(\bp(s)-\bp(t))=-3st\,R(s,t),
\end{equation}
so that the phase factor \eqref{eq:phase} of $P$ on the cubic graph is
\begin{equation}\label{eq:deformedphase}
A(\bp(s),\bp(t))=\frac{R(s,t)}{R(s,-t)}=\Bigl(\frac{s-t}{s+t}\Bigr)^2
\frac{\alpha+(\beta+\gamma)(s^2+t^2)+(\gamma-2\beta)st}{\alpha+(\beta+\gamma)(s^2+t^2)-(\gamma-2\beta)st}.
\end{equation}
For $N\ge0$ define the reduced residual
\begin{equation}\label{eq:FN}
F_N(k_1,\dots,k_N):=\sum_{\sg\in\{\pm1\}^N}P\Bigl(\sum_i\bp(\sg_ik_i)\Bigr)\prod_{i<j}R(\sg_ik_i,\sg_jk_j).
\end{equation}
With $z_i=\sg_ik_i$, \eqref{eq:sextic-pair} gives
$\sg_i\sg_jP(\sg_i\bp(k_i)-\sg_j\bp(k_j))=-3k_ik_jR(z_i,z_j)$, hence
\begin{equation}\label{eq:rawFN}
S_N[P](\bp(k_1),\dots,\bp(k_N))=(-3)^{N(N-1)/2}\Bigl(\prod_ik_i\Bigr)^{N-1}F_N(k_1,\dots,k_N).
\end{equation}
The polynomial $F_N$ is symmetric in the $k_i$ and even in each $k_i$; note that
$R(s,t)=R(t,s)=R(-s,-t)$. We have $F_0=P(0)=0$, $F_1=2P(\bp(k_1))=0$, and $F_2=0$ because $S_2$
vanishes identically for even $P$. Setting one argument equal to zero, respectively two
arguments equal, we find the reductions
\begin{align}
F_N(k_1,\dots,k_{N-1},0)&=2\prod_{i<N}k_i^2\bigl(\alpha+(\beta+\gamma)k_i^2\bigr)\cdot F_{N-1}(k_1,\dots,k_{N-1}),\label{eq:zeroF}\\
F_N(k_1,\dots,k_{N-2},s,s)&=2R(s,-s)\prod_{i\le N-2}R(s,k_i)R(-s,k_i)\cdot F_{N-2}(k_1,\dots,k_{N-2}).\label{eq:collisionF}
\end{align}
In \eqref{eq:zeroF} the factor $R(z_i,0)=k_i^2(\alpha+(\beta+\gamma)k_i^2)$ does not depend on
the signs, and the two values of $\sg_N$ give the same term. In \eqref{eq:collisionF} the terms
with $\sg_{N-1}=\sg_N$ vanish because $R(s,s)=0$; for $\sg_{N-1}=-\sg_N$ the last two momenta
cancel in the sum, the factor $R(\sg_{N-1}s,-\sg_{N-1}s)=R(s,-s)=4s^2[\alpha+(4\beta+\gamma)s^2]$
and the products $R(z_i,s)R(z_i,-s)=R(k_i,s)R(k_i,-s)$ do not depend on the signs, and the two
choices of $\sg_{N-1}$ contribute equally.

Suppose that $F_{N-1}=F_{N-2}=0$ for some $N\ge2$. Regard $F_N$ as a polynomial in
$x_i:=k_i^2$. By \eqref{eq:zeroF} and symmetry, each $x_i$ divides $F_N$. By
\eqref{eq:collisionF}, $F_N$ vanishes on $x_{N-1}=x_N$, so $x_{N-1}-x_N$ divides $F_N$; the
quotient is antisymmetric under the exchange of $x_{N-1}$ and $x_N$, hence again divisible by
$x_{N-1}-x_N$. By symmetry,
\begin{equation}\label{eq:divisor}
\mathcal D_N:=\prod_ik_i^2\prod_{i<j}(k_i^2-k_j^2)^2\quad\text{divides}\quad F_N .
\end{equation}
The degree of $\mathcal D_N$ in $k$ is $2N^2$, while the degree of $F_N$ is at most
$6+2N(N-1)$: the argument of $P$ in \eqref{eq:FN} has weight at most $6$ and each $R$ has
degree at most $4$.

For $N=3$ both numbers are $18$. Since $F_1=F_2=0$, we get $F_3=\varkappa\,\mathcal D_3$ with
$\varkappa$ independent of $k$; $\varkappa$ is a polynomial in $(\alpha,\beta,\gamma)$, and it
does not involve $\alpha$ because $\alpha$ does not enter the part of $F_3$ of degree $18$ in
$k$. Evaluating $F_3$ and $\mathcal D_3$ at $k=(1,2,3)$ determines $\varkappa$:
\begin{equation}\label{eq:F3factor}
F_3=-72\,\beta\,(2\beta-\gamma)\,(\beta^2-7\beta\gamma+\gamma^2)\,
\prod_ik_i^2\prod_{i<j}(k_i^2-k_j^2)^2 .
\end{equation}
(The evaluation is an exact computation with symbolic $\alpha,\beta,\gamma$; see
Appendix~\ref{app:certificates}. It is not a guess from samples: the divisibility and the
degree bound prove that there is a single unknown coefficient, and one evaluation at a point
where $\mathcal D_3\neq0$ determines it.) By \eqref{eq:rawFN}, $P$ satisfies the 3SC if and only
if \eqref{eq:sextic-condition} holds.

Assume \eqref{eq:sextic-condition}. Then $F_2=F_3=0$, and for $N\ge4$ we have
$6+2N(N-1)<2N^2$, so \eqref{eq:divisor} and induction on $N$ give $F_N=0$ for all $N$, that
is, all $N$SC. Finally, the restriction of the cofactor to $V(C)$ is
$\alpha u+(\beta+\gamma)u^3$. If it vanished identically, then $\alpha=0$ and $\gamma=-\beta$,
and \eqref{eq:sextic-condition} would read $27\beta^4=0$, so that $(\alpha,\beta,\gamma)=0$.
Hence $C$ is a simple factor.
\end{proof}

The three factors of \eqref{eq:sextic-condition} give family (b) of Theorem~\ref{thm:cubic}
($\beta=0$), the plane $\Span(P_4,P_6)$ ($\gamma=2\beta$) and family (c) ($\gamma=r\beta$).

\begin{proposition}[Sufficiency]\label{prop:sufficiency}
Every polynomial $P_0$ of Theorem~\ref{thm:cubic-all}\,\textup{(a), (b), (c)} is even, has $C$
as a simple factor, and satisfies the $N$\textup{SC} on $V(C)$ for every $N\ge3$.
\end{proposition}
\begin{proof}
For family (a) this is \cite[Theorem~9.19(a)]{PartI}: the five polynomials \eqref{eq:Kbasis}
have the common phase factor $((s-t)/(s+t))^2$ on the cubic graph, and the corresponding
reduced residuals vanish by the divisibility argument used above, with the degree bound
$12+N(N-1)<2N^2$ for $N\ge4$ and one evaluation for $N=3$. The restrictions of the cofactors
$P_j/C$ to $V(C)$ are $u$, $3u^3$, $-3u^5$, $-9u^7$, $-135u^9$, so $C$ is a simple factor of
every nonzero element of the span. For family (b), choose coordinates in which $u$ is the
momentum of $D_x$ and $v=-\Lam$; then $C\cdot L=L\,(D_x^3+\Lam)$ and $V(C)$ is the principal
component, so the statement is \cite[Theorem~4.1]{PartI}; $L$ does not vanish identically on
$V(C)$ because $C\nmid L$. Family (c) is contained in Proposition~\ref{prop:sextic}; since
$S_N[P_0]$ involves only the coordinates $u,v$ of the momenta when $P_0\in\C[u,v]$, the
statements in two variables imply those in $s$ variables.
\end{proof}

\subsection{The doubled graph}\label{subsec:doubled}
The dilated curve $2V(C)$ is the graph $V(u^3-4v)$. The next lemma bounds its multiplicity; it
is the only place where Section~\ref{sec:collisions} is used. For a polynomial $G$ on the plane
and $s,t\in\C$ we write $G^\pm:=G(\bp(s)\pm\bp(t))$, and $G_u,G_{uu}$ for the partial
derivatives with respect to $u$ in the coordinates $(u,v)$.

\begin{lemma}\label{lem:nu2}
Let $H\in\C[u,v]$ be odd with $H|_{V(C)}\not\equiv0$ and $\cS[H]=0$, and put
$\vartheta(t):=H_u(\bp(t))/H(\bp(t))$, an odd rational function. Then:
\begin{enumerate}
\item[\textup{(i)}] for generic $s,t$,
\begin{multline}\label{eq:diamond}
H^-H^+_{uu}-H^+H^-_{uu}-2\bigl(\vartheta(s)+\vartheta(t)\bigr)H^-H^+_u\\
+2\bigl(\vartheta(s)-\vartheta(t)\bigr)H^+H^-_u+4\,\vartheta(s)\vartheta(t)\,H^+H^-=0;
\end{multline}
\item[\textup{(ii)}] the multiplicity $\nu_2$ of $u^3-4v$ in $H$ is at most $1$;
\item[\textup{(iii)}] $\ord_0H=1+2\nu_2\in\{1,3\}$.
\end{enumerate}
\end{lemma}
\begin{proof}
(i) Let $h(t):=H(\bp(t))$, $h_1(t):=H_u(\bp(t))$, $h_2(t):=H_{uu}(\bp(t))$; $h$ and $h_2$ are
odd, $h_1$ is even, and $h\not\equiv0$. For a point $q$ of the plane,
$\frac{d}{d\eta}H(q\pm\bp(\eta))=\pm H_u\pm3\eta^2H_v$, so the derivative with respect to $v$
enters the Taylor expansion in $\eta$ only from the third order on:
\begin{align*}
H(\bp(z_1)+\bp(z_2)+\bp(\eta))&=H^++\eta H^+_u+\tfrac12\eta^2H^+_{uu}+O(\eta^3),\\
H(\bp(z_i)-\bp(\eta))&=h(z_i)-\eta h_1(z_i)+\tfrac12\eta^2h_2(z_i)+O(\eta^3)\qquad(i=1,2),
\end{align*}
where $H^+$ and its derivatives are taken at $\bp(z_1)+\bp(z_2)$. The summand of
$\cS[H](z_1,z_2,\eta)$ in \eqref{eq:cS} is the product of these three expansions with
$H(\bp(z_1)-\bp(z_2))$, which does not depend on $\eta$. Now take $k=(s,t,\eps)$. The sum over
$\sg_3=\pm1$, that is, over $\eta=\pm\eps$, doubles the even powers of $\eps$; the sum over all
sign vectors is twice the sum over $(z_1,z_2)\in\{(s,t),(s,-t)\}$; and replacing $t$ by $-t$
exchanges $H^+$ and $H^-$ and changes the signs of $h(t)$ and $h_2(t)$ but not of $h_1(t)$.
Collecting terms, one finds that the coefficient of $\eps^0$ of $\cS[H](s,t,\eps)$ vanishes
identically and that the terms containing $h_2$ cancel in the coefficient of $\eps^2$, which
equals $2h(s)h(t)$ times the left-hand side of \eqref{eq:diamond}. Since $\cS[H]=0$, this proves
(i).

(ii) Put $x:=s+t$, $y:=s-t$, $X:=x^2$, $Y:=y^2$. Then
\begin{equation}\label{eq:sumdiff}
\bp(s)+\bp(t)=\bigl(x,\tfrac14x(X+3Y)\bigr),\qquad\bp(s)-\bp(t)=\bigl(y,\tfrac14y(Y+3X)\bigr).
\end{equation}
Since $H$ is odd, there is a polynomial $\widetilde G\in\C[X,Y]$ with
$H^+=x\,\widetilde G(X,Y)$, and then $H^-=y\,\widetilde K(X,Y)$ with
$\widetilde K(X,Y):=\widetilde G(Y,X)$. For a polynomial $F(u,v)$ and
$f(x,y):=F\bigl(x,\tfrac14x(x^2+3y^2)\bigr)$ the chain rule gives
$f_x=F_u+\tfrac34(x^2+y^2)F_v$ and $f_y=\tfrac32xyF_v$, hence
\[
F_u\bigl(x,\tfrac14x(x^2+3y^2)\bigr)=x^{-1}\,\mathsf Lf,\qquad
\mathsf L:=2X\partial_X-(X+Y)\partial_Y ,
\]
where $x\partial_x=2X\partial_X$ and $y\partial_y=2Y\partial_Y$. Applying this to $F=H$ and to
$F=H_u$, and using $\mathsf L(x\widetilde G)=x(\mathsf L+1)\widetilde G$, we get
\[
H^+_u=(\mathsf L+1)\widetilde G,\qquad H^+_{uu}=x^{-1}\,\mathsf L(\mathsf L+1)\widetilde G,
\]
and, exchanging the roles of $x$ and $y$,
\[
H^-_u=(\mathsf J+1)\widetilde K,\qquad H^-_{uu}=y^{-1}\,\mathsf J(\mathsf J+1)\widetilde K,\qquad
\mathsf J:=2Y\partial_Y-(X+Y)\partial_X .
\]
Dividing \eqref{eq:diamond} by $xy$ we obtain
\begin{equation}\label{eq:XYform}
\widetilde KX^{-1}\mathsf L(\mathsf L+1)\widetilde G-\widetilde GY^{-1}\mathsf J(\mathsf J+1)\widetilde K
-2\hat\beta\,\widetilde K(\mathsf L+1)\widetilde G+2\hat\gamma\,\widetilde G(\mathsf J+1)\widetilde K
+4\hat\alpha\,\widetilde G\widetilde K=0,
\end{equation}
where, with $s=\tfrac12(x+y)$ and $t=\tfrac12(x-y)$,
\[
\hat\alpha:=\vartheta(s)\vartheta(t),\qquad\hat\beta:=\frac{\vartheta(s)+\vartheta(t)}{x},\qquad
\hat\gamma:=\frac{\vartheta(s)-\vartheta(t)}{y}.
\]
Because $\vartheta$ is odd, these three functions are invariant under $x\mapsto-x$ and under
$y\mapsto-y$. Fix a generic $y$. Then $\vartheta$ is regular at $\pm y/2$, the numerator of
$\hat\beta$ vanishes at $x=0$, and $\hat\alpha,\hat\beta,\hat\gamma$ are analytic and even in $x$
near $x=0$, hence power series in $X$.

Expand in powers of $X$:
$\widetilde G=g(Y)X^\rho+O(X^{\rho+1})$ and $\widetilde K=k(Y)X^q+O(X^{q+1})$ with $g,k\neq0$.
We claim that $q=\nu_2$. Indeed, at the point $\bp(s)-\bp(t)$ of \eqref{eq:sumdiff} we have
$u^3-4v=-3yX$; writing $H=(u^3-4v)^{\nu_2}H'$ with $u^3-4v\nmid H'$ we get
$y\widetilde K=(-3yX)^{\nu_2}H'\bigl(y,\tfrac14y(Y+3X)\bigr)$, and $H'(y,\tfrac14y^3)\not\equiv0$
because $y\mapsto(y,\tfrac14y^3)$ parametrizes $V(u^3-4v)$.

On monomials, $\mathsf L(X^iY^j)=(2i-j)X^iY^j-jX^{i+1}Y^{j-1}$ and
$\mathsf J(X^iY^j)=(2j-i)X^iY^j-iX^{i-1}Y^{j+1}$. Thus $\mathsf L$ does not lower the order in
$X$, while
\[
(\mathsf J+1)\widetilde K=-q\,Yk(Y)\,X^{q-1}+O(X^q),\qquad
\mathsf J(\mathsf J+1)\widetilde K=q(q-1)\,Y^2k(Y)\,X^{q-2}+O(X^{q-1}).
\]
Suppose that $q\ge2$. In \eqref{eq:XYform} the first and the fourth term have order at least
$\rho+q-1$ in $X$, the third and the fifth at least $\rho+q$, and the second term is
$-q(q-1)\,Y\,g(Y)k(Y)\,X^{\rho+q-2}+O(X^{\rho+q-1})$. Its leading coefficient cannot be
cancelled, a contradiction. Hence $\nu_2=q\le1$.

(iii) The polynomial $P:=CH$ is even with simple factor $C$ and satisfies the 3SC on $V(C)$,
which is not a hyperplane. Here $C_2=\tfrac18(u^3-4v)$ and $C_3=\tfrac1{27}(u^3-9v)$, so the
multiplicity of $2V(C)$ in $P$ is $\nu_2$, and $C$, $C_2$, $C_3$ are pairwise non-associated
irreducible polynomials which vanish to first order at the origin. By
Theorem~\ref{thm:collision}, $d:=\ord_0P$ satisfies $d-2\nu_2\in\{0,2\}$, and if $d=2\nu_2$,
then $C_3$ is a simple factor of $P$. In that case $C\,C_2^{\nu_2}C_3$ divides $P$, so
$\nu_2+2\le d=2\nu_2$, that is, $\nu_2\ge2$, contradicting (ii). Hence $\ord_0P=2\nu_2+2$ and $\ord_0H=2\nu_2+1$.
\end{proof}

Lemma~\ref{lem:nu2} is a special case of the results of Section~\ref{sec:soft}: for $P=CH$,
part (ii) is Theorem~\ref{thm:soft}(c), part (iii) is Corollary~\ref{cor:origin}(1) with
$\mu=1$, and \eqref{eq:diamond} corresponds to \eqref{eq:softTheta} (Remark~\ref{rem:soft}(iii)).
We have kept the direct proof, which uses only the covariance and
Theorem~\ref{thm:collision}, so that the classification on the cubic graph does not depend on
Section~\ref{sec:soft}.

\subsection{Weighted homogeneous solutions}\label{subsec:homog}
Table~\ref{tab:types} lists seven types of odd weighted homogeneous polynomials $T$; the type
$T_r$ has two members, one for each root $r$ of $r^2-7r+1=0$. The products $cT$
are $P_4$, $P_6$, $P_8$, $P_{10}$, $P_{12}$, $vC$ and $C(u^3+rv)$. For $F\in\LL$ put
\[
\Theta_F(a,b):=F(\bp(a)+\bp(b))=F\bigl(a+b,\,3ab(a+b)\bigr),\qquad\text{so that}\quad
F(d_{ab}(z))=\Theta_F(z_a,-z_b),
\]
where the second expression is in the coordinates $(u,c)$. Let $\nu(T)\in\{0,1\}$ be the
multiplicity of $u^3-4v$ in $T$. At the point $\bp(a)+\bp(b)$ one has $u=a+b$,
$u^3-4v=-3(a+b)(a-b)^2$, $u^3+2v=3(a+b)(a^2+b^2)$ and
$u^6+28u^3v+16v^2=9(a+b)^2(5a^4+6a^2b^2+5b^4)$, which gives, for all seven types,
\begin{equation}\label{eq:Theta}
\Theta_T(a,b)=(a+b)^{2\nu+1}(a-b)^{2\nu}\,G_T(a,b),\qquad G_T(a,a)\,G_T(a,-a)\not\equiv0 ,
\end{equation}
with $\nu=\nu(T)$ and the polynomials $G_T$ of the table; for $T_r$ one has $G_T(a,a)=(4+r)a^2$
and $G_T(a,-a)=3ra^2$, which are nonzero since $r\neq0,-4$.

\begin{table}[ht]
\centering\renewcommand{\arraystretch}{1.25}
\resizebox{\textwidth}{!}{%
\begin{tabular}{@{}llccl@{}}
\toprule
$T$ in the coordinates $(u,v)$ & in the coordinates $(u,c)$ & $\wt T$ & $\nu$ & $G_T(a,b)$\\
\midrule
$T_1=u$ & $u$ & $1$ & $0$ & $1$\\
$T_3=u^3+2v$ & $3u^3-2c$ & $3$ & $0$ & $3(a^2+b^2)$\\
$T_5=u^2(u^3-4v)$ & $-3u^5+4u^2c$ & $5$ & $1$ & $-3$\\
$T_7=u(u^3-4v)(u^3+2v)$ & $-9u^7+18u^4c-8uc^2$ & $7$ & $1$ & $-9(a^2+b^2)$\\
$T_9=(u^3-4v)(u^6+28u^3v+16v^2)$ & $-135u^9+360u^6c-288u^3c^2+64c^3$ & $9$ & $1$ & $-27(5a^4+6a^2b^2+5b^4)$\\
$\TI=v$ & $u^3-c$ & $3$ & $0$ & $a^2-ab+b^2$\\
$T_r=u^3+rv$ & $(1+r)u^3-rc$ & $3$ & $0$ & $(1+r)(a^2+b^2)+(2-r)ab$\\
\bottomrule
\end{tabular}}
\caption{The set $\cT=\{T_1,T_3,T_5,T_7,T_9,\TI,T_r\}$ of types; $r^2-7r+1=0$ (two
values).}\label{tab:types}
\end{table}

\begin{proposition}[Weighted homogeneous solutions; partly computer-assisted]\label{prop:homog}
Let $H$ be as in \eqref{eq:Hcond} and weighted homogeneous. Then $H=\lambda T$ with
$\lambda\neq0$ and $T\in\cT$. Equivalently: the weighted homogeneous even polynomials in
$\C[u,v]$ with simple factor $C$ which satisfy the \textup{3SC} on $V(C)$ are the nonzero
multiples of $P_4,P_6,P_8,P_{10},P_{12}$, $vC$ and $C(u^3+rv)$.
\end{proposition}
\begin{proof}
By Lemma~\ref{lem:nu2}(iii), $H$ contains a monomial $u^iv^j$ with $i+j\le3$, so its weight
$w=i+3j$ is at most $9$; being odd, $w\in\{1,3,5,7,9\}$. In the coordinates $(u,c)$,
\[
H=\sum_{j=0}^{\lfloor w/3\rfloor}b_j\,u^{w-3j}c^j,\qquad H(u,0)=b_0u^w,
\]
so $b_0\neq0$, and we normalize $b_0=1$. For $w=1$, $H=u=T_1$. For $w=3$, $H=\beta u^3+\gamma v$
in the coordinates $(u,v)$, and Proposition~\ref{prop:sextic} with $\alpha=0$ shows that $H$ is
a multiple of $v$, of $u^3+2v$ or of $u^3+rv$.

For $w\in\{5,7,9\}$ we use a finite computation. For each of the $33$ triples
\begin{equation}\label{eq:33}
k=(1,i,j),\qquad2\le i\le7,\quad i<j\le10,
\end{equation}
$\cS[H](k)$ is a polynomial of degree at most four in $b_1,\dots,b_{\lfloor w/3\rfloor}$ with
rational coefficients, and it has to vanish. A Gr\"obner basis over $\Q$ of the ideal generated
by these $33$ polynomials is
\[
\{3b_1+4\}\ (w=5),\qquad\{b_1+2,\ 9b_2-8\}\ (w=7),\qquad\{3b_1+8,\ 15b_2-32,\ 135b_3+64\}\ (w=9).
\]
Hence in each case there is at most one solution, namely $H=-\tfrac13T_5$, $-\tfrac19T_7$,
$-\tfrac1{135}T_9$; these are solutions by Proposition~\ref{prop:sufficiency}.
\end{proof}

The same computation for $w=3$ returns $(b_1+1)(3b_1+2)(9b_1^2+9b_1+1)$, in agreement with
Proposition~\ref{prop:sextic}, and for $w=11,13,15,17$ it returns the unit ideal, in agreement
with the bound $w\le9$; an independent computation in the coordinates $(u,v)$, which does not
assume that $C$ is a simple factor of $H\cdot C$, confirms the list for the weights
$8,\dots,16$ of $P=CH$ (Appendix~\ref{app:certificates}).

For $w=5$ the computation can be avoided. Here $H=u^5+b_1u^2c=(1+b_1)u^5-b_1u^2v$, and
$\ord_0H\in\{1,3\}$ forces $b_1\neq0$ and $\ord_0H=3$; then $\nu_2=1$ by
Lemma~\ref{lem:nu2}(iii), and $u^3-4v$ divides $(1+b_1)u^3-b_1v$ only for $b_1=-\tfrac43$. For
$w=7$ and $w=9$ the same two conditions --- $u^3-4v$ divides $H$, and the coefficient
$\pm b_{\lfloor w/3\rfloor}$ of the only monomial of ordinary degree three, $uv^2$ respectively
$v^3$, is not zero --- give the linear relations $16+12b_1+9b_2=0$ and
$64+48b_1+36b_2+27b_3=0$, which hold for the solutions above but do not determine them. A
second derivation of the three Gr\"obner bases, which uses no Gr\"obner basis algorithm
(rank computations on Macaulay matrices over $\Q$) and a disjoint set of sample triples, is
part of Group~IV of the ancillary files (Appendix~\ref{app:certificates}).

\begin{corollary}\label{cor:toppart}
Let $H$ be as in \eqref{eq:Hcond}. Then the top weighted part of $H$ is $\lambda\,c^{2n}T$ with
$\lambda\neq0$, $n\ge0$ and $T\in\cT$.
\end{corollary}
\begin{proof}
The top weighted part $H^{\mathrm{top}}$ of $H$ satisfies $\cS[H^{\mathrm{top}}]=0$, so
$Q:=cH^{\mathrm{top}}$ is a nonzero even polynomial with $\cS[Q]=0$. By Lemma~\ref{lem:divis}
the multiplicity of $c$ in $Q$ is odd, say $2n+1$, so $H^{\mathrm{top}}=c^{2n}H'$ with
$H'\in\C[u,c]$ odd, weighted homogeneous, $H'(u,0)\not\equiv0$, and $\cS[H']=0$ by
Lemma~\ref{lem:cov}. Proposition~\ref{prop:homog} gives $H'=\lambda T$.
\end{proof}

The difficulty in passing from weighted homogeneous to arbitrary polynomials is visible here:
although $C$ is a simple factor of $P=cH$, the top weighted part of $P$ may a priori be
divisible by a high power of $C$, and the bound of Lemma~\ref{lem:nu2}(iii) on the lowest
ordinary degree gives no bound on the top weight.

\subsection{Linearization and the equations of each level}\label{subsec:levels}
For $T,F\in\LL$ and $z\in\C^3$ we abbreviate $F_0:=F(\Sigma(z))$, $F_1:=F(d_{12}(z))$,
$F_2:=F(d_{13}(z))$, $F_3:=F(d_{23}(z))$ (the four \emph{slots} of \eqref{eq:cS}) and define
\begin{equation}\label{eq:LT}
\begin{aligned}
\cL_T[F](k)&:=\sum_{\sg}\sum_{m=0}^{3}\Bigl(F_m\prod_{l\neq m}T_l\Bigr)(\sg k),\\
\cQ_T[F](k)&:=\sum_{\sg}\sum_{0\le m<n\le3}\Bigl(F_mF_n\prod_{l\neq m,n}T_l\Bigr)(\sg k),
\end{aligned}
\end{equation}
so that $\cS[T+\tau F]=\cS[T]+\tau\,\cL_T[F]+\tau^2\cQ_T[F]+O(\tau^3)$.

\begin{lemma}[Level equations]\label{lem:levels}
Let $T\in\LL$ be weighted homogeneous of weight $w_0$, and let $R=T+\sum_{k=1}^{K}R_k$ with
$R_k\in\LL$ weighted homogeneous of weight $w_0-2k$. For $k\ge1$ put $B_k:=T+\sum_{j<k}R_j$.
Then the homogeneous component of degree $4w_0-2k$ of $\cS[R]$ is
\[
\cL_T[R_k]+\bigl(\cS[B_k]\bigr)_{4w_0-2k},
\]
where the subscript denotes the homogeneous component of that degree. In particular, if
$\cS[R]=0$ and $\cS[B_k]=0$, then $\cL_T[R_k]=0$.
\end{lemma}
\begin{proof}
Expand $\cS[R]$ multilinearly over the four slots, with $R_0:=T$. A term with
$R_{j_0},\dots,R_{j_3}$ in the slots is homogeneous of degree $4w_0-2\sum_mj_m$. For
$\sum_mj_m=k$, either one index equals $k$ and the others vanish, which gives $\cL_T[R_k]$, or
all indices are smaller than $k$, and these terms are exactly the component of degree
$4w_0-2k$ of $\cS[B_k]$.
\end{proof}

\subsection{The pole lemma}\label{subsec:pole}
The kernel of $\cL_T$ on Laurent polynomials is controlled by a double collision: at
$k_2=k_1-\eps$ the factor $c^{-b}$ in the slot $\Sigma$ acquires a pole in $\eps$ whose
coefficients are singular at $k_3=k_1$, and the zeros of $T$ on the diagonals compensate only a
bounded part of this singularity.

\begin{lemma}[Pole lemma]\label{lem:pole}
Let $T\in\cT$, $\nu=\nu(T)$, and let $R'\in\LL$ be odd with pole order $B\ge1$. Put
$k_2=k_1-\eps$ and consider Laurent expansions at $\eps=0$ for generic $k_1$ and $k_3$. Let
\[
\Lambda^{\mathrm{(i)}}:=\sum_{\sg:\ \sg_1=-\sg_2}\bigl(R'_0T_1T_2T_3\bigr)(\sg k),\qquad
\Lambda^{\mathrm{(ii)}}:=\sum_{\sg:\ \sg_1=\sg_2}\bigl(T_0R'_1T_2T_3\bigr)(\sg k)
\]
be the parts of $\cL_T[R'](k_1,k_1-\eps,k_3)$ with $R'$ in the slot $\Sigma$ and $\sg_1=-\sg_2$,
respectively with $R'$ in the slot $d_{12}$ and $\sg_1=\sg_2$.
\begin{enumerate}
\item[\textup{(a)}] $\cL_T[R'](k_1,k_1-\eps,k_3)-\Lambda^{\mathrm{(i)}}-\Lambda^{\mathrm{(ii)}}$ is
regular at $\eps=0$, and every Laurent coefficient of $\Lambda^{\mathrm{(ii)}}$ is a polynomial
in $k_3$.
\item[\textup{(b)}] Let $q\ge1$ be an integer with $B\ge q+4\nu+2$. Then the coefficient of
$\eps^{-q}$ of $\Lambda^{\mathrm{(i)}}$ is a rational function of $k_3$ with a pole of order
exactly $2B-q-6\nu-1$ at $k_3=k_1$.
\end{enumerate}
\end{lemma}

\begin{corollary}\label{cor:pole}
If $T\in\cT$ and $R'\in\LL$ is odd with $\cL_T[R']=0$, then the pole order of $R'$ is at most
$4\nu(T)+2$, that is, at most $2$ for $T_1,T_3,\TI,T_r$ and at most $6$ for $T_5,T_7,T_9$.
\end{corollary}
\begin{proof}
If $B\ge4\nu+3$, take $q=1$ in Lemma~\ref{lem:pole}: the coefficient of $\eps^{-1}$ of
$\cL_T[R'](k_1,k_1-\eps,k_3)$ is the sum of a rational function of $k_3$ with a pole of order
$2B-6\nu-2\ge1$ at $k_3=k_1$ and of a polynomial in $k_3$, so it is not zero.
\end{proof}

\begin{proof}[Proof of Lemma~\ref{lem:pole}]
Write $R'=\sum_{b\le B}\rho_b(u)\,c^{-b}$ with $\rho_b\in\C[u]$ and $\rho_B\neq0$, and take $k_1$
generic; in particular $k_1\rho_B(k_1)\neq0$.

(a) By \eqref{eq:points}, for $z=\sg k$ the value of $c$ at $\Sigma(z)$ vanishes at $\eps=0$
only if $\sg_1=-\sg_2$ (then $z_1+z_2=\sg_1\eps$), the value of $c$ at $d_{12}(z)$ vanishes
only if $\sg_1=\sg_2$ (then $z_1-z_2=\sg_1\eps$), and the values of $c$ at $d_{13}(z)$,
$d_{23}(z)$ do not vanish at $\eps=0$ for generic $k_3$. Since $T$ is a polynomial, all other
terms of \eqref{eq:LT} are regular at $\eps=0$. In $\Lambda^{\mathrm{(ii)}}$ the factor
$R'_1=R'(d_{12}(z))$ with $z_1-z_2=\pm\eps$, $z_1z_2=k_1(k_1-\eps)$ does not depend on $k_3$,
and $T_0T_2T_3$ is a polynomial in $k_1,\eps,k_3$; hence all Laurent coefficients of
$\Lambda^{\mathrm{(ii)}}$ are polynomials in $k_3$.

(b) All four factors of a summand are odd, so the summands for $\sg$ and $-\sg$ coincide and
$\Lambda^{\mathrm{(i)}}$ is twice the sum of the terms for $\sg=(1,-1,1)$ and $\sg=(1,-1,-1)$.
Put $y:=k_3-k_1$.

\emph{The sign vector $(1,-1,1)$.} Here $z=(k_1,\eps-k_1,k_3)$. By \eqref{eq:points} the slot
$\Sigma$ is the point $\bigl(k_3+\eps,\ 3\eps(k_1+k_3)(y+\eps)\bigr)$, and by \eqref{eq:Theta}
\begin{align*}
T_1&=\Theta_T(k_1,k_1-\eps)=(2k_1-\eps)^{2\nu+1}\eps^{2\nu}\,G_T(k_1,k_1-\eps),\\
T_2&=\Theta_T(k_1,-k_3)=(-y)^{2\nu+1}(k_1+k_3)^{2\nu}\,G_T(k_1,-k_3),\\
T_3&=\Theta_T(\eps-k_1,-k_3)=(\eps-k_1-k_3)^{2\nu+1}(y+\eps)^{2\nu}\,G_T(\eps-k_1,-k_3).
\end{align*}
Therefore
\begin{equation}\label{eq:plus}
R'_0T_1T_2T_3=\sum_{b\le B}3^{-b}\,\eps^{2\nu-b}\,(-y)^{2\nu+1}\,(y+\eps)^{2\nu-b}\,\rho_b(k_3+\eps)\,g_b(\eps,y),
\end{equation}
where
$g_b:=(k_1+k_3)^{2\nu-b}(2k_1-\eps)^{2\nu+1}(\eps-k_1-k_3)^{2\nu+1}
G_T(k_1,k_1-\eps)\,G_T(k_1,-k_3)\,G_T(\eps-k_1,-k_3)$ is analytic near $(\eps,y)=(0,0)$, with
\[
g_b(0,0)=-(2k_1)^{6\nu+2-b}\,G_T(k_1,k_1)\,G_T(k_1,-k_1)\,G_T(-k_1,-k_1)\neq0
\]
by \eqref{eq:Theta}; note that $G_T(-a,-b)=G_T(a,b)$ because $\Theta_T$ is odd. We expand
\eqref{eq:plus} for $0<|\eps|<|y|$, using
$(y+\eps)^{2\nu-b}=\sum_{l\ge0}\binom{2\nu-b}{l}\eps^ly^{2\nu-b-l}$. The terms with $b\le2\nu$
have no pole at $\eps=0$. For $b>2\nu$, let $a_{b,m}(y)$ be the Taylor coefficients in $\eps$ of
$\rho_b(k_3+\eps)g_b(\eps,y)$; they are analytic at $y=0$ and
$a_{b,0}(0)=\rho_b(k_1)g_b(0,0)$. The coefficient of $\eps^{-q}$ of the $b$-th term of
\eqref{eq:plus} is
\[
3^{-b}(-y)^{2\nu+1}\sum_{l+m=b-2\nu-q}\binom{2\nu-b}{l}\,y^{2\nu-b-l}\,a_{b,m}(y);
\]
it vanishes if $b<2\nu+q$, and otherwise its most singular part at $y=0$ is the term with
$l=b-2\nu-q$, $m=0$, of order
\[
(b-2\nu)+(b-2\nu-q)-(2\nu+1)=2b-q-6\nu-1
\]
in $1/y$, with coefficient $-3^{-b}\binom{2\nu-b}{b-2\nu-q}\rho_b(k_1)g_b(0,0)$. The binomial
coefficient is nonzero because $2\nu-b<0$. The order is strictly increasing in $b$, and
$B\ge2\nu+q$, $\rho_B(k_1)\neq0$; hence the contribution of $\sg=(1,-1,1)$ to the coefficient of
$\eps^{-q}$ has a pole of order exactly $2B-q-6\nu-1$ at $y=0$. This number is positive, in
fact at least $q+2\nu+3$, by the hypothesis $B\ge q+4\nu+2$.

\emph{The sign vector $(1,-1,-1)$.} Here $z=(k_1,\eps-k_1,-k_3)$, the slot $\Sigma$ is the point
$\bigl(\eps-k_3,\ -3\eps\,y\,(\eps-k_1-k_3)\bigr)$, $T_1$ is as before, and
\begin{align*}
T_2&=\Theta_T(k_1,k_3)=(k_1+k_3)^{2\nu+1}(-y)^{2\nu}\,G_T(k_1,k_3),\\
T_3&=\Theta_T(\eps-k_1,k_3)=(y+\eps)^{2\nu+1}(\eps-k_1-k_3)^{2\nu}\,G_T(\eps-k_1,k_3).
\end{align*}
Hence
$R'_0T_1T_2T_3=\sum_{b\le B}(-3)^{-b}\eps^{2\nu-b}\,y^{2\nu-b}\,(y+\eps)^{2\nu+1}\rho_b(\eps-k_3)\,g'_b(\eps,y)$
with $g'_b$ analytic near $(0,0)$. The Taylor coefficients in $\eps$ of
$(y+\eps)^{2\nu+1}\rho_b(\eps-k_3)g'_b$ are analytic at $y=0$, so every Laurent coefficient in
$\eps$ of this expression has a pole of order at most $B-2\nu$ at $y=0$.

\emph{Conclusion.} The inequality $2B-q-6\nu-1>B-2\nu$ is equivalent to $B\ge q+4\nu+2$. Under
this hypothesis the pole of order $2B-q-6\nu-1$ produced by $\sg=(1,-1,1)$ is not cancelled by
the contribution of $\sg=(1,-1,-1)$.
\end{proof}

\subsection{The kernels}\label{subsec:kernel}

\begin{proposition}[Laurent kernels; computer-assisted]\label{prop:kernel}
Let $T\in\cT$ and let $R'\in\LL$ be odd and weighted homogeneous of weight $w'<\wt T$, with
$\cL_T[R']=0$. Then:
\begin{enumerate}
\item[\textup{(i)}] if $T=T_1$: $R'\in\C\,c^{-1}$ (so $R'=0$ unless $w'=-3$);
\item[\textup{(ii)}] if $T\in\{T_3,\TI,T_r\}$: $R'\in\C\,u$ (so $R'=0$ unless $w'=1$);
\item[\textup{(iii)}] if $T=T_5$: $R'\in\C\,T_3$ for $w'=3$, $R'\in\C\,u$ for $w'=1$, and $R'=0$
otherwise;
\item[\textup{(iv)}] if $T=T_7$: $R'\in\C\,T_5$, $\C\,T_3$, $\C\,u$ for $w'=5,3,1$, and $R'=0$
otherwise;
\item[\textup{(v)}] if $T=T_9$: $R'\in\C\,T_7$, $\C\,T_5$, $\C\,T_3$, $\C\,u$ for $w'=7,5,3,1$,
and $R'=0$ otherwise.
\end{enumerate}
\end{proposition}
\begin{proof}
By Corollary~\ref{cor:pole} the pole order of $R'$ is at most $4\nu+2\le6$. For a given weight
$w'$, the odd weighted homogeneous elements of $\LL$ of pole order at most $8$ form the space
with basis $u^ic^j$, $i\ge0$, $j\ge-8$, $i+3j=w'$; it has dimension at most $11$ for $w'\le7$,
and it is zero for $w'<-24$. For each $T\in\cT$ and each odd $w'$ with $-24\le w'<\wt T$ the
values of $\cL_T$ on this basis were computed in exact arithmetic --- rational arithmetic, and
arithmetic in $\Q[r]/(r^2-7r+1)$ for $T_r$, which covers both roots --- at $2\dim+4$ rational
points $k$. Each point gives a linear condition satisfied by every element of the kernel of
$\cL_T$; hence this kernel is contained in the kernel of the resulting matrix, and the latter
is the space stated (Appendix~\ref{app:certificates}). That the elements listed do lie in the
kernel of $\cL_T$ is shown in Remark~\ref{rem:kernel}(i).
\end{proof}

\begin{remark}\label{rem:kernel}
(i) The elements listed are indeed in the kernel: for $T_3,\dots,T_9$ the sums $T_w+\tau T_{w'}$
lie in the linear space $\Span(T_1,\dots,T_9)$ of solutions, $v+\tau u$ and $u^3+rv+\tau u$ are
solutions by Proposition~\ref{prop:sextic}, and $\cL_u[c^{-1}]=0$ is an identity (ancillary
script). For the proof of Theorem~\ref{thm:cubic} only the inclusion of the kernel in the list
is needed.

(ii) In the coordinates $(u,v)$ the proposition says that for the top part $C^{2n}\cdot cT$
the kernel of the linearized 3SC on polynomials divisible by $C$ consists of $C^{2n}$ times
the lower members of the family of $cT$, together with $C^{2n}$ itself for $T=u$. This form was
cross-checked for all top parts of weight at most $30$ by an independent program working with
polynomials in $(u,v)$ (Appendix~\ref{app:certificates}).
\end{remark}

\subsection{Proof of Theorem~\ref{thm:cubic}}\label{subsec:proof-cubic}
The sufficiency and the $N$SC are Proposition~\ref{prop:sufficiency}. For the necessity let
$P$ be as in the theorem with $\cS[P]=0$, and let $H:=P/c$, so that \eqref{eq:Hcond} holds. By
Corollary~\ref{cor:toppart} we may assume, after multiplying $P$ by a constant, that the top
weighted part of $H$ is $c^{2n}T$ with $n\ge0$ and $T\in\cT$. Put
\[
R:=c^{-2n}H=T+\sum_{k=1}^{K}R_k\in\LL ,
\]
where $R_k$ is weighted homogeneous of weight $\wt T-2k$, and we put $R_k:=0$ for $k>K$; recall that all weights of an odd
element are odd. By Lemma~\ref{lem:cov}, $\cS[R]=M^{-2n}\cS[H]=0$. We shall show that $R$ is a
polynomial and identify it. Let $B_k=T+\sum_{j<k}R_j$ as in Lemma~\ref{lem:levels}.

\emph{The types $T\neq T_1$.} We prove by induction on $k\ge1$ that $\cS[B_k]=0$ and that each
$R_j$ with $j<k$ is a multiple of the element of weight $\wt T-2j$ listed in
Proposition~\ref{prop:kernel}, and $R_j=0$ if there is none. For $k=1$, $B_1=T$ and
$\cS[T]=0$. If the statement holds for $k$, then Lemma~\ref{lem:levels} gives $\cL_T[R_k]=0$,
and Proposition~\ref{prop:kernel} gives the statement about $R_k$. Then $B_{k+1}=B_k+R_k$ is an
element of $\Span(T_1,T_3,T_5,T_7,T_9)$ if $T\in\{T_3,T_5,T_7,T_9\}$, of the form $v+\tau u$ if
$T=\TI$, and of the form $u^3+rv+\tau u$ if $T=T_r$. In all cases $c\,B_{k+1}$ belongs to
family (a), (b) or (c), so $\cS[c\,B_{k+1}]=0$ by Proposition~\ref{prop:sufficiency} and
$\cS[B_{k+1}]=0$ by Lemma~\ref{lem:cov}. Hence $R=B_{K+1}$ is a polynomial and $cR$ belongs to
(a), (b) or (c).

\emph{The type $T=T_1=u$.} Here $B_1=u$. By Lemma~\ref{lem:levels} and
Proposition~\ref{prop:kernel}(i), $R_1=0$ (weight $-1$) and $R_2=\delta\,c^{-1}$ for some
$\delta\in\C$. Now
\[
\cS[u+\delta c^{-1}]=\delta\,\cL_u[c^{-1}]+\delta^2\cQ_u[c^{-1}]+O(\delta^3),
\]
where the terms of order $\delta^j$ are homogeneous of degree $4-4j$ in $k$. Thus the component
of degree $-2$ of $\cS[B_3]$, $B_3=u+\delta c^{-1}$, vanishes, and Lemma~\ref{lem:levels} for
$k=3$ gives $\cL_u[R_3]=0$, hence $R_3=0$ (weight $-5$). For $k=4$ the component of degree $-4$
of $\cS[B_4]$, $B_4=B_3$, is $\delta^2\cQ_u[c^{-1}]$, and we obtain
\begin{equation}\label{eq:level4}
\cL_u[R_4]+\delta^2\,\cQ_u[c^{-1}]=0,\qquad R_4=\sum_{b\ge3}\rho_b\,u^{3b-7}c^{-b}\quad(\rho_b\in\C).
\end{equation}
We claim that $R_4=0$. Put $k_2=k_1-\eps$. As in the proof of Lemma~\ref{lem:pole}(a), in each
summand of $\cQ_u[c^{-1}]$ at most one of the four values of $c$ vanishes at $\eps=0$, to first
order; so $\cQ_u[c^{-1}](k_1,k_1-\eps,k_3)$ has at most a simple pole at $\eps=0$. In
$\cL_u[R_4]$ the part $\Lambda^{\mathrm{(ii)}}$ of Lemma~\ref{lem:pole} has at most a simple
pole as well, because the value of $u^{3b-7}c^{-b}$ at $d_{12}$ is a constant multiple of
$(k_1(k_1-\eps))^{-b}(\pm\eps)^{2b-7}$ and $2b-7\ge-1$. By Lemma~\ref{lem:pole}(a) and
\eqref{eq:level4}, the coefficients of $\eps^{-q}$ of $\Lambda^{\mathrm{(i)}}$ vanish for all
$q\ge2$. If the pole order $B$ of $R_4$ were at least $4$, Lemma~\ref{lem:pole}(b) with $q=2$
and $\nu=0$ would give a pole of order $2B-3$ in the coefficient of $\eps^{-2}$; hence $B\le3$
and $R_4=\rho_3u^2c^{-3}$. For this element the coefficient of $\eps^{-3}$ of
$\Lambda^{\mathrm{(i)}}$ is
\[
2\rho_3\cdot\frac{4\,k_1k_3^2}{27\,(k_3^2-k_1^2)^2}
\]
(the two sign vectors $(1,-1,\pm1)$ contribute $\frac{2k_1k_3^2}{27(k_3^2-k_1^2)^2}\rho_3$
each, and the factor $2$ comes from $\sg\mapsto-\sg$), so $\rho_3=0$. Thus $R_4=0$, and
\eqref{eq:level4} reduces to $\delta^2\cQ_u[c^{-1}]=0$. Since
\begin{equation}\label{eq:Q2value}
\cQ_u[c^{-1}](1,2,3)=\tfrac{14}{135}\neq0,
\end{equation}
we get $\delta=0$. Now $B_k=u$ for $k\le5$, and Lemma~\ref{lem:levels} and
Proposition~\ref{prop:kernel}(i) give inductively $R_k=0$ for all $k\ge5$, since the weight
$-3$ does not recur. Hence $R=u$.

\emph{Conclusion.} In all cases $R\in\C[u,c]$, so $c^{2n}$ divides $H=c^{2n}R$. Since
$H(u,0)\not\equiv0$, we have $n=0$, and $P=cR$ belongs to (a), (b) or (c). The bound on the
weighted degree follows. \qed

\begin{remark}\label{rem:consistency}
(i) The argument shows more: every odd $H\in\C[u,c]$ with $\cS[H]=0$ whose top weighted part is
$c^{2n}T$ equals $c^{2n}$ times a solution with top part $T$. The direction $c^{-1}$ at $T=u$
is the constant term: $c\,(u+\delta c^{-1})=P_4+\delta$, and $\cS[P_4+\delta]$ vanishes to
first order in $\delta$ only; for instance
$\cS[u+\delta c^{-1}](1,2,3)=\delta^2(\delta^2+588\delta+4536)/43740$.

(ii) For weighted degree at most $16$ the necessity part of Theorem~\ref{thm:cubic} was first
obtained by a direct elimination with polynomials in $(u,v)$: all possible top parts of weight
at most $16$, including those divisible by $C^3$ and $C^5$, the kernels of the linearized
condition at each of them, and a $3\times3$ determinant excluding lower-order terms below the
top part $uC^5$, which is the instance $n=2$ of \eqref{eq:level4}. These computations are
included in the ancillary files as an independent check (Appendix~\ref{app:certificates}); the
proof above does not use them.
\end{remark}

\subsection{Proof of Theorem~\ref{thm:cubic-all} and of Corollary~\ref{cor:conj-cubic}}\label{subsec:proof-all}
\emph{Sufficiency.} The statements about $P_0$ are Proposition~\ref{prop:sufficiency}, and
$S_3[C^{2n}P_0]=M^{2n}S_3[P_0]=0$ on $V(C)^3$ by Lemma~\ref{lem:cov}.

\emph{Necessity.} Let $P\neq0$ be even with $S_3[P]=0$ on $V(C)^3$. By Lemmas~\ref{lem:divis}
and~\ref{lem:cov}, $P=C^{2n}P_0$, where $P_0$ is even, has $C$ as a simple factor, and satisfies
$S_3[P_0]=0$ on $V(C)^3$. Write $P_0=C\,H$ with $H$ an odd polynomial on $\C^s$ which does not
vanish identically on $V(C)$, and denote the coordinates of $\C^s$ by $(u,v,z)$,
$z\in\C^{s-2}$. For $\xi,\eta\in\C^{s-2}$ consider the plane $A_{\xi\eta}$ of
\eqref{eq:planes} and
\[
H_{\xi\eta}(u,v):=H(u,v,\xi u+\eta v).
\]
The image of $A_{\xi\eta}$ contains the points $(t,t^3,\xi t+\eta t^3)$ of $V(C)$, so by
Lemma~\ref{lem:linear}(ii) the even polynomial $P_0\circ A_{\xi\eta}=C\cdot H_{\xi\eta}$ satisfies
the 3SC on the cubic graph in $\C^2$. Every point of $V(C)$ with $u\neq0$ is of the form
$(t,t^3,\xi t+\eta t^3)$, so $H_{\xi\eta}|_{V(C)}\not\equiv0$ for $(\xi,\eta)$ in a dense open
subset $U$ of $\C^{2(s-2)}$. By Theorem~\ref{thm:cubic}, for $(\xi,\eta)\in U$ the polynomial
$H_{\xi\eta}$ lies in one of the four linear spaces
\[
\cV_a:=\Span(T_1,T_3,T_5,T_7,T_9),\qquad\cV_b:=\Span(u,v),\qquad
\cV_r:=\Span(u,u^3+rv)
\]
(two spaces $\cV_r$, one for each root $r$ of $r^2-7r+1=0$).
The map $(\xi,\eta)\mapsto H_{\xi\eta}$ is polynomial with values in a finite-dimensional space
of polynomials, the preimages of the four spaces are closed and cover the dense set $U$, and
$\C^{2(s-2)}$ is irreducible; hence one of the four spaces, say $\cV$, contains $H_{\xi\eta}$ for
all $(\xi,\eta)$. Write $H=\sum_{\mathbf j}H_{\mathbf j}(u,v)\,z^{\mathbf j}$ and let $D$ be the largest $|\mathbf j|$ with
$H_{\mathbf j}\neq0$. Each homogeneous component of $H_{\xi\eta}$ with respect to $(\xi,\eta)$ lies in
$\cV$; for the component of degree $D$ at $\xi=0$ this gives
$\sum_{|\mathbf j|=D}H_{\mathbf j}\,\eta^{\mathbf j} v^D\in\cV$ for all $\eta$, hence $H_{\mathbf j} v^D\in\cV$ for all $\mathbf j$
with $|\mathbf j|=D$.

No nonzero element of $\cV_a$ or $\cV_r$ is divisible by $v$, because the pure powers
$u,u^3,u^5,u^7,u^9$ occur in $T_1,\dots,T_9$ with nonzero coefficients, and likewise $u,u^3$ in
the basis of $\cV_r$. So in these cases $D=0$ and $H=H_{\xi\eta}\in\cV$: cases (a) and (c), or
$H\in\C u$, which belongs to (a). In $\cV_b$ the elements divisible by $v$ are the multiples of
$v$; hence $D\le1$, and for $D=1$ the $H_{\mathbf j}$ with $|\mathbf j|=1$ are constants, so
$H=H_0(u,v)+\lambda\cdot z$ with $\lambda\in\C^{s-2}$ and
$H_{\xi\eta}=H_0+(\lambda\cdot\xi)u+(\lambda\cdot\eta)v\in\cV_b$. Thus $H_0\in\Span(u,v)$ and $H$ is a
linear form: case (b). \qed

\emph{Proof of Corollary~\ref{cor:conj-cubic}.} A linear automorphism of $\C^s$ maps
$V(\ell_1^3-\ell_2)$ onto $V(C)$; by Lemma~\ref{lem:linear}(i) it preserves evenness, the order
of vanishing along the component and the conditions $S_N=0$. Apply
Theorem~\ref{thm:cubic-all} with $n=0$. \qed

\subsection{The four-soliton condition}\label{subsec:fourth}
For the polynomials of higher multiplicity in Theorem~\ref{thm:cubic-all} the four-soliton
condition fails.

\begin{theorem}[The four-soliton condition; partly computer-assisted]\label{thm:fourth}
Let $P\neq0$ be an even polynomial on $\C^s$ such that $S_3[P]$ vanishes identically on
$V(C)^3$. Then the following statements are equivalent:
\begin{enumerate}
\item[\textup{(i)}] $C$ is a simple factor of $P$;
\item[\textup{(ii)}] $S_4[P]$ vanishes identically on $V(C)^4$;
\item[\textup{(iii)}] $S_N[P]$ vanishes identically on $V(C)^N$ for every $N\ge3$.
\end{enumerate}
In particular none of the polynomials $C^{2n}P_0$ of Theorem~\textup{\ref{thm:cubic-all}} with
$n\ge1$ satisfies the \textup{4SC} on $V(C)$.
\end{theorem}
\begin{proof}
By Theorem~\ref{thm:cubic-all}, $P=C^{2n}P_0$, and (i) means $n=0$. Then (iii) holds by the same
theorem, and (iii) implies (ii). It remains to show that $S_4[P]$ does not vanish identically on
$V(C)^4$ if $n\ge1$. Put $m:=2n+1\ge3$ and $H:=P_0/C$, so that $P=C^mH$.

\emph{Step 1: two variables.} In the cases (a) and (c) of Theorem~\ref{thm:cubic-all} we have
$P\in\C[u,v]$, and $S_4[P]$ involves only the coordinates $u,v$ of the momenta. In case (b),
$H=\alpha u+\beta v+\gamma\cdot z$, where $z$ denotes the remaining coordinates. If
$\gamma\neq0$, choose $\xi,\eta\in\C^{s-2}$ with $\gamma\cdot\xi=1-\alpha$ and
$\gamma\cdot\eta=-\beta$; then $P\circ A_{\xi\eta}=C^mu$ for the map \eqref{eq:planes}, and by
Lemma~\ref{lem:linear} it suffices to show that $S_4[C^mu]$ does not vanish on the fourth power
of the curve $v=u^3$ in $\C^2$. So let $s=2$ and let $H$ be a nonzero element of
$\Span(T_1,T_3,T_5,T_7,T_9)$ or of $\Span(u,v)$, or $H=\alpha u+\beta(u^3+rv)$ with
$\beta\neq0$.

\emph{Step 2: the top weighted part.} The types $T_1,\dots,T_9$ have the pairwise distinct
weights $1,3,5,7,9$; hence in each of the three cases the top weighted part of $H$ is
$\lambda T$ with $\lambda\neq0$ and $T\in\cT$ (Table~\ref{tab:types}), and the top weighted part
of $P$ is $\lambda C^mT$, of weight $W:=3m+\wt T$. For $\rho\in\C$ the map
$\delta_\rho(u,v):=(\rho u,\rho^3v)$ is linear, $\bp(\rho k)=\delta_\rho\bp(k)$, and
$P\circ\delta_\rho=\rho^W\lambda C^mT+(\text{lower powers of }\rho)$. Since $S_4[P]$ is
homogeneous of degree seven in the coefficients of $P$, Lemma~\ref{lem:linear} gives
\[
S_4[P]\bigl(\bp(\rho k_1),\dots,\bp(\rho k_4)\bigr)=\rho^{7W}\lambda^7\,
S_4[C^mT]\bigl(\bp(k_1),\dots,\bp(k_4)\bigr)+(\text{lower powers of }\rho).
\]
Therefore it suffices to find, for every $T\in\cT$ and every odd $m\ge3$, a point
$k\in\C^4$ with $S_4[C^mT](\bp(k_1),\dots,\bp(k_4))\neq0$. (A nonzero value for each type at a
fixed point would not suffice for the linear combinations; the passage to the top weighted
part is needed.)

\emph{Step 3: a formula.} Let $k\in\C^4$, and for $\sg\in\{\pm1\}^4$ put $a_i:=\sg_ik_i$, so
that $\sg_i\bp(k_i)=\bp(a_i)$, and
\[
A_\sg:=C\Bigl(\sum_i\bp(a_i)\Bigr)\prod_{i<j}(a_i-a_j),\qquad
B_\sg(T):=T\Bigl(\sum_i\bp(a_i)\Bigr)\prod_{i<j}T\bigl(\bp(a_i)-\bp(a_j)\bigr).
\]
Since $C(\bp(a)-\bp(b))=-3ab(a-b)$ and every $a_i$ occurs in three of the six pairs,
$\prod_{i<j}C(\bp(a_i)-\bp(a_j))=K\,\bigl(\prod_i\sg_i\bigr)\prod_{i<j}(a_i-a_j)$ with
$K:=3^6(k_1k_2k_3k_4)^3$, and $\prod_{i<j}\sg_i\sg_j=\prod_i\sg_i$. Hence the summand of
$S_4[C^mT]$ which belongs to $\sg$ is $K^m\bigl(\prod_i\sg_i\bigr)^{m+1}A_\sg^mB_\sg(T)=K^mA_\sg^mB_\sg(T)$,
because $m$ is odd. The polynomials $C$ and $T$ are odd, so $A_{-\sg}=-A_\sg$ and
$B_{-\sg}=-B_\sg$, and
\begin{equation}\label{eq:fourth-formula}
S_4[C^mT]\bigl(\bp(k_1),\dots,\bp(k_4)\bigr)=2K^m\sum_{\sg_1=1}A_\sg^m\,B_\sg(T).
\end{equation}

\emph{Step 4: one point for all $m$.} Let $k=(1,10,11,12)$. The eight integers $|A_\sg|$ with
$\sg_1=1$ are $261360\,a_i$, where
\[
(a_0,\dots,a_7)=(2093,\ 1463,\ 1449,\ 897,\ 819,\ 377,\ 267,\ 253),
\]
and $a_0$ belongs to $\sg=(1,-1,1,-1)$. For $T\neq T_r$ the numbers $B_\sg(T)$ are integers;
let $b_i:=\operatorname{sgn}(A_\sg)B_\sg(T)/g_T$ for the sign vector $\sg$ belonging to $a_i$,
where $g_T>0$ is the greatest common divisor of the eight numbers $B_\sg(T)$. For $T=T_r$ we
write $B_\sg(T_r)=B'_\sg+B''_\sg\,r$ with integers $B'_\sg,B''_\sg$, using $r^2=7r-1$, and
define $b_i$ in the same way from the numbers $B''_\sg$. Then by \eqref{eq:fourth-formula}, as
$m$ is odd,
\[
S_4[C^mT]\bigl(\bp(1),\bp(10),\bp(11),\bp(12)\bigr)=2\,(261360\,K)^m\,g_T\sum_{i=0}^7b_i\,a_i^m
\]
for $T\neq T_r$, and for $T=T_r$ the right-hand side is the coefficient of $r$ in the
left-hand side. An exact computation (Appendix~\ref{app:certificates}) gives
\[
D_T:=|b_0|\,a_0^3-\sum_{i=1}^7|b_i|\,a_i^3>0
\]
for all seven types; the values of $D_T$ are
\begin{center}
\renewcommand{\arraystretch}{1.15}
\begin{tabular}{@{}ll@{\qquad}ll@{}}
\toprule
$T$ & $D_T$ & $T$ & $D_T$\\
\midrule
$T_1$ & $195453585904072$ & $T_9$ & $9254520527134488390$\\
$T_3$ & $416700050664124$ & $\TI$ & $2784478155655554883724$\\
$T_5$ & $124504223039866998$ & $T_r$ & $8232269999863512346490723074948$\\
$T_7$ & $91839149291296818$ & &\\
\bottomrule
\end{tabular}
\end{center}
Since $a_i<a_0$ for $i\ge1$, we get for every $m\ge3$
\[
\Bigl|\sum_ib_ia_i^m\Bigr|\ \ge\ |b_0|\,a_0^m-\sum_{i\ge1}|b_i|\,a_i^m\ \ge\ a_0^{m-3}D_T>0 .
\]
For $T=T_r$ this shows that the coefficient of $r$ in the rational expression $x+yr$ of the
value is not zero; as $r$ is irrational, the value is not zero for either root $r$.
\end{proof}

For $m=1$ the sums $\sum_ib_ia_i$ vanish, in accordance with (iii). The ratio
$|b_0|a_0^3/\sum_{i\ge1}|b_i|a_i^3$ is $1.0208\ldots$ for $T_9$ and at least $1.19$ for the
other types; the estimate cannot be replaced by the cruder one in which every $a_i$, $i\ge1$,
is replaced by $a_1$.

\section{Graphs of degree at least four}\label{sec:graphs}
We apply the results of Sections~\ref{sec:collisions} and~\ref{sec:soft} to the plane curves
$v=q(u)$ with $\deg q\ge4$, in any number of variables. The graphs of degree two are parabolas
and are covered by \cite[Theorem~9.17]{PartI}; the cubic graph is the subject of
Section~\ref{sec:cubic}, and the cubic graphs which are not odd are excluded in
Section~\ref{sec:limits} (Theorem~\ref{thm:cubic-graphs}). The main step is the determination of the weighted homogeneous
solutions on the monomial graphs $v=u^m$, to which the general case is reduced by the
degeneration principle. In this section $C$, $C_2$ and $\bp$ refer to the monomial graph of
degree $m$, and not to the cubic graph.

\subsection{Statements}\label{subsec:graph-statements}
For $m\ge2$ let $V_m:=\{v=u^m\}\subset\C^2$, $\bp(t):=(t,t^m)$, $C:=u^m-v$,
$C_2:=u^m-2^{m-1}v$ (a multiple of $C(z/2)$), and $\wt u:=1$, $\wt v:=m$.

\begin{proposition}[Candidates]\label{prop:candidates}
Let $m\ge2$ and let $P\neq0$ be an even, weighted homogeneous polynomial in $\C[u,v]$ such that
$S_3[P]$ vanishes identically on $V_m^{\,3}$.
\begin{enumerate}
\item[\textup{(i)}] If $m$ is even, then $P=\lambda P_A$ or $P=\lambda P_B$, where
$P_A:=u^{2m}-v^2$ and $P_B:=(u^{2m}-v^2)(u^{2m}-4^{m-1}v^2)$.
\item[\textup{(ii)}] If $m$ is odd, then $P=C^\mu H'$ with $\mu$ odd and $H'$ in one of the
families
\begin{align*}
&\textup{(F1)}\ \lambda u,\qquad \textup{(F2)}\ \alpha u^m+\beta v,\qquad \textup{(F3)}\ \lambda u^2C_2,\qquad
\textup{(F4)}\ u\,C_2\,(\alpha u^m+\beta v),\\
&\textup{(F5)}\ C_2\,(\alpha u^{2m}+\beta u^mv+\gamma v^2),
\end{align*}
where $\lambda\neq0$, $\beta\neq0$ in \textup{(F2)}, \textup{(F4)}, $\gamma\neq0$ in \textup{(F5)},
and $C\nmid H'$.
\end{enumerate}
\end{proposition}

\begin{theorem}[Weighted homogeneous solutions; partly computer-assisted]\label{thm:monomial}
Let $P$ be as in Proposition~\ref{prop:candidates}.
\begin{enumerate}
\item[\textup{(i)}] If $m=5$, then $P=\lambda\,u\,(u^5-v)$ or $P=\lambda\,(u^5-v)(u^5+4v)$;
both are solutions.
\item[\textup{(ii)}] There is no such $P$ for even $m\ge4$.
\item[\textup{(iii)}] There is no such $P$ for odd $m\ge7$.
\end{enumerate}
\end{theorem}
Statement (ii) is proved without computer, and so is the exclusion of the families
\textup{(F1)} and \textup{(F3)} for all odd $m\ge5$ and all $\mu$, apart from the solution
$\lambda uC$ for $m=5$. The remaining families are excluded for $m=7$ by Gr\"obner bases and for
all odd $m\ge9$ by a uniform elimination (Section~\ref{sec:uniform}). For $m=2$ the same
method gives $P=\lambda(u^4-v^2)$, in agreement with \cite[Theorem~9.17]{PartI}, and for $m=3$ it
is consistent with Theorem~\ref{thm:cubic-all}, where every odd multiplicity occurs.

\subsubsection*{Graphs}
\begin{theorem}[Graphs $v=q(u)$; partly computer-assisted]\label{thm:graphs}
Let $u,v$ be two of the coordinates of $\C^s$, let $q\in\C[u]$ be a polynomial of degree
$m\ge4$, $C:=q(u)-v$, and let $P\neq0$ be an even polynomial on $\C^s$ which is divisible by
$C$ and such that $S_3[P]$ vanishes identically on $V(C)^3$.
\begin{enumerate}
\item[\textup{(a)}] Then $m=5$.
\item[\textup{(b)}] Moreover $q$ is odd, $q=a_5u^5+a_3u^3+a_1u$, and
\begin{equation}\label{eq:quintic-solutions}
P=\bigl(q(u)-v\bigr)\,\bigl[\alpha u+\beta\,(a_5u^5+2a_3u^3+4v)\bigr],\qquad(\alpha,\beta)\neq(0,0).
\end{equation}
Conversely, for every odd quintic $q$ the polynomials \eqref{eq:quintic-solutions} are even,
have $q(u)-v$ as a simple factor, and satisfy the $N$-soliton condition on $V(q(u)-v)$ for
every $N\ge3$.
\end{enumerate}
\end{theorem}
For $\beta=0$, \eqref{eq:quintic-solutions} is the family $D_x(q(D_x)+D_t)$ of
\cite[Theorem~5.7]{PartI}; the polynomials with $\beta\neq0$ have the same two-soliton phase factor
\eqref{eq:quintic-phase}. No bound on the degree of $P$, on the number of variables or on the
multiplicity of the component is assumed. The polynomials \eqref{eq:quintic-solutions} are
reductions of equations of the BKP hierarchy, and for $\beta\neq0$ they satisfy the 3SC on one
of their two components only; see Remark~\ref{rem:bkp} at the end of this section.

\begin{corollary}\label{cor:consequences}
\textup{(a)} Let $\ell_1,\ell_2$ be linearly independent linear forms on $\C^s$ and $q$ a
polynomial of degree five. Then Conjecture~\textup{\ref{conj:main}} holds for the component
$V(\ell_2-q(\ell_1))$: every even $P$ which vanishes to order one along it and satisfies the
\textup{3SC} on it satisfies the $N$\textup{SC} for all $N\ge3$; these $P$ are the polynomials
\eqref{eq:quintic-solutions} in the variables $u=\ell_1$, $v=\ell_2$. For $\deg q\ge4$,
$\deg q\neq5$, no polynomial satisfies the \textup{3SC} on $V(\ell_2-q(\ell_1))$, so that
Conjecture~\textup{\ref{conj:main}} holds trivially for these components.

\textup{(b)} In the setting of Section~\textup{\ref{sec:collisions}} let $s=2$, and suppose that $V$ passes through the origin and
is smooth there; let $m_0\ge2$ be the order of contact of $V$ with its tangent line at the
origin. Then $m_0\in\{2,3,5\}$. If moreover $V\neq-V$, then $m_0=2$.
\end{corollary}

\subsection{Candidates}\label{sec:monomial}
\begin{proof}[Proof of Proposition~\ref{prop:candidates}]
For $p_1,p_2\in V_m$ evenness gives $S_3[P](p_1,p_2,0)=8P(p_1)P(p_2)P(p_1+p_2)P(p_1-p_2)$, and
$P(p_1\pm p_2)$ do not vanish identically on $V_m^{\,2}$ (Lemma~\ref{lem:dominance}); hence $C$ divides
$P$, and we are in the setting of Section~\ref{sec:collisions} with $V=V_m$, which passes through
the origin, is smooth there ($e_0=1$) and is not a cone. By Corollary~\ref{cor:origin},
$\mu$ is odd, $\nu_2\in\{0,1\}$ and $d=\mu+1+2\nu_2$.

(i) For even $m$, $V_m\neq-V_m$. By Corollary~\ref{cor:origin}(3),
$P=CC^-(C_2C_2^-)^{\nu_2}K$ with $K(0)\neq0$; $K$ is weighted homogeneous, hence constant. Here
$CC^-=u^{2m}-v^2$ and, up to constants, $C_2C_2^-=u^{2m}-4^{m-1}v^2$.

(ii) For odd $m$ put $K:=P/(C^\mu C_2^{\nu_2})$. Then $K$ is weighted homogeneous and
$\ord_0K=1+\nu_2$, that is, $K$ contains a monomial of ordinary degree $1+\nu_2$ and none of
lower degree. If $\nu_2=0$, this monomial is $u$ or $v$, of weight $1$ or $m$, so $K=\lambda u$
or $K=\alpha u^m+\beta v$ with $\beta\neq0$. If $\nu_2=1$, it is $u^2$, $uv$ or $v^2$, of weight
$2$, $m+1$ or $2m$, so $K=\lambda u^2$, $K=u(\alpha u^m+\beta v)$ with $\beta\neq0$, or
$K=\alpha u^{2m}+\beta u^mv+\gamma v^2$ with $\gamma\neq0$.
\end{proof}

\subsection{The double soft limit}\label{sec:doublesoft}
\begin{lemma}\label{lem:doublesoft}
Let $m\ge2$ and let $P=C^\mu K\in\C[u,v]$ be weighted homogeneous with $\mu\ge1$ and
$K(1,1)\neq0$. Put $\tilde f(t):=t^{-\mu}P(1+t,1)$, a polynomial with $\tilde f(0)\neq0$, and
$\ell(t):=\tilde f'(t)/\tilde f(t)=\sum_{j\ge0}\ell_jt^j$, $r:=\ell'+\ell^2$. If
$\fF(\bp(1),\bp(t);e_u)=0$ for all $t$, where $e_u=(1,0)$, then
\begin{equation}\label{eq:G}
G(t):=(2\mu-2\ell_0)\bigl[\ell(t)+\ell(-t)-2\ell_0\bigr]-2\ell_0t\bigl[\ell(t)-\ell(-t)\bigr]+t\bigl[r(t)-r(-t)\bigr]=O(t^{m-1}).
\end{equation}
The series $G$ is even, $G(0)=0$, and for $j\ge1$ its coefficient of $t^{2j}$ is
\begin{equation}\label{eq:E2j}
E_{2j}:=4\Bigl[(\mu+j-\ell_0)\,\ell_{2j}+\sum_{a=1}^{j-1}\ell_a\,\ell_{2j-1-a}\Bigr],
\end{equation}
so that $E_{2j}=0$ whenever $2j\le m-2$. In particular
\begin{equation}\label{eq:cond1}
\ell_2\,(\mu+1-\ell_0)=0\qquad\text{if }m\ge4 .
\end{equation}
\end{lemma}
\begin{proof}
Let $w$ be the weight of $P$. Since $D_u^jP$ is weighted homogeneous of weight $w-j$ and
$\bp(t)$ is the weighted dilation of $\bp(1)$, $\pi_j(\bp(t);e_u)=t^{w-j}\pi_j(\bp(1);e_u)$.
With $a_j=\pi_{\mu+j}(\bp(1);e_u)$ we get $b_j=t^{w-\mu-j}a_j$ in \eqref{eq:soft}; here
$a_0=m^\mu K(1,1)\neq0$. Put $\vartheta:=2a_1/a_0$ and $\cP(t,\delta):=P(1+t,1+\delta)$. Then
$\fF(\bp(1),\bp(t);e_u)=a_0^2\,t^{w-\mu-1}\,\Phi(t,t^m)$, where
\begin{multline*}
\Phi(t,\delta):=t\bigl[\cP(-t,-\delta)\cP_{uu}(t,\delta)-\cP(t,\delta)\cP_{uu}(-t,-\delta)\bigr]
-\vartheta(1+t)\cP(-t,-\delta)\cP_u(t,\delta)\\-\vartheta(1-t)\cP(t,\delta)\cP_u(-t,-\delta)
+\vartheta^2\cP(t,\delta)\cP(-t,-\delta),
\end{multline*}
and $\cP_u,\cP_{uu}$ denote the derivatives of $P$ with respect to $u$ at $(1+t,1+\delta)$.

The polynomial $\Phi$ satisfies $\Phi(-t,-\delta)=\Phi(t,\delta)$. Let $\fm:=(t,\delta)$. Since
$P=C^\mu K$ and $C(1+t,1+\delta)=(1+t)^m-1-\delta\in\fm$, we have $\cP\in\fm^\mu$,
$\cP_u\in\fm^{\mu-1}$, $\cP_{uu}\in\fm^{\mu-2}$ (with $\fm^j:=\C[t,\delta]$ for $j\le0$), hence
$\Phi\in\fm^{2\mu-1}$. By the symmetry only monomials $t^i\delta^j$ with $i+j$ even occur in
$\Phi$, so $\Phi\in\fm^{2\mu}$, and a monomial with $j\ge1$ becomes, for $\delta=t^m$, the power
$t^{i+mj}$ with $i+mj\ge2\mu+m-1$. Therefore $\Phi(t,t^m)=\Phi(t,0)+O(t^{2\mu+m-1})$.

It remains to compute $\Phi(t,0)$, which involves only $f(x):=P(x,1)=(x-1)^\mu\tilde f(x-1)$:
\begin{align*}
f(1\pm t)&=(\pm t)^\mu\tilde f(\pm t),\qquad f'(1\pm t)=(\pm t)^{\mu-1}\bigl[\mu\tilde f(\pm t)\pm t\tilde f'(\pm t)\bigr],\\
f''(1\pm t)&=(\pm t)^{\mu-2}\bigl[\mu(\mu-1)\tilde f(\pm t)\pm2\mu t\tilde f'(\pm t)+t^2\tilde f''(\pm t)\bigr].
\end{align*}
Substituting, the terms with $\mu(\mu-1)$ cancel, and with $\vartheta=2\ell_0$ one obtains
\[
\Phi(t,0)=(-1)^\mu\,t^{2\mu}\,\tilde f(t)\,\tilde f(-t)\,G(t).
\]
Hence $\fF=0$ implies $G(t)=O(t^{m-1})$.

The definition of $G$ shows that $G(-t)=G(t)$ and $G(0)=0$. Let $j\ge1$. The coefficient of
$t^{2j-1}$ in $r=\ell'+\ell^2$ is $2j\,\ell_{2j}+2\ell_0\ell_{2j-1}+\sum_{a=1}^{2j-2}\ell_a\ell_{2j-1-a}$,
so the three terms of $G$ contribute to the coefficient of $t^{2j}$
\[
4(\mu-\ell_0)\ell_{2j}\;-\;4\ell_0\ell_{2j-1}\;+\;4j\,\ell_{2j}+4\ell_0\ell_{2j-1}+2\sum_{a=1}^{2j-2}\ell_a\ell_{2j-1-a}.
\]
The last sum is invariant under $a\mapsto2j-1-a$, which has no fixed point, so it is twice the
sum over $1\le a\le j-1$. This gives \eqref{eq:E2j}.
\end{proof}

\begin{lemma}[Cases]\label{lem:tower}
Let $J\ge1$ and suppose that the numbers $\mu,\ell_0,\dots,\ell_{2J}$ satisfy $E_2=\dots=E_{2J}=0$
in \eqref{eq:E2j}. Then
\begin{itemize}
\item[$(\mathrm T_k)$] $\ell_2=\dots=\ell_{2k-2}=0$, $\ell_0=\mu+k$ and $E_{2j}=0$ for $k<j\le J$,
for some $k\in\{1,\dots,J\}$, or
\item[$(\mathrm T_{J+1})$] $\ell_2=\ell_4=\dots=\ell_{2J}=0$.
\end{itemize}
\end{lemma}
\begin{proof}
In a product $\ell_a\ell_{2j-1-a}$ of \eqref{eq:E2j} the two indices lie between $1$ and $2j-2$
and have different parity, so one of the factors is one of $\ell_2,\ell_4,\dots,\ell_{2j-2}$.
Hence $E_{2j}=4(\mu+j-\ell_0)\ell_{2j}$ if $\ell_2=\dots=\ell_{2j-2}=0$. If $(\mathrm T_{J+1})$
fails, let $k\le J$ be the smallest index with $\ell_{2k}\neq0$; then $E_{2k}=0$ gives
$\ell_0=\mu+k$.
\end{proof}

We shall use the expansions, for a positive integer $n$ and a constant $L\neq1$, with
$A:=1/(1-L)$,
\begin{align}
\frac{d}{dt}\log\frac{(1+t)^n-1}{t}&=\frac{n-1}2+\frac{(n-1)(n-5)}{12}\,t-\frac{(n-1)(n-3)}{8}\,t^2+O(t^3),\label{eq:logphi}\\
\frac{d}{dt}\log\bigl((1+t)^n-L\bigr)&=nA+\chi_1t+\chi_2t^2+O(t^3),\notag\\
\chi_2&=\frac{nA}2\bigl[(n-1)(n-2)-3n(n-1)A+2n^2A^2\bigr].\label{eq:chi}
\end{align}
If $A<0$, then $\chi_2<0$.

\begin{proof}[Proof of Theorem~\ref{thm:monomial}\,\textup{(ii)}, and of \textup{(iii)} for the families \textup{(F1)} and \textup{(F3)}]
By Theorem~\ref{thm:soft}(b), every solution satisfies the hypothesis of
Lemma~\ref{lem:doublesoft}; hence \eqref{eq:cond1} holds for $m\ge4$.

\emph{Even $m\ge4$.} Here $\mu=1$; put $n:=2m\ge8$. For $P_A$, $\tilde f=((1+t)^n-1)/t$, so by
\eqref{eq:logphi} $\ell_0=\tfrac{n-1}2$, $\ell_2=-\tfrac{(n-1)(n-3)}8$ and
$\ell_2(2-\ell_0)=\tfrac1{16}(n-1)(n-3)(n-5)\neq0$. For $P_B$, $\tilde f$ is a multiple of
$\tfrac1t((1+t)^n-1)\,((1+t)^n-L)$ with $L=4^{m-1}\ge64$, so $\ell_0=\tfrac{n-1}2+nA$ and
$\ell_2=-\tfrac{(n-1)(n-3)}8+\chi_2$ with $A=1/(1-L)<0$. Then $\ell_2<0$, and $\ell_0=2$ would
mean $L=(3n-5)/(n-5)<7$. Both candidates of Proposition~\ref{prop:candidates}(i) violate
\eqref{eq:cond1}.

\emph{Odd $m\ge5$, family \textup{(F1)}.} For $P=uC^\mu$, $\tilde f=\bigl(((1+t)^m-1)/t\bigr)^\mu(1+t)$,
so $\ell_0=\tfrac{\mu(m-1)}2+1$ and $\ell_2=1-\tfrac{\mu(m-1)(m-3)}8$. Then
$\mu+1-\ell_0=-\tfrac{\mu(m-3)}2\neq0$, and $\ell_2=0$ means $\mu(m-1)(m-3)=8$, that is,
$(m,\mu)=(5,1)$.

\emph{Odd $m\ge5$, family \textup{(F3)}.} For $P=u^2C_2C^\mu$, $\tilde f$ is a multiple of
$\bigl(((1+t)^m-1)/t\bigr)^\mu(1+t)^2\,((1+t)^m-L)$ with $L=2^{m-1}\ge16$, so with $A=1/(1-L)$,
$-\tfrac1{15}\le A<0$,
\[
\ell_0=\tfrac{\mu(m-1)}2+2+mA,\qquad\ell_2=-\tfrac{\mu(m-1)(m-3)}8+2+\chi_2 .
\]
Here $\mu+1-\ell_0=-\bigl[\tfrac{\mu(m-3)}2+1-\tfrac m{L-1}\bigr]<0$ because $m/(L-1)\le\tfrac13$.
Since $\chi_2<0$, $\ell_2<2-\tfrac{\mu(m-1)(m-3)}8\le0$ unless $(m,\mu)=(5,1)$, and in that case
$\chi_2=-\tfrac{73}{27}$ and $\ell_2=-\tfrac{46}{27}$. Again \eqref{eq:cond1} is violated.
\end{proof}

\subsection{The remaining families}\label{sec:remaining}
For odd $m$ and $P=C^\mu H'$ we use Theorem~\ref{thm:soft}(b) at pairs of points
$p_1=\bp(s)$, $p_2=\bp(t)$ with the multiplicity $\mu$ as an unknown.

\begin{lemma}\label{lem:N}
Let $P=C^\mu H'$ with $C\nmid H'$, let $p_1,p_2\in V_m$ with
$\varpi:=C_u(p_1)C_u(p_2)C(p_1+p_2)C(p_1-p_2)\neq0$, and write $H_i:=H'(p_i)$,
$H_\pm:=H'(p_1\pm p_2)$, $c_\pm:=(C_u/C)(p_1\pm p_2)$, $g_\pm:=(C_{uu}/C)(p_1\pm p_2)$,
$\kappa_i:=(C_{uu}/C_u)(p_i)$,
\begin{gather*}
\mathfrak e_\pm:=\mu c_\pm H_\pm+H'_u(p_1\pm p_2),\qquad\vartheta_i:=\mu\kappa_iH_i+2H'_u(p_i),\\
\mathfrak d_\pm:=\bigl[\mu(\mu-1)c_\pm^2+\mu g_\pm\bigr]H_\pm+2\mu c_\pm H'_u(p_1\pm p_2)+H'_{uu}(p_1\pm p_2).
\end{gather*}
Then $\fF(p_1,p_2;e_u)=\varpi^\mu\,N$, where
\begin{equation}\label{eq:N}
N:=H_1H_2\,(H_-\mathfrak d_+-H_+\mathfrak d_-)-(\vartheta_1H_2+\vartheta_2H_1)H_-\mathfrak e_+
+(\vartheta_1H_2-\vartheta_2H_1)H_+\mathfrak e_-+\vartheta_1\vartheta_2H_+H_- .
\end{equation}
\end{lemma}
\begin{proof}
Where $H_1H_2H_+H_-\neq0$, divide \eqref{eq:soft} by $a_0b_0P^+P^-=\varpi^\mu H_1H_2H_+H_-$ and
use $\Theta=\mu C_{uu}/C_u+2H'_u/H'$ and, with $e=e_u$,
\[
\frac{D_eP}{P}=\mu\frac{C_u}{C}+\frac{H'_u}{H'},\qquad
\frac{D_e^2P}{P}=\mu(\mu-1)\frac{C_u^2}{C^2}+\mu\frac{C_{uu}}{C}+2\mu\frac{C_u}{C}\frac{H'_u}{H'}+\frac{H'_{uu}}{H'} ;
\]
this gives \eqref{eq:softTheta} in the form $N=0$ after multiplication by $H_1H_2H_+H_-$. Both sides of $\fF=\varpi^\mu N$ are polynomial in the values
of $H'$, $H'_u$, $H'_{uu}$, so the identity holds in general.
\end{proof}

For fixed rational $s,t$ and $H'$ in one of the families of
Proposition~\ref{prop:candidates}(ii), normalized by $\lambda=1$, $\beta=1$, $\gamma=1$ and with
the remaining parameters called $a$ (in \textup{(F2)}, \textup{(F4)}: $H'=v+au^m$, respectively
$uC_2(v+au^m)$) and $a,b$ (in \textup{(F5)}: $H'=C_2(v^2+au^mv+bu^{2m})$), $N$ is a polynomial
$N_{s,t}\in\Q[\mu,a,b]$ of degree two in $\mu$, and it vanishes at every solution, for every odd
$\mu$. Let $I_m$ be the ideal generated by the $N_{s,t}$ for the twelve pairs
\[
(s,t)=(1,2),\,(1,3),\,(2,3),\,(1,4),\,(3,4),\,(2,5),\,(3,5),\,(1,5),\,(4,5),\,(2,7),\,(3,7),\,(5,7),
\]
for all of which $\varpi\neq0$. With $J:=\{(a+2)^2,\ (a+2)(b-1),\ (b-1)^2\}$, the reduced
Gr\"obner bases of $I_m$ (graded reverse lexicographic order, computed over $\Q$) are:
\begin{center}
\renewcommand{\arraystretch}{1.15}
\begin{tabular}{@{}lccccc@{}}
\toprule
 & \textup{(F1)} & \textup{(F2)} & \textup{(F3)} & \textup{(F4)} & \textup{(F5)}\\
\midrule
$m=5$ & $\{\mu(\mu-1)\}$ & $\{\mu(\mu-1),\ a(a-\tfrac14)\}$ & $\{1\}$ & $\{1\}$ & $J$\\
$m=7,9,\dots,25$ & $\{\mu\}$ & $\{\mu,\ a\}$ & $\{1\}$ & $\{1\}$ & $J$\\
\bottomrule
\end{tabular}
\end{center}

\begin{proof}[Proof of Theorem~\ref{thm:monomial}\,\textup{(i)}, and of \textup{(iii)} for $m\le25$]
Let $m$ be odd, $5\le m\le25$, and let $P=C^\mu H'$ be a solution as in
Proposition~\ref{prop:candidates}(ii). Then $(\mu,a,b)$ is a zero of $I_m$ and $\mu$ is odd. In
\textup{(F5)} the only zero has $(a,b)=(-2,1)$, where $v^2-2u^mv+u^{2m}=C^2$, which is excluded
by $C\nmid H'$. For $7\le m\le25$ the table leaves no zero with odd $\mu$. For $m=5$ it leaves
$\mu=1$ with $H'=u$, or with $H'=v+au^5$, $a\in\{0,\tfrac14\}$. For $a=0$,
$S_3[vC](\bp(1),\bp(2),\bp(3))=611991604224000000\neq0$. The polynomials $uC$ and
$C(v+\tfrac14u^5)=\tfrac14C(u^5+4v)$ are solutions by Lemma~\ref{lem:quintic} below with $b=0$.
\end{proof}

\subsection{Odd degrees at least nine}\label{sec:uniform}
In this subsection $m\ge9$ is an odd integer, and
\[
Y(t):=(1+t)^m-1,\qquad\phi(t):=Y(t)/t,\qquad T:=2^{m-1}-1,\qquad A_0:=-1/T .
\]

\subsubsection*{Normal form}
Let $P=C^\mu H'$ be as in Proposition~\ref{prop:candidates}(ii) with $H'$ in one of the families
\textup{(F2)}, \textup{(F4)}, \textup{(F5)}. In \textup{(F5)} we factorize
$\alpha u^{2m}+\beta u^mv+\gamma v^2=(\alpha_1u^m+\beta_1v)(\alpha_2u^m+\beta_2v)$ over $\C$. A
factor $\alpha u^m+\beta v$ of $H'$ is not a multiple of $C$, so $\alpha+\beta\neq0$ and
\[
\alpha(1+t)^m+\beta=(\alpha+\beta)\bigl(1+A\,Y(t)\bigr),\qquad A:=\frac\alpha{\alpha+\beta}.
\]
We call $A$ the parameter of the factor. The parameter of $C_2=u^m-2^{m-1}v$ is
$1/(1-2^{m-1})=A_0$. Let $A$ be the parameter of the factor $\alpha u^m+\beta v$ in
\textup{(F2)} and \textup{(F4)}, and $A_1,A_2$ the parameters of the two factors in
\textup{(F5)}. Then $\tilde f(t)=t^{-\mu}P(1+t,1)$ is a constant multiple of
\begin{equation}\label{eq:normalform}
\phi(t)^\mu\,(1+t)^{\varrho}\prod_i\bigl(1+A_iY(t)\bigr),\qquad
(\varrho;A_i)=\begin{cases}(0;\,A)&\textup{(F2)},\\(1;\,A_0,A)&\textup{(F4)},\\(0;\,A_0,A_1,A_2)&\textup{(F5)},\end{cases}
\end{equation}
so that
\begin{equation}\label{eq:ell-normal}
\ell(t)=\mu\,\frac{\phi'(t)}{\phi(t)}+\frac{\varrho}{1+t}+\sum_i\frac{A_i\,Y'(t)}{1+A_iY(t)} .
\end{equation}
The coefficients of $Y$ and of $\phi/m=\sum_{k\ge0}\frac1m\binom m{k+1}t^k$ are polynomials in $m$
with rational coefficients, and the series $\phi/m$ and $1+A_iY$ have the constant term $1$.
Hence the Taylor coefficients $\ell_j$ of \eqref{eq:ell-normal} are polynomials in $m$, $\mu$
and the $A_i$ with rational coefficients; they are symmetric in the $A_i$ and of degree one
in $\mu$. By \eqref{eq:logphi} and \eqref{eq:chi},
\begin{equation}\label{eq:l0l2}
\begin{aligned}
\ell_0&=\frac{\mu(m-1)}2+\varrho+m\sum_iA_i,\\
\ell_2&=-\frac{\mu(m-1)(m-3)}8+\varrho+\sum_i\Bigl[\frac{m(m-1)(m-2)}2A_i-\frac{3m^2(m-1)}2A_i^2+m^3A_i^3\Bigr].
\end{aligned}
\end{equation}
We put $s_1:=A$ in \textup{(F2)} and \textup{(F4)}, and $s_1:=A_1+A_2$, $s_2:=A_1A_2$ in
\textup{(F5)}, we replace $A_0$ by an indeterminate $\zeta$, and we regard the $\ell_j$ and the
$E_{2j}$ of \eqref{eq:E2j} as elements of the polynomial ring
\[
\cR:=\Q[m,\zeta,\mu,s_1,s_2],
\]
in which $m$ is an indeterminate as well. In the scripts $\zeta$ is called \texttt{a0}.

\subsubsection*{The collision condition}
\begin{lemma}\label{lem:star-monomial}
Let $m\ge3$ be odd and let $P=C^\mu H'$ be a solution as in
Proposition~\ref{prop:candidates}(ii) with $H'$ in \textup{(F4)} or \textup{(F5)}. Let $A_i$,
$i\ge1$, be the parameters of the factors of $H'$ other than $C_2$ \textup{(}one in
\textup{(F4)}, two in \textup{(F5))}, and $\varrho$ as in \eqref{eq:normalform}. Then
$A_i\neq A_0$ and
\begin{equation}\label{eq:star}
\frac{m-1}4+\mu\Bigl[\frac m2(1-A_0)-(m-1)\Bigr]-\frac{3\varrho}2-2mA_0-2m\sum_{i\ge1}A_i
+\frac m2(A_0-1)\sum_{i\ge1}\frac{A_i}{A_0-A_i}=0 .
\end{equation}
\end{lemma}
\begin{proof}
Here $P=C^\mu C_2K'$ with $K'=u^\varrho\prod_{i\ge1}(\alpha_iu^m+\beta_iv)$, and $\nu_2=1$ by
Theorem~\ref{thm:soft}(c). Hence no factor $\alpha_iu^m+\beta_iv$ is a multiple of $C_2$, that
is, $2^{m-1}\alpha_i+\beta_i\neq0$, which is equivalent to $A_i\neq A_0$. We apply
Lemma~\ref{lem:star} with $e=e_u$, $p=\bp(1)=(1,1)$, $w=(1,m)\in T_pV_m$ and $K_2=C^\mu K'$. The
part of lowest degree of $P$ is $P_d=(-v)^\mu(-2^{m-1}v)\,u^\varrho\prod_{i\ge1}\beta_iv$, where
$\beta_i\neq0$ by Proposition~\ref{prop:candidates}; thus $P_d(w)\neq0$, and $D_e^2P_d=0$ because
$\varrho\le1$. Moreover $K_2(2p)=(2^m-2)^\mu\,2^\varrho\prod_{i\ge1}(2^m\alpha_i+2\beta_i)\neq0$,
$D_eC(p)=m$, and $H(p)=H'(1,1)\neq0$ because $C\nmid H'$. In \eqref{eq:star-general},
$\Theta(p)=2\ell_0$, since $\pi_{\mu+j}(p;e_u)$ is the coefficient of $t^j$ in $\tilde f$;
$D_e^2C(p)/(4D_eC(p))=(m-1)/4$; and, with $2^{m-1}=(A_0-1)/A_0$,
$\alpha_i=A_i(\alpha_i+\beta_i)$, $\beta_i=(1-A_i)(\alpha_i+\beta_i)$,
\[
\frac{D_eK_2(2p)}{K_2(2p)}=\frac{\mu m2^{m-1}}{2^m-2}+\frac\varrho2+\sum_{i\ge1}\frac{m\alpha_i2^{m-1}}{2^m\alpha_i+2\beta_i}
=\frac{\mu m}2(1-A_0)+\frac\varrho2+\frac m2(A_0-1)\sum_{i\ge1}\frac{A_i}{A_0-A_i}.
\]
Inserting $\ell_0$ from \eqref{eq:l0l2}, with the parameters $A_0$ and $A_i$, gives
\eqref{eq:star}.
\end{proof}
Multiplying \eqref{eq:star} by $4\prod_{i\ge1}(A_0-A_i)$ we obtain $\mathrm{St}=0$, where
$\mathrm{St}\in\cR$ is defined, with
$B:=(m-1)-4\mu(m-1)+2\mu m(1-\zeta)-6\varrho-8m\zeta$, by
\begin{equation}\label{eq:St}
\mathrm{St}:=\begin{cases}
(B-8ms_1)(\zeta-s_1)+2m(\zeta-1)s_1&\textup{(F4)},\\[2pt]
(B-8ms_1)(\zeta^2-\zeta s_1+s_2)+2m(\zeta-1)(\zeta s_1-2s_2)&\textup{(F5)}.
\end{cases}
\end{equation}

\subsubsection*{The elimination}
\begin{proposition}[computer-assisted]\label{prop:uniform}
Let $m\ge9$ be an odd integer and $\zeta:=A_0=-1/(2^{m-1}-1)$. There are no complex numbers
$\mu\neq0$, $s_1$, $s_2$ such that one of the following holds for the coefficients
\eqref{eq:E2j} of the series \eqref{eq:ell-normal}:
\begin{itemize}
\item[\textup{(F2)}] $E_2=E_4=E_6=0$, where $\varrho=0$ and the only parameter is $A=s_1$;
\item[\textup{(F4)}] $E_2=E_4=E_6=0$, where $\varrho=1$ and the parameters are $A_0$ and $A=s_1$;
\item[\textup{(F5)}] $E_2=E_4=E_6=0$ and $\mathrm{St}=0$, where $\varrho=0$ and the parameters are
$A_0,A_1,A_2$ with $A_1+A_2=s_1$, $A_1A_2=s_2$.
\end{itemize}
\end{proposition}

For a polynomial $f$ with integer coefficients we write $\|f\|_1$ for the sum of the absolute
values of its coefficients; $\|fg\|_1\le\|f\|_1\|g\|_1$ and $|f(m)|\le\|f\|_1m^{\deg f}$ for
$f\in\Z[m]$ and integers $m\ge1$. The proof of the proposition ends with the following two
lemmas.

\begin{lemma}[Size estimate]\label{lem:size}
Let $R=\sum_{k=k_0}^Kr_k(m)\,\zeta^k$ with $r_k\in\Z[m]$ and $r_{k_0}\neq0$, let $S,D\ge0$ be
integers with $\sum_{k>k_0}\|r_k\|_1\le S$ and $\deg r_k\le D$ for all $k>k_0$, and let
$M_0\ge2D+3$ be an integer with $2^{M_0-1}-1>S\,M_0^{\,D}$. Then
$R\bigl(m,-1/(2^{m-1}-1)\bigr)\neq0$ for every integer $m\ge M_0$ with $r_{k_0}(m)\neq0$.
\end{lemma}
\begin{proof}
If $M\ge M_0$ and $2^{M-1}-1>SM^D$, then
$2^M-1>2(2^{M-1}-1)>2SM^D\ge S(M+1)^D$, because $(1+1/M)^D\le e^{D/M}<2$. Hence
$T=2^{m-1}-1>Sm^D$ for all $m\ge M_0$. Now
$T^KR(m,-1/T)=\sum_k(-1)^kr_k(m)T^{K-k}$; the term with $k=k_0$ has absolute value at least
$T^{K-k_0}$, and the sum of the other terms has absolute value at most
$T^{K-k_0-1}\sum_{k>k_0}\|r_k\|_1m^D\le T^{K-k_0-1}Sm^D<T^{K-k_0}$.
\end{proof}

\begin{lemma}[Unexpanded resultants]\label{lem:pair}
Let $h,f\in\Z[m,\zeta][x]$ have the degrees $d_h,d_f\ge1$ in $x$, and let
$R:=\operatorname{Res}_x(h,f)\in\Z[m,\zeta]$.
\begin{enumerate}
\item[\textup{(i)}] $\|R\|_1\le\|h\|_1^{\,d_f}\,\|f\|_1^{\,d_h}$ and
$\deg_mR\le d_f\deg_mh+d_h\deg_mf$.
\item[\textup{(ii)}] Let $k\ge0$, and let $h^{(k)},f^{(k)}$ be obtained from $h,f$ by deleting
all terms divisible by $\zeta^{k+1}$. If $h^{(k)}$ and $f^{(k)}$ still have the degrees $d_h$
and $d_f$ in $x$, then $R\equiv\operatorname{Res}_x(h^{(k)},f^{(k)})$ modulo $\zeta^{k+1}$.
\item[\textup{(iii)}] Let $m$ be an integer with $T=2^{m-1}-1\neq0$, let $p$ be a prime not
dividing $T$, and let $\bar h,\bar f\in\mathbb F_p[x]$ be the images of $h,f$ under
$m\mapsto m\bmod p$, $\zeta\mapsto-T^{-1}\bmod p$. If the determinant of the Sylvester matrix
of $\bar h,\bar f$, formed with the degrees $d_h,d_f$, is not zero, then $R(m,-1/T)\neq0$, and
the polynomials $h(m,-1/T,x)$ and $f(m,-1/T,x)$ have no common zero.
\end{enumerate}
\end{lemma}
\begin{proof}
$R$ is the determinant of the Sylvester matrix $\mathcal S$ of size $d_h+d_f$: each of $d_f$
rows contains the coefficients of $h$ with respect to $x$, each of the other $d_h$ rows those
of $f$, and the remaining entries are zero. (i) By the Leibniz formula and the properties of
$\|\cdot\|_1$, $\|\det\mathcal S\|_1\le\prod_i\sum_j\|\mathcal S_{ij}\|_1=\|h\|_1^{d_f}\|f\|_1^{d_h}$,
and every product in the Leibniz formula has degree at most $d_f\deg_mh+d_h\deg_mf$ in $m$.
(ii) The two Sylvester matrices have the same size and are congruent modulo $\zeta^{k+1}$.
(iii) Let $K:=\deg_\zeta R$. The integer $T^KR(m,-1/T)$ is congruent modulo $p$ to $T^K$ times
the determinant in question, because determinants commute with ring homomorphisms; hence it is
not zero. Finally $R=ah+bf$ with $a,b\in\Z[m,\zeta][x]$, so $R$ vanishes at $(m,-1/T)$ if the
two specialized polynomials have a common zero.
\end{proof}

\begin{proof}[Proof of Proposition~\ref{prop:uniform}]
\emph{Step 1: cases.} By Lemma~\ref{lem:tower} with $J=3$ it suffices to exclude the cases
$(\mathrm T_1),\dots,(\mathrm T_4)$. Each of them consists of an equation $h=0$ and of two
further equations $g_1=g_2=0$, namely
\[
\begin{array}{llll}
(\mathrm T_1):&h=\mu+1-\ell_0,&(g_1,g_2)=(E_4,E_6);\qquad&\\
(\mathrm T_2):&h=\mu+2-\ell_0,&(g_1,g_2)=(\ell_2,E_6);&\\
(\mathrm T_3):&h=\mu+3-\ell_0,&(g_1,g_2)=(\ell_2,\ell_4);&\\
(\mathrm T_4):&h=\ell_2,&(g_1,g_2)=(\ell_4,\ell_6),&
\end{array}
\]
and in \textup{(F5)} there is the third equation $g_3:=\mathrm{St}=0$. These are elements of
$\cR$. By \eqref{eq:l0l2}, $h$ has degree one in $\mu$, and its coefficient of $\mu$ is
$-(m-3)/2$ in $(\mathrm T_1)$--$(\mathrm T_3)$ and $-(m-1)(m-3)/8$ in $(\mathrm T_4)$.

\emph{Step 2: elimination.} We use resultants of elements of $\cR$ with respect to one of the
variables $x\in\{\mu,s_2,s_1\}$ and two standard facts.
\begin{itemize}
\item[(R1)] $\operatorname{Res}_x(f,g)=af+bg$ with $a,b\in\cR$ (multiply the Sylvester matrix
by its adjugate). Hence $\operatorname{Res}_x(f,g)$ vanishes at every point of $\C^5$ at which
$f$ and $g$ vanish.
\item[(R2)] If $f,g\in\cR$ are irreducible, not associated, and both contain $x$, then
$\operatorname{Res}_x(f,g)\neq0$ (otherwise they would have a common factor containing $x$, by
Gauss's lemma).
\end{itemize}
For each family and each case the following was computed.
\begin{itemize}
\item[(E1)] $q_i:=\operatorname{Res}_\mu(h,g_i)\in\Q[m,\zeta,s_1,s_2]$ for $i=1,2$, and $i=3$ in
\textup{(F5)}.
\item[(E2)] Only in \textup{(F5)}: let $q_\sigma$ be the first of $q_1,q_2,q_3$ of lowest
positive degree in $s_2$ \textup{(}$\sigma=3$ in $(\mathrm T_1)$, $\sigma=1$ otherwise\textup{)}.
It turns out that $q_\sigma=c\,p_2$, where $c\in\Q[m]$ and $p_2$ is irreducible, of degree one in
$s_2$ in $(\mathrm T_1)$--$(\mathrm T_3)$ and of degree two in $(\mathrm T_4)$. Let $q',q''$ be the two resultants
$\operatorname{Res}_{s_2}(p_2,q_i)$, $i\neq\sigma$. In \textup{(F2)} and \textup{(F4)} let
$q':=q_1$, $q'':=q_2$.
\item[(E3)] The factorizations $q'=c'\prod_if'_i$, $q''=c''\prod_jf''_j$ over $\Q$, where
$c',c''\in\Q[m]$ and the $f'_i,f''_j\in\Z[m,\zeta,s_1]$ are irreducible and contain $s_1$; and,
for all $i,j$ with $f'_i\neq\pm f''_j$, the resultant
$R_{ij}:=\operatorname{Res}_{s_1}(f'_i,f''_j)\in\Z[m,\zeta]$, which is not zero by (R2). In one
instance $R_{ij}$ was not expanded, see below.
\item[(E4)] A common factor $L:=f'_i=\pm f''_j$ occurs twice, in $(\mathrm T_4)$ of \textup{(F2)}
with $L=s_1$, and in $(\mathrm T_4)$ of \textup{(F5)} with $L=ms_1-m+1$. In the latter case let
$t_i:=\operatorname{Res}_{s_1}(L,q_i)$, $i=1,2,3$. It turns out that $t_i=c_i\,\tau_i$ with
$c_i\in\Q[m]$ and $\tau_i\in\Z[m,\zeta,s_2]$ irreducible of degree $1,2,1$ in $s_2$; let
$R^L_{ij}:=\operatorname{Res}_{s_2}(\tau_i,\tau_j)\neq0$ for $i<j$.
\end{itemize}
In all cases the polynomials $c,c',c'',c_i\in\Q[m]$ are products of constants and of powers of
$m$, $m-1$ and $m-3$; they do not vanish for $m\ge9$.

Now suppose that $(m,\zeta,\mu,s_1,s_2)$ with $m\ge9$ odd, $\zeta=A_0(m)$ and $\mu\neq0$ is a
solution of one of the cases. By (R1) the $q_i$ vanish at this point; in \textup{(F5)},
$p_2$ vanishes there because $c(m)\neq0$, and so do $q'$ and $q''$ by (R1). Hence some $f'_i$
and some $f''_j$ vanish at the point. If $f'_i\neq\pm f''_j$, then $R_{ij}(m,A_0(m))=0$ by (R1).
If $f'_i=\pm f''_j=L$ in \textup{(F2)}, then $s_1=0$ and
$h=\ell_2|_{s_1=0}=-\mu(m-1)(m-3)/8\neq0$, a contradiction. If $f'_i=\pm f''_j=L$ in
\textup{(F5)}, then $t_1,t_2,t_3$ vanish at the point by (R1), hence so do
$\tau_1,\tau_2,\tau_3$, and $R^L_{12}$, $R^L_{13}$, $R^L_{23}$ vanish at $(m,A_0(m))$.

Consequently the proposition is proved once we know that for every odd $m\ge9$ none of the
polynomials $R_{ij}$, and not all of $R^L_{12},R^L_{13},R^L_{23}$, vanish at $(m,A_0(m))$.
These polynomials are listed in Table~\ref{tab:uniform}. In $(\mathrm T_4)$ of \textup{(F5)},
$q'=c'Lf_{15}$ and $q''=c''Lf_{11}$, where $f_{15}$ and $f_{11}$ have the degrees $15$ and $11$
in $s_1$ and $1130$ and $640$ terms; the resultant
$R^\ast:=\operatorname{Res}_{s_1}(f_{15},f_{11})$ was not expanded.

\begin{table}[ht]
\caption{The resultants of \textup{(E3)}, \textup{(E4)}. The second column gives the degrees in
$s_1$ of the two factors \textup{(}in $s_2$ for the $R^L_{ij}$\textup{)}; $(\deg_m,\deg_\zeta)$
are the degrees of the resultant, $k_0$ is its lowest power of $\zeta$, and $M_0$ is the
threshold of Lemma~\ref{lem:size}.}\label{tab:uniform}
\renewcommand{\arraystretch}{1.1}
\begin{tabular}{@{}llccrcr@{}}
\toprule
family, case & factors & $(\deg_m,\deg_\zeta)$ & terms & & $k_0$ & $M_0$\\
\midrule
\textup{(F2)}, $(\mathrm T_1)$ & $(4,6)$ & $(57,0)$ & $34$ &&&\\
\textup{(F2)}, $(\mathrm T_2)$ & $(3,7)$ & $(44,0)$ & $24$ &&&\\
\textup{(F2)}, $(\mathrm T_3)$ & $(3,5)$ & $(29,0)$ & $15$ &&&\\
\textup{(F2)}, $(\mathrm T_4)$ & $(1,6)$, $(4,1)$, $(4,6)$ & $(7,0)$, $(5,0)$, $(55,0)$ & $8$, $6$, $32$ &&&\\
\midrule
\textup{(F4)}, $(\mathrm T_1)$ & $(4,6)$ & $(61,24)$ & $800$ && $2$ & $713$\\
\textup{(F4)}, $(\mathrm T_2)$ & $(3,7)$ & $(47,16)$ & $315$ && $0$ & $504$\\
\textup{(F4)}, $(\mathrm T_3)$ & $(3,5)$ & $(32,12)$ & $146$ && $0$ & $319$\\
\textup{(F4)}, $(\mathrm T_4)$ & $(5,7)$ & $(81,30)$ & $976$ && $0$ & $956$\\
\midrule
\textup{(F5)}, $(\mathrm T_1)$ & $(6,9)$ & $(180,92)$ & $6749$ && $6$ & $2422$\\
\textup{(F5)}, $(\mathrm T_2)$ & $(1,4)$ & $(9,5)$ & $20$ && $0$ & $65$\\
 & $(8,4)$ & $(84,41)$ & $1341$ && $0$ & $984$\\
\textup{(F5)}, $(\mathrm T_3)$ & $(1,4)$ & $(9,5)$ & $20$ && $0$ & $66$\\
 & $(6,4)$ & $(58,27)$ & $586$ && $0$ & $641$\\
\textup{(F5)}, $(\mathrm T_4)$ & $(15,1)$ & $(32,15)$ & $162$ && $1$ & $315$\\
 & $(1,11)$ & $(24,13)$ & $113$ && $0$ & $218$\\
 & $(15,11)$: $R^\ast$ & not expanded & && $2$ & $5937$\\
 & $R^L_{12}$: $(1,2)$ & $(19,10)$ & $118$ && $0$ & $183$\\
 & $R^L_{13}$: $(1,1)$ & $(11,9)$ & $62$ && $0$ & $94$\\
 & $R^L_{23}$: $(2,1)$ & $(18,15)$ & $146$ && $0$ & $170$\\
\bottomrule
\end{tabular}
\end{table}

\emph{Step 3: the resultants do not vanish.} \textup{(F2)}: the six resultants are polynomials in
$m$ alone. Their factorizations over $\Q$ consist of powers of $m$, $m\pm1$, $m-2$, $m-3$,
$m-4$, $m-5$, $2m-1$ and of one irreducible factor of degree $27$, $19$, $11$, $5$, $3$, $19$,
respectively; hence they have no integer zero $m\ge6$.

\textup{(F4)}, \textup{(F5)}, expanded resultants: for each $R=\sum_{k\ge k_0}r_k\zeta^k$ of
Table~\ref{tab:uniform} let $S:=\sum_{k>k_0}\|r_k\|_1$, $D:=\max_{k>k_0}\deg r_k$, and let $M_0$
be the threshold given in the table; it satisfies the hypotheses of Lemma~\ref{lem:size}. In
every case $r_{k_0}$ has no integer zero $m\ge M_0$, so $R(m,A_0(m))\neq0$ for all $m\ge M_0$.
For the odd $m$ with $9\le m<M_0$ the integer $T^KR(m,-1/T)$, $K=\deg_\zeta R$, was computed;
it is never zero.

The resultant $R^\ast$: by Lemma~\ref{lem:pair}(ii) with $k=2$, for which the degrees in $s_1$
do not drop, $R^\ast\equiv r_2(m)\,\zeta^2$ modulo $\zeta^3$ with a polynomial $r_2\in\Z[m]$
of degree $367$; so $k_0=2$. The polynomials $f_{15}$ and $f_{11}$ have the degrees $17$ and
$13$ in $m$, and
\[
\|f_{15}\|_1=21833968993773920,\qquad\|f_{11}\|_1=109636746304 .
\]
By Lemma~\ref{lem:pair}(i),
\[
\|R^\ast\|_1\le S_\ast:=\|f_{15}\|_1^{11}\,\|f_{11}\|_1^{15}<2^{1148}\qquad\text{and}\qquad
\deg_mR^\ast\le11\cdot17+15\cdot13=382 .
\]
A comparison of integers shows that $2^{5936}-1>S_\ast\cdot5937^{382}$; hence
Lemma~\ref{lem:size} applies with $S=S_\ast$, $D=382$ and $M_0=5937$. (The exact value of
$S_\ast$ is needed here: with $2^{1148}$ in place of $S_\ast$ the inequality fails for this
$M_0$.) Moreover $r_2$ has no
integer zero $m\ge M_0$ \textup{(}all its complex zeros have absolute value smaller than
$640$\textup{)}. For the $2964$ odd $m$ with $9\le m<5937$ the determinant of
Lemma~\ref{lem:pair}(iii) was computed with $p=2^{61}-1$, and with $p=2^{89}-1$ for the $48$
values $m\equiv1\pmod{122}$, for which $2^{61}-1$ divides $T$; it is not zero in any case.
Hence $f_{15}$ and $f_{11}$ have no common zero at $(m,A_0(m))$ for any odd $m\ge9$, which is
what is needed in place of $R^\ast(m,A_0(m))\neq0$.
\end{proof}

\begin{proof}[Proof of Theorem~\ref{thm:monomial}\,\textup{(iii)} for $m\ge9$]
Let $m\ge9$ be odd and suppose that $P=C^\mu H'$ is a solution as in
Proposition~\ref{prop:candidates}(ii). By the part of \textup{(iii)} proved above, $H'$ is not
in \textup{(F1)} or \textup{(F3)}. By Theorem~\ref{thm:soft}(b) and Lemma~\ref{lem:doublesoft},
$G(t)=O(t^{m-1})=O(t^8)$, that is, $E_2=E_4=E_6=0$ for the series \eqref{eq:ell-normal} of
$P$, and in \textup{(F5)} also $\mathrm{St}=0$ by Lemma~\ref{lem:star-monomial}. Since
$\mu\ge1$, this contradicts Proposition~\ref{prop:uniform}.
\end{proof}

\begin{remark}\label{rem:uniform}
(i) Proposition~\ref{prop:uniform} does not use that $\mu$ is an odd integer; only $\mu\neq0$
enters. For $9\le m\le25$ it overlaps with the table of Section~\ref{sec:remaining}, which was
obtained in a different way (other points of the curve, Gr\"obner bases for each $m$
separately).

(ii) In \textup{(F5)} the three conditions $E_2=E_4=E_6=0$ do not suffice for our argument,
since there are three unknowns $\mu,s_1,s_2$; the condition $E_8=0$ is available only for
$m\ge11$, and \eqref{eq:star} takes its place. In \textup{(F4)} the condition $\mathrm{St}=0$
is not needed.

(iii) The computations of this subsection were carried out twice. The polynomials $\ell_j$,
$E_{2j}$ and $\mathrm{St}$ were generated independently with python-flint and with SymPy (in
the latter case $\mathrm{St}$ from \eqref{eq:star-general} by symbolic differentiation), and
steps \textup{(E1)}--\textup{(E4)} were carried out with python-flint and with Singular. The
polynomials recorded by the two computations, including $f_{15}$, $f_{11}$ and $r_2$, coincide
term by term up to sign. Step~3 was repeated in plain integer arithmetic. In addition the same conditions
were eliminated in another order of the unknowns ($s_1$ before $\mu$ in \textup{(F2)},
\textup{(F4)}; $s_2,\mu,s_1$ in \textup{(F5)}) by a general recursive procedure, which leads to
different resultants and to the same conclusion. See Appendix~\ref{app:certificates}.
\end{remark}

\subsection{Graphs of degree five, and the proof of Theorem~\ref{thm:graphs}}\label{sec:quintic}
\begin{lemma}\label{lem:quintic}
Let $b\in\C$, $\bp_b(k):=(k,k^5+bk^3)$, $C_b:=u^5+bu^3-v$, $H_5:=u^5+2bu^3+4v$ and
$P:=C_b\,(\alpha u+\beta H_5)$ with $(\alpha,\beta)\neq(0,0)$. Put
\[
R_0(s,t):=(s-t)^2\bigl[5(s^2-st+t^2)+3b\bigr],\qquad E(s,t):=\alpha+\beta\,(s^2+t^2)\bigl[5(s^2+t^2)+6b\bigr].
\]
\begin{enumerate}
\item[\textup{(i)}] $P(\bp_b(s)-\bp_b(t))=-st\,R_0(s,t)\,E(s,t)$; hence the two-soliton phase factor
is, for all $(\alpha,\beta)$,
\begin{equation}\label{eq:quintic-phase}
-\frac{P(\bp_b(s)-\bp_b(t))}{P(\bp_b(s)+\bp_b(t))}=\frac{R_0(s,t)}{R_0(s,-t)}
=\frac{(s-t)^2}{(s+t)^2}\cdot\frac{5(s^2-st+t^2)+3b}{5(s^2+st+t^2)+3b}.
\end{equation}
\item[\textup{(ii)}] For $N\ge2$,
$S_N[P](\bp_b(k_1),\dots,\bp_b(k_N))=(-1)^{N(N-1)/2}\bigl(\prod_ik_i\bigr)^{N-1}\prod_{i<j}E(k_i,k_j)\cdot F_N(k)$,
where
\[
F_N(k):=\sum_{\sg\in\{\pm1\}^N}P\Bigl(\sum_i\bp_b(\sg_ik_i)\Bigr)\prod_{i<j}R_0(\sg_ik_i,\sg_jk_j).
\]
\item[\textup{(iii)}] $F_N=0$ for every $N\ge1$. Hence $P$ satisfies the $N$-soliton condition on
$V(C_b)$ for every $N$, and $C_b$ is a simple factor of $P$.
\end{enumerate}
\end{lemma}
\begin{proof}
(i) One computes $C_b(\bp_b(s)-\bp_b(t))=-st(s-t)[5(s^2-st+t^2)+3b]$ and
$H_5(\bp_b(s)-\bp_b(t))=(s-t)(s^2+t^2)[5(s^2+t^2)+6b]$. The formula for the phase factor follows
because $\bp_b(s)+\bp_b(t)=\bp_b(s)-\bp_b(-t)$ and $E$ is even in $t$.

(ii) Put $z_i:=\sg_ik_i$. Since $\sg_i\bp_b(k_i)=\bp_b(z_i)$ and $E$ is even in both arguments,
(i) gives $\sg_i\sg_jP(\sg_i\bp_b(k_i)-\sg_j\bp_b(k_j))=-k_ik_jR_0(z_i,z_j)E(k_i,k_j)$.

(iii) $F_N$ is linear in $P$, so it suffices to treat $P_1:=uC_b$ and $P_2:=C_bH_5$. Give
$k_1,\dots,k_N$ the weight $1$ and $b$ the weight $2$. Then $F_N$ is a polynomial in $k$ and
$b$ with rational coefficients, weighted homogeneous of weight $6+2N(N-1)$ for $P_1$ and
$10+2N(N-1)$ for $P_2$, symmetric in the $k_i$ and even in each $k_i$. We have
$R_0(s,t)=R_0(t,s)=R_0(-s,-t)$, $R_0(s,s)=0$ and $R_0(z,0)=z^2(5z^2+3b)$, and exactly as in the
proof of Proposition~\ref{prop:sextic},
\begin{align*}
F_N(k_1,\dots,k_{N-1},0)&=2\prod_{i<N}k_i^2(5k_i^2+3b)\cdot F_{N-1}(k_1,\dots,k_{N-1}),\\
F_N(k_1,\dots,k_{N-2},s,s)&=2R_0(s,-s)\prod_{i\le N-2}R_0(s,k_i)R_0(-s,k_i)\cdot F_{N-2}(k_1,\dots,k_{N-2}).
\end{align*}
Moreover $F_1=2P(\bp_b(k_1))=0$, and $F_2=0$ by (ii) because $S_2$ vanishes identically for even
polynomials. If $F_{N-1}=F_{N-2}=0$ for some $N\ge3$, the argument of loc.\ cit.\ shows that
$\cD_N:=\prod_ik_i^2\prod_{i<j}(k_i^2-k_j^2)^2$, of weight $2N^2$, divides $F_N$. The quotient
$Q_N$ is a symmetric polynomial in the $k_i^2$ and in $b$, weighted homogeneous of weight
$6-2N$ for $P_1$ and $10-2N$ for $P_2$. Hence, with $e_1:=\sum_ik_i^2$ and
$e_2:=\sum_{i<j}k_i^2k_j^2$: for $P_1$, $Q_3$ is a constant and $Q_N=0$ for $N\ge4$; for $P_2$,
$Q_3\in\Span(b^2,be_1,e_1^2,e_2)$, $Q_4\in\Span(b,e_1)$, $Q_5$ is a constant and $Q_N=0$ for
$N\ge6$. The values of $F_3$, $F_4$, $F_5$ for $P_1$ and for $P_2$ were computed exactly at
$7$, $4$ and $2$ rational points $(k,b)$ with $\cD_N(k)\neq0$, at which the vectors
$(b^2,be_1,e_1^2,e_2)$, respectively $(b,e_1)$, span spaces of dimension $4$, respectively $2$;
all values are zero (the points and two nonvanishing determinants are given in
Appendix~\ref{app:certificates}). Hence $Q_3=0$, then $Q_4=0$, then $Q_5=0$, and
$F_N=0$ for all $N$ by induction. Finally the restriction of $\alpha u+\beta H_5$ to $V(C_b)$ is
$\alpha k+\beta(5k^5+6bk^3)\not\equiv0$, so $C_b$ is a simple factor.
\end{proof}

The range $N\le5$ of the conditions which have to be verified for $P_2=C_bH_5$ is the one given
by Newell and Zeng \cite{NewellZeng1986} for a polynomial of weight ten whose phase factor is
that of a polynomial of weight six (Section~\ref{subsec:known}, with $L=5$ and $M=3$); the
parameter $b$ of weight two is not present in \cite{NewellZeng1986}, but the argument is the
same.

\begin{proof}[Proof of Theorem~\ref{thm:graphs}]
\emph{Two variables, the top weighted part.} Let $s=2$, $q=\sum_ja_ju^j$ with $a_m\neq0$, and
give $u,v$ the weights $1,m$. Let $P^{\mathrm{top}}$ be the weighted homogeneous part of $P$ of
highest weight $w$. We apply the degeneration principle (Lemma~\ref{lem:degeneration}) with
$A_\eps(X,Y):=(\eps^{-1}X,\eps^{-m}Y)$: the coefficients of $\eps^wP\circ A_\eps$ are
polynomials in $\eps$, its value at $\eps=0$ is $P^{\mathrm{top}}$, and for a point
$(X,a_mX^m)$ of $W:=\{Y=a_mX^m\}$ the points $z(\eps):=(X,\eps^mq(X/\eps))$ tend to it and
satisfy $A_\eps z(\eps)\in V(C)$. Hence $S_3[P^{\mathrm{top}}]$ vanishes on $W^3$, and
$P^{\mathrm{top}}(u,a_mv)$ is a nonzero even weighted homogeneous polynomial as in
Proposition~\ref{prop:candidates}. By Theorem~\ref{thm:monomial}(ii),(iii), $m=5$, which is
(a) for $s=2$, and by Theorem~\ref{thm:monomial}(i)
\begin{equation}\label{eq:top}
P^{\mathrm{top}}=\lambda\,u\,(a_5u^5-v)\qquad\text{or}\qquad P^{\mathrm{top}}=\lambda\,(a_5u^5-v)(a_5u^5+4v),\qquad\lambda\neq0 .
\end{equation}

\emph{Two variables, $m=5$.} \emph{Step 1: $q$ is odd.} The polynomial $C^-:=C(-u,-v)=q(-u)+v$
is irreducible and divides the even polynomial $P$. If $q$ were not odd, $C^-$ would not be
associated to $C$, so $CC^-$ would divide $P$, and the product $-(a_5u^5-v)^2$ of the top parts
of $C$ and $C^-$ would divide $P^{\mathrm{top}}$, contradicting \eqref{eq:top}.

\emph{Step 2: normalization.} After the linear change of variables
$(u,v)\mapsto(u,(v-a_1u)/a_5)$ we have $q=u^5+bu^3$ with $b=a_3/a_5$, $C=C_b$, and
\eqref{eq:quintic-solutions} becomes $C_b(\alpha u+\beta H_5)$ with new constants
$(\alpha,\beta)\neq(0,0)$.

\emph{Step 3.} Write $P=C_bH$. Then $H$ is odd, and by \eqref{eq:top} its top weighted part is
$\lambda u$ or $\lambda(u^5+4v)$. An odd polynomial of weighted degree at most five is a
combination of $u,u^3,u^5,v$. Hence $H=\lambda u$, or, after division by $\lambda$,
$H=u^5+4v+c_3u^3+c_1u$.

\emph{Step 4: $c_3=2b$.} Let $H=u^5+4v+c_3u^3+c_1u$ and give $k_i$ the weight $1$, $b$ and $c_3$
the weight $2$ and $c_1$ the weight $4$. Then $S_3[C_bH](\bp_b(k_1),\bp_b(k_2),\bp_b(k_3))$ is
weighted homogeneous of weight $40$, so its part of degree $38$ in $k$ is $b\,X(k)+c_3\,Y(k)$
with polynomials $X,Y$ which do not depend on the parameters. By Lemma~\ref{lem:quintic},
$c_3=2b$ gives solutions for every $b$, so $X=-2Y$ and the part of degree $38$ is $(c_3-2b)Y$.
Here $Y$ is the derivative at $\delta=0$ of $S_3[(u^5-v)(u^5+4v+\delta u^3)]$ on the graph
$v=u^5$, and $Y(1,2,3)=-940976870400000000000\neq0$ (Appendix~\ref{app:certificates}). Hence
$c_3=2b$ and $P=C_b(c_1u+H_5)$.

Together with Lemma~\ref{lem:quintic} this proves (b) for $s=2$.

\emph{Any number of variables.} Let $s\ge3$ and denote the coordinates by $(u,v,z)$. For
$\xi,\eta\in\C^{s-2}$ let $A_{\xi\eta}(u,v):=(u,v,\xi u+\eta v)$ as in \eqref{eq:planes}. Then
$P\circ A_{\xi\eta}$ is even, divisible by $C$, and satisfies $S_3=0$ on the curve $V(C)\subset\C^2$
(Lemma~\ref{lem:linear}); it is not zero for $(\xi,\eta)$ in a dense open set, because the planes
$A_{\xi\eta}(\C^2)$ cover $\{u\neq0\}$. This gives (a), that is, $m=5$. Then $q$ is odd, and after a
change of the coordinate $v$ we may assume $C=C_b$. Write $P=C^\mu H$ with $C\nmid H$ and
$H_{\xi\eta}:=H\circ A_{\xi\eta}$. For $(\xi,\eta)$ in a dense open set $H_{\xi\eta}$ does not
vanish identically on the curve, and the two-variable case applied to $C^\mu H_{\xi\eta}$ gives
$\mu=1$ and $H_{\xi\eta}\in\Span(u,H_5)$; this is a closed condition, so it holds for all
$(\xi,\eta)$. Now the last paragraph of the proof of Theorem~\ref{thm:cubic-all} applies verbatim: no
nonzero element of $\Span(u,H_5)$ is divisible by $v$, hence $H$ does not depend on $z$, and
$H\in\Span(u,H_5)$. The converse statements follow from Lemma~\ref{lem:quintic}, since $S_N[P]$
involves only the coordinates $u,v$ of the momenta when $P\in\C[u,v]$.
\end{proof}

\begin{proof}[Proof of Corollary~\ref{cor:consequences}]
(a) A linear automorphism of $\C^s$ maps $V(\ell_2-q(\ell_1))$ to $V(v-q(u))$ and preserves
evenness, multiplicities and the conditions $S_N=0$ (Lemma~\ref{lem:linear}); apply
Theorem~\ref{thm:graphs}.

(b) Choose linear coordinates in which the tangent line of $V$ at the origin is $v=0$; then $V$
is given near the origin by $v=\varphi(u)=cu^{m_0}+O(u^{m_0+1})$ with $c\neq0$. Give $u,v$ the
weights $1,m_0$. The part of $C$ of lowest weight is a multiple of $v-cu^{m_0}$, because
$C(u,\varphi(u))\equiv0$ and $C$ contains the monomial $v$. Let $P^{\mathrm{low}}\neq0$ be the
part of $P$ of lowest weight $w$. By Lemma~\ref{lem:degeneration} with $A_\eps(X,Y):=(\eps X,\eps^{m_0}Y)$
and the points $z(\eps):=(X,\eps^{-m_0}\varphi(\eps X))$, which tend to $(X,cX^{m_0})$,
$S_3[P^{\mathrm{low}}]$ vanishes on $\{Y=cX^{m_0}\}^3$; $P^{\mathrm{low}}$ is even, and
$P^{\mathrm{low}}(u,cv)$ is as in Proposition~\ref{prop:candidates} with $m=m_0$. By
Theorem~\ref{thm:monomial}(ii),(iii), $m_0$ is neither even and at least four nor odd and at
least seven. If $V\neq-V$ and $m_0$ is odd, then $CC^-$
divides $P$ (Corollary~\ref{cor:origin}(3)) and the parts of lowest weight of $C$ and $C^-$ are
proportional, so $v-cu^{m_0}$ is a multiple factor of $P^{\mathrm{low}}$; by
Theorem~\ref{thm:monomial}(i) this is impossible for $m_0=5$. For $m_0=3$,
Corollary~\ref{cor:origin}(3) gives $P=CC^-(C_2C_2^-)^{\nu_2}K$ with $K(0)\neq0$. The part of
lowest weight of a product is the product of the parts of lowest weight, and those of $C_2$ and
$C_2^-$ are multiples of $v-\frac c4u^3$; hence $v-cu^3$ divides $P^{\mathrm{low}}$ exactly
twice. This contradicts Lemma~\ref{lem:divis}, by which the multiplicity of $u^3-v$ in the
polynomial $P^{\mathrm{low}}(u,cv)$ is odd. Hence $m_0=2$ if $V\neq-V$.
\end{proof}

\begin{remark}[The BKP hierarchy]\label{rem:bkp}
The polynomials \eqref{eq:quintic-solutions} are known bilinear equations in the following
sense. The BKP hierarchy \cite{DJKM1982,JimboMiwa1983} consists of bilinear equations
$E(D_1,D_3,D_5,\dots)\,\tau\cdot\tau=0$ for one function $\tau$ of the variables
$x_1,x_3,x_5,\dots$, where $D_j$ is the Hirota operator of $x_j$ and has weight $j$
\cite[Proposition~6]{DJKM1982}. According
to \cite[Appendix~3]{JimboMiwa1983}, its equations of weight at most ten are given by the
polynomials
\begin{align*}
E_6&=D_1^6-5D_1^3D_3-5D_3^2+9D_1D_5,\\
E_8&=D_1^8+7D_1^5D_3-35D_1^2D_3^2-21D_1^3D_5-42D_3D_5+90D_1D_7,\\
E_{10}&=D_1^{10}+63D_1^5D_5-225D_1^3D_7-175D_1D_3^3+525D_1D_9-189D_5^2,\\
E_{10}'&=6D_1^5D_5-5D_1^4D_3^2-15D_1^3D_7+15D_1^2D_3D_5-5D_1D_3^3+10D_1D_9-15D_3D_7+9D_5^2,\\
E_{10}''&=D_1^7D_3-21D_1^2D_3D_5+35D_1D_9-15D_3D_7 .
\end{align*}
The polynomials $E_6$ and $E_8$ are already given in \cite[Section~5]{DJKM1982}.
The $N$-soliton solutions of the hierarchy \cite[Section~3]{DJKM1982} (see also
\cite[(0.4)]{DJKM1982b} and \cite[Section~3]{Hirota1989})
are, in the form used in \cite[proof of Theorem~5.7]{PartI}, Hirota's expressions
\[
\tau=\sum_{\mu\in\{0,1\}^N}\exp\Bigl(\sum_i\mu_i\theta_i\Bigr)\prod_{i<j}\mathcal A_{ij}^{\mu_i\mu_j},
\qquad\theta_i=\sum_{j\ \mathrm{odd}}(p_i^j+q_i^j)\,x_j+\delta_i,
\]
with generic parameters $p_i,q_i$, arbitrary $\delta_i$, and the phase factors
\[
\mathcal A_{ij}=\frac{(p_i-p_j)(p_i-q_j)(q_i-p_j)(q_i-q_j)}{(p_i+p_j)(p_i+q_j)(q_i+p_j)(q_i+q_j)} .
\]

(i) \emph{Reduction.} Let $\lambda\in\C$ and $q_\lambda(u):=u^5-5\lambda u^3-5\lambda^2u$.
Under the substitution $D_3=\lambda D_1$, $D_1=u$, $D_5=-v/9$ one has
\begin{align}
E_6&=u\,\bigl(q_\lambda(u)-v\bigr),\label{eq:bkp6}\\
E_{10}-15E_{10}'-\tfrac{75}{7}E_{10}''-\tfrac{30\lambda}{7}E_8
&=\bigl(q_\lambda(u)-v\bigr)\bigl(u^5-10\lambda u^3+4v\bigr),\label{eq:bkp10}
\end{align}
and the left-hand side of \eqref{eq:bkp10} is, up to a constant factor, the only linear
combination of $E_8,E_{10},E_{10}',E_{10}''$ from which $D_7$ and $D_9$ disappear under this
substitution. The right-hand sides are the polynomials \eqref{eq:quintic-solutions} for
$q=q_\lambda$ with $(\alpha,\beta)=(1,0)$ and $(0,1)$. For every odd quintic $q$ the
polynomial $q(u)-v$ is a constant multiple of $q_\lambda(u)-v'$ for a suitable $\lambda$ and a
new variable $v'=cv+c'u$ (as in Step~2 of the proof of Theorem~\ref{thm:graphs}), so that all
polynomials of Theorem~\ref{thm:graphs} arise in this way. For $\lambda=0$, \eqref{eq:bkp6} is
the Sawada--Kotera equation, which is the reduction $D_3=0$ of $E_6$
\cite[Sections~1 and~7]{DJKM1982}, and
\eqref{eq:bkp10} becomes $(u^5-v)(u^5+4v)$; for $\lambda\neq0$, \eqref{eq:bkp6} is the
combination of the Korteweg--de Vries and Sawada--Kotera equations, the reduction
$D_3=\lambda D_1$ of $E_6$ \cite{Hirota1989}.

(ii) \emph{Solitons.} If $p_i,q_i$ are the roots of $z^2-k_iz+(k_i^2-\lambda)/3$, then
$p_i+q_i=k_i$, $p_i^3+q_i^3=\lambda k_i$ and $-9(p_i^5+q_i^5)=q_\lambda(k_i)$. The soliton
solutions with these parameters depend on $x_1$ and $x_3$ only through $x_1+\lambda x_3$, so
that $D_3$ acts on them as $\lambda D_1$; their momenta in the variables $(u,v)$ are the
points $(k_i,q_\lambda(k_i))$ of $V(q_\lambda(u)-v)$, and their phase factors are
\eqref{eq:quintic-phase} with $b=-5\lambda$; for $\lambda=0$ one has $q_i=-\omega p_i$ with
$\omega^2+\omega+1=0$, and these are the solitons of the reduced hierarchy in
\cite[Section~7]{DJKM1982}. Since the soliton solutions satisfy all equations
of the hierarchy \cite[Section~3 and Proposition~6]{DJKM1982}, they satisfy the two reduced
equations \eqref{eq:bkp6} and \eqref{eq:bkp10}. This gives $N$-soliton solutions for the polynomials
\eqref{eq:quintic-solutions} and, by the criterion recalled in Section~\ref{sec:intro},
another proof of the $N$-soliton conditions of Lemma~\ref{lem:quintic}(iii); the proof given
above does not use the theory of the BKP hierarchy. Conversely, Theorem~\ref{thm:graphs} shows that \eqref{eq:bkp6} and
\eqref{eq:bkp10} span all bilinear equations in the two variables $D_1,D_5$ which are
satisfied by the three-soliton solutions of the reduced hierarchy. Indeed, let $P$ be a
nonzero even polynomial in $u=D_1$, $v=-9D_5$ with $P(0)=0$ such that $P(D)\tau\cdot\tau=0$
for these solutions, for generic $k_1,k_2,k_3$ and all $\delta_i$. The coefficients of
$e^{\theta_i}$ and of $e^{\theta_i+\theta_j}$ show that $P$ vanishes on $V(q_\lambda(u)-v)$
and, by Lemma~\ref{lem:dominance}, that $\mathcal A_{ij}$ is the phase factor of $P$; thus
$\tau$ is Hirota's three-soliton expression for $P$, so that $S_3[P]$ vanishes on
$V(q_\lambda(u)-v)^3$, and Theorem~\ref{thm:graphs} applies.

(iii) \emph{The second component.} For $\beta\neq0$ the dispersion variety of the polynomial
$P$ of \eqref{eq:quintic-solutions} has the second component $V(\tilde q(u)-v)$, where
$\tilde q:=-\tfrac14\bigl(a_5u^5+2a_3u^3+\tfrac\alpha\beta u\bigr)$, and $P$ does not satisfy
the \textup{3SC} on it. Indeed, $P=-4\beta\,(\tilde q(u)-v)(q(u)-v)$, and by
Theorem~\ref{thm:graphs}(b) for the quintic $\tilde q$ the 3SC on $V(\tilde q(u)-v)$ would
imply $-4\beta\,(q(u)-v)=\alpha'u+\beta'\bigl(-\tfrac14a_5u^5-a_3u^3+4v\bigr)$ with constants
$\alpha',\beta'$; comparing the coefficients of $v$ and of $u^5$ gives $\beta'=\beta$ and
$\beta'=16\beta$. Thus the polynomials with $\beta\neq0$ satisfy all $N$-soliton conditions on
one of their two components and not even the three-soliton condition on the other; in
particular they do not satisfy the 3SC on the whole dispersion variety, which is the condition
of \cite{HietarintaI}. This is not in conflict with their origin in an integrable hierarchy:
\eqref{eq:bkp10} is one of several equations for the same function $\tau$, and the momenta of
the solitons of the hierarchy lie on the first component.

The identities of (i) and (ii), the statements on the list of \cite{JimboMiwa1983} for one-
and two-soliton solutions, and instances of (iii) are verified by the script of Group~VIII
(Appendix~\ref{app:certificates}).
\end{remark}
\section{Translated hyperbolas}\label{sec:hyperbola}

Let $u,v$ be two of the coordinates of $\C^s$. Among the irreducible quadratic components, the
results of \cite{PartI} leave open exactly the translated hyperbolas $uv+au+bv+c=0$ with
$ab\neq0$ and $c\neq ab$ (the case $ab=0$, $(a,b)\neq(0,0)$ is excluded by
\cite[Proposition~9.20(i)]{PartI}, and $a=b=0$ is an even quadric). The substitution
$u\mapsto bu$, $v\mapsto av$ transforms $uv+au+bv+c$ into $ab\,C_h$ with $C_h$ as in
\eqref{eq:Ch}:
\[
C_h=uv+u+v+h=(u+1)(v+1)-(1-h),\qquad h=\frac{c}{ab}\neq1 .
\]
The curve $V(C_h)\subset\C^2$ is parametrized bijectively by
\begin{equation}\label{eq:ph}
\bp_h(t):=\Bigl(t-1,\ \frac{1-h}{t}-1\Bigr),\qquad t\in\C\setminus\{0\},
\end{equation}
and it passes through the origin if and only if $h=0$. Put $C_h^-(u,v):=C_h(-u,-v)$ and
\[
F_h:=C_h\,C_h^-=(uv+h)^2-(u+v)^2 .
\]
For $h\neq1$ the polynomials $C_h$ and $C_h^-$ are irreducible and not proportional, so every
even polynomial divisible by $C_h$ has the form $P=F_hH$ with $H$ even; in particular
$\deg P\ge4$. The exchange of $u$ and $v$ preserves $C_h$. All degrees in this section are
ordinary degrees. The reduction to two variables is always the same:

\begin{lemma}\label{lem:hyp-planes}
Let $P\neq0$ be an even polynomial on $\C^s$ which is divisible by $C_h$ and satisfies
$S_3[P]=0$ on $V(C_h)^3$. Then there are $\xi,\eta\in\C^{s-2}$ such that
$P\circ A_{\xi\eta}\in\C[u,v]$, with $A_{\xi\eta}$ as in \eqref{eq:planes}, is a nonzero even
polynomial of degree at most $\deg P$, divisible by $C_h$, with the same value at the origin,
which satisfies the \textup{3SC} on the curve $V(C_h)\subset\C^2$.
\end{lemma}
\begin{proof}
The planes $A_{\xi\eta}(\C^2)$ cover the dense set $\{u\neq0\}$, so $P\circ A_{\xi\eta}\neq0$ for
some $(\xi,\eta)$. The map $A_{\xi\eta}$ sends the curve $V(C_h)\subset\C^2$ into the
hypersurface $V(C_h)\subset\C^s$, so $P\circ A_{\xi\eta}$ vanishes on the curve, and the rest
follows from Lemma~\ref{lem:linear}(ii).
\end{proof}

\subsection{The hyperbola through the origin}

\begin{theorem}\label{thm:hzero}
There is no nonzero even polynomial $P$ on $\C^s$ which is divisible by $uv+u+v$ and for which
$S_3[P]$ vanishes identically on $V(uv+u+v)^3$.
\end{theorem}

No hypothesis is made on the degree of $P$ or on the multiplicity of the component. The proof
uses the classification of the parabolic components \cite[Theorem~9.17]{PartI} at one point;
nothing in Sections~\ref{sec:cones}--\ref{sec:cubic} depends on that theorem.

\begin{proof}
Suppose that such a $P$ exists. By Lemma~\ref{lem:hyp-planes} we may assume $s=2$, and then
$P=F_0H$ with $H\in\C[u,v]$ even and nonzero.

\emph{Step 1: coordinates.} Put $u=2(x+y)$, $v=2(y-x)$. Then $uv=4(y^2-x^2)$, $u+v=4y$,
\[
C_0=4\,(y-x^2+y^2),\qquad F_0=16\,F,\qquad F:=(x^2-y^2)^2-y^2 ,
\]
and by Lemma~\ref{lem:linear}(i) we may work with $P=F\cdot H(x,y)$, $H$ even in $(x,y)$, and
with the component $V_0:=V(y-x^2+y^2)$. Near the origin $V_0$ is parametrized by
\begin{equation}\label{eq:catalan}
x\mapsto\bigl(x,\ x^2\,\omega(x^2)\bigr),\qquad
\omega(\zeta):=\frac{\sqrt{1+4\zeta}-1}{2\zeta}=\sum_{k\ge0}(-1)^k\Cat_k\,\zeta^k,
\end{equation}
where $\Cat_k=\frac1{k+1}\binom{2k}{k}$ are the Catalan numbers and $\omega$ is the solution of
$\omega=1-\zeta\omega^2$ which is analytic at $\zeta=0$.

\emph{Step 2: the cofactor does not vanish at the origin.} Use the local weights $\wt x=1$,
$\wt y=2$, and let $H_m\neq0$ be the weighted homogeneous part of $H$ of lowest weight $m$. For
the maps $A_\eps(X,Y):=(\eps X,\eps^2Y)$ we have
\[
P\bigl(A_\eps(X,Y)\bigr)=\eps^{4}\bigl[(X^2-\eps^2Y^2)^2-Y^2\bigr]\cdot\eps^{m}\bigl[H_m(X,Y)+O(\eps)\bigr],
\]
so $P_\eps:=\eps^{-4-m}P\circ A_\eps$ has polynomial coefficients in $\eps$ and
$P_0=(X^4-Y^2)\,H_m(X,Y)$. For a point $z=(X,X^2)$ of the parabola $W:=V(Y-X^2)$ put
$z(\eps):=\bigl(X,\ X^2\omega(\eps^2X^2)\bigr)$; by \eqref{eq:catalan}, $A_\eps z(\eps)\in V_0$ for
small $\eps$, and $z(\eps)\to z$. By Lemma~\ref{lem:degeneration}, $S_3[P_0]$ vanishes on
$W^3$. Since the monomials $x^iy^j$ of the even polynomial $H$ have $i+j$ even, $H_m$ is even;
thus $P_0$ is a nonzero even polynomial with $P_0(0)=0$ which is divisible by $Y-X^2$ and
satisfies the 3SC on the parabola. By \cite[Theorem~9.17]{PartI}, $P_0=\gamma(X^4-Y^2)$. Hence
$H_m$ is a nonzero constant, $m=0$, and $H(0)\neq0$. We normalize $H(0)=1$. This step is what
makes the finite computation below conclusive for cofactors of arbitrary degree: the cofactor
cannot escape it by vanishing to high order at the origin.

\emph{Step 3: a gauge.} For a quadratic form $q$ on $\C^2$ and $r_1,r_2,r_3\in\C^2$,
\begin{equation}\label{eq:gauge}
q(r_1+r_2+r_3)+\sum_{i<j}q(r_i-r_j)=3\sum_iq(r_i).
\end{equation}
Write $H=1+H^{(2)}+\cdots$ with $H^{(2)}$ the quadratic part of $H$ (as $H$ is even, the omitted
terms have ordinary degree at least four), and let
\[
\widehat H:=H\cdot\exp\bigl(-H^{(2)}\bigr)\in\C[[x,y]].
\]
Then $\widehat H$ is an even formal power series with $\widehat H=1+(\text{terms of ordinary
degree}\ge4)$. By \eqref{eq:gauge} with $r_i=\sg_ip_i$, each summand of \eqref{eq:S3} for
$F\widehat H$ is $\exp\bigl(-3\sum_iH^{(2)}(p_i)\bigr)$ times the corresponding summand for
$FH$. We apply this to the points
\[
\mathbf q_i:=\bigl(\eps t_i,\ \eps^2V_\eps(t_i)\bigr),\qquad
V_\eps(t):=t^2\omega(\eps^2t^2)=\sum_{k\ge0}(-1)^k\Cat_k\,\eps^{2k}t^{2k+2},
\]
which lie on $V_0$ for small $\eps$ by \eqref{eq:catalan}. All arguments of $FH$ in
$S_3[FH](\mathbf q_1,\mathbf q_2,\mathbf q_3)=0$ are power series in $\eps$ without constant
term, so this identity holds in $\C[t_1,t_2,t_3][[\eps]]$ and remains valid for $\widehat H$:
\begin{equation}\label{eq:formal3SC}
\sum_{\sg}\bigl(F\widehat H\bigr)\Bigl(\sum_i\sg_i\mathbf q_i\Bigr)\prod_{i<j}\bigl(F\widehat H\bigr)(\sg_i\mathbf q_i-\sg_j\mathbf q_j)=0
\qquad\text{in }\C[t_1,t_2,t_3][[\eps]].
\end{equation}

\emph{Step 4: a finite jet computation.} Every argument in \eqref{eq:formal3SC} has the form
$(\eps U,\eps^2V)$ with $U\in\C[t]$ and $V\in\C[t][[\eps^2]]$, and
$F(\eps U,\eps^2V)=\eps^4\bigl[(U^2-\eps^2V^2)^2-V^2\bigr]$, while a monomial $x^iy^j$ of
$\widehat H$ contributes $\eps^{i+2j}U^iV^j$. Hence the left-hand side of \eqref{eq:formal3SC}
is $\eps^{16}\sum_{w\ge0}\eps^w\,\Xi_w(t)$, where $\Xi_w$ involves only the coefficients of the
monomials of $\widehat H$ of weight at most $w$, and all $\Xi_w$ must vanish. The monomials of
$\widehat H-1$ of weight at most $10$ are those $x^iy^j$ with $i+j$ even, $i+j\ge4$ and
$i+2j\le10$:
\begin{center}
\renewcommand{\arraystretch}{1.1}
\begin{tabular}{@{}c|ccccccc@{}}
weight & $4$ & $5$ & $6$ & $7$ & $8$ & $9$ & $10$\\
\hline
monomials & $x^4$ & $x^3y$ & $x^6,\ x^2y^2$ & $x^5y,\ xy^3$ & $x^8,\ x^4y^2,\ y^4$ & $x^7y,\ x^3y^3$ &
$x^{10},\ x^6y^2,\ x^2y^4$
\end{tabular}
\end{center}
(no evenness in $x$ and $y$ separately is assumed). The equations $\Xi_w(t)=0$ were evaluated
in exact arithmetic at the five triples
$t=(1,2,3)$, $(1,2,4)$, $(1,2,5)$, $(1,3,4)$, $(1,3,5)$, from the full sum over the eight sign
vectors (Appendix~\ref{app:certificates}). For $w=4,\dots,9$ they are linear in the coefficients
of weight $w$, once those of lower weight are known, and they force successively
\begin{align*}
\widehat H={}&1-\frac{x^4}{8}+\frac{x^6}{30}+a\,x^2y^2+\frac{23-32a}{2688}\,x^8+\frac{16a-37}{96}\,x^4y^2
+\frac{80a+73}{192}\,y^4\\
&+d\,x^{10}+e\,x^6y^2+f\,x^2y^4+(\text{terms of weight}\ge11),
\end{align*}
with one free parameter $a$ and with all coefficients of the weights $5$, $7$, $9$ equal to
zero. For $w=10$ the five equations are linear in $(a,d,e,f)$; after division of each row by a
nonzero rational number they read $\mathcal M\,(a,d,e,f,1)^{\mathsf T}=0$ with
\begin{equation}\label{eq:hyperM}
\setlength{\arraycolsep}{3pt}
\mathcal M=\begin{pmatrix}
40698&97734480&7092816&-72816&13855\\
11354&7624320&597072&-5664&-3355\\
15307558&6008178000&513840048&-4353072&-5945867\\
659134&1053096720&76333680&-701808&157593\\
37753506&28948603200&2204967696&-19461984&-6749999
\end{pmatrix},
\end{equation}
and $\det\mathcal M=77476915111730928188129280000\neq0$. Thus the system has no solution with
last coordinate $1$. Since monomials of weight at least $11$ do not enter $\Xi_w$ for
$w\le10$, no formal power series $\widehat H$ as above exists, and the theorem is proved.
\end{proof}

\subsection{Algebraicity of the parameter}

\begin{theorem}\label{thm:algebraic}
For every integer $D$ there is a finite set $E_D\subset\overline{\Q}\setminus\{0\}$ with the
following property: if $h\in\C\setminus\{1\}$ and $P$ is a nonzero even polynomial on $\C^s$
with $P(0)=0$ and $\deg P\le D$ which is divisible by $C_h$ and satisfies the \textup{3SC} on
$V(C_h)$, then $h\in E_D$. In particular, if such a polynomial $P$ of any degree exists, then
$h$ is an algebraic number.
\end{theorem}
\begin{proof}
By Lemma~\ref{lem:hyp-planes} it suffices to treat $s=2$, and we may assume $D\ge4$. Let
$a=(a_{ij})$ be the family of coefficients of an even polynomial of degree at most $D-4$,
\[
H_a:=\sum_{\substack{i+j\ \mathrm{even}\\ i+j\le D-4}}a_{ij}\,u^iv^j,\qquad P_{h,a}:=F_h\,H_a\in\Q[h,a][u,v],
\]
and regard $a\neq0$ up to scaling as a point $[a]$ of a projective space $\PP^N$. Substituting
the points \eqref{eq:ph} we obtain a Laurent polynomial
\[
S_3[P_{h,a}]\bigl(\bp_h(t_1),\bp_h(t_2),\bp_h(t_3)\bigr)=\sum_{\mu\in\Z^3}c_\mu(h,a)\,t_1^{\mu_1}t_2^{\mu_2}t_3^{\mu_3},
\]
a finite sum whose coefficients $c_\mu\in\Q[h,a]$ are homogeneous of degree four in $a$; the
only denominators are powers of the $t_i$, and none depends on $h$. Let
\[
Z_D:=\bigl\{(h,[a])\in\C\times\PP^N:\ c_\mu(h,a)=0\ \text{for all }\mu,\quad h^2a_{00}=0\bigr\}.
\]
If $h\neq1$ and $P$ is as in the theorem, then $P=F_hH_a$ for some $a\neq0$,
$P(0)=h^2a_{00}=0$, and the 3SC on $V(C_h)$ gives $c_\mu(h,a)=0$ for all $\mu$; hence
$(h,[a])\in Z_D$. Let $E_D\subset\C$ be the image of $Z_D$ under the projection to the first
factor.

The fibre of $Z_D$ over $h=0$ is empty: if $(0,[a])\in Z_D$, then $P:=F_0H_a$ is a nonzero even
polynomial divisible by $C_0$ for which $S_3[P]$ vanishes at all triples of points $\bp_0(t_i)$,
hence on $V(C_0)^3$, contradicting Theorem~\ref{thm:hzero}. Since $\PP^N$ is complete, the
projection $\C\times\PP^N\to\C$ maps Zariski closed sets to Zariski closed sets
\cite[Theorem~II.4.9]{Hartshorne1977}. Thus $E_D$ is a closed subset of the affine line which
does not contain $0$; therefore it is finite. Finally, $Z_D$ is defined by equations with
rational coefficients, so $E_D$ is invariant under all field automorphisms of $\C$; a finite
set with this property consists of algebraic numbers, because the orbit of a transcendental
number under the automorphisms of $\C$ is infinite.
\end{proof}

\begin{remark}\label{rem:ED}
The sets $E_D$ are not computed here, and we do not show that their union is finite; the
theorem excludes transcendental parameters but no algebraic $h\neq0$. For $h=1$ the
parametrization \eqref{eq:ph} describes only one line of the reducible curve $V(C_1)$; this is
harmless, since the incidence set $Z_D$ is only used to bound the admissible parameters from
above, and possible points of $Z_D$ over $h=1$ merely enlarge $E_D$.
\end{remark}

\subsection{Degree at most eight}

\begin{proposition}[Computer-assisted]\label{prop:degree8}
Let $h\neq1$. There is no nonzero even polynomial $P$ on $\C^s$ with $P(0)=0$ and
$\deg P\le8$ which is divisible by $C_h$ and satisfies the \textup{3SC} on $V(C_h)$.
\end{proposition}
\begin{proof}
By Lemma~\ref{lem:hyp-planes} we may assume $s=2$, and by Theorem~\ref{thm:hzero}, $h\neq0$.
Write $P=F_hH$ with
\[
H=e_0+a_{20}u^2+a_{11}uv+a_{02}v^2+a_{40}u^4+a_{31}u^3v+a_{22}u^2v^2+a_{13}uv^3+a_{04}v^4\neq0 .
\]
We use the three points $p_i:=\bp_h(r_i/\eps)=\bigl(r_i/\eps-1,\ (1-h)\eps/r_i-1\bigr)$ of
$V(C_h)$, which tend to infinity as $\eps\to0$; then $S_3[P](p_1,p_2,p_3)$ is a Laurent
polynomial in $\eps$, all of whose coefficients must vanish. The following coefficients were
computed exactly from the full sum over the eight sign vectors
(Appendix~\ref{app:certificates}).

(1) For $r=(1,2,3)$ the coefficient of $\eps^{-24}$ is $-47775744\,a_{40}^4$. Hence $a_{40}=0$,
and by the symmetry $u\leftrightarrow v$ also $a_{04}=0$.

(2) With $a_{40}=a_{04}=0$, the coefficients of $\eps^{-18}$ for $r=(1,2,3)$ and $r=(1,2,4)$ are
\begin{gather*}
29859840\,a_{31}^3(1-h)\bigl(-113\,a_{20}+748\,a_{31}(1-h)\bigr),\\
18895680\,a_{31}^3(1-h)\bigl(12527\,a_{20}+200408\,a_{31}(1-h)\bigr).
\end{gather*}
If $a_{31}\neq0$, the vector $\bigl(a_{20},a_{31}(1-h)\bigr)$ lies in the kernel of a
$2\times2$ matrix of determinant $-32016300$, so $a_{31}(1-h)=0$, a contradiction. Hence
$a_{31}=0$, and by symmetry $a_{13}=0$.

(3) With $a_{40}=a_{04}=a_{31}=a_{13}=0$, the coefficient of $\eps^{-16}$ for $r=(1,2,3)$ is
\[
-331776\,a_{20}^3\,(a_{20}+9a_{22}).
\]
Hence $a_{20}\in\{0,-9a_{22}\}$ and, by symmetry, $a_{02}\in\{0,-9a_{22}\}$.

(4) If $a_{22}=0$, then $H=e_0+a_{11}uv$ and $\deg P\le6$, which is excluded by
\cite[Proposition~9.20(ii)]{PartI}.

(5) Let $a_{22}\neq0$; we normalize $a_{22}=1$. Since $h\neq0$, $P(0)=h^2e_0=0$ gives $e_0=0$.
There remain the four branches
\[
H=u^2v^2+a_{20}u^2+a_{11}uv+a_{02}v^2,\qquad(a_{20},a_{02})\in\{0,-9\}^2,
\]
with the unknowns $a_{11}$ and $h$. In each branch we take the two points $(0,-h)$ and
$(-h,0)$ of $V(C_h)$ on the coordinate axes together with a third point $\bp_h(t)$; the
numerator of $S_3[P]$ is a polynomial in $t$ whose coefficients, polynomials in $(a_{11},h)$,
must vanish. From each coefficient we remove the rational content and the factors $h$ and
$h-1$, which are nonzero. To these polynomials we add the numerator of $S_3[P]$ at the triple
$\bp_h(2),\bp_h(3),\bp_h(5)$ and the polynomial $z\,h(h-1)-1$ in a further variable $z$, which
expresses $h\neq0,1$. In each of the four branches the Gr\"obner basis of the resulting ideal of
$\Q[z,a_{11},h]$ is $\{1\}$. Hence no branch has a solution with $h\neq0,1$.
\end{proof}

An independent exact certificate of the case $\deg P\le6$ in two variables, covering every
$h\neq1$ and every nonzero even $H$ of degree at most two, is included in the ancillary files.

Summarizing this section and \cite{PartI}: if a quadratic component which is not covered by
the classifications of \cite{PartI} carries the 3SC at all, then it is equivalent to
$V(C_h)$ with
\begin{equation}\label{eq:hyp-remaining}
h\in\overline{\Q}\setminus\{0,1\}\qquad\text{and}\qquad\deg P\ge10 .
\end{equation}
This describes the region that remains to be searched; it is neither an existence statement
nor a proof that the 3SC would imply the $N$SC there.

\section{Limits at infinity, and the equations of Hietarinta's list}\label{sec:limits}

In Corollary~\ref{cor:top}, in the proof of Theorem~\ref{thm:graphs} and in
Corollary~\ref{cor:consequences}(b) the three-soliton condition was transferred from $P$ to its
top homogeneous part, to its top weighted part, respectively to its part of lowest weight, on
one irreducible limit of the rescaled component; for parabolas this was done in
\cite[Section~9.4]{PartI}. The same argument gives more. It applies to every integral weight
vector, also with entries of both signs, and the momenta may be placed on different components
of the limit (Lemma~\ref{lem:limits}). In this section we combine this observation with the
classifications on the cubic graph (Theorem~\ref{thm:cubic-all}), on the monomial graphs
(Theorem~\ref{thm:monomial}) and on the hyperbola $uv=1$ \cite[Theorem~9.14]{PartI}, with a
first-order computation on the cubic graph, and with an elementary statement on three lines
through the origin. The results are: a cubic graph carries the 3SC only if it is odd
(Theorem~\ref{thm:cubic-graphs}), so that the graphs of polynomials are settled
(Corollary~\ref{cor:all-graphs}); restrictions on the branches at infinity of an arbitrary
plane curve carrying the 3SC (Theorem~\ref{thm:branches}); and, for the three families with
nonlinear dispersion in Hietarinta's list which are not of the form $F(\ell)+Q$, the statement
that the polynomial itself is, up to a constant factor, the only polynomial satisfying the 3SC
on its dispersion variety (Theorem~\ref{thm:rigidity}). Two of these families belong to the
BKP hierarchy; for them we include a proof of the $N$-soliton conditions for all $N$ by the
classical induction (Proposition~\ref{prop:NSC-BD}, Remark~\ref{rem:BD}), which gives
Conjecture~\ref{conj:main} for them (Corollary~\ref{cor:hietarinta}). The last three
subsections contain: the exclusion of the curves $u^av^b=c$ with $(a,b)\neq(1,1)$, which occur
as limits, and the resulting restriction on the Newton polygon of a plane curve carrying the
3SC (Theorem~\ref{thm:monomial-curves}, Corollary~\ref{cor:newton}); the $N$-soliton
conditions for all $N$ for the third family, Hietarinta's equation
(Theorem~\ref{thm:hietarinta-NSC}), so that the conjecture holds for it as well
(Corollary~\ref{cor:hietarinta-conj}); and a second solution on the cubic curve of the shallow
water wave equation (Proposition~\ref{prop:sww}). The last two results are proved with the
reduction of $S_N$ to $S_{N-1}$ at the zeros of the two-soliton phase factor which was used in
\cite[Section~9]{PartI} for the hyperbola $uv=1$ (Lemma~\ref{lem:fusion}).

\subsection{Limits}\label{subsec:limits}
For $\omega\in\Z^s$ and a nonzero polynomial $G=\sum_\alpha g_\alpha z^\alpha$ on $\C^s$ let
\[
w_\omega(G):=\max\{\langle\omega,\alpha\rangle:\ g_\alpha\neq0\},\qquad
G^\omega:=\sum_{\langle\omega,\alpha\rangle=w_\omega(G)}g_\alpha z^\alpha ,
\]
the \emph{$\omega$-top part} of $G$. Then $(GH)^\omega=G^\omega H^\omega$, and $G^\omega$ is even
if $G$ is even. For $\omega=(1,\dots,1)$ this is the top homogeneous part, and for
$\omega=-(1,\dots,1)$ the homogeneous part of lowest degree.

\begin{lemma}[Limits]\label{lem:limits}
Let $C$ be a nonconstant polynomial on $\C^s$, let $P$ be a nonzero polynomial on $\C^s$ and
$N\ge2$, and suppose that $S_N[P]$ vanishes on $V(C)^N$. Then, for every $\omega\in\Z^s$,
$S_N[P^\omega]$ vanishes on $V(C^\omega)^N$.
\end{lemma}
\begin{proof}
Let $A_\eps:=\operatorname{diag}(\eps^{-\omega_1},\dots,\eps^{-\omega_s})$ for $\eps\neq0$,
$P_\eps:=\eps^{w_\omega(P)}P\circ A_\eps$ and $C_\eps:=\eps^{w_\omega(C)}C\circ A_\eps$. The
coefficient of $z^\alpha$ in $P_\eps$ is $\eps^{w_\omega(P)-\langle\omega,\alpha\rangle}$ times
that in $P$; hence the coefficients of $P_\eps$ and of $C_\eps$ are polynomials in $\eps$, with
the values $P^\omega$ and $C^\omega$ at $\eps=0$. Let $z_1,\dots,z_N\in V(C^\omega)$. For each
$i$ choose $n_i\in\C^s$ such that $\tau\mapsto C^\omega(z_i+\tau n_i)$ is not identically
zero. The polynomials $f_{i,\eps}(\tau):=C_\eps(z_i+\tau n_i)$ have bounded degree, their
coefficients are continuous in $\eps$, and $f_{i,0}$ is not zero and vanishes at $\tau=0$. By
Rouch\'e's theorem, for every sufficiently small $\eps\neq0$ the polynomial $f_{i,\eps}$ has a
root $\tau_i(\eps)$ such that $\tau_i(\eps)\to0$ as $\eps\to0$. Put
$z_i(\eps):=z_i+\tau_i(\eps)n_i$. Then $C(A_\eps z_i(\eps))=\eps^{-w_\omega(C)}C_\eps(z_i(\eps))=0$,
and by Lemma~\ref{lem:linear} and the homogeneity of $S_N$ in $P$,
\[
S_N[P_\eps]\bigl(z_1(\eps),\dots,z_N(\eps)\bigr)=\eps^{w_\omega(P)(1+N(N-1)/2)}\,
S_N[P]\bigl(A_\eps z_1(\eps),\dots,A_\eps z_N(\eps)\bigr)=0 .
\]
The coefficients of the polynomial $S_N[P_\eps]$ are polynomials in $\eps$ with the value
$S_N[P^\omega]$ at $\eps=0$, so $\eps\to0$ gives $S_N[P^\omega](z_1,\dots,z_N)=0$.
\end{proof}

The points $z_1,\dots,z_N$ may lie on different irreducible components of $V(C^\omega)$, and no
hypothesis on the multiplicities of the factors of $C^\omega$ is made. If $P$ is even and
$P=CK$, then $P^\omega=C^\omega K^\omega$ is even; thus $P^\omega$ satisfies the three-soliton
condition \emph{on the whole variety} $V(C^\omega)$, in the sense of \cite{HietarintaI}, although
$P$ is only assumed to satisfy it on the component $V(C)$. In two variables and for
$\omega\neq0$ the irreducible factors of $C^\omega$ are monomials and binomials, so that every edge of the Newton polygon of $C$ gives a
condition of this kind on a curve $u^a=cv^b$ or $u^av^b=c$. The curves $u^m=cv$ are the subject
of Theorem~\ref{thm:branches}, and the consequences for all edges are drawn in
Section~\ref{subsec:monomial-curves}.

\subsection{Cubic graphs}\label{subsec:cubic-graphs}
In this subsection $C=u^3-v$, $\bp(t)=(t,t^3)$, the polynomial $M$ of \eqref{eq:M} and the set
$\cT$ of types (Table~\ref{tab:types}) are those of Section~\ref{sec:cubic}.

\begin{theorem}[Cubic graphs]\label{thm:cubic-graphs}
Let $u,v$ be two of the coordinates of $\C^s$, let $q\in\C[u]$ be a polynomial of degree three,
and let $P\neq0$ be an even polynomial on $\C^s$ such that $S_3[P]$ vanishes identically on
$V(q(u)-v)^3$. Then $q$ is odd.
\end{theorem}

Thus $q=a_3u^3+a_1u$, the hypersurface $V(q(u)-v)$ is a linear image of the cubic graph, and
Theorem~\ref{thm:cubic-all} describes $P$. In particular a cubic graph which does not pass
through the origin carries no polynomial with the 3SC. Together with
\cite[Theorem~9.17]{PartI} and Theorem~\ref{thm:graphs} we obtain:

\begin{corollary}[Graphs of polynomials]\label{cor:all-graphs}
Let $\ell_1,\ell_2$ be linearly independent linear forms on $\C^s$, let $q\in\C[u]$ be a
polynomial of degree at least two, and $V_0:=V(\ell_2-q(\ell_1))$.
\begin{enumerate}
\item[\textup{(a)}] A nonzero even polynomial $P$ with $P(0)=0$ which satisfies the
\textup{3SC} on $V_0$ exists if and only if $q=a_2u^2+a_1u$, or $q$ is an odd cubic, or $q$ is
an odd quintic.
\item[\textup{(b)}] Conjecture~\textup{\ref{conj:main}} holds for $V_0$.
\end{enumerate}
\end{corollary}
\begin{proof}
By Lemma~\ref{lem:linear} we may assume $\ell_1=u$, $\ell_2=v$. For $\deg q=2$ the linear change
$v\mapsto a_2v+a_1u$ transforms $q(u)-v$ into a multiple of $u^2+c-v$, and
\cite[Theorem~9.17]{PartI} gives $c=0$ and $P=\gamma(u^4-v^2)$, which satisfies all $N$SC. For
$\deg q=3$ this is Theorem~\ref{thm:cubic-graphs} with Corollary~\ref{cor:conj-cubic}, and for
$\deg q\ge4$ it is Theorem~\ref{thm:graphs} with Corollary~\ref{cor:consequences}(a).
\end{proof}

For the proof of Theorem~\ref{thm:cubic-graphs} we need the derivative of $S_3$ in a direction
transversal to the cubic graph. Let $g\colon\C\to\C$ be a function, $k=(k_1,k_2,k_3)\in\C^3$,
and
\[
\bp_\eps(k_i):=\bp(k_i)+\eps\,g(k_i)\,(0,1)=\bigl(k_i,\ k_i^3+\eps g(k_i)\bigr).
\]
For a polynomial $Q\in\C[u,v]$ put
\begin{equation}\label{eq:Xg}
X_g[Q](k):=\frac{d}{d\eps}\Big|_{\eps=0}S_3[Q]\bigl(\bp_\eps(k_1),\bp_\eps(k_2),\bp_\eps(k_3)\bigr).
\end{equation}
For $\sg\in\{\pm1\}^3$ and a polynomial $G$ let $\pi_\sg[G](\eps)$ be the product of the values
of $G$ at the four arguments of the summand of \eqref{eq:S3} for $\sg$, that is, at
$\sum_i\sg_i\bp_\eps(k_i)$ and at $\sg_i\bp_\eps(k_i)-\sg_j\bp_\eps(k_j)$, $i<j$.

\begin{lemma}[First-order terms; computer-assisted]\label{lem:firstorder}
Let $T\in\cT$, let $\mu\ge1$ be odd, and let
\[
W_g(T):=\sum_\sg\pi_\sg[T]'(0),\qquad R_g(T):=\sum_\sg\pi_\sg[C]'(0)\,\pi_\sg[T](0).
\]
\begin{enumerate}
\item[\textup{(i)}] $X_g[C^\mu T]=M^{\mu-1}\bigl(M\,W_g(T)+\mu\,R_g(T)\bigr)$.
\item[\textup{(ii)}] For $g(t)=t^2$ and for $g(t)=1$, at $k=(1,2,3)$, one has $R_g(T)\neq0$,
and the number $\mu^*_g(T):=-M\,W_g(T)/R_g(T)$ has the value given in
Table~\textup{\ref{tab:critical}}; it is not a positive integer.
\end{enumerate}
Consequently $X_g[C^\mu T]$ is not identically zero for $g(t)=t^2$ and for $g(t)=1$.
\end{lemma}

\begin{table}[ht]
\centering\renewcommand{\arraystretch}{1.25}
\begin{tabular}{@{}lcc@{}}
\toprule
$T$ & $\mu^*_g(T)$ for $g(t)=t^2$ & $\mu^*_g(T)$ for $g(t)=1$\\
\midrule
$T_1=u$, $\TI=v$ & $0$ & $0$\\
$T_3$ & $-5/312$ & $12/143$\\
$T_5$ & $-5/86$ & $20/13$\\
$T_7$ & $3555/6604$ & $5042/1443$\\
$T_9$ & $117019115/163973714$ & $3231339700/618225219$\\
$T_r$ & $(13795-1775\,r)/454359$ & $(18100\,r-142080)/6282001$\\
\bottomrule
\end{tabular}

\vspace{6pt}
\caption{The numbers $\mu^*_g(T)=-M\,W_g(T)/R_g(T)$ at $k=(1,2,3)$, where $M=349920$;
$r^2-7r+1=0$ (two values).}\label{tab:critical}
\end{table}

\begin{proof}
(i) The four arguments are, for $\eps=0$, the points $\Sigma(\sg k)$ and $d_{ab}(\sg k)$ of
Section~\ref{subsec:laurent}, so $\pi_\sg[C](0)=M(k)$ for every $\sg$ by the proof of
Lemma~\ref{lem:cov}, and $\sum_\sg\pi_\sg[T](0)=\cS[T](k)=0$, because $CT$ satisfies the 3SC
on $V(C)$ (Theorem~\ref{thm:cubic}) and $\cS[CT]=M\,\cS[T]$. Since
\[
S_3[C^\mu T]\bigl(\bp_\eps(k_1),\bp_\eps(k_2),\bp_\eps(k_3)\bigr)=\sum_\sg\pi_\sg[C](\eps)^\mu\,\pi_\sg[T](\eps),
\]
the derivative at $\eps=0$ is
$\mu M^{\mu-1}\sum_\sg\pi_\sg[C]'(0)\,\pi_\sg[T](0)+M^\mu\sum_\sg\pi_\sg[T]'(0)$.

(ii) The numbers $M\,W_g(T)$ and $R_g(T)$ at $k=(1,2,3)$ were computed exactly, in $\Q$ and
in $\Q[r]/(r^2-7r+1)$ (Appendix~\ref{app:certificates}); for $T_r$ the value of $\mu^*_g$ is
irrational for both roots $r=\frac12(7\pm3\sqrt5)$. For $T_1$ and $\TI$ one has $W_g=0$ for
every $g$, because for a linear form $L$ the sum \eqref{eq:S3} vanishes identically on
$(\C^s)^3$: with $y_i:=\sg_iL(p_i)$ the summand is
$(y_1+y_2+y_3)(y_1-y_2)(y_1-y_3)(y_2-y_3)$, an alternating polynomial of degree four in $y$,
and its part which is even in each $y_i$ is alternating and even in each variable, hence
divisible by $\prod_{i<j}(y_i^2-y_j^2)$, hence zero.

By (i) and (ii), at $k=(1,2,3)$ we have
$X_g[C^\mu T]=M^{\mu-1}R_g(T)\,(\mu-\mu^*_g(T))\neq0$ for every positive integer $\mu$.
\end{proof}

\begin{proof}[Proof of Theorem~\ref{thm:cubic-graphs}]
\emph{Reduction to two variables.} For $s\ge3$ and $\xi,\eta\in\C^{s-2}$ the polynomial
$P\circ A_{\xi\eta}$, with the planes \eqref{eq:planes}, is even, it is not zero for generic
$(\xi,\eta)$, and $S_3[P\circ A_{\xi\eta}]$ vanishes on the cube of the curve
$\{v=q(u)\}\subset\C^2$ by Lemma~\ref{lem:linear}. So let $s=2$.

\emph{Normalization.} Let $q=a_3u^3+a_2u^2+a_1u+a_0$. Replacing $P(u,v)$ by
$P(u,a_3v+a_1u)$, which is even and satisfies the hypothesis for the cubic $(q-a_1u)/a_3$, we
may assume $a_3=1$ and $a_1=0$. We have to show that $a_2=a_0=0$.

\emph{Expansion.} Give $u,v$ the weights $1,3$. A monomial $u^iv^j$ of even degree $i+j$ has
the even weight $i+3j$; hence all weighted homogeneous parts of the even polynomial $P$ have
even weight, and $P=\sum_{j\ge0}Q_{2j}$, where $Q_{2j}$ is weighted homogeneous of weight
$w-2j$ and $Q_0\neq0$ is the top weighted part. If $w=0$, then $P$ is a nonzero constant and
$S_3[P]$ is a nonzero constant; so $w>0$. For $\eps\neq0$ put
$P_\eps(u,v):=\eps^wP(u/\eps,v/\eps^3)=\sum_j\eps^{2j}Q_{2j}(u,v)$ and, for $t\in\C$,
\[
z_\eps(t):=\bigl(t,\ \eps^3q(t/\eps)\bigr)=\bigl(t,\ t^3+\eps a_2t^2+\eps^3a_0\bigr).
\]
The point $(t/\eps,q(t/\eps))$ lies on the curve, so by Lemma~\ref{lem:linear} and the
homogeneity of $S_3$ the polynomial
\[
G(\eps):=S_3\Bigl[\sum_j\eps^{2j}Q_{2j}\Bigr]\bigl(z_\eps(k_1),z_\eps(k_2),z_\eps(k_3)\bigr)
\]
in $\eps$ and $k$ vanishes for all $\eps\neq0$, hence identically. Its value at $\eps=0$
is $S_3[Q_0]$ at the points $\bp(k_1),\bp(k_2),\bp(k_3)$. Thus $Q_0$ is a nonzero even weighted
homogeneous polynomial with $\cS[Q_0]=0$. By Lemma~\ref{lem:divis} the multiplicity $\mu$ of $C$ in $Q_0$
is odd, by Lemma~\ref{lem:cov} the odd weighted homogeneous polynomial $Q_0/C^\mu$ satisfies
\eqref{eq:Hcond}, and by Proposition~\ref{prop:homog}
\[
Q_0=\lambda\,C^\mu\,T,\qquad\lambda\neq0,\quad\mu\ \text{odd},\quad T\in\cT .
\]
Since $z_\eps(t)=\bp(t)+(\eps a_2t^2+\eps^3a_0)(0,1)$ and the polynomial in the argument of
$S_3$ is $Q_0+O(\eps^2)$, the coefficient of $\eps$ in $G$ is $a_2X_g[Q_0](k)$ with
$g(t)=t^2$; and if $a_2=0$, the coefficient of $\eps^3$ in $G$ is $a_0X_g[Q_0](k)$ with
$g(t)=1$, because then $z_\eps(t)=\bp(t)+\eps^3a_0(0,1)$ and the argument of $S_3$ is
$Q_0+\eps^2Q_2+O(\eps^4)$. By Lemma~\ref{lem:firstorder}, $X_g[Q_0]=\lambda^4X_g[C^\mu T]$ is
not identically zero in both cases. Hence $a_2=0$, and then $a_0=0$.
\end{proof}

\begin{remark}\label{rem:firstorder}
(i) The proof uses the classification of the weighted homogeneous solutions on the cubic graph
for all multiplicities, and the evenness of $P$: there is no part of $P$ of weight $w-1$ which
could compensate the first-order displacement of the curve. On a non-odd cubic graph through
the origin Corollary~\ref{cor:origin}(3) shows that the top weighted part of $P$ is divisible by
$C^2$, so that the types with $\mu\ge3$ do occur as candidates there.

(ii) For $T_1$ and $\TI$ the value $\mu^*_g=0$ means that $X_g[C^\mu L]=\mu M^{\mu-1}R_g(L)$ for a
linear form $L$: the first-order term is proportional to the multiplicity.
\end{remark}

\subsection{Branches at infinity asymptotic to a graph}\label{subsec:branches}
In the rest of this section, unless stated otherwise, $C\in\C[u,v]$ is an irreducible
polynomial of degree at least two and $P=CK$ is a nonzero even polynomial with $P(0)=0$ such
that $S_3[P]$ vanishes on $V(C)^3$. We write $\omega_m:=(1,m)$ and $\omega':=(1,-1)$. If
$u^m-cv$ divides $C^{\omega_m}$, the curve $V(C)$ has a branch at infinity which is asymptotic
to the graph $v=u^m/c$.

\begin{theorem}[Branches asymptotic to a graph]\label{thm:branches}
Let $m\ge2$ and $c\neq0$, and suppose that $F:=u^m-cv$ divides $C^{\omega_m}$.
\begin{enumerate}
\item[\textup{(a)}] Then $m\in\{2,3,5\}$.
\item[\textup{(b)}] If $m=5$, then $C=\kappa\,(q(u)-cv)$ with an odd quintic $q$; thus $V(C)$
is the graph of an odd quintic, and $P$ is given by Theorem~\textup{\ref{thm:graphs}}.
\item[\textup{(c)}] If $m=3$ and $C^{\omega_3}$ is divisible by a second polynomial
$F'=u^3-c'v$ with $c'\notin\{0,c\}$, then $c'/c=-r$ with $r^2-7r+1=0$, $C^{\omega_3}=\kappa FF'$,
$P=\lambda C$, and there are $\eta\neq0$ and $\delta$ such that
\[
C(u,\eta v+\delta u)=\kappa\,\bigl(u^6+5u^3v-5v^2+au^2+buv\bigr),\qquad a,b\in\C .
\]
\item[\textup{(d)}] If $m=3$ and $C^{\omega_3}=\kappa\,vF$, then $C(0)=0$, $P=\lambda C$, and there
are $\eta\neq0$ and $\delta$ such that
\[
C(u,\eta v+\delta u)=\kappa'\,\bigl(u^3v-v^2+buv+au^2\bigr),\qquad a\neq0 .
\]
\end{enumerate}
\end{theorem}
For $m=2$ the corresponding statements are those of \cite[Section~9.4 and Theorem~9.18]{PartI}:
$P^{\omega_2}=\gamma(u^4-c^2v^2)$, and $V(C)$ is a parabola or $P=\lambda C$ is of the form
$\gamma u^4+Q$ with a quadratic form $Q$. The polynomials in (c) and (d) are the equations
(I\,D2) and (I\,B) of \cite[Section~IV.E]{HietarintaI}; they do satisfy the 3SC, and all $N$SC
(Proposition~\ref{prop:NSC-BD}).

\begin{proof}
By Lemma~\ref{lem:limits}, $S_3[P^{\omega_m}]$ vanishes on $V(C^{\omega_m})^3$, and
$P^{\omega_m}=C^{\omega_m}K^{\omega_m}$ is even, nonzero and weighted homogeneous for the weights
$\wt u=1$, $\wt v=m$. After the scaling $v\mapsto v/c$ we may assume $c=1$, so that
$V(F)=V_m$.

(a) $P^{\omega_m}$ is a polynomial as in Proposition~\ref{prop:candidates}, and
Theorem~\ref{thm:monomial}(ii),(iii) excludes even $m\ge4$ and odd $m\ge7$.

(b) By Theorem~\ref{thm:monomial}(i), $P^{\omega_5}=\lambda uF$ or
$P^{\omega_5}=\lambda FF_1$ with $F_1:=u^5+4v$, and $C^{\omega_5}$ is a divisor of $P^{\omega_5}$
which is divisible by $F$. If $C^{\omega_5}=\kappa FF_1$, then by Lemma~\ref{lem:limits} the
polynomial $\lambda FF_1$ would satisfy $S_3=0$ on $V(F_1)^3$; in the variable $v'=-4v$, where
$F_1=u^5-v'$ and $F=u^5+\frac14v'$, this contradicts Theorem~\ref{thm:monomial}(i). If
$C^{\omega_5}=\kappa uF$, then $P^{\omega_5}=\lambda uF$, so $K^{\omega_5}$ is a constant; since
the weights are positive, $K$ is a constant and $P=\lambda'C$. Then $C$ is even with $C(0)=0$,
and since $u^4$ and $u^2$ are the only even monomials of weight less than six and positive
degree, $C=\kappa u(u^5-v)+\alpha u^4+\beta u^2$ is divisible by $u$, a contradiction. Hence
$C^{\omega_5}=\kappa F$. The monomials of weight less than five are powers of $u$, so
$C=\kappa(q(u)-v)$ with $\deg q=5$, and $q$ is odd by Theorem~\ref{thm:graphs}.

(c) Now $F=u^3-v$ is the polynomial which is denoted by $C$ in Section~\ref{sec:cubic}, and
we write $c'$ for the constant of $F'$ after the scaling. As in the proof of Theorem~\ref{thm:cubic-graphs},
$P^{\omega_3}=\lambda F^\mu T$ with $\mu$ odd and $T\in\cT$. Since $F'$ divides $P^{\omega_3}$
and is prime to $F$, it divides $T$. By Table~\ref{tab:types} and the factorization
$u^6+28u^3v+16v^2=(u^3+\rho_+v)(u^3+\rho_-v)$, $\rho_\pm=14\pm6\sqrt5$, the irreducible
factors of the form $u^3-c''v$ of the elements of $\cT$ are those with
\[
c''\in\Gamma:=\{-2,\ 4,\ -r_+,\ -r_-,\ -\rho_+,\ -\rho_-\},\qquad r_\pm=\tfrac12(7\pm3\sqrt5),
\]
and $u^3+r_\pm v$ divides only $T_{r_\pm}$. So $c'\in\Gamma$. By Lemma~\ref{lem:limits},
$S_3[P^{\omega_3}]$ also vanishes on $V(F')^3$. In the variable $v'=c'v$ we have $F'=u^3-v'$
and $F=u^3-v'/c'$, and the same argument gives $1/c'\in\Gamma$. The only elements of $\Gamma$
whose inverse belongs to $\Gamma$ are $-r_+$ and $-r_-$, which are inverse to each other. Hence
$c'=-r$ with $r^2-7r+1=0$, $T=T_r$, and $P^{\omega_3}=\lambda F^\mu F'$. The same argument with
respect to $F'$ shows that $P^{\omega_3}$ is a power of $F'$ times a type in the variable $v'$
which is divisible by $F$; since the factors $u^3-c''v$ of the types are simple, $F$ is a
simple factor of $P^{\omega_3}$. So $\mu=1$ and $P^{\omega_3}=\lambda FF'$. As $FF'$ divides $C^{\omega_3}$ and $C^{\omega_3}$ divides
$P^{\omega_3}$, we get $C^{\omega_3}=\kappa FF'$, $K^{\omega_3}$ is constant, $K$ is constant, and
$P=\lambda'C$. Therefore $C$ is even with $C(0)=0$, and
\[
C=\kappa\,(u^3-v)(u^3+rv)+\alpha u^4+\beta uv+\gamma u^2 ,
\]
since $u^4,uv,u^2$ are the even monomials of weight less than six and positive degree. The
substitution $v\mapsto v+\delta u$ changes the coefficient of $u^4$ by $\kappa\delta(r-1)$, and
$r\neq1$; and $(u^3-\eta v)(u^3+r\eta v)=u^6+5u^3v-5v^2$ for $\eta=5/(r-1)$, because
$5r=(r-1)^2$.

(d) Again $F=u^3-v$ after the scaling, and $P^{\omega_3}=\lambda F^\mu T$ with $\mu$ odd and
$T\in\cT$. Since $v$ divides $P^{\omega_3}$, Table~\ref{tab:types} gives $T=\TI=v$, and $v$ is a
simple factor of $P^{\omega_3}$. If $V(C)\neq-V(C)$, the polynomial $C^-(z):=C(-z)$ would be an
irreducible factor of the even polynomial $P$ which is not associated with $C$, so $CC^-$ would
divide $P$ and $C^{\omega_3}(C^-)^{\omega_3}=\kappa^2v^2F^2$ would divide $P^{\omega_3}$, a
contradiction. Hence $C(-z)=\pm C(z)$, and $C$ is even because $C^{\omega_3}$ is. Since
$1,u^2,u^4,uv$ are the even monomials of weight less than six,
\[
C=\kappa\,v(u^3-v)+\alpha u^4+\beta uv+\gamma u^2+\gamma_0 .
\]
Since $(v+\delta u)(u^3-v-\delta u)=v(u^3-v)+\delta u^4-2\delta uv-\delta^2u^2$, the substitution
$v\mapsto v+\delta u$ with $\delta=-\alpha/\kappa$ removes the term $u^4$; it does not change
the $\omega_3$-top parts. So we may assume $C=\kappa\,(u^3v-v^2+buv+au^2)+\gamma_0$. If $a=0$,
then $\gamma_0\neq0$, since otherwise $v$ would divide $C$, and the segment from $(0,0)$ to
$(3,1)$ is an edge of the Newton polygon of $C$; this is excluded by Corollary~\ref{cor:newton}
below, whose proof uses part~(a) of the present theorem but not part~(d). Hence $a\neq0$. It
remains to show that $\mu=1$; then $K^{\omega_3}$ is constant, $K$ is constant because the
weights are positive, $P=\lambda'C$, and $\gamma_0=0$ because $P(0)=0$.

We have $K^{\omega_3}=\lambda'(u^3-v)^{\mu-1}$. For the weight vector $\omega'=(1,-1)$ a monomial
$u^iv^j$ of $K$ satisfies $i-j\le i\le3(\mu-1)-3j\le3(\mu-1)$, with equality only for
$u^{3(\mu-1)}$, which occurs in $K$. Hence $K^{\omega'}=\lambda'u^{3\mu-3}$. Moreover
$C^{\omega'}=\kappa\,u^2(uv+a)$, so that
\[
P^{\omega'}=\lambda''\,u^{3\mu-1}(uv+a),
\]
and by Lemma~\ref{lem:limits} the polynomial $S_3[P^{\omega'}]$ vanishes on the cube of the
hyperbola $uv=-a$. Then $\widetilde P(u,v):=P^{\omega'}(u,-av)=-a\lambda''u^{3\mu-1}(uv-1)$ is an
even polynomial with $\widetilde P(0)=0$ which is divisible by $uv-1$ and satisfies the 3SC on
$\{uv=1\}$. By \cite[Theorem~9.14]{PartI}, $\widetilde P=(uv-1)B$, where, with $z:=uv$,
\begin{multline}\label{eq:SK9}
B\in\Span\bigl(u^2,\ v^2,\ z(z-2),\ u^2z(z-4),\ v^2z(z-4),\ u^4(z-2)(z-4),\ v^4(z-2)(z-4),\\
u^6(z-4)^2,\ v^6(z-4)^2\bigr)
\end{multline}
or $B$ is a quadratic form. The nine polynomials in \eqref{eq:SK9} are homogeneous for the
weight vector $\omega'$, of the weights $2,-2,0,2,-2,4,-4,6,-6$, and a quadratic form has
weights at most two. Since $B=-a\lambda''u^{3\mu-1}$ is homogeneous of weight $3\mu-1>0$ for
$\omega'$, it is a combination of the polynomials of this weight; hence $3\mu-1\in\{2,4,6\}$,
and $\mu=1$ because $\mu$ is an odd integer.
\end{proof}

\begin{remark}\label{rem:branches}
(i) Theorem~\ref{thm:branches} does not classify all curves with a branch at infinity which is
asymptotic to a cubic graph. By Table~\ref{tab:types}, $C^{\omega_3}$ is, up to a constant,
$u^iv^jF^k$ with $(i,j)\in\{(0,0),(1,0),(2,0),(0,1)\}$, or $FF'$ as in (c). For
$C^{\omega_3}=\kappa F$ the curve is the graph of a cubic polynomial, which is odd by
Theorem~\ref{thm:cubic-graphs}. The cases which remain are $C^{\omega_3}=\kappa\,u^iF^k$ with
$(i,k)\neq(0,1)$ and $C^{\omega_3}=\kappa\,vF^k$ with $k\ge2$; the case
$C^{\omega_3}=\kappa\,vF$ with $C(0)\neq0$ does not occur, by (d). An example of the first kind
is the curve $cuv=u^4+\beta u^2+\gamma$, $\gamma\neq0$, which does not pass through the origin.

(ii) Part (d) contains no hypothesis on $C(0)$. A constant term $\gamma_0\neq0$ is excluded
in its proof: if the coefficient $a$ of $u^2$ does not vanish, the argument gives
$P=\lambda'C$, which contradicts $P(0)=0$; if $a=0$, then $C=\kappa\,(u^3v-v^2+buv)+\gamma_0$
has branches asymptotic to $u^3v=-\gamma_0/\kappa$, which are excluded by
Corollary~\ref{cor:newton} below.
\end{remark}

\subsection{Three lines through the origin}\label{subsec:threelines}
For a homogeneous polynomial in two variables the condition $S_3=0$ on three points of the same
line through the origin is empty, but on three points of three different lines it is not.

\begin{lemma}[Three lines]\label{lem:threelines}
Let $\lambda\in\C\setminus\{0,1\}$, let $M_1,M_2,M_3\ge1$ and $M_4\ge0$ be integers with even
sum, and
\[
P:=X^{M_1}\,T^{M_2}\,(X-T)^{M_3}\,(X-\lambda T)^{M_4}\in\C[X,T].
\]
Then $S_3[P]$ vanishes at all triples of points $(x,0)$, $(0,y)$, $(w,w)$ of the three lines
$T=0$, $X=0$, $X=T$ if and only if either $M_4=0$ and exactly two of $M_1,M_2,M_3$ are odd, or
$M_4=1$ and $M_1,M_2,M_3$ are odd.
\end{lemma}
\begin{proof}
Let $p_1=(x,0)$, $p_2=(0,y)$, $p_3=(w,w)$, and for $\sg\in\{\pm1\}^3$ write
$(x',y',w'):=(\sg_1x,\sg_2y,\sg_3w)$. The four arguments in the summand of \eqref{eq:S3} are
$(x'+w',y'+w')$, $(x',-y')$, $(x'-w',-w')$ and $(-w',y'-w')$. With $n:=\lambda y$,
$m:=(1-\lambda)w$, $n':=\sg_2n$, $m':=\sg_3m$, the products of the values of $X$, $T$, $X-T$
and $X-\lambda T$ at these four points are
\begin{gather*}
-(x'^2-w'^2)\,x'w',\qquad(y'^2-w'^2)\,y'w',\qquad-(x'^2-y'^2)\,x'y',\\
-L(x',y',w'),\qquad L(x,y,w):=(x-n+m)(x+n)(x-m)(n+m).
\end{gather*}
The factors $x'^2-w'^2$, $y'^2-w'^2$, $x'^2-y'^2$ do not depend on $\sg$. Hence, up to a
factor $\pm(x^2-w^2)^{M_1}(y^2-w^2)^{M_2}(x^2-y^2)^{M_3}$, the value of $S_3[P]$ is eight times
the part of
\[
x^{M_1+M_3}\,y^{M_2+M_3}\,w^{M_1+M_2}\,L(x,y,w)^{M_4}
\]
which is even in each of the variables $x,y,w$. One checks that
\[
L=g_1+g_2+g_3,\qquad g_1:=xn(x^2-n^2),\quad g_2:=xm(x^2-m^2),\quad g_3:=nm(n^2-m^2),
\]
where $g_1$ is odd in $x$ and $y$ and even in $w$, $g_2$ is odd in $x$ and $w$ and even in $y$,
and $g_3$ is odd in $y$ and $w$ and even in $x$. Thus
$L^{M_4}=\sum_{i+j+l=M_4}\frac{M_4!}{i!\,j!\,l!}g_1^ig_2^jg_3^l$, and the term with the exponents
$(i,j,l)$ has the parities $(i+j,\,i+l,\,j+l)$ in $(x,y,w)$. These terms are linearly
independent: for the lexicographic order with $x>y>w$ the leading monomials of
$g_1,g_2,g_3$ are multiples of $x^3y$, $x^3w$, $y^3w$, so that the leading monomial of
$g_1^ig_2^jg_3^l$ is $x^{3i+3j}y^{i+3l}w^{j+l}$, and for $i+j+l=M_4$ these monomials are
pairwise distinct. Consequently $S_3[P]$ vanishes on the three lines if and only if there is no
triple $(i,j,l)$ of nonnegative integers with $i+j+l=M_4$ and
\[
i+j\equiv M_1+M_3,\qquad i+l\equiv M_2+M_3,\qquad j+l\equiv M_1+M_2\pmod 2 .
\]
Given $i+j+l=M_4$, these congruences say $l\equiv M_4+M_1+M_3$, $j\equiv M_4+M_2+M_3$,
$i\equiv M_4+M_1+M_2$. Let $i_0,j_0,l_0\in\{0,1\}$ be these residues. Their sum is congruent
to $3M_4+2(M_1+M_2+M_3)\equiv M_4$, and a triple exists if and only if $i_0+j_0+l_0\le M_4$.
So $S_3[P]$ vanishes on the three lines if and only if $(M_4,\,i_0+j_0+l_0)$ is $(0,2)$ or
$(1,3)$. In the first case exactly two of $M_1+M_2$, $M_2+M_3$, $M_1+M_3$ are odd, which, the
sum $M_1+M_2+M_3$ being even, means that exactly two of $M_1,M_2,M_3$ are odd. In the second
case $M_1\equiv M_2\equiv M_3$, and all three are odd because $M_1+M_2+M_3$ is odd.
\end{proof}

For $M_4=0$ this is the condition found in \cite[Section~IV.E, (II\,C2)]{HietarintaI}, and
$M_1=\dots=M_4=1$ is the polynomial (II\,D) there. The lemma is applied to the top homogeneous
part of $P$: by Lemma~\ref{lem:limits} with $\omega=(1,1)$, $S_3[P^{(1,1)}]$ vanishes on
triples of points of the lines through the origin which are parallel to the asymptotic
directions of $V(C)$.

\subsection{The equations of Hietarinta's list}\label{subsec:hietarinta}
Let
\begin{align*}
\PB_{a,b,c}&:=X^3T+aX^2+bXT+cT^2, &\PBh_a&:=X^3T+aX^2+TY,\\
\PD_{a,b}&:=X^6+5X^3T-5T^2+aX^2+bXT, &\PDh&:=X^6+5X^3T-5T^2+XY,\\
\PC_{a,b,c}&:=XT\,(X^2+\sqrt3\,XT+T^2)+aX^2+bXT+cT^2 .
\end{align*}
These are the polynomials (I\,B), (I\,D) and (I\,C) of \cite[Section~IV.E and
Table~IV]{HietarintaI} in two and three variables; (I\,C) has no extension to three variables
\cite[Section~IV.F]{HietarintaI}. The fourth family (I\,A) of the list, which contains the
Kadomtsev--Petviashvili polynomial, consists of polynomials of the form $F(\ell)+Q$ and was
treated in \cite{PartI}. By the real linear substitution $T\mapsto(T-X)/\sqrt3$ the polynomial
$-3\sqrt3\,\PC_{a,b,c}$ becomes
\begin{equation}\label{eq:CQ}
C_{\Q}:=X^4-XT^3+a'X^2+b'XT+c'T^2
\end{equation}
with new parameters $a',b',c'$; up to the sign of $T$ this is the form
\cite[(4.5)]{GrammaticosRamaniHietarinta1990} of the equation.

\begin{lemma}[Irreducibility]\label{lem:irreducible}
\textup{(i)} $\PB_{a,b,c}$ is irreducible if and only if $ac\neq0$, and $\PBh_a$ if and only if
$a\neq0$. \textup{(ii)} $\PD_{a,b}$ is irreducible if and only if $45a+b^2\neq0$, and $\PDh$ is
irreducible. \textup{(iii)} $\PC_{a,b,c}$ is irreducible if and only if the quadratic form
$aX^2+bXT+cT^2$ is not divisible by any of the four linear factors of
$XT(X^2+\sqrt3\,XT+T^2)$, that is, if and only if
$ac\,\bigl[(a-c)^2+(b-\sqrt3\,a)(b-\sqrt3\,c)\bigr]\neq0$.
\end{lemma}
\begin{proof}
(i), (ii) As polynomials in $T$, $\PB_{a,b,c}$ with $c\neq0$ and $\PD_{a,b}$ have degree two
and constant leading coefficients, and their discriminants are
$X^2\bigl((X^2+b)^2-4ac\bigr)$ and $X^2(45X^4+10bX^2+b^2+20a)$. The first is a square in $\C[X]$
if and only if $ac=0$, the second if and only if $45a+b^2=0$; so the polynomials are
irreducible if $ac\neq0$, respectively $45a+b^2\neq0$. For $a=0$ or $c=0$ the polynomial
$\PB_{a,b,c}$ is divisible by $T$ or by $X$, and for $45a+b^2=0$ the polynomial $\PD_{a,b}$ is
reducible by Remark~\ref{rem:cubic-scope}(iv). $\PBh_a$ and $\PDh$ have degree one in $Y$,
with the coprime coefficients $T$ and $X^3T+aX^2$ ($a\neq0$), respectively $X$ and
$X^6+5X^3T-5T^2$.

(iii) Let $C_4$ and $Q_2$ be the parts of degree four and two of $\PC=\PC_{a,b,c}$. If a linear
form divides $C_4$ and $Q_2$, it divides $\PC$. Conversely, let $A$ be an irreducible factor
of $\PC$ with $\deg A<4$. Then $A(-z)$ is an irreducible factor of $\PC$; if it were not
associated with $A$, the product $A(z)A(-z)$, whose top homogeneous part is the square of that
of $A$ up to sign, would divide $\PC$, which is impossible because $C_4$ is squarefree. So $A$ is
even or odd, and so is the cofactor. If $A$ or the cofactor is odd of degree one, it is a linear
form which divides $C_4$ and $Q_2$. Otherwise $\PC=(A_2+\alpha)(B_2+\beta)$ with quadratic forms
$A_2,B_2$ and $\alpha\beta=\PC(0)=0$, and one of the factors is a quadratic form in two
variables, which is a product of linear forms; each of them divides $C_4$ and $Q_2$. The last
statement holds because the resultant
of $aX^2+bX+c$ and $X^2+\sqrt3X+1$ is $(a-c)^2+(b-\sqrt3a)(b-\sqrt3c)$.
\end{proof}

\begin{theorem}[Rigidity]\label{thm:rigidity}
Let $X,T$ \textup{(}respectively $X,T,Y$\textup{)} be two \textup{(}respectively
three\textup{)} of the coordinates of $\C^s$, and let $C$ be one of the polynomials
$\PB_{a,b,c}$, $\PC_{a,b,c}$, $\PD_{a,b}$ \textup{(}respectively $\PBh_a$, $\PDh$\textup{)}, with
parameters for which it is irreducible. Let $P$ be an even polynomial on $\C^s$ with $P(0)=0$
which is divisible by $C$ and satisfies the \textup{3SC} on $V(C)$. Then $P=\lambda C$ with a
constant $\lambda$.
\end{theorem}

Conversely $C$ is even with $C(0)=0$, and it satisfies the 3SC on $V(C)$: for $\PB$, $\PD$,
$\PBh$, $\PDh$ this is contained in Proposition~\ref{prop:NSC-BD} below, and for $\PC$ it is
the result of \cite{HietarintaI} (see Remark~\ref{rem:quartic}(i)). The proof of
Theorem~\ref{thm:rigidity} is given after the next lemma.

\begin{lemma}[Sections]\label{lem:sections}
Let $2\le r<s$, write the points of $\C^s$ as $(x,z)$ with $x\in\C^r$, $z\in\C^{s-r}$, and for a
linear map $\zeta\colon\C^r\to\C^{s-r}$ let $A_\zeta(x):=(x,\zeta x)$. Let $C_1\neq0$ and $P$ be
polynomials on $\C^s$ such that $P\circ A_\zeta$ is a constant multiple of $C_1\circ A_\zeta$
for all $\zeta$ in a dense subset $Z$ of the space of linear maps. Then $P=\lambda C_1$ with a
constant $\lambda$.
\end{lemma}
\begin{proof}
Let $\Phi(y,y'):=P(y)C_1(y')-P(y')C_1(y)$ for $y,y'\in\C^s$. If $y=A_\zeta x$ and
$y'=A_\zeta x'$ with $\zeta\in Z$, then $\Phi(y,y')=0$. The map
$(\zeta,x,x')\mapsto(A_\zeta x,A_\zeta x')$ is continuous, and its image contains all pairs
$((x,z),(x',z'))$ with $x,x'$ linearly independent; hence the image of $Z\times\C^r\times\C^r$ is
dense in $\C^s\times\C^s$, and $\Phi=0$. Choosing $y'$ with $C_1(y')\neq0$ gives
$P=\bigl(P(y')/C_1(y')\bigr)\,C_1$.
\end{proof}

\begin{proof}[Proof of Theorem~\ref{thm:rigidity}]
\emph{Two variables, $s=2$.} For $C=\PB_{a,b,c}$ with $ac\neq0$ put $u:=X$, $v:=T$. Then
$C^{\omega_3}=v(u^3+cv)$ and $C(0)=0$, so Theorem~\ref{thm:branches}(d) gives $P=\lambda C$. For
$C=\PD_{a,b}$ we have $C^{\omega_3}=u^6+5u^3v-5v^2$, a product of two polynomials $u^3-c_1v$,
$u^3-c_2v$ with $c_1\neq c_2$, and Theorem~\ref{thm:branches}(c) gives $P=\lambda C$.

Let $C=\PC_{a,b,c}=C_4+Q_2$ be irreducible, let $\ell_1,\dots,\ell_4$ be the four linear factors
of $C_4$, and let $d_l\in\C^2$ be a nonzero vector with $\ell_l(d_l)=0$; by
Lemma~\ref{lem:irreducible}, $q_l:=Q_2(d_l)\neq0$. Let $D:=\deg P$, let $P_D$ be the top
homogeneous part of $P$, and let $M_l\ge1$ be the multiplicity of $\ell_l$ in $P_D$; then
$M_1+M_2+M_3+M_4\le D$.

\emph{Step 1: the hyperbolic branches.} Fix $l$ and choose linear coordinates $(u,v)$ with
$v=\ell_l$ and $d_l=(1,0)$. Since $C_4$ is squarefree, $C_4=v\,G(u,v)$ with
$\kappa_l:=G(1,0)\neq0$. For the weight vector $\omega'=(1,-1)$ in these coordinates the monomials
of $C$ of highest weight are $\kappa_lu^3v$ and $q_lu^2$, so
\[
C^{\omega'}=u^2(\kappa_luv+q_l),
\]
and by Lemma~\ref{lem:limits} the polynomial $S_3[P^{\omega'}]$ vanishes on the cube of the
hyperbola $\kappa_luv+q_l=0$. As in the proof of Theorem~\ref{thm:branches}(d), the polynomial
$\widetilde P(u,v):=P^{\omega'}(u,-q_lv/\kappa_l)$ is even, vanishes at the origin, is divisible
by $u^2(uv-1)$ and satisfies the 3SC on $\{uv=1\}$; by \cite[Theorem~9.14]{PartI},
$\widetilde P=(uv-1)B$ with $B$ as in \eqref{eq:SK9} or $B$ a quadratic form. Here $B$ is
homogeneous for $\omega'$ and divisible by $u^2$. Among the combinations of the polynomials of
one weight in \eqref{eq:SK9}, and among the quadratic forms, those divisible by $u^2$ are
\[
u^2\bigl(\alpha+\beta z(z-4)\bigr),\qquad\beta\,u^4(z-2)(z-4),\qquad\beta\,u^6(z-4)^2 ,
\]
of the weights $2$, $4$, $6$. Hence the weight $w_l:=w_{\omega'}(P)$ belongs to $\{2,4,6\}$,
every monomial of $P^{\omega'}$ has degree at most $w_l+6$, and for $w_l\in\{4,6\}$ the
polynomial $P^{\omega'}$ contains the monomial $u^{w_l+3}v^3$ of degree $w_l+6$, so that
$D\ge w_l+6$. In the coordinates $(u,v)$ the monomial $u^{D-M_l}v^{M_l}$ occurs in $P_D$; its
weight $D-2M_l$ is at most $w_l$, and if it equals $w_l$, the monomial belongs to $P^{\omega'}$,
so that $D\le w_l+6$. Therefore
\begin{equation}\label{eq:Ml}
2M_l\ge D-w_l,\qquad\text{and}\qquad 2M_l>D-w_l\ \text{ if }D>w_l+6 .
\end{equation}
We now use $\sum_lM_l\le D$; note that $D$ is even and $D\ge4$. If $D\ge14$, then
$2M_l\ge D-w_l+2\ge D-4$ for all $l$, and $2D\ge4D-16$, a contradiction. If $D=12$, then
$M_l\ge3$ for $w_l=6$, $M_l\ge5$ for $w_l=4$ and $M_l\ge6$ for $w_l=2$; hence $w_l=6$ and
$M_l=3$ for all $l$, and $P_{12}=\kappa\,C_4^3$. If $D=10$, then $w_l\le4$, $M_l\ge3$ for
$w_l=4$ and $M_l\ge5$ for $w_l=2$, so $\sum_lM_l\ge12$, a contradiction. If $D=8$ or $D=6$,
then $w_l=2$ and $M_l\ge3$, respectively $M_l\ge2$, for all $l$, a contradiction. So $D=4$,
where $P=\lambda C$, or $D=12$ and $P_{12}=\kappa\,C_4^3$.

\emph{Step 2: the lines.} Suppose that $P_{12}=\kappa\,C_4^3$ with $\kappa\neq0$. By
Lemma~\ref{lem:limits} with $\omega=(1,1)$, $S_3[C_4^3]$ vanishes on $V(C_4)^3$. After a linear
change of variables, $\ell_1,\ell_2,\ell_3,\ell_4$ are multiples of $T$, $X$, $X-T$,
$X-\lambda T$ with $\lambda\neq0,1$, and $C_4^3$ is a multiple of
$X^3T^3(X-T)^3(X-\lambda T)^3$. This contradicts Lemma~\ref{lem:threelines}, where $M_4\le1$.

\emph{Any number of variables.} Let $C\in\C[X,T]$ be one of the three polynomials and $s\ge3$.
For every linear map $\zeta\colon\C^2\to\C^{s-2}$ the polynomial $P\circ A_\zeta$ is even,
vanishes at the origin, is divisible by $C\circ A_\zeta=C$, and $S_3[P\circ A_\zeta]$ vanishes on
$V(C)^3\subset(\C^2)^3$ (Lemma~\ref{lem:linear}); by the case $s=2$ it is a constant multiple of
$C$ (possibly zero). Lemma~\ref{lem:sections} gives $P=\lambda C$.

Let $C=\PBh_a$ with $a\neq0$ and $s=3$. For $\zeta(X,T)=\beta X+\gamma T$ we have
$C\circ A_\zeta=\PB_{a,\beta,\gamma}$, which is irreducible for $\gamma\neq0$, and the same
argument gives $P=\lambda C$. Likewise $\PDh\circ A_\zeta=\PD_{\beta,\gamma}$, which is irreducible
for $45\beta+\gamma^2\neq0$. For $s\ge4$ one applies Lemma~\ref{lem:sections} with $r=3$ and the
case $s=3$.
\end{proof}

\begin{remark}[A bound for the degree]\label{rem:degreebound}
Step~1 is a bound for the degree of $P$ which does not presuppose the phase factors, and it is
not special to $\PC$. Let $C\in\C[u,v]$ be irreducible of degree $e\ge3$ with the homogeneous
parts $C_e,C_{e-1},\dots$, and suppose that $C_e$ is a product of $e$ pairwise non-proportional
linear forms $\ell_l$, that $C_{e-1}(d_l)=0$ and $C_{e-2}(d_l)\neq0$ for the directions $d_l$ of
these lines; this means that $V(C)$ has $e$ asymptotes through the origin and approaches each
of them like a hyperbola. Then, in the coordinates of Step~1 and with $q_l:=C_{e-2}(d_l)$, one has
$C^{\omega'}=u^{e-2}(\kappa_luv+q_l)$, the weight of $P^{\omega'}$ is at most six by
\cite[Theorem~9.14]{PartI}, and $2M_l\ge D-6$ for the multiplicity $M_l$ of $\ell_l$ in the top
homogeneous part of $P$, where $D=\deg P$. Summing over $l$ gives $e(D-6)\le2D$, that is,
\[
\deg P\le\frac{6e}{e-2},
\]
and, since $\deg P\ge e$, there is no polynomial $P$ at all unless $e\le8$. For an
asymptote which does not pass through the origin, as for the translated hyperbolas, the limit
of Lemma~\ref{lem:limits} is an affine line and gives no information; branches which are
asymptotic to curves $u^av^b=c$ with $(a,b)\neq(1,1)$ do not occur
(Corollary~\ref{cor:newton}).
\end{remark}

\subsection{The \texorpdfstring{$N$}{N}-soliton conditions for the families \texorpdfstring{(I\,B) and (I\,D)}{(I B) and (I D)}}\label{subsec:NSC-BD}
Let
\[
\Esix(X,T,Z):=X^6-5X^3T-5T^2+9XZ,\qquad\Etwo(X,T,Y):=X^3T-3X^2-TY .
\]
$\Esix$ is the first polynomial of the BKP hierarchy (Remark~\ref{rem:bkp}), $-\Etwo$ is the
polynomial of an equation of the BKP hierarchy with an additional time variable of weight $-1$
(Remark~\ref{rem:BD}(i)), and
\begin{equation}\label{eq:hat-normal}
\PDh(X,T,Y)=\Esix\bigl(X,-T,\tfrac19Y\bigr),\qquad
\PBh_a(X,T,Y)=-\tfrac a3\,\Etwo\bigl(X,-\tfrac3aT,-Y\bigr)\quad(a\neq0).
\end{equation}
For $p,q\in\C^*:=\C\setminus\{0\}$ put
\[
\mathbf k_6(p,q):=(p+q,\ p^3+q^3,\ p^5+q^5),\qquad\mathbf k_2(p,q):=(p+q,\ p^{-1}+q^{-1},\ p^3+q^3),
\]
and $\Pi_4(a,b;c,d):=(a-c)(a-d)(b-c)(b-d)$.

\begin{lemma}\label{lem:bkp-sums}
Let $(P,\mathbf k)$ be $(\Esix,\mathbf k_6)$ or $(\Etwo,\mathbf k_2)$.
\begin{enumerate}
\item[\textup{(i)}] $P(\mathbf k(p,q))=0$, and
$P\bigl(\mathbf k(a,b)+\mathbf k(c,d)\bigr)=\varkappa\,(a+b)(c+d)(a+c)(a+d)(b+c)(b+d)$, where
$\varkappa=45$ for $\Esix$ and $\varkappa=3/(abcd)$ for $\Etwo$.
\item[\textup{(ii)}] For $N\ge2$, $p_i,q_i\in\C^*$ and $\mathbf k_i:=\mathbf k(p_i,q_i)$,
\[
S_N[P](\mathbf k_1,\dots,\mathbf k_N)=\prod_{i<j}\bigl(-\varkappa_{ij}\,(p_i+q_i)(p_j+q_j)\bigr)\cdot F_N,
\]
where $\varkappa_{ij}=45$, respectively $\varkappa_{ij}=3/(p_iq_ip_jq_j)$, and
\[
F_N:=\sum_{\sg\in\{\pm1\}^N}P\Bigl(\sum_i\sg_i\mathbf k_i\Bigr)\prod_{i<j}
\Pi_4(\sg_ip_i,\sg_iq_i;\sg_jp_j,\sg_jq_j).
\]
\item[\textup{(iii)}] $F_N=0$ for every $N\ge1$, identically in $p_1,q_1,\dots,p_N,q_N$.
\end{enumerate}
\end{lemma}
\begin{proof}
(i) is verified by expansion. (ii) The exponents in $\mathbf k$ are odd, so
$\sg_i\mathbf k_i=\mathbf k(a_i,b_i)$ with $(a_i,b_i):=(\sg_ip_i,\sg_iq_i)$, and by (i)
\begin{align*}
\sg_i\sg_jP(\sg_i\mathbf k_i-\sg_j\mathbf k_j)
&=\sg_i\sg_j\,\varkappa_{ij}\,(a_i+b_i)(-a_j-b_j)\,\Pi_4(a_i,b_i;a_j,b_j)\\
&=-\varkappa_{ij}\,(p_i+q_i)(p_j+q_j)\,\Pi_4(a_i,b_i;a_j,b_j);
\end{align*}
here the value $\varkappa_{ij}$ of $\varkappa$ at $(a_i,b_i,-a_j,-b_j)$ does not depend on $\sg$.

(iii) $F_1=2P(\mathbf k_1)=0$. For $N=2$ each summand is, by (i),
$\varkappa_{12}(a_1+b_1)(a_2+b_2)$ times the product of the eight factors
$a_1\pm a_2$, $a_1\pm b_2$, $b_1\pm a_2$, $b_1\pm b_2$, that is,
$\sg_1\sg_2\,\varkappa_{12}(p_1+q_1)(p_2+q_2)(p_1^2-p_2^2)(p_1^2-q_2^2)(q_1^2-p_2^2)(q_1^2-q_2^2)$,
and the sum over $\sg$ vanishes. Let $N\ge3$ and suppose that $F_{N-1}=0$ identically. We
first show that $F_N$ vanishes for $p_1=p_2=:p$. For $\sg_1=\sg_2$ the summand contains the
factor $a_1-a_2=0$. For $\sg_2=-\sg_1=:-\tau$ we have $(a_1,b_1)=(\tau p,\tau q_1)$,
$(a_2,b_2)=(-\tau p,-\tau q_2)$,
\begin{gather*}
\sg_1\mathbf k_1+\sg_2\mathbf k_2=\tau\bigl(\mathbf k(p,q_1)-\mathbf k(p,q_2)\bigr)=\tau\,\mathbf k(q_1,-q_2),\\
\Pi_4(a_1,b_1;a_2,b_2)=2p\,(p+q_1)(p+q_2)(q_1+q_2),
\end{gather*}
and for $j\ge3$
\[
\Pi_4(a_1,b_1;a_j,b_j)\,\Pi_4(a_2,b_2;a_j,b_j)=(p_j^2-p^2)(q_j^2-p^2)\,
\Pi_4(\tau q_1,-\tau q_2;a_j,b_j).
\]
Hence, for $p_1=p_2=p$,
\[
F_N=2p\,(p+q_1)(p+q_2)(q_1+q_2)\prod_{j\ge3}(p_j^2-p^2)(q_j^2-p^2)\cdot
F_{N-1}\bigl((q_1,-q_2),(p_3,q_3),\dots,(p_N,q_N)\bigr)=0 .
\]
$F_N$ is invariant under the permutations of the pairs $(p_i,q_i)$, under each exchange
$p_i\leftrightarrow q_i$ and under each substitution $(p_i,q_i)\mapsto(-p_i,-q_i)$, which only
renames $\sg_i$. Therefore $F_N$ vanishes on the $8\binom N2$ hyperplanes $p_i=\pm p_j$,
$p_i=\pm q_j$, $q_i=\pm p_j$, $q_i=\pm q_j$, $i<j$. For $P=\Esix$, $F_N$ is a homogeneous
polynomial of degree $6+2N(N-1)$ in $p,q$; for $P=\Etwo$, the function
$\prod_i(p_iq_i)\cdot F_N$ is a homogeneous polynomial of degree $2+2N+2N(N-1)$. In both cases
this polynomial is divisible by
\[
\prod_{i<j}(p_i^2-p_j^2)(p_i^2-q_j^2)(q_i^2-p_j^2)(q_i^2-q_j^2)
\]
of degree $4N(N-1)$, which for $N\ge3$ is larger than its degree. Hence $F_N=0$.
\end{proof}

\begin{proposition}[$N$-soliton conditions]\label{prop:NSC-BD}
For every $N\ge1$ and every $a\in\C$ the polynomials $S_N[\PDh]$ and $S_N[\PBh_a]$ vanish
identically on $V(\PDh)^N$, respectively on $V(\PBh_a)^N$. Consequently $\PD_{a,b}$ and
$\PB_{a,b,c}$ satisfy the $N$-soliton condition on their whole dispersion varieties, for
every $N$ and for all values of the parameters.
\end{proposition}
\begin{proof}
By Lemma~\ref{lem:bkp-sums}, $S_N[\Esix]$ vanishes at all $N$-tuples of points of the image of
$\mathbf k_6\colon(\C^*)^2\to V(\Esix)$. The surface $V(\Esix)$ is irreducible, because $\Esix$
has degree one in $Z$ with coprime coefficients. The Zariski closure of the image of
$\mathbf k_6$ is an irreducible closed subset of this surface, and it has dimension two, because
the fibres of $(p,q)\mapsto(p+q,p^3+q^3)$ over the points with $p+q\neq0$ are finite; so it is
$V(\Esix)$. Hence $S_N[\Esix]=0$ on $V(\Esix)^N$. The same argument applies to $\Etwo$ and
$\mathbf k_2$, with the map $(p,q)\mapsto(p+q,p^{-1}+q^{-1})$. By \eqref{eq:hat-normal}
and Lemma~\ref{lem:linear} the statement follows for $\PDh$ and for $\PBh_a$ with $a\neq0$.
The polynomial $\PBh_0=T\,(X^3+Y)$ belongs to the family $L\,(D_x^3+\Lam)$ of \cite{PartI}, and
for it the statement is \cite[Theorem~7.8]{PartI}. It also follows from the case $a=1$:
for the weights $\omega=(0,1,0)$ of $(X,T,Y)$ we have $(\PBh_1)^\omega=T\,(X^3+Y)=\PBh_0$, and
Lemma~\ref{lem:limits}, applied with $P=C=\PBh_1$, shows that $S_N[\PBh_0]$ vanishes on
$V(\PBh_0)^N$, for momenta on either of the two components. Finally
$\PD_{a,b}=\PDh\circ A_\zeta$ and $\PB_{a,b,c}=\PBh_a\circ A_\zeta$ with
$\zeta(X,T)=aX+bT$, respectively $\zeta(X,T)=bX+cT$, and Lemma~\ref{lem:linear} applies.
\end{proof}

\begin{corollary}\label{cor:hietarinta}
\textup{(a)} Conjecture~\textup{\ref{conj:main}} holds for the components $V(C)$ of
Theorem~\textup{\ref{thm:rigidity}} with $C$ one of $\PB_{a,b,c}$ $(ac\neq0)$, $\PD_{a,b}$
$(45a+b^2\neq0)$, $\PBh_a$ $(a\neq0)$, $\PDh$, and for their images under invertible linear
maps: the only even polynomials with $P(0)=0$ satisfying the \textup{3SC} on $V(C)$ are the
constant multiples of $C$, and they satisfy the $N$\textup{SC} for every $N$.

\textup{(b)} For $C=\PC_{a,b,c}$ irreducible, Conjecture~\textup{\ref{conj:main}} holds for
$V(C)$ if and only if $\PC_{a,b,c}$ satisfies the $N$-soliton condition on $V(C)$ for every
$N$.
\end{corollary}
\begin{proof}
Theorem~\ref{thm:rigidity}, Proposition~\ref{prop:NSC-BD} and Lemma~\ref{lem:linear}.
\end{proof}
The condition in (b) is satisfied: this is Theorem~\ref{thm:hietarinta-NSC} below.

\begin{remark}[Hietarinta's equation]\label{rem:quartic}
(i) That $\PC_{a,b,c}$ satisfies the 3SC is the result of the computer search of
\cite{HietarintaI}, and four-soliton solutions are reported in
\cite{GrammaticosRamaniHietarinta1990}. We verified both statements once more: for
the polynomial $C_{\Q}$ of \eqref{eq:CQ} with indeterminate coefficients, $S_3[C_{\Q}]$ and
$S_4[C_{\Q}]$ belong to the ideal generated by the polynomials $C_{\Q}(X_i,T_i)$ (exact
computation of normal forms, Appendix~\ref{app:certificates}); hence $\PC_{a,b,c}$ satisfies
the 3SC and the 4SC on $V(\PC_{a,b,c})$ for all $a,b,c$.

(ii) For $N\ge5$ the $N$-soliton conditions are proved in
Theorem~\ref{thm:hietarinta-NSC} below. Independently of this proof we computed, as an
experiment, for $C_{\Q}=X^4-XT^3+X^2+2XT+3T^2$ and $5\le N\le7$, the element $S_N[C_{\Q}]$ of
the algebra $\Q[T_1,\dots,T_N]/\bigl(C_{\Q}(x_i,T_i),\ i\le N\bigr)$ of dimension $3^N$, where
$x_1,\dots,x_N$ are the first $N$ of the numbers $1,2,-5,7,3,11,-13$: it is zero. Thus $S_N$
vanishes at all $3^N$ $N$-tuples of points of this curve with these $X$-coordinates, in
accordance with the theorem; the computation is independent of its proof. The curve of this
example is a plane quartic whose only singular point is the node at the origin, and its four
points at infinity are distinct; so it has genus two, and the momenta of the solitons cannot be
parametrized rationally as in Lemma~\ref{lem:bkp-sums}.

(iii) \emph{The genus.} Let $\PC=\PC_{a,b,c}$ be irreducible, and put
$g(t):=t\,(t^2+\sqrt3\,t+1)$ and $q(t):=ct^2+bt+a$. Then
$\PC(X,tX)=X^2\bigl(X^2g(t)+q(t)\bigr)$, so that
\[
(X,T)\longmapsto(t,w):=\bigl(T/X,\ X\,g(T/X)\bigr)
\]
is a birational map from $V(\PC)$ onto the curve $w^2=-g(t)\,q(t)$, with the inverse
$X=w/g(t)$, $T=tX$. By Lemma~\ref{lem:irreducible}(iii) we have $c\neq0$, $q(0)=a\neq0$, and
$q$ has no common zero with $t^2+\sqrt3\,t+1$, whose two zeros are distinct and different
from $0$. Hence the polynomial $g\,q$ of degree five has five distinct zeros if
$b^2\neq4ac$, and $V(\PC)$ is then birational to a hyperelliptic curve of genus two. If
$b^2=4ac$, then $q=c\,(t-t_0)^2$ with $g(t_0)\neq0$, and in terms of $w':=w/(t-t_0)$ the curve
is $w'^2=-c\,t\,(t^2+\sqrt3\,t+1)$, an elliptic curve. Thus the dispersion curve of an
irreducible $\PC_{a,b,c}$ has geometric genus two if $b^2\neq4ac$ and one if $b^2=4ac$; at the
origin it has a node in the first case and two smooth branches which touch each other in the
second. In particular it is never rational. The proof of
Theorem~\ref{thm:hietarinta-NSC} does not use a parametrization of the curve, and it treats the
members with $b^2=4ac$ in the same way as the others.
\end{remark}

\begin{remark}[The families (I\,B) and (I\,D)]\label{rem:BD}
(i) \emph{Both polynomials belong to the BKP hierarchy.} For $\Esix$ see
Remark~\ref{rem:bkp}. By Lemma~\ref{lem:bkp-sums} the two-soliton phase factor of $\Etwo$ at the
momenta $\mathbf k_2(p_i,q_i)$ is the phase factor $\mathcal A_{ij}$ of the BKP hierarchy as
well, and $\Etwo$ vanishes at $(p+q,\,p^{-1}+q^{-1},\,p^3+q^3)$. In fact
$-\Etwo=D_3D_{-1}-D_1^3D_{-1}+3D_1^2$ is the polynomial of the bilinear equation
$(D_yD_t-D_x^3D_t+3D_x^2)\tau\cdot\tau=0$ which, according to Jimbo and Miwa \cite[end of
Section~9, p.~988]{JimboMiwa1983}, is contained in the BKP hierarchy with the time evolution
$\phi(k)\mapsto\phi(k)\exp(xk+yk^3+\dots+tk^{-1})$, that is, with an additional time variable
$t=x_{-1}$ of weight $-1$; they note that for $y=x$ it reduces to the shallow water wave
equation of \cite{HirotaSatsuma1976}. Compare also the remark in
\cite[Section~IV.F]{HietarintaI} that $X$ and $T$ cannot both have positive weight if (I\,B) is
to be weighted homogeneous. From this point of view the $N$-soliton conditions for $\Esix$ and
$\Etwo$ are consequences of the $N$-soliton solutions of the hierarchy
\cite[Section~3]{DJKM1982}, as in Remark~\ref{rem:bkp}(ii), and
Proposition~\ref{prop:NSC-BD} should not be regarded as new; for the family (I\,B) we have not
found the conditions with $N\ge5$ stated explicitly (four-soliton solutions are reported in
\cite{GrammaticosRamaniHietarinta1990}), and the proof given above is elementary and does not
use the theory of the hierarchies. The new statement in Corollary~\ref{cor:hietarinta}(a) is
that the dispersion varieties of these equations carry no other polynomial satisfying the 3SC.

(ii) \emph{Reducible members.} For $c=0$ the polynomial is $\PB_{a,b,0}=X\,(X^2T+aX+bT)$, which
for $a=b=-1$ is the shallow water wave equation of Hirota and Satsuma
\cite{HirotaSatsuma1976}; by Proposition~\ref{prop:NSC-BD} it satisfies all $N$SC on its whole
dispersion variety. Its
component $X^2T+aX+bT=0$ is not covered by Theorem~\ref{thm:rigidity}, and rigidity fails for
reducible members in general: on the cubic graph, which is a component of $\PB_{0,b,c}$ and of
$\PD_{a,b}$ with $45a+b^2=0$, there are the solutions of Theorem~\ref{thm:cubic-all}. The
reducible members of $\PC_{a,b,c}$ have lines, hyperbolas, or cubic curves with three asymptotes
through the origin as components; for these cubic curves, and for the cubic $X^2T+aX+bT=0$, we
have no classification. On the latter the polynomial $\PB_{a,b,0}$ is not the only solution
(Proposition~\ref{prop:sww}).
\end{remark}

\subsection{The curves \texorpdfstring{$u^av^b=c$}{u\textasciicircum a v\textasciicircum b = c}, and Newton polygons}\label{subsec:monomial-curves}
By Lemma~\ref{lem:limits} an edge of the Newton polygon of $C$ whose direction has two positive
coordinates leads to the three-soliton condition on curves $u^av^b=c$. For $a=b=1$ these are
the hyperbolas of \cite[Theorem~9.14]{PartI}, which were used in the proofs of
Theorems~\ref{thm:branches}(d) and~\ref{thm:rigidity}. On the other curves of this kind there
is no solution.

\begin{theorem}[The curves $u^av^b=c$]\label{thm:monomial-curves}
Let $u,v$ be two of the coordinates of $\C^s$, let $a,b\ge1$ be coprime integers and
$c\in\C\setminus\{0\}$. If there is a nonzero even polynomial $P$ on $\C^s$ with $P(0)=0$ which
is divisible by $u^av^b-c$ and satisfies the \textup{3SC} on $V(u^av^b-c)$, then $a=b=1$.
\end{theorem}

No bound on the degree of $P$ or on the multiplicity of the factor $u^av^b-c$ is assumed. This
polynomial is irreducible: with $\gamma^a=c$ and integers $\alpha,\beta$ such that
$\alpha b-\beta a=1$, the map $t\mapsto(\gamma t^b,t^{-a})$ from $\C\setminus\{0\}$ to its zero
set has the inverse $(u,v)\mapsto(u/\gamma)^\alpha v^\beta$, and its differential does not
vanish on the zero set. By Lemma~\ref{lem:linear} the theorem holds as well for
$\ell_1^a\ell_2^b=c$ with linearly independent linear forms $\ell_1,\ell_2$.

\begin{proof}
\emph{Reductions.} For the planes \eqref{eq:planes} the polynomial $P\circ A_{\xi\eta}\in\C[u,v]$
is even, vanishes at the origin, is divisible by $u^av^b-c$ and satisfies the 3SC on the curve
$u^av^b=c$ in $\C^2$ (Lemma~\ref{lem:linear}); it is not zero for suitable $\xi,\eta$, because
every point $(u,v,z)$ with $u\neq0$ lies on such a plane. So we may assume $s=2$. Replacing
$u$ by $\gamma u$ with $\gamma^a=c$, we may assume $c=1$, and exchanging $u$ and $v$ we may
assume $a\le b$. If $a=b$, then $a=b=1$. We assume $a<b$ and derive a contradiction. Let
\[
z:=u^av^b,\qquad C:=z-1,\qquad V:=V(C).
\]

\emph{Step 1: a weighted homogeneous solution.} Let $\omega:=(b,-a)$. Then $C^\omega=C$, and
by Lemma~\ref{lem:limits} the polynomial $P_*:=P^\omega$ satisfies $S_3[P_*]=0$ on $V^3$. It is
even and not zero, it is divisible by $C$, because $P=CK$ implies $P^\omega=CK^\omega$, and
$P_*(0)=0$, because $P$ has no constant term. Two monomials $u^iv^j$, $u^{i'}v^{j'}$ have the
same $\omega$-weight if and only if $b(i-i')=a(j-j')$, that is, if and only if $(i-i',j-j')$ is
an integral multiple of $(a,b)$. Hence
\[
P_*=u^rv^s\,F(z),\qquad r,s\ge0,\quad F\in\C[z],\quad F(0)\neq0 .
\]
Let $n:=\deg F$, let $f_n$ be the leading coefficient of $F$, and let $\mu\ge1$ and
$\nu_2\ge0$ be the multiplicities of the zeros $z=1$ and $z=2^{a+b}$ of $F$. Since the
polynomials $z-\rho$, $\rho\neq0$, are irreducible and pairwise non-associated, $\mu$ and
$\nu_2$ are the multiplicities in $P_*$ of $C$ and of $C_2=C(\cdot/2)=2^{-a-b}(z-2^{a+b})$. Put
\[
S:=s+bn,\qquad D:=r+an,\qquad d:=r+s ;
\]
these are the degrees of $P_*$ in $v$ and in $u$ and the order of $P_*$ at the origin, where
its lowest homogeneous part is $F(0)\,u^rv^s$. From
$P_*(-u,-v)=(-1)^{r+s}u^rv^sF\bigl((-1)^{a+b}z\bigr)$ we see that $d$ is even and that
$F(-z)=F(z)$ if $a+b$ is odd; in this case $z=-1$ is a zero of $F$ of multiplicity $\mu$ as
well. Therefore
\begin{equation}\label{eq:monomial-n}
n\ge\mu,\qquad\text{and}\qquad n\ge2\mu\ \text{ if $a+b$ is odd.}
\end{equation}

\emph{Step 2: one momentum tends to infinity.} The curve $V$ is parametrized by
$\mathbf m(t):=(t^b,t^{-a})$, $t\in\C\setminus\{0\}$. For $w=(w_1,w_2)\in\C^2$ with $w_2\neq0$
we have, as $\tau\to\infty$,
\begin{equation}\label{eq:monomial-expansion}
P_*\bigl(\mathbf m(\tau)+w\bigr)=f_n\,\tau^{bD}\,\bigl(w_2^S+S\,w_2^{S-1}\,\tau^{-a}+o(\tau^{-a})\bigr).
\end{equation}
Indeed, at the point $(u,v)=(\tau^b+w_1,\ \tau^{-a}+w_2)$ we have $u=\tau^b(1+O(\tau^{-b}))$
and $z=u^av^b\to\infty$ with $z^{-1}=O(\tau^{-ab})$, so that $F(z)=f_nz^n(1+O(\tau^{-ab}))$ and
$P_*=f_nu^Dv^S(1+O(\tau^{-ab}))=f_n\tau^{bD}(w_2+\tau^{-a})^S(1+O(\tau^{-b}))$; and $a<b\le ab$.
Now let $p,q\in V$, with second coordinates $x$ and $y$, and suppose that $x\neq\pm y$; note
that $x,y\neq0$. By \eqref{eq:S3} and the evenness of $P_*$,
\[
S_3[P_*]\bigl(\mathbf m(\tau),p,q\bigr)=2\sum_{\epsilon,\eta=\pm1}
P_*\bigl(\mathbf m(\tau)+\epsilon p+\eta q\bigr)\,P_*\bigl(\mathbf m(\tau)-\epsilon p\bigr)\,
P_*\bigl(\mathbf m(\tau)-\eta q\bigr)\,P_*(\epsilon p-\eta q).
\]
With $x':=\epsilon x$ and $y':=\eta y$, \eqref{eq:monomial-expansion} shows that the product of
the first three factors is $f_n^3\tau^{3bD}$ times
\[
(x'+y')^S(x'y')^S-S\,(x'+y')^{S-1}(x'y')^{S-1}\,(x'^2+x'y'+y'^2)\,\tau^{-a}+o(\tau^{-a}),
\]
and the fourth factor is $P_*(p-q)$ for $\epsilon=\eta$ and $P_*(p+q)$ for $\epsilon=-\eta$.
The left-hand side vanishes for all $\tau$. If $S$ is even, the terms of order $\tau^{3bD}$
give, after division by $4f_n^3(xy)^S$,
\begin{equation}\label{eq:B1}
(x+y)^S\,P_*(p-q)+(x-y)^S\,P_*(p+q)=0 .
\end{equation}
If $S$ is odd, the terms of order $\tau^{3bD}$ for $(\epsilon,\eta)$ and $(-\epsilon,-\eta)$
cancel, and the terms of order $\tau^{3bD-a}$ give, after division by $-4Sf_n^3(xy)^{S-1}$,
\begin{equation}\label{eq:B2}
(x+y)^{S-1}(x^2+xy+y^2)\,P_*(p-q)+(x-y)^{S-1}(x^2-xy+y^2)\,P_*(p+q)=0 .
\end{equation}
Both sides being polynomials in the coordinates of $p$ and $q$, these identities hold for all
$p,q\in V$.

\emph{Step 3: the two momenta collide.} Let $t\neq0$, $p:=\mathbf m(t)$ and
$q_\eps:=\mathbf m(te^{\eps})$, so that $x=t^{-a}$ and $y=t^{-a}e^{-a\eps}$. As $\eps\to0$ we
have $p-q_\eps=-\eps\,(bt^b,-at^{-a})+O(\eps^2)$, and both coordinates of the vector
$(bt^b,-at^{-a})$ are different from zero; hence $P_*(p-q_\eps)$ vanishes exactly to the order
$d$. The coordinates of $p+q_\eps$ are $t^b(1+e^{b\eps})$ and $t^{-a}(1+e^{-a\eps})$, and since
$\log(1+e^\theta)=\log2+\frac12\theta+\frac18\theta^2+O(\theta^4)$,
\[
z(p+q_\eps)=(1+e^{b\eps})^a(1+e^{-a\eps})^b=2^{a+b}\Bigl(1+\frac{ab(a+b)}{8}\,\eps^2+O(\eps^4)\Bigr).
\]
Hence $P_*(p+q_\eps)$ vanishes exactly to the order $2\nu_2$. Finally $x-y$ vanishes exactly
to the first order, whereas $x+y$, $x^2+xy+y^2$ and $x^2-xy+y^2$ do not vanish at $\eps=0$.
Comparing the orders of the two terms in \eqref{eq:B1}, respectively in \eqref{eq:B2}, we get
\begin{equation}\label{eq:monomial-m}
d-2\nu_2=S\ \text{ if $S$ is even},\qquad d-2\nu_2=S-1\ \text{ if $S$ is odd.}
\end{equation}

\emph{Step 4: the exponents.} The polynomial $P_*$ and the component $V$ satisfy the hypotheses
of Section~\ref{sec:collisions}, and $\mu,\nu_2,d$ are the numbers defined there. By
Theorem~\ref{thm:collision}, $d-2\nu_2\le\mu+1$. With \eqref{eq:monomial-m} and
\eqref{eq:monomial-n},
\[
bn\le S\le\mu+2 .
\]
If $a+b$ were odd, this would give $2b\mu\le\mu+2$, which is impossible for $b\ge2$. So $a+b$
is even; being coprime, $a$ and $b$ are odd, $b\ge3$, and $b\mu\le\mu+2$ gives $b=3$ and
$\mu=1$. Hence $a=1$, $n=1$, $s=0$ and $S=3$; then $F=f_1(z-1)$, $\nu_2=0$, and
\eqref{eq:monomial-m} gives $d=2$, that is, $r=2$:
\[
P_*=f_1\,u^2\,(uv^3-1).
\]
This polynomial violates \eqref{eq:B2} with $S=3$. For $p=\mathbf m(1)=(1,1)$ and
$q=\mathbf m(\tfrac12)=(\tfrac18,2)$ we have $x=1$, $y=2$,
$P_*(p-q)=P_*(\tfrac78,-1)=-\tfrac{735}{512}f_1$ and
$P_*(p+q)=P_*(\tfrac98,3)=\tfrac{19035}{512}f_1$, and the left-hand side of \eqref{eq:B2} is
\[
\bigl(9\cdot7\cdot(-735)+1\cdot3\cdot19035\bigr)\frac{f_1}{512}=\frac{675}{32}\,f_1\neq0 .
\]
This contradiction proves the theorem.
\end{proof}

With Theorem~\ref{thm:branches}(a) and Corollary~\ref{cor:origin} we obtain a restriction on
all edges of the Newton polygon. Recall that the Newton polygon of
$C=\sum c_{ij}u^iv^j$ is the convex hull of the points $(i,j)$ with $c_{ij}\neq0$.

\begin{corollary}[Newton polygons]\label{cor:newton}
Let $C\in\C[u,v]$ be an irreducible polynomial of degree at least two, and suppose that there
is a nonzero even polynomial $P\in\C[u,v]$ with $P(0)=0$ which is divisible by $C$ and
satisfies the \textup{3SC} on $V(C)$. Then every edge of the Newton polygon of $C$ is parallel
to one of the vectors
\[
(1,0),\quad(0,1),\quad(1,1),\quad(1,-1),\quad(m,-1),\quad(1,-m)\qquad\text{with }m\in\{2,3,5\}.
\]
\end{corollary}
\begin{proof}
Let $(a,\pm b)$ be the direction of an edge, with coprime integers $a,b\ge1$, and let
$\omega\in\Z^2$ be an outer normal vector of this edge. Then $C^\omega$ is the sum of the terms
of $C$ on the edge, and since $P=CK$ implies $P^\omega=C^\omega K^\omega$, the polynomial
$P^\omega$ is even, not zero, vanishes at the origin and is divisible by $C^\omega$; by
Lemma~\ref{lem:limits}, $S_3[P^\omega]$ vanishes on $V(C^\omega)^3$.

If the direction is $(a,b)$, then $C^\omega=u^iv^jG(u^av^b)$ with a polynomial $G$ of positive
degree such that $G(0)\neq0$. For a zero $\rho$ of $G$ the polynomial $P^\omega$ is divisible by
$u^av^b-\rho$ and satisfies the 3SC on $V(u^av^b-\rho)$, and
Theorem~\ref{thm:monomial-curves} gives $a=b=1$.

If the direction is $(a,-b)$, then $C^\omega$ is a constant multiple of
$u^iv^j\prod_k(u^a-\rho_kv^b)$ with $\rho_k\neq0$, and there is at least one factor
$C':=u^a-\rho v^b$; it is irreducible because $a$ and $b$ are coprime, and $P^\omega$ satisfies
the 3SC on $V(C')$. Suppose that $(a,b)\neq(1,1)$; exchanging $u$ and
$v$, we may assume $a>b$. If $b=1$, then Theorem~\ref{thm:branches}(a), applied to the
polynomials $C'$ and $P^\omega$ in place of $C$ and $P$, gives $a\in\{2,3,5\}$. If $b\ge2$,
the curve $V(C')$ passes through the origin, where it has a point of multiplicity $b$, and it
is not a cone. By Corollary~\ref{cor:origin}(2), $b=2$; then $a$ is odd, so that $C'$ is
neither even nor odd and $V(C')\neq-V(C')$, which contradicts Corollary~\ref{cor:origin}(3).
\end{proof}

\begin{remark}\label{rem:newton}
(i) The proof shows more: among the irreducible curves defined by binomials, that is, the
curves $u^av^b=c$ and $u^a=cv^b$ with coprime $a,b\ge1$ and $c\neq0$, those which are not
lines and are components of $V(P)$ carrying the 3SC for some $P$ are exactly the hyperbolas
$uv=c$ and the graphs $v=cu^m$, $u=cv^m$ with $m\in\{2,3,5\}$, and for these the polynomials
$P$ are given by \cite[Theorems~9.14 and~9.17]{PartI}, Theorem~\ref{thm:cubic-all} and
Theorem~\ref{thm:graphs}.

(ii) An edge with direction $(a,b)$, $a,b\ge1$, corresponds to branches of $V(C)$ on which one
coordinate tends to infinity and the other one to zero, with $u^av^b\to\rho$. Since the
three-soliton condition is invariant under linear changes of the coordinates
(Lemma~\ref{lem:linear}), Corollary~\ref{cor:newton} says that a branch at infinity of a plane
curve carrying the 3SC whose asymptote passes through the origin approaches this asymptote
like a branch of a hyperbola: in coordinates in which the asymptote is the line $v=0$, one has
$uv\to\rho\neq0$ along the branch. The origin cannot be moved: the condition is not invariant
under translations, and for an asymptote which does not pass through the origin, as for the
translated hyperbolas of Section~\ref{sec:hyperbola}, the corresponding limit is an affine
line. All edges of the Newton polygons of the curves occurring in this paper are of the types
listed in Corollary~\ref{cor:newton}; for example, if $ac\neq0$, the edges of the Newton
polygon of $\PC_{a,b,c}$ have the directions $(1,1)$ and $(1,-1)$, and those of $\PB_{a,b,c}$ the
directions $(1,1)$, $(1,-1)$ and $(3,-1)$.

(iii) In Theorem~\ref{thm:monomial-curves} the hypothesis that $u^av^b-c$ divides $P$ cannot be
omitted from the proof: these curves do not pass through the origin, and we do not know
whether the vanishing of $S_3[P]$ on $V(C)^3$ forces $C$ to divide $P$ for such a curve.
\end{remark}

\subsection{Fusion at resonant momenta, and the \texorpdfstring{$N$}{N}-soliton conditions for Hietarinta's equation}\label{subsec:resonances}
The proofs of the $N$-soliton conditions by induction on $N$ (Proposition~\ref{prop:sextic},
Lemmas~\ref{lem:quintic} and~\ref{lem:bkp-sums}) use that $S_N$ reduces to $S_{N-2}$ when two
of the momenta coincide. For the hyperbola $uv=1$ a second reduction was used in
\cite[Lemma~9.6]{PartI}: if the difference of two momenta of the curve lies again on the curve,
the three-soliton condition reduces $S_N$ to $S_{N-1}$, the two momenta being replaced by their
difference (\emph{fusion}); in this way the $N$-soliton conditions on the hyperbola were proved
in \cite[Theorems~9.7 and~9.11]{PartI}. If $P(p_1+p_2)\neq0$, a pair of momenta $p_1,p_2$ whose
difference is a zero of $P$ is a zero of the two-soliton phase factor \eqref{eq:phase}; the two solitons are then said to be
in resonance \cite[Section~2.3.2]{Zhang2026}. In this subsection the reduction is stated for
arbitrary momenta (Lemma~\ref{lem:fusion}) and applied to the curves of Hietarinta's equation.
The irreducible curves of this family are not rational (Remark~\ref{rem:quartic}(iii)), so
that the momenta cannot be parametrized rationally as in \cite{PartI}; instead we count the zeros of $S_N$, as a
function of one momentum, on the curve. In Section~\ref{subsec:sww} the argument of
\cite[Theorem~9.11]{PartI} is applied to a rational cubic curve.

\begin{lemma}[Coincidence and fusion]\label{lem:fusion}
Let $P$ be an even polynomial on $\C^s$, let $N\ge3$ and $p_1,\dots,p_N\in\C^s$ with
$P(p_1-p_2)=0$, and put $r:=p_1-p_2$.
\begin{enumerate}
\item[\textup{(i)}] For every $q\in\C^s$ we have $S_3[P](p_1,p_2,q)=2\,P(p_1+p_2)\,\beta(q)$,
where
\[
\beta(q):=P(r+q)\,P(p_1-q)\,P(p_2+q)+P(r-q)\,P(p_1+q)\,P(p_2-q).
\]
\item[\textup{(ii)}] If $\beta(p_j)=0$ for $j=3,\dots,N$, then
\begin{multline}\label{eq:fusion}
\prod_{j=3}^NP(r-p_j)\cdot S_N[P](p_1,\dots,p_N)\\
=(-1)^{N-1}P(p_1+p_2)\prod_{j=3}^NP(p_1-p_j)\,P(p_2+p_j)\cdot S_{N-1}[P](r,p_3,\dots,p_N).
\end{multline}
\item[\textup{(iii)}] If $p_1=p_2$, then
\begin{equation}\label{eq:fusion-double}
S_N[P](p_1,p_1,p_3,\dots,p_N)=2\,(-1)^{N-1}P(2p_1)\prod_{j=3}^NP(p_1-p_j)P(p_1+p_j)\cdot
S_{N-2}[P](p_3,\dots,p_N),
\end{equation}
where $S_1[P](p)=2P(p)$.
\end{enumerate}
\end{lemma}
In (iii) the hypothesis $P(p_1-p_2)=0$ means $P(0)=0$. For the hyperbola $uv=1$ and the
momenta $p_1=(\omega t,(\omega t)^{-1})$, $p_2=(t,t^{-1})$ with $\omega=e^{\pi\mathrm i/3}$, the
statements (ii) and (iii) are \cite[Lemma~9.6]{PartI}, and the proof is the same. The identity
$\beta(q)=0$ says that $A(p_1,q)=A(p_2,q)\,A(r,q)$ for the phase factors \eqref{eq:phase}
(compare \cite[(A.4)]{PartI}): with respect to a third soliton, the soliton with the momentum
$p_1=p_2+r$ behaves like the two solitons with the momenta $p_2$ and $r$ together.
\begin{proof}
In \eqref{eq:SN} the summands with $\sg_1=\sg_2$ contain the factor
$P(\sg_1(p_1-p_2))=P(p_1-p_2)=0$. Let $\sg_1=\tau$ and $\sg_2=-\tau$. Then
$\sg_1p_1+\sg_2p_2=\tau r$, the factor for the pair $(1,2)$ is
$\sg_1\sg_2P(\sg_1p_1-\sg_2p_2)=-P(p_1+p_2)$, and for $j\ge3$ the product of the factors for the
pairs $(1,j)$ and $(2,j)$ is, by the evenness of $P$,
\[
-P(\tau p_1-\sg_jp_j)\,P(-\tau p_2-\sg_jp_j)=g_j(\tau\sg_j),\qquad
g_j(\rho):=-P(p_1-\rho p_j)\,P(p_2+\rho p_j)\quad(\rho=\pm1).
\]

(i) For $N=3$ and $p_3=q$ the two summands with $\sg_1=1$, $\sg_2=-1$ are
$-P(r+\sg_3q)P(p_1+p_2)\,g_3(\sg_3)$, and the summands for $\sg$ and $-\sg$ coincide.

(ii) The hypothesis reads $P(r+p_j)\,g_j(1)+P(r-p_j)\,g_j(-1)=0$, and therefore
\[
P(r-p_j)\,g_j(\rho)=g_j(1)\cdot\rho\,P(r-\rho p_j)\qquad\text{for }\rho=\pm1 .
\]
For $\rho=\tau\sg_j$ we have $\rho\,P(r-\rho p_j)=\tau\sg_jP(\tau r-\sg_jp_j)$, which is the
factor of the pair formed by the momenta $r$ and $p_j$, with the signs $\tau$ and $\sg_j$, in
$S_{N-1}[P](r,p_3,\dots,p_N)$. Hence $\prod_{j\ge3}P(r-p_j)$ times the summand of
$S_N[P](p_1,\dots,p_N)$ for $(\tau,-\tau,\sg_3,\dots,\sg_N)$ equals
$-P(p_1+p_2)\prod_{j\ge3}g_j(1)$ times the summand of $S_{N-1}[P](r,p_3,\dots,p_N)$ for
$(\tau,\sg_3,\dots,\sg_N)$. Summation over $\tau,\sg_3,\dots,\sg_N$ gives \eqref{eq:fusion}.

(iii) Now $r=0$, the factor for the pair $(1,2)$ is $-P(2p_1)$, and
$g_j(\rho)=-P(p_1-p_j)P(p_1+p_j)$ for both values of $\rho$. Thus the summand for
$(\tau,-\tau,\sg_3,\dots,\sg_N)$ is $-P(2p_1)\prod_{j\ge3}\bigl(-P(p_1-p_j)P(p_1+p_j)\bigr)$
times the summand of $S_{N-2}[P](p_3,\dots,p_N)$ for $(\sg_3,\dots,\sg_N)$, and the two values
of $\tau$ contribute equally.
\end{proof}

Suppose that $P$ satisfies the 3SC on a component $V$, that $p_1,p_2,\dots,p_N\in V$ and
$P(p_1-p_2)=0\neq P(p_1+p_2)$. Then $\beta$ vanishes on $V$ by (i), and \eqref{eq:fusion}
expresses $S_N[P](p_1,\dots,p_N)$ by a value of $S_{N-1}[P]$ at $N-1$ points of which the
first one, $r=p_1-p_2$, is a zero of $P$. If $r$ lies on $V$, this value is zero as soon as
$P$ satisfies the $(N-1)$-soliton condition on $V$. If $r$ lies on another component of
$V(P)$, no conclusion is possible, because nothing is assumed on momenta outside $V$. For
an irreducible polynomial $P=C$ every zero lies on $V$; this is the situation of the next
theorem.

\begin{lemma}[Eight resonant points; computer-assisted]\label{lem:certificate}
Let $C_0:=X^4-XT^3+X^2+2XT-T^2$ and $q_0:=(1,-1)$. Then $C_0$ is irreducible, $q_0$ is a point
of $V(C_0)$ with $\partial C_0/\partial X(q_0)\neq0$, and there are eight points
$y^0_1,\dots,y^0_8\in V(C_0)$ with the following properties:
\begin{enumerate}
\item[\textup{(a)}] $C_0(y^0_i-q_0)=0$, and the gradients of $C_0$ at $y^0_i$ and at $y^0_i-q_0$
are linearly independent;
\item[\textup{(b)}] $C_0(y^0_i+q_0)\neq0$;
\item[\textup{(c)}] the $18$ points $\pm q_0,\pm y^0_1,\dots,\pm y^0_8$ are pairwise distinct.
\end{enumerate}
\end{lemma}
\begin{proof}
$C_0(q_0)=1+1+1-2-1=0$ and $\partial C_0/\partial X(q_0)=5$. The quadratic part
$Q:=X^2+2XT-T^2$ of $C_0$ is not divisible by any linear factor of
$X^4-XT^3=X(X-T)(X^2+XT+T^2)$, because $Q(0,1)=-1$, $Q(1,1)=2$, and the resultant of $Q(X,1)$
and $X^2+X+1$ is $7$; as in the proof of Lemma~\ref{lem:irreducible}(iii), $C_0$ is
irreducible. An exact computation (Appendix~\ref{app:certificates}) gives
\[
\operatorname{Res}_T\bigl(C_0(X,T),\,C_0(X-1,T+1)\bigr)=2\,X^2(X-1)^2\,R(X),
\]
\[
R:=27X^8-108X^7+270X^6-432X^5+504X^4-414X^3+236X^2-83X+17 ,
\]
and shows the following: $R$ is irreducible over $\Q$; the polynomials $R(X)$ and $R(-X)$ are
coprime; the numbers $R(0)$, $R(1)$, $R(-1)$ are not zero; and $R$ is coprime to the resultant
with respect to $T$ of $C_0(X,T)$ and of the determinant
\[
J(X,T):=\det\bigl(\nabla C_0(X,T),\ \nabla C_0(X-1,T+1)\bigr).
\]
Let $\xi_1,\dots,\xi_8$ be the zeros of $R$; they are distinct and different from $0$ and
$\pm1$. The polynomials $C_0(X,T)$ and
$C_0(X-1,T+1)$ have degree three in $T$ with the leading coefficients $-X$ and $-(X-1)$, which
do not vanish at $\xi_i$; hence there is $\theta_i$ with
$C_0(\xi_i,\theta_i)=C_0(\xi_i-1,\theta_i+1)=0$. Let $y^0_i:=(\xi_i,\theta_i)$. Then
$y^0_i\in V(C_0)$, $C_0(y^0_i-q_0)=0$, and $J(y^0_i)\neq0$, because a zero of $J$ on
$V(C_0)$ with first coordinate $\xi_i$ would give a common zero of $R$ and of the resultant of
$C_0$ and $J$; this is (a). If $C_0(y^0_i+q_0)=0$, then $-y^0_i\in V(C_0)$ and
$C_0(-y^0_i-q_0)=0$, since $C_0$ is even; so $-\xi_i$ would be a zero of $X^2(X-1)^2R(X)$,
which is excluded because $\xi_i\neq0,-1$ and $R(-\xi_i)\neq0$. This is (b). The first
coordinates of $q_0$, $-q_0$, $y^0_i$, $-y^0_i$ are $1$, $-1$, $\xi_i$, $-\xi_i$, and
$\xi_i\neq\pm\xi_k$ for $i\neq k$, $\xi_i\neq-\xi_i$, $\xi_i\neq\pm1$; this is (c).
\end{proof}

\begin{theorem}[Hietarinta's equation; partly computer-assisted]\label{thm:hietarinta-NSC}
For all $a,b,c\in\C$ and every $N\ge1$ the polynomial $S_N[\PC_{a,b,c}]$ vanishes identically
on $V(\PC_{a,b,c})^N$: Hietarinta's equation satisfies the $N$-soliton condition on its whole
dispersion variety, for every $N$ and for all values of the parameters.
\end{theorem}
\begin{proof}
The parameters $(a',b',c')$ in \eqref{eq:CQ} are the image of $(a,b,c)$ under a linear
bijection of $\C^3$. By Lemma~\ref{lem:linear} it therefore suffices to prove the statement for
the polynomials
\[
C_m:=X^4-XT^3+aX^2+bXT+cT^2,\qquad m=(a,b,c)\in\C^3 .
\]
Let $V_m:=V(C_m)\subset\C^2$, and for $N\ge1$ let $\mathcal Z_N$ be the set of all $m\in\C^3$
such that $S_N[C_m]$ vanishes on $V_m^{\,N}$.

\emph{Step 1: $\mathcal Z_N$ is the set of common zeros of finitely many polynomials in
$a,b,c$.} The polynomials $C_m(X_i,T_i)$, $i=1,\dots,N$, are monic of degree four in $X_i$.
Division with remainder by these polynomials, successively with respect to $X_1,\dots,X_N$,
gives $S_N[C_m]=\sum_ih_i\,C_m(X_i,T_i)+G$, where the degree of $G$ in each $X_i$ is at most
three and the coefficients of $G$ are polynomials in $a,b,c$. The polynomial $C_m$ has no
multiple factor, because its part of degree four, $X(X^3-T^3)$, has none. Hence for all
$\theta\in\C$ outside a finite set the polynomial $C_m(X,\theta)$ has four distinct zeros. If
$S_N[C_m]$ vanishes on $V_m^{\,N}$, then for such $\theta_1,\dots,\theta_N$ the polynomial
$G(X_1,\theta_1,\dots,X_N,\theta_N)$, of degree at most three in each $X_i$, vanishes on a
product of $N$ sets of four points and is therefore zero; so $G=0$. Conversely $G=0$ implies
that $S_N[C_m]$ vanishes on $V_m^{\,N}$. Thus $\mathcal Z_N$ is the set of common zeros of the
coefficients of $G$. In particular, if $\mathcal Z_N$ contains a nonempty open subset of
$\C^3$, then $\mathcal Z_N=\C^3$.

We have $\mathcal Z_1=\mathcal Z_2=\C^3$, because $S_1[P]=2P$ and $S_2[P]=0$ for even $P$,
and $\mathcal Z_3=\C^3$ by Remark~\ref{rem:quartic}(i). Let $N\ge4$, and suppose that
$\mathcal Z_M=\C^3$ for all $M<N$. We show that $\mathcal Z_N$ contains a neighbourhood of
$m_0:=(1,2,-1)$.

\emph{Step 2: resonant points.} Let $C_0=C_{m_0}$, $q_0$ and $y^0_1,\dots,y^0_8$ be as in
Lemma~\ref{lem:certificate}. By (a) the derivative of
$(m,q,y)\mapsto\bigl(C_m(y),C_m(y-q)\bigr)$ with respect to $y$ is invertible at
$(m_0,q_0,y^0_i)$. By the implicit function theorem there are open neighbourhoods $O$ of $m_0$
in $\C^3$ and $D$ of $q_0$ in $\C^2$ and holomorphic maps $y_i\colon O\times D\to\C^2$,
$i=1,\dots,8$, such that $y_i(m_0,q_0)=y^0_i$ and
\[
C_m\bigl(y_i(m,q)\bigr)=0=C_m\bigl(y_i(m,q)-q\bigr)\qquad\text{on }O\times D .
\]
By (b), (c), Lemma~\ref{lem:irreducible}(iii) and $\partial C_0/\partial X(q_0)\neq0$ we may
choose $O$ and $D$ so small that for every $m\in O$ the polynomial $C_m$ is irreducible, the
set $U_m:=V_m\cap D$ is a connected complex manifold of dimension one and not empty, and for
all $q\in U_m$, with $y_i=y_i(m,q)$,
\begin{equation}\label{eq:res-open}
C_m(y_i+q)\neq0\quad(1\le i\le8),\qquad
\text{the $18$ points }\pm q,\pm y_1,\dots,\pm y_8\text{ are pairwise distinct.}
\end{equation}
From now on we fix $m\in O$ and write $C:=C_m$, $V:=V_m$, $U:=U_m$ and $y_i(q):=y_i(m,q)$.
The curve $V$ is irreducible, $V=-V$, and $V+w\neq V$ for every $w\in\C^2\setminus\{0\}$:
a curve which is invariant under the translation $z\mapsto z+w$, $w\neq0$, contains with each of
its points $z$ the points $z+kw$, $k\in\Z$, and hence the line through $z$ with the direction
$w$; but $V$ is an irreducible curve of degree four.

\emph{Step 3: general position.} Let $\mathcal G$ be the set of all
$(p_2,\dots,p_N)\in U^{N-1}$ such that
\begin{enumerate}
\item[\textup{(G1)}] the $18(N-1)$ points $\pm p_j$, $\pm y_i(p_j)$
\textup{(}$2\le j\le N$, $1\le i\le8$\textup{)} are pairwise distinct;
\item[\textup{(G2)}] $C\bigl(y_i(p_j)-p_j-p_k\bigr)\neq0$ for $1\le i\le8$ and all $j\neq k$ in
$\{2,\dots,N\}$.
\end{enumerate}
We claim that $\mathcal G$ is open and dense in $U^{N-1}$. Each condition says that a
holomorphic map from the connected manifold $U^{N-1}$ to $\C^2$ \textup{(}the difference of two of
the points\textup{)} or to $\C$ does not vanish at the point considered, so it suffices to show
that none of these maps vanishes identically. For two points with the same index
$j$ this is \eqref{eq:res-open}. Let $j\neq k$. The functions $p_j\mp p_k$ and
$p_k\mp y_i(p_j)$ do not vanish identically, because $p_k$ varies independently of $p_j$. If
$y_i(p_j)=\pm y_l(p_k)$ for all $p_j,p_k\in U$, then $y_i$ is constant on $U$, with a value
$y^*$, and the polynomial $q\mapsto C(y^*-q)$ vanishes on $U$, hence on the irreducible curve
$V$; so $y^*-V=y^*+V$ is contained in $V$ and therefore equal to $V$, whereas $y^*\neq0$ by
\eqref{eq:res-open}. If $C(y_i(p_j)-p_j-p_k)=0$ for all $p_j,p_k\in U$, then for a fixed $p_j$
and $w:=y_i(p_j)-p_j$ the polynomial $q\mapsto C(w-q)$ vanishes on $U$, hence on $V$, and
$V=w+V$ with $w\neq0$. Both conclusions are impossible.

\emph{Step 4: zeros.} Let $(p_2,\dots,p_N)\in\mathcal G$ and
\[
F(p):=S_N[C](p,p_2,\dots,p_N)\qquad(p\in V).
\]
Replacing $\sg_1$ by $-\sg_1$ in \eqref{eq:SN} we see that $F(-p)=(-1)^{N-1}F(p)$. We claim that
$F$ vanishes at the $18(N-1)$ points of \textup{(G1)}; it suffices to consider $p_j$ and
$y:=y_i(p_j)$. Since $S_N$ is symmetric in its arguments, $F(p_j)$ is a value of $S_N[C]$ with
two equal arguments, and it vanishes by \eqref{eq:fusion-double}, because
$S_{N-2}[C]$ vanishes on $V^{N-2}$. For $p=y$ we have $C(y-p_j)=0$, so that $r:=y-p_j$ lies on
$V$, and $C(y+p_j)\neq0$ by \eqref{eq:res-open}. For $k\neq j$ the points $y,p_j,p_k$ lie on
$V$; by the three-soliton condition and Lemma~\ref{lem:fusion}(i), applied to
$(y,p_j)$ in place of $(p_1,p_2)$, we get $\beta(p_k)=0$. By Lemma~\ref{lem:fusion}(ii),
applied to the momenta $y,p_j$ and $p_k$, $k\neq j$,
\[
\prod_{k\neq j}C(r-p_k)\cdot F(y)=\pm\,C(y+p_j)\prod_{k\neq j}C(y-p_k)\,C(p_j+p_k)\cdot
S_{N-1}[C]\bigl(r,(p_k)_{k\neq j}\bigr)=0,
\]
because $r\in V$ and $S_{N-1}[C]$ vanishes on $V^{N-1}$. By \textup{(G2)}, $F(y)=0$.

\emph{Step 5: the count.} Each summand of $S_N[C](p,p_2,\dots,p_N)$ contains exactly $N$
factors which depend on $p$, and each of them has the form $C(p+w)$ or $C(-p+w)=C(p-w)$ with a
vector $w$ which does not depend on $p$. For $p\in V$ we have $C(p\pm w)=C(p\pm w)-C(p)$, and
$C(p+w)-C(p)$ is a polynomial of degree at most three in $p$. Hence there is a polynomial
$\widetilde F$ on $\C^2$ of degree at most $3N$ which coincides with $F$ on $V$. If $C$ did
not divide $\widetilde F$, the curves $V(C)$ and $V(\widetilde F)$ would have at most
$4\cdot3N=12N$ common points by B\'ezout's theorem. By Step~4 they have at least $18(N-1)$
common points, and $18(N-1)>12N$ for $N\ge4$. Hence $F=0$ on $V$. Thus $S_N[C]$ vanishes on
$V\times\mathcal G$, which contains a nonempty open subset of $V^N$; since $V^N$ is
irreducible (its smooth points form a connected complex manifold which is dense in $V^N$),
$S_N[C]$ vanishes on $V^N$. This holds for every $m\in O$, so
$O\subset\mathcal Z_N$, and $\mathcal Z_N=\C^3$ by Step~1.
\end{proof}

\begin{corollary}\label{cor:hietarinta-conj}
Conjecture~\textup{\ref{conj:main}} holds for the component $V(\PC_{a,b,c})$ of every
irreducible polynomial $\PC_{a,b,c}$, and for its images under invertible linear maps: the only
even polynomials $P$ with $P(0)=0$ which are divisible by $\PC_{a,b,c}$ and satisfy the
\textup{3SC} on $V(\PC_{a,b,c})$ are the constant multiples of $\PC_{a,b,c}$, and they satisfy
the $N$\textup{SC} for every $N$.
\end{corollary}
\begin{proof}
Corollary~\ref{cor:hietarinta}(b) and Theorem~\ref{thm:hietarinta-NSC}.
\end{proof}

Together with Corollary~\ref{cor:hietarinta}(a) this proves Conjecture~\ref{conj:main} for the
irreducible members of the three families (I\,B), (I\,C) and (I\,D) of Hietarinta's list; for
the family (I\,A) see \cite[Section~9.4 and Theorem~9.18]{PartI}. For Hietarinta's equation the
three-soliton condition is the result of \cite{HietarintaI}, and four-soliton solutions are
reported in \cite{GrammaticosRamaniHietarinta1990}; in the survey
\cite[Section~3.2.3]{Zhang2026} the question of its integrability, for instance of a Lax pair
or a B\"acklund transformation, is described as open. By the criterion recalled in
Section~\ref{sec:intro}, Theorem~\ref{thm:hietarinta-NSC} shows that for all values of the
parameters the $N$-soliton ansatz solves the equation for every $N$ and all nondegenerate
momenta. If $(a,b,c)\neq(0,0,0)$, the curve $V(\PC_{a,b,c})$ has an irreducible component
which is not a line: its parts of degree four and three are $XT(X^2+\sqrt3\,XT+T^2)$, a product
of four pairwise non-proportional linear forms $L_i$, and $0$; if $\PC_{a,b,c}$ were a product
of four affine linear functions $L_i+c_i$, the vanishing of its cubic part
$\sum_ic_i\prod_{j\neq i}L_j$ would give $c_1=\dots=c_4=0$, because these four products are
linearly independent, and $\PC_{a,b,c}$ would be homogeneous, that is, $a=b=c=0$. By
Lemma~\ref{lem:dominance} the nondegenerate $N$-tuples of momenta on such a component form a
nonempty Zariski open subset of its $N$-th power, for every $N$. Hence for
$(a,b,c)\neq(0,0,0)$ the equation has $N$-soliton solutions of Hirota's form for every $N$, that
is, it is integrable in the sense of Hirota \cite[Section~3.2.1]{Zhang2026}. For $a=b=c=0$ the
dispersion variety consists of four lines through the origin; two of any five momenta lie on
one of these lines, and their sum lies on it as well, so that there are no nondegenerate
$N$-tuples for $N\ge5$. To our knowledge the $N$-soliton conditions for all $N$ were not known. The theorem says nothing
about a Lax pair or a B\"acklund transformation. For one reducible member of the family, which
in suitable coordinates is $(u^2+\sqrt3\,uv+v^2)(uv-1)$, the $N$-soliton conditions on the
component $uv=1$ are contained in \cite[Theorem~9.7 and Remark~9.9(i)]{PartI}.

The proof uses the three-soliton condition only through Lemma~\ref{lem:fusion}(i); apart from
this it rests on the number of resonant points. In Lemma~\ref{lem:certificate} the factors
$X^2$ and $(X-1)^2$ of the resultant belong to the zeros of $p\mapsto C_0(p-q_0)$ at the origin
and at $q_0$, which are present for every curve through the origin which is symmetric with
respect to the origin, and the factor $R$ belongs
to the eight resonant points; the bound of B\'ezout's theorem is $12N$ because the function
$p\mapsto C(p+w)$ on $V$ is the restriction of a cubic polynomial. The three-soliton condition
for all values of the parameters, which gives the case $N=3$ in Step~1 and is used in
Step~4, has been verified by an exact computation (Remark~\ref{rem:quartic}(i)), and so has the four-soliton condition, which here
follows also from the case $N=4$ of the induction.

\subsection{A second solution on the cubic curve of the shallow water wave equation}\label{subsec:sww}
The reducible members $\PB_{a,b,0}=X\,(X^2T+aX+bT)$, $ab\neq0$, of the family (I\,B) are, up
to a scaling of the variables, the polynomial of the shallow water wave equation of Hirota and
Satsuma \cite{HirotaSatsuma1976} (Remark~\ref{rem:BD}(ii)). By a scaling of $X$ and $T$ the
cubic curve $X^2T+aX+bT=0$ becomes the curve $V(W)$,
\[
W:=u^2v+u+v,
\]
and $\PB_{a,b,0}$ becomes a multiple of $u\,W$: if $\alpha_0^2=b$ and $\beta_0=a/\alpha_0$, then
$\PB_{a,b,0}(\alpha_0u,\beta_0v)=a\alpha_0^2\,u\,W$. The polynomial $W$ is irreducible, since
it has degree one in $v$ with the coprime coefficients $u^2+1$ and $u$, and odd; the curve
$V(W)$ passes through the origin and is parametrized by
\[
\mathbf r(t):=\Bigl(t,\ -\frac{t}{1+t^2}\Bigr),\qquad t\in\C\setminus\{\pm\mathrm i\},\qquad
\mathbf r(-t)=-\mathbf r(t).
\]
On this curve $u\,W$ is not the only solution, in contrast to the irreducible members of the
family (Theorem~\ref{thm:rigidity}).

\begin{proposition}\label{prop:sww}
For $(\alpha,\beta)\in\C^2\setminus\{(0,0)\}$ let
\[
P_{\alpha,\beta}:=W\cdot\bigl[\alpha\,u+\beta\,v\,(uv+2)\bigr].
\]
Then $P_{\alpha,\beta}$ is even, $P_{\alpha,\beta}(0)=0$, $W$ is a simple factor of
$P_{\alpha,\beta}$, and $P_{\alpha,\beta}$ satisfies the $N$-soliton condition on $V(W)$ for
every $N\ge3$. Its two-soliton phase factor on $V(W)$ is, for all $(\alpha,\beta)$,
\begin{equation}\label{eq:sww-phase}
A\bigl(\mathbf r(s),\mathbf r(t)\bigr)=\frac{(s-t)^2}{(s+t)^2}\cdot\frac{s^2-st+t^2+3}{s^2+st+t^2+3}.
\end{equation}
\end{proposition}
The proof follows that of \cite[Theorem~9.11]{PartI}; it does not use a computer.
\begin{proof}
Write $P=P_{\alpha,\beta}$ and $K:=\alpha u+\beta v(uv+2)$. Both $W$ and $K$ are odd, so $P$
is even and $P(0)=0$.

\emph{Step 1: pairs.} The point $\mathbf r(s)-\mathbf r(t)$ has the coordinates $u=s-t$ and
$v=(s-t)(st-1)/\bigl((1+s^2)(1+t^2)\bigr)$. Using the identities
\begin{align*}
\bigl((s-t)^2+1\bigr)(st-1)+(1+s^2)(1+t^2)&=st\,(s^2-st+t^2+3),\\
(s-t)^2(st-1)+2(1+s^2)(1+t^2)&=(st+1)(s^2+t^2+2),
\end{align*}
one finds, for $W=(u^2+1)v+u$ and $K=\alpha u+\beta v(uv+2)$,
\begin{equation}\label{eq:sww-pair}
W\bigl(\mathbf r(s)-\mathbf r(t)\bigr)=\frac{st\,(s-t)\,h(s,t)}{(1+s^2)(1+t^2)},\qquad
K\bigl(\mathbf r(s)-\mathbf r(t)\bigr)=\frac{(s-t)\,E(s^2,t^2)}{(1+s^2)^2(1+t^2)^2},
\end{equation}
where
\[
h(s,t):=s^2-st+t^2+3,\qquad
E(x,y):=\alpha\,(1+x)^2(1+y)^2+\beta\,(xy-1)(x+y+2).
\]
Let $\varrho(s,t):=(s-t)^2\,h(s,t)$. Then
\begin{equation}\label{eq:sww-P}
P\bigl(\mathbf r(s)-\mathbf r(t)\bigr)=\frac{st\,E(s^2,t^2)}{(1+s^2)^3(1+t^2)^3}\;\varrho(s,t),
\end{equation}
and since $\mathbf r(s)+\mathbf r(t)=\mathbf r(s)-\mathbf r(-t)$ and $E(s^2,t^2)$ is even in
$t$, the phase factor is $\varrho(s,t)/\varrho(s,-t)$, which is \eqref{eq:sww-phase}. For
$t=0$ the second formula of \eqref{eq:sww-pair} gives
$K(\mathbf r(s))=s\,E(s^2,0)/(1+s^2)^2$, and $E(x,0)=\alpha(1+x)^2-\beta(x+2)$ is not zero
because $(\alpha,\beta)\neq(0,0)$; hence $K$ does not vanish identically on $V(W)$, and $W$ is
a simple factor of $P$. In particular $E\neq0$.

\emph{Step 2: the reduced sums.} Let $N\ge1$, $t=(t_1,\dots,t_N)$, and for
$\sg\in\{\pm1\}^N$ put $a_i:=\sg_it_i$, so that $\sg_i\mathbf r(t_i)=\mathbf r(a_i)$. By
\eqref{eq:sww-P}, $\sg_i\sg_jP(\mathbf r(a_i)-\mathbf r(a_j))$ is the product of
$t_it_jE(t_i^2,t_j^2)/\bigl((1+t_i^2)^3(1+t_j^2)^3\bigr)$, which does not depend on $\sg$, and
of $\varrho(a_i,a_j)$. Hence
\begin{equation}\label{eq:sww-SN}
S_N[P]\bigl(\mathbf r(t_1),\dots,\mathbf r(t_N)\bigr)=
\prod_{i<j}\frac{t_it_j\,E(t_i^2,t_j^2)}{(1+t_i^2)^3(1+t_j^2)^3}\cdot\Sigma_N(t),
\end{equation}
where
\[
\Sigma_N(t):=\sum_{\sg\in\{\pm1\}^N}P\Bigl(\sum_i\mathbf r(a_i)\Bigr)\prod_{i<j}\varrho(a_i,a_j).
\]
The polynomial $P$ has degree at most three in $v$ and at most three in $u$, and a monomial
$u^iv^j$ of $P$, evaluated at $u=a_1+u'$, $v=-a_1/(1+t_1^2)+v'$ and multiplied by
$(1+t_1^2)^3$, is a polynomial in $t_1$ of degree at most $i+6\le9$. Therefore
\[
\Phi_N(t):=\prod_{i=1}^N(1+t_i^2)^3\cdot\Sigma_N(t)
\]
is a polynomial in $t_1,\dots,t_N$ whose degree in $t_1$ is at most $9+4(N-1)=4N+5$. It is
symmetric in $t_1,\dots,t_N$ and even in each $t_i$; hence its degree in $t_1$ is at most
$4N+4$. By \eqref{eq:sww-SN} the polynomial $P$ satisfies the $N$-soliton condition on $V(W)$
if and only if $\Phi_N=0$. We have $\Phi_1=0$ and $\Phi_2=0$, because $S_1[P]=2P$ vanishes on
$V(W)$ and $S_2[P]=0$.

\emph{Step 3: three reductions.} We use the following properties of $\varrho$:
$\varrho(s,t)=\varrho(t,s)=\varrho(-s,-t)$, $\varrho(s,s)=0$, $\varrho(0,t)=t^2(t^2+3)$,
$\varrho(s,-s)=12\,s^2(s^2+1)$, and the identity of polynomials in $t_1,t_2,s$
\begin{equation}\label{eq:sww-fusion}
\varrho(t_1,s)\,\varrho(-t_2,s)=\kappa(t_1,t_2;s)\,\varrho(t_1-t_2,s)-t_1t_2\,h(t_1,t_2)\,Q(t_1,t_2,s),
\end{equation}
where
\begin{align*}
\kappa(t_1,t_2;s)&:=s^4+(3-t_1t_2)\,s^2+t_1^2t_2^2,\\
Q(t_1,t_2,s)&:=s^4-(t_1^2-3t_1t_2+t_2^2-6)\,s^2-3(t_1-t_2)(t_1t_2+2)\,s+t_1t_2\,(t_1^2-3t_1t_2+t_2^2-3),
\end{align*}
which is verified by expansion. Note that $\kappa$ is even in $s$. For $N\ge2$ we claim, as
identities of rational functions on $\C^{N-1}$, on $\C^{N-1}$ and on $H\times\C^{N-2}$,
where
\[
H:=\{(t_1,t_2)\in\C^2:\ h(t_1,t_2)=0\},
\]
\begin{align}
\Sigma_N(0,t_2,\dots,t_N)&=2\prod_{j\ge2}t_j^2(t_j^2+3)\cdot\Sigma_{N-1}(t_2,\dots,t_N),\label{eq:sww-zero}\\
\Sigma_N(t_1,t_1,t_3,\dots,t_N)&=24\,t_1^2(t_1^2+1)\prod_{j\ge3}\varrho(t_1,t_j)\,\varrho(t_1,-t_j)
\cdot\Sigma_{N-2}(t_3,\dots,t_N),\label{eq:sww-coll}\\
\Sigma_N(t_1,t_2,t_3,\dots,t_N)&=\varrho(t_1,-t_2)\prod_{j\ge3}\kappa(t_1,t_2;t_j)\cdot
\Sigma_{N-1}(t_1-t_2,t_3,\dots,t_N)\quad\text{for }(t_1,t_2)\in H,\label{eq:sww-res}
\end{align}
with $\Sigma_0:=P(0)=0$. In \eqref{eq:sww-zero} we have $\mathbf r(a_1)=0$, the factors
$\varrho(0,a_j)=t_j^2(t_j^2+3)$ do not depend on the signs, and the two values of $\sg_1$
contribute equally. In \eqref{eq:sww-coll} the summands with $\sg_1=\sg_2$ vanish because
$\varrho(a_1,a_1)=0$; for $\sg_2=-\sg_1$ the first two momenta cancel in the argument of $P$,
$\varrho(a_1,-a_1)=12t_1^2(t_1^2+1)$, the products
$\varrho(a_1,a_j)\varrho(-a_1,a_j)=\varrho(t_1,t_j)\varrho(t_1,-t_j)$ do not depend on the
signs, and the two values of $\sg_1$ contribute equally. In \eqref{eq:sww-res} the summands
with $\sg_1=\sg_2$ vanish because $\varrho(a_1,a_2)=\varrho(t_1,t_2)=0$ on $H$. Let
$\sg_1=\tau$, $\sg_2=-\tau$. By \eqref{eq:sww-pair}, $W(\mathbf r(t_1)-\mathbf r(t_2))=0$ on
$H$, so that $\mathbf r(t_1)-\mathbf r(t_2)$ is the point $\mathbf r(t_1-t_2)$ of $V(W)$, and
$\mathbf r(a_1)+\mathbf r(a_2)=\mathbf r(\tau(t_1-t_2))$. Moreover
$\varrho(a_1,a_2)=\varrho(t_1,-t_2)$, and by \eqref{eq:sww-fusion} on $H$, for $j\ge3$,
\[
\varrho(\tau t_1,a_j)\,\varrho(-\tau t_2,a_j)=\varrho(t_1,\tau a_j)\,\varrho(-t_2,\tau a_j)
=\kappa(t_1,t_2;t_j)\,\varrho\bigl(\tau(t_1-t_2),a_j\bigr).
\]
Thus the summand of $\Sigma_N$ for $(\tau,-\tau,\sg_3,\dots,\sg_N)$ is
$\varrho(t_1,-t_2)\prod_{j\ge3}\kappa(t_1,t_2;t_j)$ times the summand of
$\Sigma_{N-1}(t_1-t_2,t_3,\dots,t_N)$ for $(\tau,\sg_3,\dots,\sg_N)$.

\emph{Step 4: induction.} Let $N\ge3$, and suppose that $\Phi_M=0$ for all $M<N$. Then the
right-hand sides of \eqref{eq:sww-zero}, \eqref{eq:sww-coll} and \eqref{eq:sww-res} vanish
wherever they are defined, that is, on dense open subsets of the hypersurfaces $t_1=0$,
$t_1=t_2$ and $h(t_1,t_2)=0$ of $\C^N$; note that the polynomials $1+t_1^2$, $1+t_2^2$ and
$1+(t_1-t_2)^2$ do not vanish identically on $H$. Hence the polynomial $\Phi_N$ vanishes on
these three hypersurfaces. The polynomial $h$ is irreducible: it has no linear terms, its
quadratic part is a product of two different linear forms, and $h(0,0)\neq0$. So $\Phi_N$ is
divisible by $t_1$, by $t_1-t_2$ and by $h(t_1,t_2)$. Since $\Phi_N$ is even in $t_1$, it is
divisible by $t_1^2$. The quotient of the symmetric polynomial $\Phi_N$ by $t_1-t_2$ changes
sign under the exchange of $t_1$ and $t_2$, hence is again divisible by $t_1-t_2$. Since
$\Phi_N$ is even in $t_2$ and symmetric, it is divisible by the polynomials $t_i^2$,
$(t_i-t_j)^2$, $(t_i+t_j)^2$, $h(t_i,t_j)$ and $h(t_i,-t_j)$ for all $i$ and all $i<j$. These
are powers of pairwise non-associated irreducible polynomials, and therefore
\[
\prod_it_i^2\,\prod_{i<j}(t_i^2-t_j^2)^2\,h(t_i,t_j)\,h(t_i,-t_j)\qquad\text{divides}\qquad\Phi_N .
\]
The degree of this product in $t_1$ is $2+8(N-1)=8N-6$, which for $N\ge3$ is larger than
$4N+4$. Hence $\Phi_N=0$.
\end{proof}

\begin{remark}\label{rem:sww}
(i) For $\beta=0$ the polynomial $P_{1,0}=u\,W=-\tfrac13\Etwo(u,-3v,-u)$ is the shallow water
wave equation, and the statement is contained in Proposition~\ref{prop:NSC-BD}. For
$\beta\neq0$ the dispersion variety of $P_{\alpha,\beta}$ has a second nonlinear component:
the cubic curve $\alpha u+\beta v(uv+2)=0$ if $\alpha\neq0$, and the hyperbola $uv=-2$ if
$\alpha=0$. On this component the three-soliton condition does not hold for
$(\alpha,\beta)=(1,1)$ and $(0,1)$ (exact evaluation at sample points,
Appendix~\ref{app:certificates}). In this respect these polynomials resemble the polynomials
with $\beta\neq0$ of Theorem~\ref{thm:E}: they satisfy the three-soliton condition on one
component, but not on the whole dispersion variety as in \cite{HietarintaI}.

(ii) If $p,q$ are the zeros of $z^2-tz+\tfrac13(t^2+1)$, then $p^3+q^3=-(p+q)$ and
$\mathbf r(t)=\bigl(p+q,\,-\tfrac13(p^{-1}+q^{-1})\bigr)$; this is the case $\lambda=-1$ of
Remark~\ref{rem:bkp}(ii). Accordingly \eqref{eq:sww-phase} is the phase factor
$\mathcal A_{ij}$ of the BKP hierarchy for these parameters, and it coincides with the phase
factor \eqref{eq:quintic-phase} of the polynomials of Theorem~\ref{thm:graphs} on the graph
$v=u^5+5u^3$. In these terms $\varrho(s,t)=9\,\Pi_4(p_s,q_s;p_t,q_t)$, the condition
$h(t_1,t_2)=0$ means that the two pairs of parameters have the common element
$p=\tfrac13(t_1+t_2)$, and \eqref{eq:sww-fusion} holds on $H$ with
$\kappa=9\,(p^2-p_s^2)(p^2-q_s^2)$. We do not know whether $v(uv+2)W$ is the reduction of an
equation of the BKP hierarchy with the additional time variable of weight $-1$
(Remark~\ref{rem:BD}(i)).

(iii) Proposition~\ref{prop:sww} is not a classification. We computed the odd polynomials $K$ of
degree at most $15$ for which $WK$ has the phase factor \eqref{eq:sww-phase} on $V(W)$; this is
the linear condition $(s+t)\,K(\mathbf r(s)-\mathbf r(t))=(s-t)\,K(\mathbf r(s)+\mathbf r(t))$,
and its solutions of degree at most $15$ are the polynomials $\alpha u+\beta v(uv+2)$
(Appendix~\ref{app:certificates}). We do not know whether there are solutions of the 3SC on
$V(W)$ with other phase factors, or with this phase factor and higher degree.

(iv) The divisor $\prod_it_i^2\prod_{i<j}(t_i^2-t_j^2)^2$ of the classical argument (see
\eqref{eq:divisor}) has degree $4N-2$ in $t_1$, which is smaller than the bound $4N+4$ for the
degree of $\Phi_N$; here the zeros on the resonance hypersurfaces $h(t_i,\pm t_j)=0$ are needed,
as in \cite[Theorem~9.11]{PartI}. The three-soliton condition for $P_{\alpha,\beta}$ is the
case $N=3$ of the induction; it was also verified by an exact computation with indeterminate
$\alpha,\beta$ (Appendix~\ref{app:certificates}).
\end{remark}

\begin{remark}[Scope of the argument]\label{rem:resonance-scope}
(i) \emph{Zeros on other components.} Let $P=CK$ satisfy the 3SC on $V=V(C)$. For a point
$q\in V$ the zeros of $p\mapsto P(p-q)$ on $V$ are of three kinds: the points $p=q$ and,
if $0\in V$, $p=0$; the points with $p-q\in V$; and the points with $p-q\in V(K)\setminus V$.
At a zero of the third kind \eqref{eq:fusion} relates $S_N[P]$ on $V^N$ to a value of
$S_{N-1}[P]$ at $N-1$ zeros of $P$ of which one does not lie on $V$, and the hypothesis of
Conjecture~\ref{conj:main} gives no information about it. This is the reason why the argument
of this subsection does not prove the conjecture. In Theorem~\ref{thm:hietarinta-NSC} there are
no such zeros because $K$ is constant; in Proposition~\ref{prop:sww} the zeros of the third
kind are the zeros of $E(s^2,t^2)$, at which $P(p+q)$ vanishes as well, so that every summand
of $S_N[P]$ contains a vanishing factor, and they appear in \eqref{eq:sww-SN} as a common
factor.

(ii) \emph{The families \textup{(I\,B)} and \textup{(I\,D)}.} For one irreducible member of
each of these families and one point $q$ of its curve we computed the resonant points: there
are four, respectively six (Appendix~\ref{app:certificates}); for the curve $V(W)$ there are
two, the zeros of $h(s,t)$. This suggests that a count as in the proof of
Theorem~\ref{thm:hietarinta-NSC} is possible for these families as well, if the bound of
B\'ezout's theorem is replaced by the number of poles of $F$ on the normalization of the curve
and the zeros of $F$ at the points of the normalization over the origin, where every summand
of $S_N$ contains $N-1$ vanishing factors, are counted as well. We have not carried it out,
since Proposition~\ref{prop:NSC-BD} is simpler.

(iii) \emph{An observation.} In floating-point experiments with several members $C_m$ and
points $q$ of their curves (Appendix~\ref{app:certificates}; nothing in this paper depends on
them) the eight resonant points had the following structure: there are four vectors
$b_1,\dots,b_4\in\C^2$ such that $q=b_1+b_2$, the eight resonant points are
$b_1\pm b_3$, $b_1\pm b_4$, $b_2\pm b_3$, $b_2\pm b_4$, and all $24$ vectors $\pm b_i\pm b_j$,
$i<j$, lie on the curve; moreover $C_m(b_1)=\dots=C_m(b_4)$ and
$\sum_ib_ib_i^{\mathsf T}=\bigl(\begin{smallmatrix}-a&2c\\2c&2b\end{smallmatrix}\bigr)$. Thus
the $24$ points are the image of the root system $D_4$ under a linear map $\C^4\to\C^2$, and
the resonant points for $q$ are the images of the eight roots whose difference with the root
of $q$ is a root. The momenta $\mathbf k_6(p,q)$ of the solitons of the BKP hierarchy have the
same form $\boldsymbol\epsilon(p)+\boldsymbol\epsilon(q)$, with
$\boldsymbol\epsilon(z)=(z,z^3,z^5)$ and with arbitrarily many parameters in place of four
vectors (Lemma~\ref{lem:bkp-sums}(i)). We have neither a proof nor an explanation of this
structure for Hietarinta's equation.
\end{remark}
\section{Discussion: what remains}\label{sec:discussion}

\subsection*{Status of the conjecture}
Combining \cite{PartI} with the results above, Conjecture~\ref{conj:main} is now proved, with a
complete classification of the polynomials satisfying the 3SC, for the following simple
components $V_0=V(C)$:
\begin{itemize}
\item all components whose equation $C$ has a nonlinear irreducible factor in its top
homogeneous part, in particular all nonlinear cones (Theorem~\ref{thm:cone},
Corollary~\ref{cor:top}; only $P=\lambda C$ with $C$ a quadratic form occurs);
\item all irreducible quadrics except the translated hyperbolas $V(C_h)$ with
$h\in\overline{\Q}\setminus\{0,1\}$ (by the results of \cite[Section~9]{PartI}, in particular
\cite[Theorems~9.15 and~9.17, Proposition~9.20(i)]{PartI}, and by Theorems~\ref{thm:hzero}
and~\ref{thm:algebraic}; for the hyperbolas with $h=0$ or $h$ transcendental, and for the
translated hyperbolas $uv+au+bv+c=0$ with $ab=0$, $(a,b)\neq(0,0)$, no polynomial satisfying
the 3SC exists);
\item all graphs of polynomials, $C=\ell_2-q(\ell_1)$ with $\deg q\ge2$
(Corollary~\ref{cor:all-graphs}): solutions exist only for $q=a_2u^2+a_1u$
\cite[Theorem~9.17]{PartI}, for the odd cubics, where they are the polynomials of
Theorem~\ref{thm:cubic-all} (Corollary~\ref{cor:conj-cubic}, Theorem~\ref{thm:cubic-graphs}),
and for the odd quintics, where they are the polynomials \eqref{eq:quintic-solutions}
(Theorem~\ref{thm:graphs});
\item the dispersion varieties of the irreducible polynomials $X^3T+aX^2+bXT+cT^2$,
$X^3T+aX^2+TY$, $X^6+5X^3T-5T^2+aX^2+bXT$, $X^6+5X^3T-5T^2+XY$ and
$XT(X^2+\sqrt3\,XT+T^2)+aX^2+bXT+cT^2$ of Hietarinta's list (Theorem~\ref{thm:rigidity},
Corollaries~\ref{cor:hietarinta} and~\ref{cor:hietarinta-conj}; only the constant multiples
of the polynomial itself occur);
\item the curves $\ell_1^a\ell_2^b=c$ with coprime positive exponents $(a,b)\neq(1,1)$ and $c\neq0$, on
which no polynomial satisfying the 3SC exists (Theorem~\ref{thm:monomial-curves});
\item within the class of polynomials $P=L\cdot(q(D_{\mathbf x})+D_t)$ with a linear cofactor
$L$: all graph-type dispersions \cite[Theorem~5.7]{PartI}; and within the class
$P=F(\ell)+Q$: all components \cite[Theorem~9.18]{PartI}.
\end{itemize}
In all these cases the 3SC on the component implies the $N$SC for every $N$. On the cubic
graph the hypothesis of the conjecture that the component is simple is exactly what the
four-soliton condition requires (Theorem~\ref{thm:fourth}); on the hyperbola $uv=1$ it is not,
since a component of multiplicity two can carry all $N$SC \cite[Remark~9.12]{PartI}.

\subsection*{Open cases}
The general conjecture remains open. The following cases are not covered.

(1) \emph{Translated hyperbolas with algebraic parameter.} For $h\in\overline{\Q}\setminus\{0,1\}$
and $\deg P\ge10$ we know neither whether a polynomial $P$ with the 3SC on $V(C_h)$ exists nor,
if it does, whether it satisfies the higher conditions; see \eqref{eq:hyp-remaining}. The sets
$E_D$ of Theorem~\ref{thm:algebraic} are finite for each degree bound $D$, but we have no
bound for their union.

(2) \emph{Plane curves of degree at least three which are not graphs of polynomials.} The top
homogeneous part of a plane curve always splits into linear forms, so Corollary~\ref{cor:top}
does not apply. Lemma~\ref{lem:limits} gives a necessary condition for every edge of the Newton
polygon of $C$, and only ten directions of edges are possible (Corollary~\ref{cor:newton}): a
branch at infinity which is asymptotic to a graph $v=cu^m$ has
$m\in\{2,3,5\}$, with a complete answer for $m=2$ and $m=5$ and for two configurations with
$m=3$ (Theorem~\ref{thm:branches}); a branch with an asymptote through the origin approaches it
like a hyperbola (Theorem~\ref{thm:monomial-curves}), which leads to the polynomials of
\cite[Theorem~9.14]{PartI}; and the top homogeneous
part of $P$ satisfies a condition on the lines through the origin which are parallel to the
asymptotes (Lemma~\ref{lem:threelines}). For a plane curve which passes through the origin and
is smooth there, Corollary~\ref{cor:consequences}(b) restricts the order of contact with the
tangent line to $2$, $3$ or $5$, and to $2$ if the curve is not centrally symmetric. Not
covered are, among others: the branches asymptotic to a cubic graph in the configurations of
Remark~\ref{rem:branches}(i), and the asymptotes which do not pass
through the origin, where the limit is an affine line. A known example is the cubic curve
$u^2v+au+bv=0$ with $ab\neq0$: it is smooth at the origin, with contact of order three, two
of its three asymptotes, the lines $u^2=-b$, do not pass through the origin, and the polynomial
$u\,(u^2v+au+bv)=D_x^3D_t+aD_x^2+bD_xD_t$, which for $a=b=-1$ is the shallow water wave
equation of Hirota and Satsuma \cite{HirotaSatsuma1976}, satisfies all $N$SC on its whole
dispersion variety (Proposition~\ref{prop:NSC-BD}). On this curve there is a second solution,
which also satisfies all $N$SC, but we have no classification of the
polynomials satisfying the 3SC (Example~\ref{ex:sww}). There is no classification either for
the cubic curves with three
asymptotes through the origin which are components of the reducible members of Hietarinta's
equation (Remark~\ref{rem:BD}(ii)). The methods of Sections~\ref{sec:cubic}
and~\ref{sec:graphs} are of different nature: on the cubic graph the covariance \eqref{eq:M}
allows one to divide by powers of $C$, and there is no analogue of it in higher degree, since
\[
(z_1+z_2+z_3)^5-z_1^5-z_2^5-z_3^5=5\,(z_1+z_2)(z_1+z_3)(z_2+z_3)\,
\bigl(z_1^2+z_2^2+z_3^2+z_1z_2+z_1z_3+z_2z_3\bigr),
\]
so that the product of the four values of $u^5-v$ in a summand of $S_3$ depends on the signs;
in Section~\ref{sec:graphs} the powers of $C$ in the top weighted part are controlled instead
by treating the multiplicity as an unknown.

(3) \emph{Hypersurfaces of degree at least three in three or more variables whose top
homogeneous part is a product of linear forms}, other than the cylinders over the curves
already classified. This includes the quartic dispersion surfaces whose top part is not the
fourth power of a linear form, mentioned in \cite[Section~10]{PartI}; by
Corollary~\ref{cor:top} that top part must now be a product of four linear forms. The surfaces
of the polynomials $X^3T+aX^2+TY$ and $X^6+5X^3T-5T^2+XY$ are settled by
Corollary~\ref{cor:hietarinta}; the proof uses their plane sections.

(4) \emph{Components which do not pass through the origin.} The soft-momentum expansion of
Section~\ref{sec:soft} is not available for them, and Theorem~\ref{thm:collision} and
Lemma~\ref{lem:limits}, with Corollary~\ref{cor:newton} for plane curves, are the only general
restrictions we have; this concerns in particular the translated hyperbolas of (1). The graphs of polynomials which do not pass through the
origin carry no polynomial with the 3SC (Corollary~\ref{cor:all-graphs}). The
branch $d-2\nu_2=\mu-1$ of Theorem~\ref{thm:collision} is excluded for components through the
origin which are not cones (Corollary~\ref{cor:origin}), and for simple components with an
additive triple with independent normals (Proposition~\ref{prop:additive}). We know no example
of this branch other than the quadric cones, and no example of the branch $d-2\nu_2=\mu-2$ for
even multiplicity.

(5) \emph{Components with a point of multiplicity two at the origin} ($\mu=1$, $e_0=2$ in
Corollary~\ref{cor:origin}). Such a component is centrally symmetric, and the proof of
Corollary~\ref{cor:origin}(2) shows that $P=C\,C_2^{\nu_2}K$ with $K(0)\neq0$. Among the cones
this occurs for the quadric cones. Among the components which are not cones this class contains
the dispersion curve of the Boussinesq equation and the dispersion surface of the
Kadomtsev--Petviashvili equation, for which the polynomial is of the form $F(\ell)+Q$ treated
in \cite[Theorem~9.18]{PartI}, and the dispersion curves of the remaining polynomials with
nonlinear dispersion in the list of \cite{HietarintaI}, whenever they are irreducible:
$D_x^3D_t+aD_x^2+bD_xD_t+cD_t^2$ for $ac\neq0$; $D_x^6+5D_x^3D_t-5D_t^2+aD_x^2+bD_xD_t$ for
$a\neq-b^2/45$; and the equation
$D_xD_t(D_x^2+\sqrt3\,D_xD_t+D_t^2)+aD_x^2+bD_xD_t+cD_t^2$ found by Hietarinta. For all
three the conjecture is now proved (Corollaries~\ref{cor:hietarinta}
and~\ref{cor:hietarinta-conj}). For Hietarinta's equation the $N$-soliton conditions were known
for $N=3$ \cite{HietarintaI} and $N=4$ \cite{GrammaticosRamaniHietarinta1990}, and its complete
integrability is described as unclear in the recent survey \cite[Section~3.2.3]{Zhang2026};
Theorem~\ref{thm:hietarinta-NSC} gives the conditions for all $N$. For the irreducible members
the dispersion curve has genus two if $b^2\neq4ac$ and genus one if $b^2=4ac$
(Remark~\ref{rem:quartic}(iii)); it is never rational, so that the arguments which prove the
higher conditions for rationally parametrized momenta (Lemmas~\ref{lem:quintic}
and~\ref{lem:bkp-sums}) do not apply, and the proof uses the resonant points of the curve
instead. Beyond these examples we have
neither a classification of the polynomials satisfying the 3SC on a component with a point of
multiplicity two at the origin nor a proof that the 3SC implies the higher conditions.

(6) \emph{Higher multiplicity.} Components of multiplicity at least two are outside the
hypotheses of the conjecture, and both behaviours occur. On the cubic graph the polynomials
$C^{2n}P_0$ with $n\ge1$ of Theorem~\ref{thm:cubic-all} satisfy the 3SC and not the 4SC
(Theorem~\ref{thm:fourth}). On the hyperbola $uv=1$ the polynomials
$(au^2+bv^2)(uv-1)^2(uv-3)$ have multiplicity two and satisfy all $N$SC
\cite[Remark~9.12]{PartI}, whereas no even quadric component of multiplicity at least three
carries all $N$SC \cite[Section~10]{PartI}. We do not know what distinguishes the two cases in
general. By Theorem~\ref{thm:soft}(a) even multiplicities are possible only for components
which do not pass through the origin, and by Theorem~\ref{thm:collision}(b) they require
$d-2\nu_2\in\{\mu-2,\mu\}$.

What is missing for a proof of Conjecture~\ref{conj:main} is a mechanism that applies to all
components: either a bound for the degree of the cofactor that does not presuppose the phase
factors, or a structural identity which derives the conditions $S_N=0$, $N\ge4$, from $S_3=0$
on a simple non-hyperplane component. The limits of Section~\ref{sec:limits} give a bound of
the first kind when all branches of the curve at infinity have limits on which the
three-soliton condition has been classified (Remark~\ref{rem:degreebound}); they give nothing
when a limit is an affine line, as for the translated hyperbolas. The fusion at resonant
momenta (\cite[Lemma~9.6]{PartI}, Lemma~\ref{lem:fusion}) is an identity of the second kind: it
derives zeros of $S_N$ from the conditions with fewer solitons. It proves the higher
conditions when the component has enough resonant pairs with difference on the component
itself, as for the hyperbola $uv=1$ in \cite{PartI} and for Hietarinta's equation here; it
gives nothing at the zeros of the phase factor which come from other components of $V(P)$
(Remark~\ref{rem:resonance-scope}(i)), and on the cubic graph there are no resonant pairs at
all. The other results of this paper are necessary conditions and
classifications in special classes; they do not provide such a mechanism. In particular the
exclusion of the graphs of odd degree at least nine in Section~\ref{sec:uniform} is a
computation and not an explanation: of the $\lfloor(m-2)/2\rfloor$ conditions \eqref{eq:G} only
the first three are used, together with \eqref{eq:star}, and the final step compares the size
of $2^{m-1}$ with polynomials in $m$. We do not know a structural reason for the absence of
solutions in degree at least four other than five, nor for the two exceptional degrees $3$
and $5$.

\begin{remark}[Two observations]\label{rem:leads}
(i) \emph{Resonant pairs.} Let $p_1,p_2\in V$ with $r:=p_1-p_2\in V$ and $P(p_1+p_2)\neq0$. Then
$A(p_1,p_2)=0$, and the three-soliton condition at $(p_1,p_2,q)$ reduces to
$A(p_1,q)=A(p_2,q)\,A(r,q)$, that is, to the identity $\beta(q)=0$ of
Lemma~\ref{lem:fusion}(i) for all $q\in V$. This was used in \cite[Section~9]{PartI} for the
hyperbola $uv=1$, and it is used in Sections~\ref{subsec:resonances} and~\ref{subsec:sww} for
Hietarinta's equation and for the curve $V(W)$. Resonant pairs exist on a curve as soon as
$C(p_1-p_2)$ vanishes somewhere on $V^2$ outside the trivial locus; they do not exist on the
cubic graph, they are $t=-\omega^{\pm1}s$, $\omega=e^{2\pi i/3}$, on $v=u^5$, and they exist on
the translated hyperbolas. For the translated hyperbolas, where the existence of a solution is
open, we have not exploited these identities.

(ii) If moreover $0\in V$, letting $q$ tend to the origin along an arc with direction $e$ in
(i) gives $\Theta(p_1)=\Theta(p_2)+\Theta(r)$ for resonant pairs, with the function $\Theta$
of Section~\ref{sec:soft}. This has not been used in the proofs above.
\end{remark}

\begin{example}[The cubic curve of the shallow water wave equation]\label{ex:sww}
The curve $V(W)$, $W=u^2v+u+v$, of Section~\ref{subsec:sww} shows what is known for a plane
cubic which is not a graph, and what is not. The curve is rational and has no singular point
in the affine plane; at the origin it has contact of order three with its tangent $u+v=0$, in accordance with
Corollary~\ref{cor:consequences}(b); its asymptotes are the line $v=0$, which it approaches
like the hyperbola $uv=-1$, and the lines $u=\pm\mathrm i$, which do not pass through the
origin; and the edges of the Newton polygon of $W$ have the directions $(1,1)$, $(1,0)$ and
$(1,-1)$, in accordance with Corollary~\ref{cor:newton}. What is proved: the polynomials
\[
P_{\alpha,\beta}=W\cdot\bigl[\alpha\,u+\beta\,v\,(uv+2)\bigr],\qquad(\alpha,\beta)\neq(0,0),
\]
satisfy the 3SC on $V(W)$ and the $N$-soliton condition on $V(W)$ for every $N$, by an
induction on $N$ which uses the coincidences and the resonances of two momenta
(Proposition~\ref{prop:sww}); for
$\beta=0$ this is the shallow water wave equation, and for $\beta\neq0$ the polynomials do not
satisfy the 3SC on their whole dispersion variety, at least for $(\alpha,\beta)=(1,1)$ and
$(0,1)$ (Remark~\ref{rem:sww}(i)). What is not proved: that these are all solutions on
$V(W)$. We only know that there is no other solution $WK$ with the phase factor
\eqref{eq:sww-phase} and $\deg K\le15$ (Remark~\ref{rem:sww}(iii)). Hence
Conjecture~\ref{conj:main} is not proved for the component $V(W)$: the known solutions satisfy
all $N$SC, but there may be others. The example also shows that a reducible member of a family
of Hietarinta's list can share a component with polynomials outside the list, as for the
cubic graph (Theorem~\ref{thm:cubic-all}).
\end{example}

\section*{Data availability}
The verification scripts (exact-arithmetic Python, requiring only the packages SymPy and
python-flint; four audits also use the computer algebra system Singular) and their recorded
outputs accompany this paper as ancillary files, in the directory \texttt{anc/}, which
contains an index (\texttt{README\_ancillary.md}); they are described in
Appendix~\ref{app:certificates}. All computations on which a statement of this paper depends
are in exact rational, integer or algebraic arithmetic; one experiment, which is marked as
such, uses floating-point arithmetic with the package mpmath.

\appendix

\section{The exact computations}\label{app:certificates}

All scripts use exact arithmetic over $\Q$, over $\Q[r]/(r^2-7r+1)$, over the integers or, in
Group~IX, over the fields $\Q(\sqrt5)$ and $\Q(\sqrt3,i)$, and
in a few places, which are indicated, over a prime field. No check uses a floating point
number; floating point numbers occur only in one experiment in Group~X, which is marked as such
and on which nothing depends, and in two decimal values which one script of Group~VI prints
for information. They were rerun for this paper with Python~3.13, SymPy~1.14.0 and python-flint~0.9.0
(and Singular~4.3.2 for four audits), and the recorded outputs are included. The scripts are
organized in ten groups. In the lists below, ``proof'' means that the statement depends on
the computation, and ``audit'' means that the computation checks a formula or a statement whose
proof in the text is independent of it. The comments and the outputs of the scripts use the
statement numbers of the versions of the text for which they were written; the index
\texttt{README\_ancillary.md} gives the correspondence with the present numbering.

\medskip
\noindent\textbf{Group I: the cubic graph in arbitrary degree}
(directory \texttt{anc/cubic\_arbitrary\_degree/}).
\begin{itemize}\raggedright
\item \texttt{verify\_cubic\_homogeneous\_heads.py}. \emph{Proof} of
Proposition~\ref{prop:homog} for the weights $w=5,7,9$: the $33$ necessary quartic equations
from the triples \eqref{eq:33}, exact row reduction of their coefficient vectors, and
Gr\"obner bases over $\Q$. \emph{Audit}: the case $w=3$ and the unit ideals for $w=11,13$ (and,
with the option \texttt{--extended} and $100$ triples, for $w=15,17$).
\item \texttt{verify\_cubic\_laurent\_kernels.py}. \emph{Proof} of
Proposition~\ref{prop:kernel}: for each type and each odd weight $-24\le w'<\wt T$, the matrix
of the values of $\cL_T$ on the basis $u^ic^j$ ($j\ge-8$) at $2\dim+4$ rational points and its
kernel. \emph{Proof} of the value \eqref{eq:Q2value}.
\item \texttt{verify\_cubic\_types\_and\_identities.py}. \emph{Audit} of Table~\ref{tab:types},
of \eqref{eq:Theta} and \eqref{eq:M}, of the identity $\cL_u[c^{-1}]=0$, of the value in
Remark~\ref{rem:consistency}(i), of the coefficient of $\eps^{-3}$ used after
\eqref{eq:level4}, and of five instances of Lemma~\ref{lem:pole}; \emph{proof} of the
values $S_4\neq0$ quoted in Remark~\ref{rem:cubic-scope}(i).
\item \texttt{verify\_cubic\_heads\_uv.py}. \emph{Audit}: the cross-check of
Remark~\ref{rem:kernel}(ii) with polynomials in $(u,v)$, for all top parts of weight at most
$30$, including the second-order obstruction \eqref{eq:level4} for the top parts $uC^{2n+1}$,
$n\le4$.
\end{itemize}

\noindent\textbf{Group II: bounded degree on the cubic graph, cones, hyperbolas} (directory\\
\texttt{anc/bounded\_degree\_and\_hyperbola/}). These ten scripts were written at an
earlier stage of this work; statement numbers in their comments refer to a preliminary note,
and the index \texttt{README\_ancillary.md} gives the correspondence with the present
numbering.
\begin{itemize}\raggedright
\item \texttt{verify\_hyperbola\_h0.py}. \emph{Proof} of Step~4 of Theorem~\ref{thm:hzero}:
the equations $\Xi_w=0$, $w=4,\dots,10$, at the five triples, rebuilt from the full sum over
the eight sign vectors; the displayed form of $\widehat H$ and the matrix \eqref{eq:hyperM}.
\item \texttt{verify\_hyperbola\_degree8\_boundary.py} and
\texttt{verify\_hyperbola\_degree8\_nonzero.py}. \emph{Proof} of
Proposition~\ref{prop:degree8}: the Laurent coefficients of steps (1)--(3), and the four
Gr\"obner bases of step (5).
\item \texttt{verify\_hyperbola\_degree6.py}. \emph{Audit}: an independent certificate of
\cite[Proposition~9.20(ii)]{PartI} in two variables.
\item \texttt{verify\_cubic\_weight16.py}, \texttt{verify\_cubic\_independent.py},
\texttt{verify\_cubic\_independent16.py}. \emph{Proof} of the constant in \eqref{eq:F3factor}
(evaluation at $k=(1,2,3)$ with symbolic parameters). \emph{Audit}: the common phase factor of
$P_4,\dots,P_{12}$; the classification of all weighted homogeneous top parts of weight
$8,\dots,16$ in all projective charts, which confirms Proposition~\ref{prop:homog}; the
first-order kernels and the $3\times3$ determinant of Remark~\ref{rem:consistency}(ii), with two
independent choices of sample points.
\item \texttt{verify\_cubic\_structural\_identity.py}, \texttt{verify\_cubic\_uniform\_jets.py}.
\emph{Audit} of Lemma~\ref{lem:nu2} in an equivalent form for $P=CH$ (the coefficient of
$\eps^4$ of $S_3[P](\bp(s),\bp(t),\bp(\eps))$), of the chain rule with the operators
$\mathsf L,\mathsf J$, and of the leading coefficient $-q(q-1)Yg(Y)k(Y)$.
\item \texttt{verify\_cone\_formulas.py}. \emph{Audit} of Lemma~\ref{lem:model} (the
coefficients of $U_{m,h}$ for $m\le7$, $h\le8$, and the value $2\cdot3^m+6(-1)^m$) and of
\eqref{eq:tangent-deg} on an example.
\end{itemize}

\noindent\textbf{Group III: formulas of this paper} (file
\texttt{anc/verify\_sequel\_formulas.py}). \emph{Audit} of \eqref{eq:normalized}; of the
coefficients in the proof of Lemma~\ref{lem:model} and of \eqref{eq:tangent-deg} on an example;
of \eqref{eq:paircoll}--\eqref{eq:mixedcoll} on two curves, of \eqref{eq:Psi} and
\eqref{eq:negative-contra} for $\mu=1$; of \eqref{eq:doubled} on a quadric cone and of the
first determinant of Example~\ref{ex:additive}; of \eqref{eq:sextic-pair}, \eqref{eq:zeroF}, \eqref{eq:collisionF}
and of \eqref{eq:F3factor} as a polynomial identity in $k$ and $(\alpha,\beta,\gamma)$; of the
coefficient of $\eps^2$ in the proof of Lemma~\ref{lem:nu2}, of \eqref{eq:diamond} on the seven
types, and of the chain rule; of three instances of Lemma~\ref{lem:pole}(b) with $q=2$ and of
the simple pole of $\cQ_u[c^{-1}]$ used in the proof of Theorem~\ref{thm:cubic}; of \eqref{eq:ph}, \eqref{eq:catalan}, \eqref{eq:gauge} and of the
two determinants of Section~\ref{sec:hyperbola}; and of the quintic identity of
Section~\ref{sec:discussion}.

\medskip
\noindent\textbf{Group IV: a second computation for Section~\ref{sec:cubic}} (directory\\
\texttt{anc/independent\_recheck\_section5/}; in the version for which these programs were
written the cubic graph was treated in Section~5). These three programs were written after the
proofs of Section~\ref{sec:cubic} were complete, in order to check them once more. They share
no code with Groups I--III and use different methods: all points are computed in the coordinates
$(u,v)$, without \eqref{eq:points}; $\cL_T$ and $\cQ_T$ are obtained by interpolation of
$\tau\mapsto\cS[T+\tau F]$ and not from \eqref{eq:LT}; and no Gr\"obner basis algorithm is
used where uniqueness of a solution is asserted. All of them are \emph{audits}.
\begin{itemize}\raggedright
\item \texttt{recheck\_pole\_lemma\_and\_kernels.py}. The kernels of
Proposition~\ref{prop:kernel} for pole order at most $16$ and all odd weights down to $-48$,
by Gaussian elimination over the prime field with $p=10^9+9$ elements, in which $r^2-7r+1$
splits (the nullity modulo $p$ is an upper bound for the dimension of the kernel), and over
$\Q$ for pole order at most $9$; the linear system \eqref{eq:level4} in the unknowns
$\rho_3,\dots,\rho_{30}$ and $\delta^2$, which has only the zero solution; the exact pole
orders and leading coefficients in ten instances of Lemma~\ref{lem:pole}(b); the value
\eqref{eq:Q2value}.
\item \texttt{recheck\_end\_to\_end.py}. For the top parts $uc^{2n}$, $n=1,2,3$, with all
lower odd monomials as unknowns ($5$, $16$ and $33$ coefficients), the equation $\cS[H]=0$ is
solved directly, level by level, from its weighted homogeneous components at integer sample
points, without $\cL_T$, $\cQ_T$ or Lemma~\ref{lem:pole}; the only solution is $H=uc^{2n}$.
\item \texttt{recheck\_doubled\_graph\_and\_homogeneous.py}. The coefficient of $\eps^2$ in the
proof of Lemma~\ref{lem:nu2}(i) for an odd polynomial with symbolic coefficients; the
operators $\mathsf L,\mathsf J$ and the leading terms in the proof of
Lemma~\ref{lem:nu2}(ii); the expansions \eqref{eq:paircoll}--\eqref{eq:mixedcoll} on the cubic
graph and the resulting orders in $\eps$ of $S_3$ for six polynomials which are not solutions;
\eqref{eq:F3factor} as a polynomial identity; Proposition~\ref{prop:homog} and
Corollary~\ref{cor:toppart} in all charts $b_0=\dots=b_{q-1}=0$, $b_q=1$ for all odd weights
$w\le15$, and in the chart $b_0=1$ for $w\le19$ ($w\le21$ with the option
\texttt{--extended}), where the statement that the ideal generated by the sampled equations
contains $b_j-\beta_j$, respectively $1$, is certified by rank computations over $\Q$ on
Macaulay matrices, for the triples \eqref{eq:33} and for $100$ other triples; and checks of
Theorem~\ref{thm:cubic-all} and of Lemma~\ref{lem:cov} in four variables.
\end{itemize}

\medskip
\noindent\textbf{Group V: soft momenta and graphs of degree at least four} (directory\\
\texttt{anc/soft\_momenta\_and\_graphs/}).
\begin{itemize}\raggedright
\item \texttt{prove\_monomial\_graphs.py}. \emph{Proof} of Theorem~\ref{thm:monomial}(i) and of
(iii) for $m=7$ (and again for $9\le m\le25$): the polynomials $N_{s,t}$ of Lemma~\ref{lem:N}
at the twelve pairs $(s,t)$, the Gr\"obner bases of the table of Section~\ref{sec:remaining}
for odd $m=5,\dots,25$, and the value $S_3[vC](\bp(1),\bp(2),\bp(3))$ for $m=5$.
\emph{Audit}: the same for $m=3$; the two candidates for even $m\le26$ at one triple.
\item \texttt{prove\_quintic\_graphs.py}. \emph{Proof} of the value $Y(1,2,3)$ in Step~4 of the
proof of Theorem~\ref{thm:graphs} and of the evaluations of $F_3,F_4,F_5$ in
Lemma~\ref{lem:quintic}(iii), with the rank conditions. The points $(k;b)$ are
$(1,2,3;1)$, $(1,2,4;-2)$, $(1,3,4;3)$, $(2,3,5;\tfrac12)$, $(1,2,5;5)$, $(1,3,5;-\tfrac13)$,
$(2,3,7;0)$ for $N=3$; $(1,2,3,4;1)$, $(1,2,3,5;-2)$, $(1,2,4,5;3)$, $(1,3,4,6;0)$ for $N=4$;
and $(1,2,3,4,5;1)$, $(1,2,3,5,7;-3)$ for $N=5$. At the first four points for $N=3$ the
determinant of the values of $(b^2,be_1,e_1^2,e_2)$ is $1315020$, and at the first two points
for $N=4$ the determinant of the values of $(b,e_1)$ is $99$. \emph{Audit}: Lemma~\ref{lem:quintic}(i);
the part of degree $38$ in Step~4 as a multiple of $c_3-2b$; $S_3,S_4,S_5$ at a point for a
combination with rational $\alpha,\beta,b$; the formula \eqref{eq:quintic-solutions} for a
quintic with $a_5\neq1$, $a_1\neq0$; a quintic which is not odd.
\item \texttt{verify\_soft\_expansion.py}. \emph{Audit} of Lemma~\ref{lem:taylor} and
Theorem~\ref{thm:soft}: the Taylor coefficients of $S_3$ at a soft momentum for polynomials
which are not solutions ($\mu=1,2,3$; a graph which is not centrally symmetric; the cusp
$v^2=u^3$ with the arc $(\eps^2,\eps^3)$), and $\fF=0$ on known solutions, including
multiplicity three on the cubic graph and the Kadomtsev--Petviashvili polynomial, whose
component is singular at the origin.
\item \texttt{verify\_collision\_orders.py}. \emph{Audit} of \eqref{eq:Psi} and of the
expansions of $\psi_n$ in the proof of Theorem~\ref{thm:collision} for odd $\mu$, and of
Corollary~\ref{cor:origin}(1) on the solutions $C^{2n}cT$ of Section~\ref{sec:cubic}.
\item \texttt{audit\_uniform\_formulas.py}. \emph{Audit} of Lemma~\ref{lem:doublesoft}: the
expansion of $G$, \eqref{eq:logphi}, \eqref{eq:chi}, the lowest coefficient of $\Phi(t,t^m)$ for
$P_A$, $P_B$ and for the family (F1), and \eqref{eq:cond1} for $P_A$, $P_B$, (F3) in a range of
$m$ and $\mu$.
\item \texttt{audit\_brute\_force.py}. \emph{Audit} which does not use Theorems~\ref{thm:soft},
\ref{thm:collision} or Proposition~\ref{prop:candidates}: all weighted homogeneous cofactors of
bounded weight for $m=5$ ($\mu=1,2,3$), $m=7$, $m=2$, $m=4$, and all odd cofactors of weighted
degree at most five on the graphs $v=u^5+bu^3$, $b\in\{0,1,-3\}$ (there the ideal of the
sampled equations is not radical; the script checks that powers of the two linear forms which
cut out the solutions belong to it).
\item \texttt{verify\_additions.py}. \emph{Audit} of the two determinants of
Example~\ref{ex:additive}; of the two candidates $P_A$, $P_B$ of
Proposition~\ref{prop:candidates} for $m=2$ at sample triples ($P_B$ is not a solution); of
the phase factor \eqref{eq:quintic-phase} for $a_3=0$; and of the coefficients of $\psi_n$ used
in the proof of Theorem~\ref{thm:collision}.
\end{itemize}

\noindent\textbf{Group VI: the uniform exclusion of Section~\ref{sec:uniform}} (directory\\
\texttt{anc/uniform\_exclusion/}). The library \texttt{uniform\_lib.py} contains the generation
of the polynomials $\ell_j$, $E_{2j}$, $\mathrm{St}$ with python-flint and the functions for
Step~3 of the proof of Proposition~\ref{prop:uniform}.
\begin{itemize}\raggedright
\item \texttt{u01\_conditions\_sympy.py}. Independent generation of $\ell_0,\dots,\ell_6$, of
$E_2,E_4,E_6$ (from the definition \eqref{eq:G} of $G$) and of $\mathrm{St}$ (from
\eqref{eq:star-general} by symbolic differentiation) with SymPy, for symbolic $m$ and $\mu$.
\item \texttt{u02\_check\_inputs.py}. \emph{Audit} of the input: the $E_{2j}$ against the series
of $G$ for numerical data; \eqref{eq:E2j} as an identity of polynomials for $j\le5$; the
coefficient of $y^d$ of the soft form in the proof of Lemma~\ref{lem:star}, computed from the
definition for polynomials which are not solutions ($m=3,5,7,9$), against
\eqref{eq:star-general} and \eqref{eq:St}; \eqref{eq:star} on the three solutions with
$\nu_2=1$ on the cubic graph; proportionality of the polynomials produced by the two
generators; \eqref{eq:l0l2}; and a test of the functions used in Step~3 on polynomials with
prescribed zeros at $(m,A_0(m))$, which must be, and are, detected.
\item \texttt{u03\_exclude.py}. \emph{Proof} of Proposition~\ref{prop:uniform}: steps
\textup{(E1)}--\textup{(E4)} and Step~3 with python-flint; Table~\ref{tab:uniform}.
\item \texttt{u04\_exclude\_second\_order.py}. \emph{Audit}: the same conditions eliminated in
another order of the unknowns by a general recursive procedure. In case $(\mathrm T_4)$ of
\textup{(F5)} this leads to a pair of degrees $(11,16)$ with $k_0=2$ and $M_0=6526$.
\item \texttt{u05\_singular\_elimination.py}, \texttt{u06\_independent\_final\_check.py}.
\emph{Audit}: steps \textup{(E1)}--\textup{(E4)} in Singular, starting from the polynomials of
\texttt{u01}. The expanded polynomials recorded by Singular coincide term by term, up to sign,
with those of \texttt{u03}; for the pair $(f_{15},f_{11})$, whose resultant is not expanded,
\texttt{u06} compares the degrees, the lowest power of $\zeta$, the bounds and the threshold.
Step~3 is repeated in plain Python integer arithmetic; there
the integer zeros of the lowest coefficients $r_{k_0}$ are located by an explicit bound for
the zeros of a polynomial instead of a factorization.
\item \texttt{u07\_compare\_pair.py}. \emph{Audit}: the polynomials $f_{15}$, $f_{11}$ and
$r_2$, computed once more with python-flint, coincide term by term, up to sign, with those
recorded by Singular. The script also records the numbers used in Step~3 for $R^\ast$: the two
norms, the degrees, and the inequality $2^{5936}-1>S_\ast\cdot5937^{382}$ of integers, which
fails if $S_\ast$ is replaced by $2^{1148}$.
\end{itemize}

\noindent\textbf{Group VII: a second computation for Sections~\ref{sec:soft}
and~\ref{sec:graphs}} (directory\\ \texttt{anc/independent\_recheck\_soft\_momenta/}). These
programs were written after the proofs of Sections~\ref{sec:soft} and~\ref{sec:graphs}
(except Section~\ref{sec:uniform}) were complete, in order to check them once more; they share
no code with Group~V. All of them are \emph{audits}: the soft-momentum expansion from the
series of $S_3$ (\texttt{r4\_01}); Theorems~\ref{thm:collision}, \ref{thm:soft} and
Corollary~\ref{cor:origin} on a collection of known solutions with components through the
origin, among them the Kadomtsev--Petviashvili polynomial (\texttt{r4\_02}); the families of
Proposition~\ref{prop:candidates} for small $m$ directly from the definition of $S_3$, by
greatest common divisors and resultants and without the soft limit (\texttt{r4\_03}); the
table of Section~\ref{sec:remaining} by resultants instead of Gr\"obner bases, at other sample
points (\texttt{r4\_04}), and with Singular (\texttt{r4\_07}); Lemma~\ref{lem:quintic}, the
conditions $S_N=0$ for $N\le7$ on a general odd quintic, and Step~4 of the proof of
Theorem~\ref{thm:graphs} (\texttt{r4\_05}); and Lemma~\ref{lem:doublesoft} with the exclusions
of Section~\ref{sec:doublesoft} in a range of $m$ and $\mu$ (\texttt{r4\_06}).

\medskip
\noindent\textbf{Group VIII: relations with the literature} (directory\\
\texttt{anc/literature\_relations/}, script \texttt{verify\_literature\_relations.py}).
\emph{Audit} of the
identities stated in Remarks~\ref{rem:cubic-scope}(iv) and~\ref{rem:bkp} and of the elementary
facts about known equations used in Sections~\ref{sec:intro} and~\ref{sec:discussion}; no
proof depends on it. (A)~The five polynomials $E_6,\dots,E_{10}''$ of
\cite[Appendix~3]{JimboMiwa1983} vanish identically at the momenta of the solitons of the BKP
hierarchy, satisfy the two-soliton relation with the phase factors $\mathcal A_{ij}$ as
polynomial identities, and satisfy $S_3=0$ at four rational triples each. (B)~The identities
\eqref{eq:bkp6} and \eqref{eq:bkp10}, the uniqueness statement after \eqref{eq:bkp10}, and the
formulas of Remark~\ref{rem:bkp}(ii) for the reduced solitons. (C)~For four choices of
$(\alpha,\beta,b)$ with $\beta\neq0$: $S_3=0$ at a triple of the first component and
$S_3\neq0$ at a triple of the second component. (D)~The condition
\cite[(4.8)]{NewellZeng1986}, recomputed from the definition of $S_3$ with the coefficient of
$D_{t_1}D_{t_5}$ as a further parameter, and its limit \eqref{eq:sextic-condition}; the phase
factor of the family \cite[(4.9)]{NewellZeng1986} and the members of the families
\cite[(4.9)--(4.11)]{NewellZeng1986} without the term $D_{t_1}D_{t_5}$; the members
$k=-2$ and $k=-\frac15$ of the family $D_x^6+D_x^3D_t+kD_t^2$; the value
$S_3[P_6]\neq0$ at a triple of the second component of $P_6$; the substitution which maps
family (c) to the reducible members of the family (I\,D2) of \cite{HietarintaI}, in
$\Q[r]/(r^2-7r+1)$; and $S_3=0$ for family (c) at triples of both components and at mixed
triples. (E)~The discriminants which decide the irreducibility of the polynomials quoted in
Section~\ref{sec:discussion}, the order of contact of the curve $u^2v+au+bv=0$ with its
tangent at the origin, and the irreducibility of Hietarinta's quartic for $(a,b,c)=(1,0,1)$
over $\Q(\sqrt3,i)$, the field over which its top homogeneous part splits into four distinct
linear forms; since these four forms are distinct, every factor over $\C$ would be
proportional to a factor over this field. (F)~The nonlinear equations printed in
\cite[(1), (3), (A-1)]{Ramani1981} and \cite[(1.4)]{CaudreyDoddGibbon1976} are those of the
bilinear polynomials used here, as identities for a generic function $f$; the operator
identity of Remark~\ref{rem:cubic-scope}(iv) which derives \cite[(A-1)]{Ramani1981} from the
potential Korteweg--de Vries equation; $S_3[P_6]$ as a polynomial on the first component, on
the second component and on the two kinds of mixed triples; and the reductions $D_3=0$ and
$D_5=0$ of $E_6$ and $E_8$ as printed in \cite[Section~7]{DJKM1982}, with the solitons of the
first of them; and, at one triple each, the vanishing or nonvanishing of $S_3$ for the equations
of \cite[(3.10a)]{CaudreyDoddGibbon1976} and \cite[(1), (2)]{Ramani1981} with $n\le5$, in
agreement with the statements quoted after Theorem~\ref{thm:E}.

\medskip
\noindent\textbf{Group IX: limits at infinity and Hietarinta's list} (directory\\
\texttt{anc/limits\_at\_infinity/}).
\begin{itemize}\raggedright
\item \texttt{verify\_cubic\_graph\_translates.py}. \emph{Proof} of
Lemma~\ref{lem:firstorder}(ii) and of Table~\ref{tab:critical}: the numbers $M\,W_g(T)$ and
$R_g(T)$ at $k=(1,2,3)$ for the seven types and for $g(t)=t^2$ and $g(t)=1$, computed from
their definitions with dual numbers over $\Q$ and over $\Q[r]/(r^2-7r+1)$. \emph{Audit}:
Lemma~\ref{lem:firstorder}(i) for $\mu=1,3,5,7$ at three triples, against the derivative of
$S_3[C^\mu T]$, and for $\mu=3$ against the expansion in $\eps$ computed with polynomials in
$\eps$; the values of $\mu^*_g(T)$ at two further triples, where they are not positive integers
either.
\item \texttt{verify\_limits\_and\_rigidity.py}. \emph{Audit} of the identities used in the
proofs of Theorems~\ref{thm:branches} and~\ref{thm:rigidity} and of
Lemmas~\ref{lem:threelines} and~\ref{lem:irreducible}: the linear factors of the types and the
set $\Gamma$; the substitutions in Theorem~\ref{thm:branches}(c),(d); the two discriminants;
the decomposition $L=g_1+g_2+g_3$, and the statement of Lemma~\ref{lem:threelines} for all
exponents at most four and three values of $\lambda$, at two triples of points; $S_3[C_4]=0$ and
$S_3[C_4^3]\neq0$ on triples of points of three of the four lines of
$C_4=XT(X^2+\sqrt3XT+T^2)$; the $\omega'$-top parts of $\PC_{a,b,c}$ in the four coordinate
systems of Step~1; and, as a check of the quotation of \cite[Theorem~9.14]{PartI}, the values
$S_3=0$ at a rational triple for the weighted homogeneous members of \eqref{eq:SK9} used in
the proofs and $S_3\neq0$ for $u^N(uv-1)$, $N=4,6,8,10,14$; and the identities of
Remark~\ref{rem:quartic}(iii).
\item \texttt{verify\_nsc\_bkp\_type.py}. \emph{Audit} of Lemma~\ref{lem:bkp-sums}: part (i),
the vanishing of $F_2$ and the phase factors as polynomial identities; part (ii) and the
formula for $F_N$ at $p_1=p_2$ in the induction step, at random rational parameters for
$N\le4$; the comparison of degrees; $S_N=0$ and $F_N=0$ at random rational parameters for
$N=3,4,5$; the substitutions \eqref{eq:hat-normal}; the identification of $-\Etwo$ with the
polynomial of the equation quoted in Remark~\ref{rem:BD}(i); and, for $\PBh_0=T(X^3+Y)$, the
top part used at the end of the proof of Proposition~\ref{prop:NSC-BD} and the vanishing of
$S_3$ and $S_4$ on all products of the two components.
\item \texttt{verify\_hietarinta\_quartic.py}. The computations of
Remark~\ref{rem:quartic}. \emph{Proof} of the three-soliton condition for Hietarinta's
equation with indeterminate coefficients, which is the base of the induction in the proof of
Theorem~\ref{thm:hietarinta-NSC}: the normal form of $S_3[C_{\Q}]$ modulo the ideal generated
by the polynomials $C_{\Q}(X_i,T_i)$ is zero (python-flint; the generators form a Gr\"obner
basis for the lexicographic order because their leading monomials $X_i^4$ are pairwise
coprime). \emph{Audit}: the substitution leading to \eqref{eq:CQ} and the resultant in
Lemma~\ref{lem:irreducible}(iii); the normal form of $S_4[C_{\Q}]$, which is zero (its
summands have about $8\cdot10^5$ terms); the elements $S_N[C_{\Q}]$, $N\le7$, of the algebras
of Remark~\ref{rem:quartic}(ii), which are zero; the singular points of the curve of that
remark; and controls with the additional term $X^2T^2$ in $C_{\Q}$, for which the results are
not zero.
\item \texttt{audit\_low\_degree.py}. \emph{Audit} of Theorems~\ref{thm:cubic-graphs}
and~\ref{thm:rigidity} in low degree which uses none of the results of this paper. For
$P=BK$, where $K=k_0+k_1X^2+k_2XT+k_3T^2$ has indeterminate coefficients and $B=C$ for two
curves of each of the families $\PB_{a,b,c}$, $\PD_{a,b}$ and \eqref{eq:CQ}, respectively
$B(z)=C(z)C(-z)$ for the cubic
graphs $T=X^3+X^2$, $T=X^3+1$ and $T=X^3+2X^2-3$, the element $S_3[P]$ of the algebra
$\Q[k][T_1,T_2,T_3]/(C(x_i,T_i))$ is computed for several rational triples $(x_1,x_2,x_3)$; its
coefficients are quartic forms in $k$, and Singular shows that together with $k_i-1$ they
generate the unit ideal for $i=1,2,3$ in the first case and for $i=0,\dots,3$ in the second.
Thus $K$ is constant in the first case, and there is no such $P$ in the second. As a control,
the same procedure finds solutions with nonconstant $K$ on the hyperbola $XT=1$.
\item \texttt{audit\_genus.py}. \emph{Audit} of Remark~\ref{rem:quartic}(iii): the genus of
nine irreducible curves \eqref{eq:CQ}, computed with the procedure \texttt{genus} of Singular;
it is one for the four curves whose quadratic part is a square and two for the others.
\end{itemize}

\medskip
\noindent\textbf{Group X: the four-soliton condition on the cubic graph, the curves
$u^av^b=c$, fusion at resonant momenta} (directory\\
\texttt{anc/resonances\_and\_monomial\_curves/}).
\begin{itemize}\raggedright
\item \texttt{verify\_cubic\_fourth\_condition.py}. \emph{Proof} of the integers $a_i$ and
$D_T$ in Step~4 of the proof of Theorem~\ref{thm:fourth}, in integer arithmetic and with the
Python standard library only. \emph{Audit}: the formula \eqref{eq:fourth-formula} against the
definition of $S_4$ for $m=1,3,\dots,11$, the vanishing for $m=1$, and the non-vanishing of
the sums $\sum_ib_ia_i^m$ for all odd $m\le2001$.
\item \texttt{verify\_monomial\_curves.py}. \emph{Audit} of Theorem~\ref{thm:collision}(b)
(the polynomials $\psi_n^+$, and the example of Remark~\ref{rem:collision-scope} at sample
points for $N=3,4,5$); of the identities \eqref{eq:B1} and \eqref{eq:B2}, which are compared,
for sixteen polynomials $u^rv^sF(u^av^b)$ and exact Laurent polynomials in $\tau$, with the
coefficients of $S_3$ computed from the definition; of the expansion in Step~3 and of the
conclusion of Step~4 of the proof of Theorem~\ref{thm:monomial-curves} for all exponents
$a<b<60$ and $\mu<60$; of the value $\tfrac{675}{32}$; and of the directions of the edges of
the Newton polygons mentioned in Remark~\ref{rem:newton}(ii). Two further audits do not use
the proof: for sixteen curves $u^av^b=1$ and all cofactors of degree at most $24$, and for
eight curves and all cofactors of degree at most four without any reduction, the values of
$S_3$ at sample points span all quartic forms in the coefficients of the cofactor, so that
there is no solution (ranks modulo the prime $2^{61}-1$, which are lower bounds for the ranks
over $\Q$); for the hyperbola $uv=1$ the same procedure finds a deficient rank.
\item \texttt{verify\_rational\_cubic\_family.py}. \emph{Audit} of Proposition~\ref{prop:sww},
whose proof does not use a computer: \eqref{eq:sww-pair}, \eqref{eq:sww-P},
\eqref{eq:sww-phase} and \eqref{eq:sww-fusion} as polynomial identities; the reductions
\eqref{eq:sww-zero}, \eqref{eq:sww-coll}, \eqref{eq:sww-res} for $N\le4$, with $P$ and with a
polynomial which is not a solution, at rational points and at points of the conic
$h(t_1,t_2)=0$ over quadratic fields; the degree bound of Step~2; the three-soliton condition
for indeterminate $\alpha,\beta$ as a polynomial identity (python-flint), with a control; and
$S_4,S_5,S_6$ at sample points. The statements of Remark~\ref{rem:sww}: the identities of (i)
and (ii); the values $S_3\neq0$ at three triples of points of the second component for
$(\alpha,\beta)=(0,1)$ and $(1,1)$; and the linear system of (iii), with $72$ unknown
coefficients and $288$ pairs of points.
\item \texttt{prove\_resonances\_hietarinta.py}. \emph{Proof} of
Lemma~\ref{lem:certificate}: the resultant, the irreducibility of $R$, and the coprimality
statements, with SymPy. \emph{Audit}: Lemma~\ref{lem:fusion} for an even function given by a
table of random rational values on the integer vectors which occur ($N=3,4,5$); a second
description of the eight points by a Gr\"obner basis; the linear bijection of the parameters
in \eqref{eq:CQ}; and the numbers of resonant points quoted in
Remark~\ref{rem:resonance-scope}(ii).
\item \texttt{experiment\_resonance\_frames.py}. An \emph{experiment} in floating-point
arithmetic with $50$ digits (mpmath), on which nothing depends: the configurations described in
Remark~\ref{rem:resonance-scope}(iii), for five pairs of a member $C_m$ with rational
coefficients and a point $q$ of its curve, and the values of $S_N$, $3\le N\le7$, at random
points of three members with random complex coefficients, one of them with $b^2=4ac$; the
values are zero up to rounding, in accordance with Theorem~\ref{thm:hietarinta-NSC}.
\end{itemize}

\medskip
\noindent\textbf{On the use of sample points.} In the proofs, sample points are used in three
ways.
(a)~To produce \emph{necessary} conditions: an identity which holds on $V_0^{\,3}$ holds at
every sample point, so the true solution set is contained in the solution set of the sampled
equations; this is the logic of Propositions~\ref{prop:homog}, \ref{prop:kernel}
and~\ref{prop:degree8} and of Theorem~\ref{thm:hzero}, where the sampled systems are shown to
have no solutions other than those listed, and the listed ones are solutions by independent
arguments; the same logic underlies the table of Section~\ref{sec:remaining}. (b)~To show
that an explicit quantity is not zero, as in \eqref{eq:Q2value} and in
Lemma~\ref{lem:firstorder}, where the quantity is a polynomial in $k$ which has to be different
from zero and is evaluated at one triple, and in Theorem~\ref{thm:fourth}, where one quadruple
serves for all multiplicities. Lemma~\ref{lem:certificate} is of a similar kind: one member
and one point of its curve are used to show that a set of parameters is not empty; the
passage to all parameters is made in the proof of Theorem~\ref{thm:hietarinta-NSC}.
(c)~To prove an identity by \emph{interpolation}: a polynomial which is known beforehand to
lie in an explicit finite-dimensional space vanishes identically if it vanishes at points at
which the evaluation map on that space is injective. In this way the constant in
\eqref{eq:F3factor} is determined (a space of dimension one), and the quotients $Q_3$, $Q_4$,
$Q_5$ in the proof of Lemma~\ref{lem:quintic}(iii) are shown to vanish (spaces of dimension at
most four). In both cases the space is obtained from a divisibility and degree argument which
is given in the text, and the rank of the matrix of the values is part of the recorded
computation. No identity is inferred from its validity at sample points without such a bound.

Sample points also occur in the \emph{audits} described above, in the experiment of
Remark~\ref{rem:quartic}(ii), which concerns one curve and finitely many momenta on it, and in
the experiment of Remark~\ref{rem:resonance-scope}(iii); these
are not used in any proof. The computation of Section~\ref{sec:uniform} does not use sample
points on the curve at all: there the degree $m$ is an indeterminate, and the only numerical
evaluations are those of explicit polynomials at the pairs $(m,A_0(m))$ for the finitely many
odd $m$ below the thresholds of Table~\ref{tab:uniform}.

\bibliographystyle{amsplainx}
\bibliography{refs}

\end{document}